\documentclass{elsart}
\usepackage{graphicx}
\usepackage{amsmath}
\usepackage{amssymb}

\newcommand{\ds}{\displaystyle}
\renewcommand{\theequation}{\arabic{section}.\arabic{equation}}

\begin{document}

\begin{frontmatter}

\title{Phase Fluctuations and \\ Pseudogap Phenomena}
\author[Kiev]{Vadim M. Loktev\thanksref{E-mail}},
\author[Pretoria]{Rachel M. Quick},
\author[Pretoria]{Sergei G. Sharapov\thanksref{Leave}}

\address[Kiev]{Bogolyubov Institute for Theoretical Physics\\
      14-b Metrologicheskaya Str., 03143 Kiev, Ukraine}

\address[Pretoria]{Department of Physics, University of Pretoria,\\
                 Pretoria 0002, South Africa}


\thanks[E-mail]{Corresponding author. E-mail: {\rm vloktev@bitp.kiev.ua}}

\thanks[Leave]{On leave of absence from Bogolyubov Institute for
Theoretical Physics, 03143 Kiev, Ukraine\\
{\em Present address:} Institut de Physique, Universit\'e de Neuch\^atel,
CH-2000, Neuch\^atel, Switzerland}

\begin{abstract}
This article reviews the current status of precursor superconducting
phase fluctuations as a possible mechanism for pseudogap formation in
high-temperature superconductors. In particular we compare this
approach which relies on the two-dimensional nature of the
superconductivity to the often used $T$-matrix approach. Starting
from simple pairing Hamiltonians we present a broad pedagogical
introduction to the BCS-Bose crossover problem. The finite
temperature extension of these models naturally leads to a
discussion of the Berezinskii-Kosterlitz-Thouless superconducting
transition and the related phase diagram including the effects
of quantum phase fluctuations and impurities. We stress the
differences between simple Bose-BCS crossover theories and the
current approach where one can have a large pseudogap region even at
high carrier density where the Fermi surface is well-defined.
The Green's function and its associated spectral function, which
explicitly show non-Fermi liquid behaviour, is constructed in
the presence of vortices. Finally different mechanisms including
quasi-particle-vortex and vortex-vortex interactions
for the filling of the gap above $T_c$ are considered.
\end{abstract}

\begin{keyword}
high-temperature superconductivity, pseudogap, phase fluctuations\\
{\em PACS}: 74.25.-q ---
General properties; correlation between physical properties
in normal and superconducting states;
74.40.+k --- Fluctuations;
74.62.Dh --- Effects of crystal defects, doping and substitution;
74.72.-h --- High-$T_{c}$ compounds
\end{keyword}

\end{frontmatter}

\tableofcontents

\section{Introduction}
\setcounter{equation}{0}

\subsection{Pseudogap Phenomena in High-$T_{c}$ superconductors}

The discovery of high temperature superconductors (HTSC)
\cite{Bednorz} revealed new problems in solid state physics in general and in
the theory of superconductivity in particular. A combination of factors,
including unusual magnetic and electronic properties,
a lowered dimensionality, closeness to the metal-insulator transition and
relatively low carrier densities, makes the construction of an
appropriate theory both difficult and far from resolved. There is
no consensus as to the correct theoretical approach. It is even unclear
as to which physical features of cuprate superconductors should
be considered as the basis for the correct theory.

One of the clearest differences between the BCS scenario of superconductivity
and superconductivity in the cuprates is the existence of a {\em
pseudogap\/}.  This is simply a depletion of the single particle spectral
weight around the Fermi level \cite{Levi,Varma.2000} (see also a recent
excellent experimental review by Timusk and Statt
\cite{Timusk}).\footnote{Note that the word ``pseudogap'' for  HTSC
was originally suggested by Friedel \cite{Friedel}, 
although in a different context. The notion of pseudogap has been used
before in other areas such as 1D conductors.} 
The earliest experiments to reveal gap-like behaviour in the normal state 
were the NMR measurements of the Knight shift (see Refs. in \cite{Timusk}) 
which probes the uniform spin susceptibility.
\begin{figure}[h]
\centering
\resizebox{0.50\textwidth}{!}{%
  \includegraphics{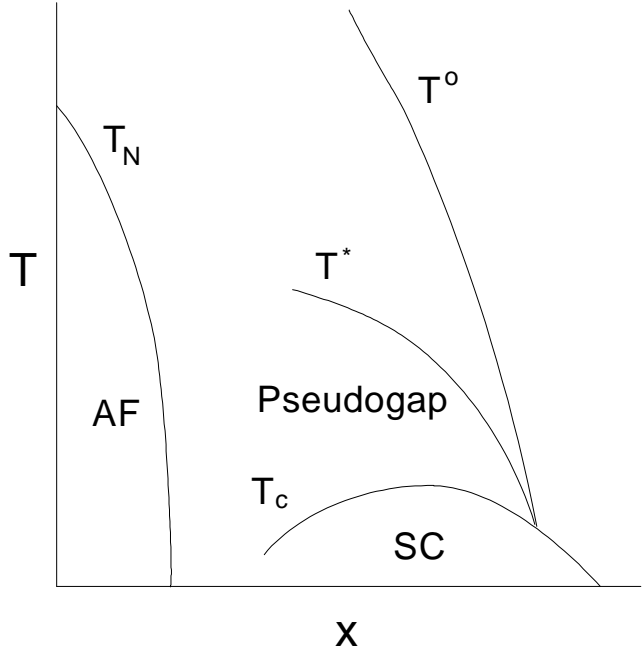}
} \caption{Qualitative phase diagram of cuprates given in terms of the
variables, temperature $T$ and the delocalized carrier concentration $x$,
taken from \protect{\cite{Timusk}}. Pseudogap crossover temperatures were
suggested from the NMR data. The crossover temperatures merge into $T_{c}$
slightly into the overdoped region of the phase diagram. } \label{fig:01}
\end{figure}
Based on the NMR data the phase diagram presented in Fig.~\ref{fig:01} was
suggested for the cuprate superconductors.

In Fig.~\ref{fig:01} there are two main phase boundaries. The first is the
transition into a long-range antiferromagnetic state at the N\'eel ordering
temperature, $T_{\mathrm{N}}$, in the very low doping $x$ regime. The second
is the transition into a superconducting state, with the maximal critical
temperature $T_{c}$ at approximately $x=0.2$ holes/CuO$_{2}$ unit cell. With
respect to the maximal $T_{c}$, under-, optimally and overdoped regimes are
distinguished.

In addition to the above-mentioned phase boundaries, two
crossover lines, $T^{\ast}$ and $T^{0}$,
are evident in these materials, even at
rather low doping levels. It must be noted, however,
that the experimental determination of these lines is still
somewhat controversial. We will return to the
discussion of how the different theoretical
approaches treat the phase diagram.

At the upper crossover temperature $T^{0}$ the Knight shift
changes behaviour. Above $T^{0}$ it is temperature independent,
while below $T^{0}$ it decreases linearly with temperature. Below the
lower crossover temperature $T^{\ast}$ the Knight shift
decreases supra-linearly with temperature. It was
suggested (see \cite{Timusk}) that these two crossover temperatures
may originate from two different physical phenomena.
In particular the lower crossover temperature $T^{\ast}$
appears to be the result of a gap appearing in the spectrum of
elementary antiferromagnetic excitations. This manifestation of
the pseudogap phenomenon was thus called a {\em spin gap\/}.

Subsequently the optical conductivity data (which has been discussed in detail
in the review \cite{Puchkov}) showed in addition a gapping of the charge
degrees of freedom.  Specific heat data (see Refs. in \cite{Timusk})
also provided evidence  that an electronic gap opens below
$T^{\ast}$. Considerable recent experimental progress in angle
resolved photo-emission spectroscopy (ARPES)
(see for example the reviews \cite{Dessau,Campuzano})
gives an experimental window on the single-particle spectral function and
other fundamental quantities such as the electron
self-energy \cite{Norman.1999}.
In these experiments one can clearly see the presence of a strong
reduction in the quasi-particle weight, or equivalently a pseudogap,
below $T^{\ast}$. Until recently most ARPES measurements
were performed for the photon energy range between 19 and 25$eV$.
The most recent data \cite{Chuang}, which was taken with a higher photon
energy of 33$eV$, is in contradiction with the previous data.
This question is now under intensive investigation
\cite{Feng}, particularly since ohmic losses
may alter the ARPES spectra \cite{Arko}.

An explanation of the pseudogap phenomenon is regarded as
one of the most important unresolved questions in the theory of
superconductivity. The ARPES and the scanning tunnelling spectroscopy
(STS) experiments (see Refs. in \cite{Timusk}) showed a smooth
crossover from the pseudo- to superconducting gap, so that in these
single-particle spectroscopies, the transition into the true
superconducting state at $T_{c}$ is barely noticeable. Moreover the
pseudo- and superconducting gaps also reveal experimentally the same
$d$-wave symmetry of the order parameter \cite{Wollman,Hardy}
(see also the reviews \cite{Scalapino.review,Timusk,Campuzano}).
We note, however, that at this stage the question of the symmetry
of the order parameter in HTSC is still under discussion \cite{Klemm.1998}.
For example, it is shown in \cite{Mesot,Gatt,Angilella} that the
observed symmetry of the order parameter cannot be fitted by only the
lowest harmonic of the $d$-wave order parameter. Furthermore, a recent
experiment with twisted Josephson junctions in the Bi-cuprates
\cite{Klemm.1999}, is in favour of an extended $s$-wave order parameter,
and has shown the absence of a $d$-wave part in the order parameter. It has
also been suggested that a transition from the $d_{x^2-y^2}$ symmetry to
the $d_{x^2-y^2} + id_{xy}$ symmetry could be induced by an external magnetic
field \cite{Laughlin.1998} and may be observed in the cuprates. Despite
remaining uncertainties in the experimental data, the existence of a
pseudogap and a corresponding crossover temperature is well established,
and the underlying physics demands an explanation.

\ack
At the outset of the review we would like to acknowledge the
people who and the institutions which have made this review
possible. In particular V.M.L. and S.G.Sh. would like to thank
E.V.~Gorbar, V.P.~Gusynin, V.A.~Miransky, I.A.~Shovkovy and
V.M.~Turkowski for their fruitful collaboration and numerous
discussions which  helped to clarify some deep questions of
low dimensional phase transitions.  Particularly we thank V.P.~Gusynin
and Yu.G.~Pogorelov for many thoughtful comments on this manuscript.
We would like to thank M.~Capezzali for sending us his PhD thesis.
R.M.Q. would like to thank M.P.~Das, D.M.~Eagles and M.~de~Llano
for very useful discussions.  V.M.L. and S.G.Sh. thank A.A.~Abrikosov,
C.C.~Almasan, D.~Ariosa, H.~Beck, M.~Oda, T.~Schneider, J.M.~Singer
and A.~Varlamov for very useful discussions, comments and sending us 
papers and preprints.  One of us (S.G.Sh.) is grateful
to the members of the Department of Physics of the University of
Pretoria for hospitality.
R.M.Q. and S.G.Sh. acknowledge the financial support of the NRF,
Pretoria. This work was also supported by the SCOPES-project 
7UKPJ062150.00/1 of the Swiss National Science Foundation 
(V.M.L. and S.G. Sh.).
Finally we would like to express our appreciation to the Referee
whose careful reading of and valuable remarks
on our manuscript undoubtedly led to a better review.

\subsection{Current theoretical explanations for the origin of the
pseudogap: magnetic, spin-density waves, stripes, etc.}
\label{sec:other}

Due to the extremely complex structural and electronic nature of
the cuprate systems \cite{Loktev.review}, and the somewhat
controversial nature of much of the current experimental data,
there are many theoretical attempts to explain pseudogap behaviour.
\begin{figure}[h]
\centering
\resizebox{0.50\textwidth}{!}{%
\includegraphics{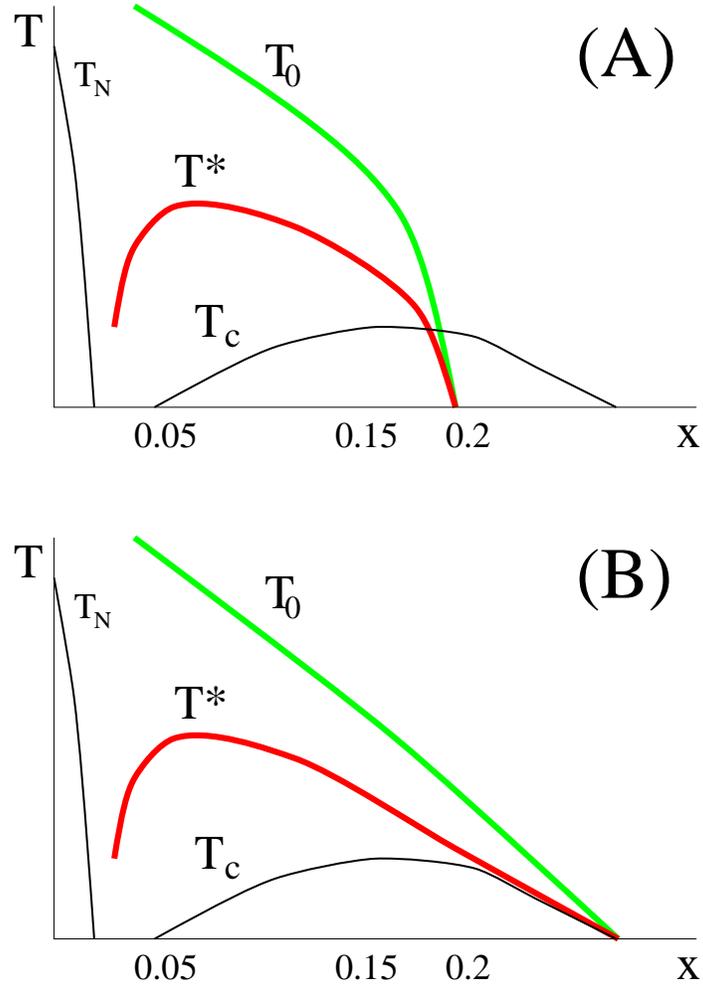} }
\caption{Theoretical patterns for the phase diagrams, taken from
\protect{\cite{Zachar}}.}
\label{fig:02}
\end{figure}
As an example, we show two theoretical phase diagrams
\cite{Zachar}. The first is based on the theory developed by
Pines and collaborators of the nearly antiferromagnetic Fermi liquid
\cite{Pines.review} (Fig.~\ref{fig:02}-A). The second is based on the
arguments of Emery and Kivelson (Fig.~\ref{fig:02}-B)
which are based on superconducting phase fluctuations
\cite{Emery.1995a,Emery.1995b} (see also \cite{Carlson}).
Both these theories are attempting to interpret the experimental phase
diagram presented in Fig.~\ref{fig:01}, and give in addition, predictions
about the less well experimentally studied regions of the diagram.
As an example, the continuation of the pseudogap lines below $T_{c}$ on
Fig.~\ref{fig:02} should be interpreted \cite{Zachar} as
``what would be present if there was no superconducting phase''. This
can be realized in practice by applying a magnetic field.

At this stage most interpretations of the NMR experiments
\cite{Gorny,Zheng} support the magnetic theory of the pseudogap
(or to be more exact the existence of a spin gap). On the other hand the
ARPES \cite{Timusk,Campuzano,Fedorov} and STS \cite{Timusk,Renner,Oda}
experiments can and have been interpreted using theories based solely on
precursor superconducting fluctuations (see Sec.~\ref{sec:precursor}).
This claim however depends on the precise way in which the experimental results
have been interpreted, and how well developed the particular theory
is for the given range of doping and/or the applied magnetic field if present.
For instance, it is argued in \cite{Zachar} that the current NMR data still do
not exclude the scenario based on the superconducting fluctuations and
that more measurements in the overdoped cuprates are needed.
An opposite point of view which supports the diagram in Fig.~\ref{fig:02}-A is
presented in a very  recent survey of the experimental data \cite{Tallon}.
However, so evident disagreement between \cite{Tallon} and \cite{Zachar} (see
also \cite{Varma.2000}) once more demonstrates vividly that there is no
consensus about the true shape of the cuprate phase diagram.

Another possible explanation relates the pseudogap to
charge- and/or spin-density waves \cite{Klemm.1998} (see also
very recent paper \cite{Krasnov} and Refs. therein).
A theory based on the phase separation in the form of a strip
structure of metallic and insulating domains with a spin gap in the
AFM regions has been suggested by Emery, Kivelson and Zachar
\cite{Emery.1997} (see also the more recent paper \cite{Emery.1999}).
Such a separation is accompanied by the formation
of a pairing amplitude through a proximity effect between
non-magnetic metallic and AFM insulating stripes, but with
large phase fluctuations.  Below $T^{\ast}$ this theory has much in
common with the more phenomenological approach based on the
2D attractive Hubbard model discussed here simply because both theories
are based on phase fluctuations. A stripe-phase quantum-critical-point
scenario for high-$T_c$ superconductors which has much in common with that of
Ref. \cite{Emery.1997} has been developed in \cite{Caprara}. $SO(5)$ symmetry
\cite{Zhang} which unifies the antiferromagnetic and superconducting
behaviour has also been used to explain pseudogap behaviour. Finally
spin-charge separation \cite{Lee} has been invoked.
This last theory also predicts \cite{Senthil.1998}, in a quasi-2D system, a
novel spin-metal phase with a nonzero spin diffusion constant at zero
temperature. It is therefore entirely conceivable that the
phase diagrams shown earlier are incomplete, and that new phases
await discovery and/or confirmation, particularly if one applies
a magnetic field (see for example \cite{Senthil.1999}).

\subsection{Precursor superconducting fluctuations
in particular phase fluctuations}
\label{sec:precursor}

As mentioned above, one major thrust suggests
that some form of precursor superconducting fluctuations or in particular
phase fluctuations \cite{Emery.1995a,Emery.1995b,Carlson} are
responsible for the pseudogap phenomenon. This assumption differs
in principle from, for example, bipolaron theory of high-$T_{c}$
superconductivity (see the review \cite{Alexandrov} and textbook
\cite{Mott}), where a very strong electron-phonon coupling is assumed and
where the bose-pairs are formed as stable units at temperatures higher
than the temperature of the superconducting (in this case superfluid)
transition.

An incoherent pair tunnelling experiment \cite{Scalapino} has been
proposed which may be able to answer whether superconducting
fluctuations are truly responsible for the pseudogap behaviour or
whether another mechanism is involved. A second test of the nature of
the condensation which comprises comparing two methods of determination
of the gap has been proposed in \cite{Deutscher}. The first method
for finding the gap is related to the single-particle excitation energy
$\Delta_{\mathrm{pair}}$, measured by ARPES or STS. The second method
is related to the energy coherence range $\Delta_{\mathrm{coh}}$,
defined by Andreev reflection.
Finally, Andreev interferometry --- the sensitivity of the tunneling
current to spatial variations in the local superconducting order at an
interface --- has also been proposed recently \cite{Goldbart} as a probe of
the spatial structure of the phase correlations in the pseudogap state.

One cannot however exclude the possibility that the pseudogap is the result of
a combination of various factors, e.g. spin and superconducting
(including phase) fluctuations\footnote{Indeed,
the anisotropy of the spin subsystem in HTSC
belongs to an easy-plane type which in turn implies that the magnetic
vectors (averaged spins) can be described by a model similar to the $XY$
model. The loss of long-range magnetic order under doping and the appearance
of magnetic correlation length $\xi_{\mathrm{mag}} \sim x^{-1/2}$
(see \cite{Loktev.review}) results in some sense in the formation
of a gap $\sim \xi_{\mathrm{mag}}^{-1}$ in the spin spectrum.
Superconducting phase fluctuations, as will be seen in
Sec.~\ref{chap:diagram}, can also be described by a $XY$-model and their
phase transition behaviour is similar.}
(see, for example,  models \cite{Onoda,Kobayashi} where the
interplay of antiferromagnetic and $d$-wave pairing fluctuations has
been recently studied).

Different authors argue that different types of superconducting
fluctuations are responsible for the pseudogap. The simplest
example of a theory which has the pseudogap/gap in the normal
state is the so-called {\em BCS-Bose crossover\/} theory.

Imagine starting from the weak coupling limit,
which is well described by BCS theory \cite{Schrieffer}, and
gradually increasing the intensity of the attraction between
the electrons. If the coupling is sufficiently strong, one
no longer needs the presence of the Fermi surface which was
essential in the BCS limit for the formation of Cooper pairs in metals
\cite{Polchinski}. The size of the Cooper pairs decreases
until they can be regarded as almost structureless singlet bose-particles.
The temperature of formation of these particles is given by the
BCS pairing temperature, $T_{\mathrm{pair}}$ which is determined
by the strength of the coupling.
However in contrast to the BCS case, these ``pre-formed'' bosons
condense into a single superfluid quantum state not at $T_{\mathrm{pair}}$,
but only at the temperature of Bose-Einstein condensation (BEC),
$T_{\mathrm{B}} \ll T_{\mathrm{pair}}$.  It is obvious that even above
$T_{\mathrm{B}}$ one will need to add extra energy
$\Delta_{\mathrm{pair}} = k_{B} T_{\mathrm{pair}}$ to break up these pairs
and this should lead to the gap-like features observed above
$T_{\mathrm{B}}$. This argument is strictly valid only in 3D.
Note that in the space of lowered (2D or 1D) dimensionality the formation of
local pairs does not demand as strong an attraction as in 3D and the
condensation process (see below) is not BEC.

There is even some evidence \cite{Mihailovic,Demsar.1998} from
time-resolved optical spectroscopy measurements of the
quasi-particle relaxation dynamics that pre-formed pairs
do exist in the underdoped YBa$_2$Cu$_3$O$_{7-\delta}$.
We shall describe the history of the BCS-Bose crossover problem in
Sec.~\ref{chap:history} and consider a number of simple 2D
models in Sec.~\ref{chap:crossover} in the review since this
problem appears to be very useful for a better understanding of
superconducting fluctuations in general.
Furthermore we believe that the BCS-Bose
crossover problem is in its own right an exciting chapter in the
history of research on superconductivity.

It is, however, rather difficult to combine the arguments
about the existence of the pre-formed local pairs
with the fact that the one observes a gapped Fermi surface in HTSC
\cite{Timusk,Campuzano}. In contrast the simple BCS-Bose crossover
models (see for the reviews \cite{Randeria.book,LSh.review})
predict its absence in precisely the Bose limit of pre-formed pairs.
It follows from these observations that the attraction between carriers in
HTSC is not strong enough to destroy the Fermi surface, so that
the pairs above $T_{c}$ appear to be short-lived
(the same is true for bipolarons \cite{Alexandrov,Mott}).
The presence of such resonant pairs above $T_{c}$ may still substantially
affect the normal state properties of the system and this effect
has been investigated in the papers of Levin et al.
\cite{Levin,Levin.quasi2D,Levin.1999}. We also note that for
spatially inhomogeneous systems it may be possible to have a
well defined Fermi surface and yet regions of low carrier densities
where pre-formed local pairs are present \cite{Demsar.1999}.

Up to this point our arguments were essentially independent of the
dimensionality of the superconductivity. Of course as mentioned above, if
one considers the 3D case, stronger attraction is necessary to reach either
the Bose limit or resonant pair regime than for the 2D system. It is known,
however, that all HTSC are highly anisotropic quasi-2D systems with an
almost 2D character of the conducting, superconducting and magnetic
properties. Indeed because the coherence length is less than a lattice
spacing in the direction perpendicular to the CuO$_2$ planes
(${\bf c}$-direction), the superconductivity in the copper oxides
takes place mainly in the weakly connected interacting CuO$_2$
layers (or their blocks).  This is the reason why pure 2D models of HTSC are
commonly accepted and studied. Undoubtedly the cuprate layers do exchange
carriers even if these layers are situated in different unit cells (blocks).
The mechanism for this transport (by coherent or incoherent (including pair)
electronic tunnelling) is not yet established (see, for example,
\cite{Puchkov}).

There is no doubt nonetheless that one must take into account the
possibility of different (for instance, direct or indirect) interlayer
hoppings to develop the full theory of HTSC.  Strictly
speaking therefore, one needs to consider quasi-2D models of these
superconductors. In practice, however, predominantly either 2D (or 3D) models
have been considered with only a few attempts to consider quasi-2D models.
This is, of course, related to the fact that the theoretical analysis
even for the relatively simple attractive model is complicated. This is due to
the necessity to go beyond the mean-field approximation, which proved good for
BCS theory, and to take into account the well developed fluctuations of the
superconducting order parameter.

It can be seen from the anisotropy of the conductivity
that the influence of the third dimension varies strongly from one
family of cuprates to the other. For example, this anisotropy
reaches $10^{5} \,-\, 10^6$ for the Bi- and Tl-based cuprates,
while its value for YBaCu$_{3}$O$_{6+\delta}$ compound is close
to $10^{2}$.  It seems therefore plausible that in the HTSC
Bi(Tl)-compounds the transport in the ${\bf c}$-direction is
incoherent, while in the Y-ones this transport is coherent
at least near the optimal doping.

These brief remarks are intended to indicate
how difficult it is to take into account the layered structure of HTSC.
As a further example, we note that the interlayer tunnelling mechanism proposed
by Anderson \cite{Anderson.book}, and based on the layered structure of the
cuprates, was considered as a leading candidate for the theory of
HTSC superconductivity until a significant discrepancy with
experiment was revealed \cite{Tsvetkov}.

One should also mention that the effect of interaction between
the layers has been recently experimentally studied in \cite{Choy}
by intercalating an organic compound into bismuth-based
cuprates.  Even though the distance between layers increased
dramatically the value of $T_{c}$ was almost unchanged
from that for the pristine material. This provides unambiguous evidence that
the superconductivity in layered cuprates is intrinsically of a 2D
nature.

The low-dimensionality of HTSC along with a relatively
small (at least in comparison to ordinary metals) carrier density
provide especially good conditions for the formation of different
types of vortex excitations.
Although the superconducting transition at $T_{c}$ itself
belongs to $d=3$ $XY$ universality class, the large anisotropy
of certain HTSC should lead to the Berezinskii-Kosterlitz-Thouless
(BKT) \cite{Berezinskii} 2D XY regime over a very wide temperature range,
until a crossover to 3D $XY$ critical behaviour over a rather narrow
temperature range (see, for example, the textbook \cite{Schneider.book}).
One also expects that the transition temperature could
well be practically unchanged \cite{Abrikosov.1997},
$T_{\rm BKT} \lesssim T_c$.
Thus, outside the transition region the low-energy
physics will be governed by the vortex fluctuations\footnote{In a
strictly 2D system the BKT transition is associated
with proliferation of unbound vortex-antivortex pairs. Weak coupling
between the planes leads to correlated motion between vortices in
adjacent planes which form 3D vortex loops close to the critical
temperature. The transition to the disordered phase is then characterized
by the appearance of vortex loops with arbitrarily large radii
\cite{Sudbo}.},
so one may expect the 2D models which we are going to
discuss below to be especially relevant for the description of
the pseudogap phase.
Thus we arrive at the picture of Emery and Kivelson
\cite{Emery.1995a,Emery.1995b} (see also earlier paper
by Doniach and Inui \cite{Doniach}) where the role of the phase
fluctuations of the superconducting order parameter, particularly
the vortex excitations is especially important.
Indeed this picture has been recently supported experimentally
\cite{Corson,Corson.low} by the measurements of the screening and
dissipation of a high-frequency electromagnetic field in Bi-cuprate
films. These measurements provide evidence for a phase-fluctuation
driven transition from the superconducting to normal state.

Without pretending to present the whole picture (this is now
practically impossible), we take
the point of view that the effect of low dimensionality is leading
to strong order-parameter phase fluctuations, and the dependence
of the conducting properties on doping, are among the key ingredients
of any proposed theory. We shall use general physical assumptions and
relatively simple models (which permit much analytical calculation) to
discuss the new properties of HTSC, including the pseudogap, based on the
assumptions given above. As we have discussed above, there is no consensus
about the origin of the pseudogap. We therefore believe that all currently
existing theories of pseudogap behaviour should be developed until their
predictions can both be experimentally tested and contrasted with the
predictions from the competing theories. Thus the main goal of this review
is to outline recent developments in the theory based on phase fluctuations.

\subsection{Outline}
This review focuses on the simplest pairing Hamiltonians but in the limit of
two dimensions and low carrier density. As such we emphasize the separation
into modulus and phase variables so essential in two dimensions. Not only are
these models the simplest prototype Hamiltonians in which to discuss phase
fluctuations and pseudogap behaviour, but as we shall demonstrate, illustrate
many of the properties of the wider class of theories based on phase
fluctuations.

In Section 2 we consider
a brief historical introduction to the BCS-Bose crossover as the simplest
theory displaying a pseudogap. In particular we critically discuss the role of
dimensionality and indicate why phase fluctuations are vital in the extreme
two dimensional limit. In Section 3 we present the BCS-Bose crossover
in a variety of models but at zero temperature where mean field theory
still applies. In particular we focus on how the inclusion of more realistic
Hamiltonians influence the crossover. The effects discussed in this section
include the role of $d$-wave pairing, the role of the retardation of the
interaction, and  the effect of interlayer couplings. We stress that the theory
of phase fluctuations is not in fact synonymous with the BCS-Bose crossover.
In Section 4 we briefly sketch the corresponding results at finite temperature
obtained using the $T$-matrix theory. These results have been derived
predominantly in three dimensions and we indicate how this approach breaks
down in the 2D limit. In Section 5 we present the BCS-Bose crossover
in two dimensions at finite temperature. The superconducting transition
temperature in this case is the Berezinskii-Kosterlitz-Thouless
transition temperature. Particular attention has been paid to the important
effects of quantum phase fluctuations, Coulomb repulsions and the observed
presence of non-magnetic impurities. In Sections 6 and 7 we construct the single
particle Green's function and the corresponding spectral density which has
been measured in recent ARPES experiments. We indicate the non-Fermi liquid
behaviour induced by phase fluctuations and discuss how various mechanisms can
``fill'' the mean-field gap to give pseudogap behaviour. Section 8 presents our
conclusions.

\section{A history of the BCS-Bose crossover problem}
\setcounter{equation}{0}
\label{chap:history}

\subsection{Early history}
The idea that the composite bosons (or local pairs as they are sometimes
called) exist and define the superconducting properties of metals
is in fact more than 10 years older than the BCS theory.  As early
as 1946, a sensational communication appeared saying that the
chemist-experimentalist Ogg had observed superconductivity in the
solution of Na in NH$_{3}$ at 77K \cite{Ogg}. It is very interesting
that the researcher made an attempt to interpret his own result in
terms of the BEC of {\em paired electrons\/}. Unfortunately, the
discovery was not confirmed and both it and his theoretical
concept, were soon completely forgotten (for the details see e.g.
\cite{Dmitrenko,Griffin}).

It is instructive to note that the history of
superconductivity has many similar examples. Probably, this explains
why Bednorz and M\"uller named their first paper "Possible high-$T_{c}$
superconductivity..." \cite{Bednorz}. And even today there appear
many unconfirmed communications about room-temperature superconducting
transitions.

A new step in the development of the local pair concept,
was taken in 1954 by Schafroth who in fact re-discovered the idea of
the electronic quasi-molecules \cite{Schafroth}. This idea was
further developed in the Schafroth, Blatt and Butler theory of
quasi-chemical equilibrium \cite{Blatt}, where superconductivity was
considered versus the BEC. Such a scenario, unfortunately, could not
compete with the BCS one due to some mathematical difficulties which
did not permit the authors to obtain the famous BCS results.
Then the triumph of the BCS theory replaced the far more obvious
concept of the local pairs and their BEC by the Cooper ones and their
instability which takes place in the necessary presence of a Fermi
surface (or more precisely a finite density of states at the Fermi level).
In contrast for the local (i.e. separated) pairs the formation is
not in principle connected to this density of electronic states. Also
unlike the local pairs, the Cooper ones are highly overlapping
in real space. More exactly there are no pairs in real
space and the Cooper pairing should be understood as a
{\em momentum\/} space pairing.

    Later experimental results reminded physicists of the existence of local
pairs. Indeed, it was Frederikse et al. \cite{Frederikse} who found
that the superconducting compound SrTiO$_{3}$ ($T_{c} \sim 0.3$K) has
a relatively low density of carriers and, moreover, this density is
controlled by Zr doping. The first deep discussion of the possibility
of the BCS-Bose crossover and electron binding above the
superconductivity transition temperature for a low density of
carriers  was carried out by Eagles \cite{Eagles.1969}
(see also his et al. relatively recent article
\cite{Eagles.1989}) in the context of SrTiO$_{3}$.

  The subsequent history of the investigation of the BCS-Bose
crossover has been often cited in the current literature. So, we
note only that the features of 3D crossover at $T = 0$ were considered
in \cite{Leggett} (see also recent analytical investigation of the
problem \cite{Marini}), and its extension to finite temperatures was
first given by Nozi\`eres and Schmitt-Rink \cite{Nozieres}.
The 2D crossover in superfluid $^{3}$He was studied in
\cite{Miyake}, where the very natural and convenient physical
parameter $\varepsilon_{b}$, the bound pair state energy, was first
used.

\subsection{Relevance to HTSC}
\label{sec:relevance}

      The discovery of HTSC rekindled the interest
in the problem of crossover and related phenomena. Let us discuss
the reasons why the crossover problem appeared to be
relevant to the general problem of the understanding of the HTSC.
Indeed, these superconducting compounds have some peculiarities
which place them much closer to the Bose  or at least to the
crossover region than the majority of low temperature superconductors.

      The ground state of the copper-oxide based materials
forms due to the  strong Hubbard repulsion resulting in antiferromagnetic
spin fluctuations in the proximity of the metal-insulator transition
(see e.g. reviews \cite{Loktev.review,Imada}). It seems
plausible that the subsequent doping of these materials results in
the appearance of weakly interacting (i.e. non-strongly correlated)
itinerant carriers (holes). However the exact nature of the ground
state including the strong electron-electron correlations in the
presence of itinerant carriers is in fact not yet understood.
We note that the experimental evidence for La$_{2-x}$Sr$_x$CuO$_4$
indicates that the unusual ``insulator-to-metal'' crossover is even
near optimal doping \cite{Boebinger}.
In addition the recent measurements of the of the transport
properties in YBa$_2$Cu$_3$O$_x$ \cite{Almasan.1999} also indicate
that the metal-insulator transition takes place close to $x=6.42$ which
corresponds to the underdoped region. Thus it is an important challenge
for theoreticians to include the metal-insulator transition
into the theory of superconductivity.
However currently most theoretical papers assume that superconductivity
in HTSC develops on the metallic side of the metal-insulator transition
and our review is not an exception.

The density $n_{f}$\footnote{One should
distinguish between the doping $x$ which is usually
connected to the chemical formula of the compound and $n_{f}$ which is defined
as the carrier density in each separate layer and because of this
can depend on the number of layers per unit cell, chemical
interactions, etc. In particular, $n_f \equiv n_{f}^{2D} = n_{f}^{3D} c$, where
$n_{f}^{3D}$ is the 3D (bulk) carrier density and $c$ is the lattice constant.}
of these itinerant holes is not as
large as in ordinary metals, so that the mean distance between them proves
to be comparable with a pair size $\xi_{\mathrm{pair}}$ or a coherence
length.\footnote{It is
good to bear in mind that the coherence length
strictly speaking is distinguished from the pair size, especially
at low carrier density, since the first one
is related to the energy coherence range $\Delta_{\mathrm{coh}}$ and the
second to the single-particle excitation energy
$\Delta_{\mathrm{pair}}$ \cite{Deutscher}
(this question will be treated in Section~\ref{sec:derivative}).}
This situation is significantly different from the conventional BCS
theory where the parameter $\xi_{\mathrm{pair}}$ greatly exceeds the mean
distance between carriers which is $\sim n_{f}^{1/2}$.  Experimentally
the dimensionless value of $k_{\mathrm{F}} \xi_{\mathrm{pair}}$
($k_{\mathrm{F}}$ is the Fermi momentum) which describes the ratio of
the pair size and the distance between carriers is about $5$ -- $20$
for HTSC while for the low-temperature superconductors it is about
$10^{3}$ -- $10^{4}$ \cite{Loktev.review,Randeria.book}.

The quantitative differences between HTSC and low temperature
superconductors were summarized on the well-known Uemura plot
\cite{Uemura}, where the superconductors are classified by the ratio
of their critical temperature, $T_{c}$ and the superfluid density
expressed in terms of the effective Fermi energy, $\epsilon_{F}$.
\begin{figure}[h]
\centering
\resizebox{0.60\textwidth}{!}{%
  \includegraphics{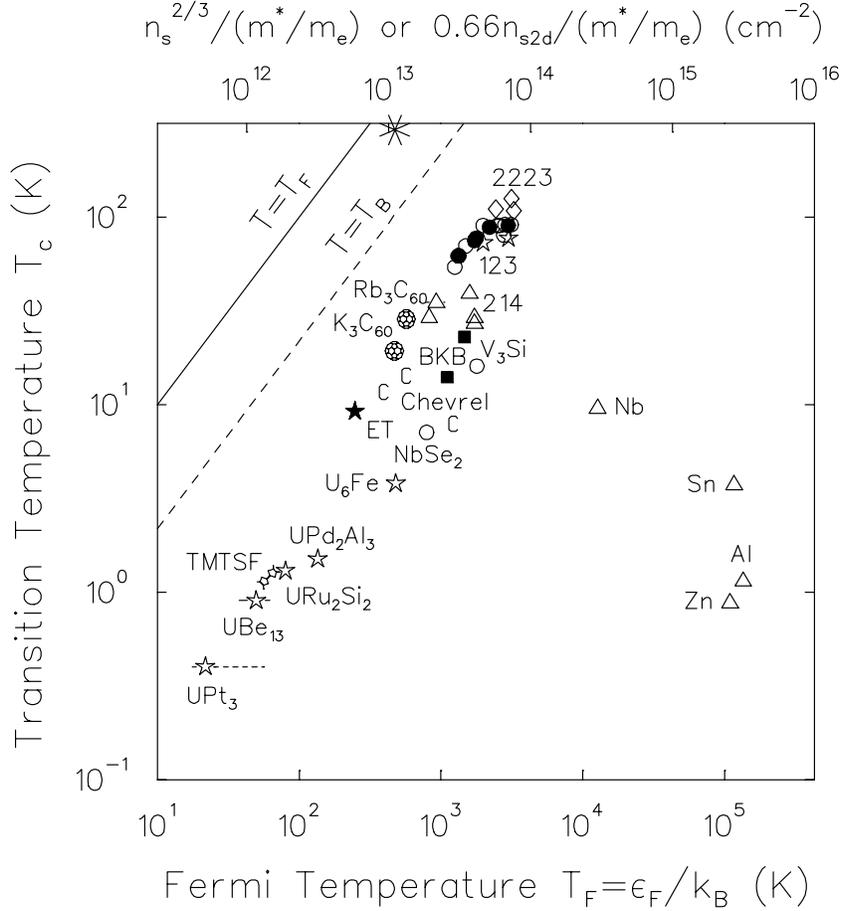}
}
\caption{A plot of $T_{c}$ versus the effective Fermi energy,
$\epsilon_{F} \sim n_{\mathrm{s2d}}/m^{\ast}$, where
$n_{\mathrm{s2d}}$ is the 2D superfluid density and $m^{\ast}$ is the effective
mass of the carriers in CuO$_2$ planes. $T_{\mathrm{B}}$ denotes BEC
temperature for non-interacting bosons with density $n_{s}/2$ and mass
$2m^{\ast}$. This plot was taken from \protect{\cite{Uemura}}.}
\label{fig:03}
\end{figure}
One can see from Fig.~\ref{fig:03}
that cuprates, organic superconductors and some other ``exotic''
superconductors have $k_{B} T_{c}/\epsilon_{F}$ as high as 0.01-0.1, much
higher than those of conventional superconductors. This value of $T_{c}$ is
however 4-5 times less than $T_{\mathrm{B}}$.  Another important issue which
follows from Fig.~\ref{fig:03} is the linear relationship $T_{c}
\sim n_{s}/m^{\ast}$.  Uemura has interpreted this dependence as
originating from BEC, but we would like to note that this dependence
may also be understood within the phase fluctuation scenario.
This linear dependence is absent in overdoped cuprates where the
depression of $T_{c}$ is associated with a decrease of the superconducting
condensate density $n_{s}$ in spite of the increasing normal-state
carrier density \cite{Budnick}.

In fact, as we will see later, the new materials are likely
to be in an intermediate regime between the Cooper pairs and the
composite bosons, at least when the doping is not large and the value
of $T_{c}$ is far from the highest possible (optimal) one.

\begin{figure}[h]
\centering
\resizebox{0.50\textwidth}{!}{%
  \includegraphics{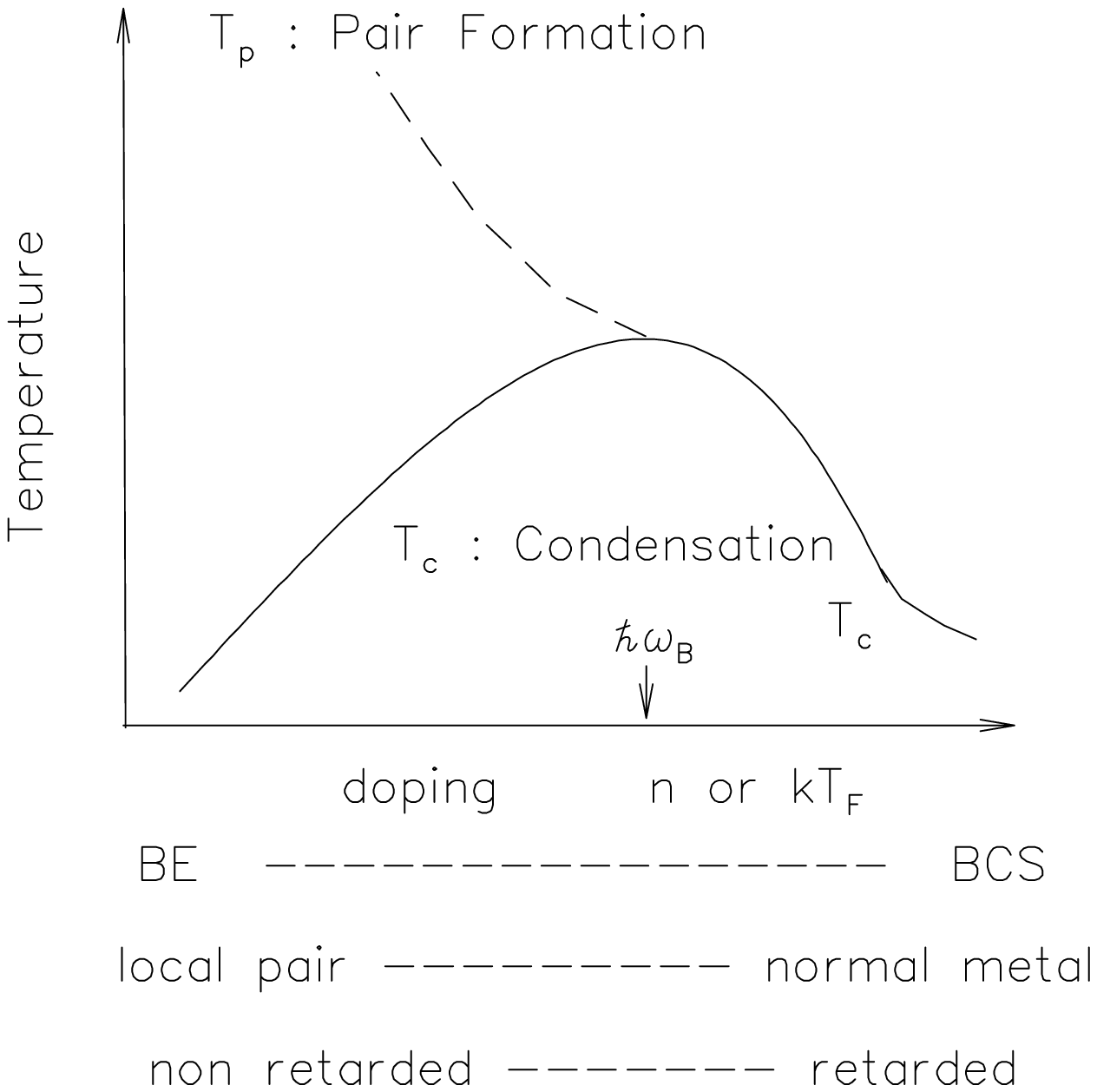}
}
\caption{Phase diagram describing BEC-BCS crossover with increasing
carrier concentration $n$. The phase diagram can be mapped to that
of cuprates assuming that the pseudogap temperature $T^{\ast}$
corresponds to the formation of normal state pairs. This diagram was taken from
\protect{\cite{Uemura}}.}
\label{fig:04}
\end{figure}

Based on the phenomenological classification presented in
Fig.~\ref{fig:03} Uemura \cite{Uemura} suggested the following
interpretation of the phase diagram from Fig.~\ref{fig:01}
in terms of evolution from BEC (in real space with non-retarded
strong interaction) to BCS condensation (in momentum space with
retarded weak interaction) shown in Fig.~\ref{fig:04}.

There is an important outstanding question as to how the picture proposed by
Uemura correlates with the ideas suggested by Emery and Kivelson
\cite{Emery.1995a,Emery.1995b} and we will discuss this point
in Sec.~\ref{sec:impure}.

\subsection{BCS-Bose crossover and pseudogap:
survey of the current literature}
\label{sec:survey}

There are so many published articles dealing with the BCS-Bose crossover
study that we are only able to mention the review papers
\cite{Randeria.book,LSh.review} where the interested reader may find more
references.

We also note the existence of the comprehensive review of Ginzburg
\cite{Ginzburg} (see also \cite{Griffin}) which presents
a historical overview of the parallel (and generally speaking independent)
development of the macro-  and microphysics of superfluidity and
superconductivity.

In the present review we focus on the more recent results
related to the crossover problem, and with a stronger accent on
the pseudogap phenomena. In most cases the attractive
(negative-$U$) 2D or 3D Hubbard model has been considered
(see the review \cite{Micnas.review} for the results
obtained prior to 1990).
In particular this model has been studied within the
``$\Phi$ derivable'' in the sense of Baym \cite{Baym},
conserving \cite{Kadanoff} and self-consistent
$T$-matrix approximation both analytically
\cite{Haussmann.1993,Tchernyshyov,Kagan} and numerically
\cite{Serene,Fresard,Micnas,Haussmann.1994,Pedersen,Schafroth.1997,Gooding,Kornilovitch,Keller}.
In all of the above papers the 2D model was considered
with the exception of \cite{Haussmann.1993,Haussmann.1994}, where the 3D model
was studied and \cite{Tchernyshyov} where the influence of the
third direction was introduced.

The non ``$\Phi$ derivable'', but still conserving $T$-matrix
approximation was considered in \cite{Levin,Levin.quasi2D,Levin.1999}.
As we have already mentioned, the pseudogap in this approach has been
related to the resonant pair scattering of correlated electrons above $T_{c}$.
This approach regards the pseudogap as a 3D phenomenon.

A comprehensive critical analysis of the self-consistent $T$-matrix
approximation which is, as we can see, widely utilized in the literature
was performed for the 3D case in \cite{Pieri}. The authors came to the
conclusion that a new class of the diagrams has to be included.
In particular, they demonstrate the importance of the correct treatment
of the residual interaction between the composite bosons in order
to regularize the strong coupling limit. It is claimed \cite{Pieri}
that the self-consistency of the fermionic Green's function in the $T$-matrix
approximation maybe less important than due care about the residual
bose-bose interaction. Their conclusion (in the 3D case) that the
$T$-matrix approximation is in fact inadequate underlines our
strong reservations about the applicability of the $T$-matrix
approximation in the 2D limit (see Sec.~\ref{sec:limitations}).

The pseudogap for the case of $d$-wave
pairing has also been studied in \cite{Nazarenko,Dagotto,Ichinomiya,Yanase}
(for the review see \cite{Randeria.review}) and quantum Monte-Carlo
simulations for the 2D attractive Hubbard model were performed in
\cite{Moreo,Trivedi,Singer,Tremblay.1998,Tremblay.1999,Tremblay.1999a}.

There is in fact a big difference between the 2D and 3D theories mentioned
here. In particular, the pseudogap obtained in the 3D approach
\cite{Levin,Levin.1999} is the result of rather strong coupling, while to
obtain the pseudogap in 2D theory one needs intermediate or even weak coupling
\cite{Tremblay.1999,Loktev.dirty}. Thus the 3D approach \cite{Levin} has far
stronger constraints on the mechanism of HTSC, than do the studies of the 2D
models mentioned here (see also the papers
\cite{Loktev.dirty,Gusynin.JETP,Sofo,Alvarez,Ariosa.1991,Ariosa.1992,Beck.1997,Ariosa.1998,Capezzali.1999,Franz,Dorsey,Gusynin.new}
which we will discuss in the subsequent chapters of the review). These show,
that almost independently of the mechanism of HTSC and the intensity of
coupling, the fluctuations in the order parameter phase contribute to the
pseudogap. Even if the origin of the pseudogap is related to some other
mechanism e.g. those discussed in Sec.~\ref{sec:other}, the preconditions for
the existence of a superconducting phase contribution to the pseudogap are so
easily satisfied that such a contribution must always be present.

It is assumed in 3D \cite{Levin,Levin.1999} and quasi-2D \cite{Levin.quasi2D}
models that $T_{c}$ is the temperature for the formation of long-range order
which must go to zero in the 2D limit in accordance with the
Coleman--Mermin--Wagner--Hohenberg (CMWH) theorem \cite{Coleman}.\footnote{We
note that the CMWH theorem has been recently revisited for the special cases
of 1D and 2D Hubbard (both attractive and repulsive) models \cite{Su}.} Indeed
the theorem prohibits the formation of a superconducting phase with {\em
homogeneous\/} order parameter (the latter related to breaking a continuous
symmetry) in 2D. This is due to the fact that the fluctuations of the order
parameter phase destroy the long-range order \cite{Rice} (see also
\cite{Miransky}). Thus even this very general argument as to why long-range
order is absent in 2D leads one to the conclusion that the phase fluctuations
play a significant role in low dimensional models of metals like HTSC.

The ideology of the 2D theories is completely different from those in 3D.
As we have just explained, in 2D there is no superconducting transition
into a state with long-range order, and the only possible kind of transition
is the BKT one \cite{Berezinskii}. (Since the physics of the BKT
transition is rather subtle, and may be unfamiliar,
we bring to the attention of the reader the following review articles
\cite{Minnhagen,Pierson} and textbooks \cite{Izyumov,Plischke,Huang}).
Thus the temperature $T_{c}$ in these models is identified with the
temperature of the BKT transition, $T_{\mathrm{BKT}}$ below which an algebraic
order is established in the system. This transition does not require
symmetry breaking and is therefore not forbidden by the CMWH
\cite{Coleman,Su} theorem. Although, due to the presence of even weak
interlayer tunnelling, the transition is of $d=3$ $XY$ type with
true long-range order, the critical temperature $T_c$
is expected to be close to the value of $T_{\mathrm{BKT}}$ calculated for
the pure 2D model \cite{Abrikosov.1997}.

An approach to the 2D single-band repulsive and attractive Hubbard
models which allows one to automatically satisfy the CMWH theorem,
and enforce the conservations laws, the Pauli principle and a number
of sum rules has been discussed in \cite{Tremblay.1997}.
We refer also to another approach \cite{Manske} developed for
the Hubbard model with a strong repulsion which results in indirect 
attraction between carriers. It is worth to mention this approach 
here since it treats the emerging superconducting phase fluctuations 
in the way very similar to the method presented in the review.
The phenomenology of superconductivity resulting from the Bose
condensation of pre-formed pairs coexisting with unpaired fermions
has been studied in \cite{Geskenbein}. It was shown that the
phenomenology describes reasonably well the data in the underdoped
Y-Ba-Cu-O family but fails to describe these compounds in the
optimally doped regime or underdoped La-Sr-Cu-O. The influence of
the existence of the fermion bound states above $T_{c}$ was also
considered in \cite{Akhiezer} and the theory of chemical
equilibrium between these bound states and single fermions
was applied in \cite{Kallio} to explain the normal state properties
of HTSC (see also \cite{deLlano} which considers the BEC of such pairs).

The large number of references cited here shows how popular
this topic is currently and the authors apologize in advance to
those whose work has been omitted.

Certainly, it is necessary to point out immediately that the
crossover phenomena in particular and the investigation of the
attractive Hubbard model in general do not (and cannot) address
the problem of the mechanism for HTSC. The problem of the mechanism for
HTSC  is very difficult and still controversial. Nevertheless the treatment
of the crossover or, more generally, the low dimensional models with
an {\em a priori\/} postulated attraction between carriers may shed light
on some of the features of HTSC.

\section{BCS-Bose crossover in 2D systems at $T=0$}
\setcounter{equation}{0}
\label{chap:crossover}

This chapter is devoted to the description of the BCS-Bose
crossover in a variety of 2D systems at $T=0$. The reason why
we decided to restrict our initial consideration to the
zero temperature limit is to avoid the complexities related to the
CMWH theorem \cite{Coleman}. Indeed, at $T=0$ the problem becomes,
due to the integration over frequency (which replaces the summation
over Matsubara frequencies for $T \neq 0$), effectively a 3D one.
Therefore, the 2D theorems \cite{Coleman} are not applicable to
this case and one can safely consider long-range order in 2D
systems at $T=0$. Thus one might expect at $T=0$ the
simplest BCS approximation with minor modifications to be
able to adequately describe the physics of the BCS-Bose crossover.

\subsection{One band continuum model with $s$-wave pairing and
non-retarded attraction}
\label{sec:one-band}

Let us introduce a continuum field theoretical model of fermions
with an attractive two-body
interaction.\footnote{It is worthwhile to
emphasize once more that the application
of a weak coupling model for the description of HTSC suggests
weak coupling for the doped carriers only. Indeed, as stated above,
HTSC compounds are antiferromagnetic insulators with a strong
onsite (Hubbard) repulsion on Cu ions that enforces an almost local
distribution of the spin density on these ions (localized fermions
with spin $S=1/2$, or copper $3d^9$ configuration). This initial system
should undoubtedly be described by strong coupling models
near half-filling ($n=1$). The doping of the system decreases
the number of electrons on the copper ions (and also on the oxygen ions,
if a $pd$-hybridization model is used), introducing the positively
charged delocalized carriers (holes) which move in the medium with a
short-range magnetic order. According to our assumption these
current-carrying holes are weakly coupled and form correspondingly
a small Fermi surface. These arguments justify to some extent
why the simplest 2D models with a direct (negative-$U$ Hubbard model)
or indirect (for example, phonon-like) attraction are considered below
to establish how the variable carrier density and the condensate phase
fluctuations affect the properties of 2D superconductivity. One
can hope that, under these circumstances, many of the physical properties
described below persist in more realistic approaches and models.}
Our goal is to consider
how the energy gap $\Delta$, the chemical potential $\mu$, the
coherence length $\xi_{\mathrm{coh}}$ and the pair size
$\xi_{\mathrm{pair}}$
change as a function of the density of bare fermions, $n_{f}$.
We shall follow here the papers \cite{LSh.review,GGL}
(see also the original paper \cite{Miyake}, the subsequent paper
and review \cite{Randeria.2D,Randeria.book}) and
discuss the difference between the coherence length and Cooper
pair size using Refs.~\cite{Pistolesi.1994,Pistolesi.1996,Quick,Carter}.

The simplest model is described by the Hamiltonian density,
which is the field-theoretical analog of the negative-$U$ Hubbard model
(see Sec.~\ref{sec:s.lattice}),
\begin{equation}
\mathcal{H} = - \psi^{\dagger}_{\sigma}(x) \left(
   \frac{\nabla^{2}}{2 m} + \mu \right) \psi_{\sigma}(x)
 - U \psi^{\dagger}_{\uparrow}(x) \psi^{\dagger}_{\downarrow}(x)
       \psi_{\downarrow}(x) \psi_{\uparrow}(x) ,
                         \label{Hamiltonian.2D}
\end{equation}
where $x \equiv {\bf r}, \tau$ (${\bf r}$ is a 2D vector and $\tau$ is the
imaginary time);
$\psi_{\sigma}(x)$ is the Fermi field; $m$ is the fermion effective
mass; $\sigma = \uparrow, \downarrow$ is the fermion spin;
$U > 0$ is the attraction constant. The chemical potential $\mu$
fixes the average density $n_{f}$ of the free (bare) carriers.
We choose units in which $\hbar = k_{\mathrm{B}} = 1$ and the
system occupies the volume $v$.

\subsubsection{Formalism}
\label{sec:formalism}
The functional integral approach is appropriate for the problem
studied here and will prove very useful in the subsequent chapters.
Since we will need the results presented here in subsequent chapters
(including the case of nonzero temperatures), we are going to use the
Matsubara thermal technique from the beginning, but will take the limit
$T=0$ in this chapter. Thus, let us consider the formalism used.

Introducing the Nambu spinors for the fermion fields
\cite{Nambu}
(see also the textbooks \cite{Schrieffer,Rickayzen.book}),
\begin{equation}
\Psi(x) = \left( \begin{array}{c}
\psi_{\uparrow}(x) \\  \psi_{\downarrow}^{\dagger}(x)
\end{array} \right), \qquad
\Psi^{\dagger}(x) = \left( \begin{array}{cc}
\psi_{\uparrow}^{\dagger}(x) \quad \psi_{\downarrow}(x)
\end{array} \right),
                            \label{Nambu.variables}
\end{equation}
one should rewrite (\ref{Hamiltonian.2D}) in the appropriate  form:
\begin{equation}
\mathcal{H} =  -\Psi^{\dagger}(x)
\left( \frac{\nabla^{2}}{2 m} + \mu \right) \tau_{3} \Psi(x)
            - U \Psi^{\dagger}(x) \tau_{+} \Psi(x)
        \Psi^{\dagger}(x) \tau_{-} \Psi(x) ,
                    \label{Hamiltonian.Nambu}
\end{equation}
where $\tau_{3}$, $\tau _{\pm} \equiv (\tau _{1} \pm i \tau _{2}) / 2$
are Pauli matrices.

       Now the partition function can be expressed through the
Hamiltonian (\ref{Hamiltonian.Nambu}) as:
\begin{equation}
Z(v, \mu, T) = \int \mathcal{ D} \Psi \mathcal{D} \Psi^{\dagger}
\exp \left\{-\int_{0}^{\beta} d \tau \int d^2  r
[\Psi^{\dagger}(x) \partial_{\tau} \Psi(x) + \mathcal{H}({\bf r})]
\right\},                    \label{partition.main}
\end{equation}
where $\beta \equiv 1/T$ and $\mathcal{D} \Psi \mathcal{D} \Psi^{\dagger}$
denotes the measure of the integration over the Grassmann variables
$\Psi$ and $\Psi^{\dagger}$, satisfying the anti periodic boundary
conditions: $\Psi(\tau, {\bf r}) = -\Psi(\tau+\beta, {\bf r})$
and $\Psi^{\dagger}(\tau, {\bf r}) =
-\Psi^{\dagger}(\tau+\beta, {\bf r})$.

      If it were possible to calculate the partition function
(\ref{partition.main}) exactly one could obtain all
thermodynamical functions from the thermodynamical potential
\begin{equation}
\Omega(v,\mu,T) = - T \ln Z(v, \mu, T).
\label{thermopotential}
\end{equation}

        Introducing the auxiliary Hubbard-Stratonovich complex scalar
field in the usual way one can represent (\ref{partition.main}) in
an exactly equivalent form (see the review \cite{Kleinert}):
\begin{eqnarray}
&& Z(v, \mu, T) =
\int \mathcal{D} \Psi \mathcal{D} \Psi^{\dagger}
\mathcal{D} \Phi \mathcal{D} \Phi^{\ast}
\exp \left\{-\int_{0}^{\beta} d \tau \int d^2  r
\left[ \frac{|\Phi(x)|^{2}}{U} +  \right. \right.
\nonumber                                       \\
&& \quad \left. \left.
\Psi^{\dagger}(x) \left[ \partial_{\tau} \hat I -
\tau_{3} \left( \frac{\nabla^{2}}{2 m} + \mu \right) -
\tau_{+} \Phi(x) - \tau_{-} \Phi^{\ast}(x)
\right]\Psi(x) \right]
\right\}.                    \label{partition.Hubbard}
\end{eqnarray}
The main virtue of this representation is the non-perturbative
introduction of the composite fields
$\Phi(x) = U \Psi^{\dagger}(x) \tau_{-} \Psi(x) = $
$U \psi_{\downarrow}(x) \psi_{\uparrow}(x)$,
$\Phi^{\ast}(x) =
U \Psi^{\dagger}(x) \tau_{+} \Psi(x) = $
$U\psi^{\dagger}_{\uparrow}(x) \psi^{\dagger}_{\downarrow}(x)$
and the possibility to develop a consistent approach.
Specifically, the expression (\ref{partition.Hubbard})
turns out to be rather convenient for studying
non-perturbative phenomena such as superconductivity.
In this case the complex Hubbard-Stratonovich field naturally
describes the order parameter arising due to the formation of
the Cooper or the local pairs. The average value of
$|\Phi|( \equiv \Delta)$ is proportional to the density of pairs,
on the one hand, and determines the gap in the one-particle
Fermi-spectrum, on the other.

        The integration over the fermion fields in
(\ref{partition.Hubbard}) can be done formally even though $\Phi$
and $\Phi^{\ast}$ depend on the spatial and temporal coordinates.
Thus, one obtains (formally exactly)
\begin{equation}
Z(v, \mu, T) =
\int \mathcal{D} \Phi \mathcal{D} \Phi^{\ast}
\exp [ - \beta \Omega(v, \mu, T, \Phi(x), \Phi^{\ast}(x))],
            \label{partition.effective}
\end{equation}
where
\begin{equation}
\beta \Omega(v, \mu, T, \Phi(x), \Phi^{\ast}(x)) =
\frac{1}{U} \int_{0}^{\beta} d \tau \int d^2  r |\Phi(x)|^{2}
- \mbox{TrLn}G^{-1} + \mbox{TrLn}G_{0}^{-1}
            \label{Effective.Action}
\end{equation}
is the one-loop effective action.
This action includes in itself a series of terms containing
derivatives with respect to $\Phi(x)$ and $\Phi^{\ast}(x)$. In
the lowest orders it corresponds to the Ginzburg-Landau effective
action.

           The operation $\mbox{Tr}$ in (\ref{Effective.Action}) is taken with
respect to the space ${\bf r}$, the imaginary time $\tau$ and
the Nambu indices. The action (\ref{Effective.Action}) is expressed
through the fermion Green's function which obeys the equation:
\begin{equation}
\left[-\hat I \partial_{\tau} +
\tau_{3} \left(\frac{\nabla^{2}}{2 m} + \mu \right)
+ \tau_{+} \Phi(\tau, {\bf r})
+ \tau_{-} \Phi^{\ast}(\tau, {\bf r}) \right]
G(\tau, {\bf r}) = \delta(\tau) \delta({\bf r})
          \label{Green.fermion}
\end{equation}
with boundary condition
\begin{equation}
G(\tau + \beta, {\bf r}) = - G(\tau, {\bf r}).
                                     \label{boundary}
\end{equation}
The free Green's function
\begin{equation}
G_{0}(\tau, {\bf r}) =
G(\tau, {\bf r})|_{\Phi, \Phi^{\ast}, \mu = 0}
                          \label{Green.free}
\end{equation}
in (\ref{Effective.Action}) is needed to provide the regularization
in the calculation of \\
$\Omega(v, \mu, T, \Phi, \Phi^{\ast})$.
The representation (\ref{partition.effective}),
(\ref{Effective.Action}) is exact, although to perform the
calculation in practice it is necessary to restrict ourselves to some
approximation. Below we shall use the assumption,
that one neglect the fluctuations of the fields
$\Phi(x)$ and $\Phi^{\ast}(x)$ and thus replace them by their
equilibrium values which will now also be denoted as $\Phi$ and
$\Phi^{\ast}$. This saddle-point approximation is in fact absolutely
equivalent to the BCS mean-field approximation and works quite
well for $T=0$ even in 2D. We note that the assumption
that $\Phi = \; \mbox{const} \neq 0$ implies that we are considering
a state with long-range order which is of course invalid for $T \neq 0$.
We will return to this problem in Chapter~\ref{chap:diagram},
where the generalization for finite $T$ will be done consistently.

The thermodynamical potential $\Omega$ as well as the
partition sum $Z$ now depend on $\Phi$ and $\Phi^{\ast}$
which play the role of the order parameter. The order parameter
appears due to the fact that one is not considering
the exact potential $\Omega(v, \mu, T)$
which contains a functional integral over the
auxiliary fields. Instead one is considering its saddle-point
approximation $\Omega(v, \mu, T, \Phi, \Phi^{\ast})$:
\begin{eqnarray}
&& \! \! \! \!
\Omega_{pot}(v, \mu, T, \Phi, \Phi^{\ast}) =
\nonumber                     \\
&& \left. \left(
\frac{1}{U}  \int d^2 r |\Phi(x)|^{2}
- T \mbox{TrLn}G^{-1} + T \mbox{TrLn}G_{0}^{-1}
\right) \right|_{\Phi = \Phi^{\ast} = const} .
            \label{Omega.Potential}
\end{eqnarray}
in which the integral over the auxiliary fields
has been replaced by a single term in which the auxiliary fields
take on their equilibrium values $\Phi$ and $\Phi^*$, and
about which one can subsequently expand to obtain better approximations.

To avoid possible misunderstanding, we shall
write down the formulae for the  Fourier transformations which are
used throughout the review. They  connect the coordinate
and momentum representations in the usual manner:
\begin{equation}
F(i \omega_{n}, {\bf k}) = \int_{0}^{\beta} d \tau \int
d^2 r F(\tau, {\bf r}) \exp(i \omega_{n} \tau - i
{\bf k} {\bf r}) \,, \label{Fourier.momentum}
\end{equation}
\begin{equation}
F(\tau, {\bf r}) =
T \sum_{n = -\infty}^{+\infty}
\int \frac{d^2 k}{(2 \pi)^{2}} F(i \omega_{n}, {\bf k})
\exp(-i \omega_{n} \tau + i {\bf k} {\bf r}),
\label{Fourier.coordinate}
\end{equation}
where $\omega_{n} = \pi T (2n + 1)$ are the fermion (odd) Matsubara
frequencies. In the case of bosons the odd frequencies
should be replaced by even ones: $\Omega_{n} = 2 \pi n T$.

         For example, the Green's function (\ref{Green.fermion})
has, in the momentum representation, the following form:
\begin{equation}
G(i \omega_{n}, {\bf k}) = -
\frac{i \omega_{n} {\hat I} + \tau_{3} \xi({\bf k})
-\tau_{+} \Phi - \tau_{-} \Phi^{\ast}}
{\omega_{n}^{2} + \xi^{2}({\bf k})  + |\Phi|^{2}},
               \label{Green.fermion.momentum}
\end{equation}
where $\xi({\bf k}) = \varepsilon({\bf k}) - \mu$ with
$\varepsilon({\bf k}) = {\bf k}^{2}/2m$,
$\Phi \equiv \langle \Phi(x) \rangle$ and
$\Phi^{\ast} \equiv \langle \Phi^{\ast}(x) \rangle$ are already taken
to be constants which represent the complex order parameter
(or more precisely, {\em the complex ordering field\/}, because in 2D at
$T \neq 0$ there is no order parameter in its usual sense).

Substituting (\ref{Green.fermion.momentum}) into
(\ref{Omega.Potential}), one arrives at (see Appendix~\ref{sec:A}
and Refs. \cite{GGL,LSh.review,Gusynin.JETP})
\begin{eqnarray}
&&\Omega_{pot}(v, \mu, T, \Phi, \Phi^{\ast})  =
\nonumber                        \\
&& \qquad
v \left\{\frac{|\Phi|^{2}}{U} \right.  -
\int \frac{d^2 k}{(2 \pi)^{2}}
\left[ 2T
\ln \cosh \frac{\sqrt{\xi^{2}({\bf k}) + |\Phi|^{2}}}{2T}
- \xi({\bf k}) \right] +
\nonumber                        \\
&& \qquad \qquad
\left. \int \frac{d^2 k}{(2 \pi)^{2}}
\left[ 2T
\ln \cosh \frac{\varepsilon({\bf k})}{2T}
- \varepsilon({\bf k}) \right] \right\} .
\label{Omega.Potential.expr}
\end{eqnarray}

\subsubsection{The Effective Action and Potential}
\label{sec:effective}

As has been noted above, it is impossible to calculate
(\ref{Effective.Action}) for $\Phi$ dependent on $x$
in general. However, if one assumes that
the gradients of $\Phi$ and $\Phi^{\ast}$ are small, the action
(\ref{Effective.Action}) can be naturally divided into kinetic and
potential parts
\begin{equation}
\Omega(v,\mu, \Phi(x),\Phi^{\ast}(x)) =
\Omega_{kin}(v, \mu, \Phi(x),\Phi^{\ast}(x)) +
\Omega_{pot}(v, \mu, \Phi,\Phi^{\ast}),
                    \label{kinetic+potential}
\end{equation}
where the effective potential has been defined by
(\ref{Omega.Potential}) (its final
expression is given by (\ref{Omega.Potential.expr})). The terms
$\Omega_{kin}(\Phi(x),\Phi^{\ast}(x))$ with derivatives in
expansion (\ref{kinetic+potential}) contain important physical
information, which we shall consider in Sec.~\ref{sec:derivative}.

      Let us return to the effective potential.
In the limit $T \to 0$ (\ref{Omega.Potential.expr}) reduces to
\begin{equation}
\Omega_{pot}(v, \mu, \Phi,\Phi^{\ast}) = v
\left[ \frac{|\Phi|^{2}}{U} - \int \frac{d^2k}{(2 \pi)^{2}}
\left(\sqrt{\xi^{2}({\bf k}) + |\Phi|^{2}} -
\xi ({\bf k}) \right) \right],
                        \label{Omega.Potential.T=0}
\end{equation}
where the terms which do not depend on $\Phi,\Phi^{\ast}$ and $\mu$
have been omitted.

      It is interesting that, by virtue of the invariance of
      the partition function (\ref{partition.Hubbard})
with respect to the phase transformation of the group $U(1)$
\begin{eqnarray}
&& \Psi(x) \to e^{i \alpha \tau_{3}} \Psi(x), \qquad
\Psi^{\dagger}(x) \to \Psi^{\dagger}(x) e^{-i \alpha \tau_{3}};
\nonumber                           \\
&& \Phi(x) \to e^{2 i \alpha} \Phi(x), \qquad
\Phi^{\ast}(x) \to  \Phi^{\ast}(x) e^{-2 i \alpha},
             \label{transformation}
\end{eqnarray}
with real $\alpha$ the potential $\Omega_{pot}(v, \mu, \Phi,\Phi^{\ast})$
(\ref{Omega.Potential}) (see also (\ref{Omega.Potential.expr}) and
(\ref{Omega.Potential.T=0})) can
only be dependent on the invariant product $\Phi^{\ast} \Phi$\footnote{There
is another transformation
(when the sign of the phase $\alpha$ is defined by the fermion spin
rather than the charge) under which the Hamiltonian (\ref{Hamiltonian.2D})
(or (\ref{Hamiltonian.Nambu})) is also invariant.
Such a transformation proves to be important for
fermion-fermion repulsion (i.e. $U<0$), or for the fermion-antifermion
(electron-hole) channel of pairing.  Apart from this difference
the formalism for the case of a repulsive interaction is
identical to that under consideration. The complete set of gauge
transformations for the Hamiltonian under consideration were originally
given by Nambu \cite{Nambu}.}.

        The analytic solution of the problem for the 2D case that we
consider here is easier than the 3D one \cite{Marini} which
is obtained in terms of special functions
(compare with Eqs.~(\ref{solution.2D.band}) and
(\ref{solution.2D}) below).
Indeed, after performing the integration over ${\bf k}$ in
(\ref{Omega.Potential.T=0}) (which can be done straightforwardly
due to the energy independence of the free fermion density of states)
one obtains
\begin{eqnarray}
&& \! \! \! \!
\Omega_{pot}(v, \mu, \Phi,\Phi^{\ast}) =  v
|\Phi|^{2} \left\{ \frac{1}{U} - \frac{m}{4 \pi} \left[ \ln \frac{W -
\mu + \sqrt{(W - \mu)^{2} + |\Phi|^{2}}} {\sqrt{\mu^{2} + |\Phi|^{2}}
- \mu} \right. \right.
\nonumber                 \\
&& + \left.
\left.  \frac{W - \mu}{W - \mu + \sqrt{(W - \mu)^{2} + |\Phi|^{2}}} +
\frac{\mu}{\sqrt{\mu^{2} + |\Phi|^{2}} - \mu} \right] \right\},
\label{Omega.Potential.2D.expr}
\end{eqnarray}
where the value $W = {\bf k}_{B}^{2}/2m$ is the conduction
bandwidth and ${\bf k}_{B}$ is the maximal (in some sense
Brillouin boundary) momentum.

\subsubsection{Main equations and analysis of solution}
\label{sec:solution}

     If the quantity $\Delta$\footnote{This is the parameter that
is responsible for the appearance of a new (ordered, or with lowered
symmetry) phase (see Sec.~\ref{sec:formalism}) which is
permissible at $T=0$.}
is defined as the average value of $|\Phi|$,
then the equation for the extremum
\begin{equation}
\left. \frac{\partial \Omega_{pot}(v, \mu, \Phi, \Phi^{\ast})}
{\partial \Phi} \right|_{\Phi = \Phi^{\ast} = \Delta} = 0,
\label{minima}
\end{equation}
yields, according to (\ref{Omega.Potential.2D.expr})
\begin{equation}
\Delta \left[ \frac{1}{U} - \frac{m}{4 \pi}
\ln \frac{W - \mu + \sqrt{(W - \mu)^{2} + \Delta^{2}}}
{\sqrt{\mu^{2} + \Delta^{2}} - \mu} \right] = 0,
             \label{gap.2D.band}
\end{equation}
while the condition
\begin{equation}
-\frac{1}{v} \left. \frac{\partial
\Omega_{pot}(v, \mu, \Phi, \Phi^{\ast})}{\partial \mu}
\right|_{\Phi = \Phi^{\ast} = \Delta} = n_{f},
                    \label{number}
\end{equation}
which sets the density of the particles in the system,
takes the form
\begin{equation}
W - \sqrt{(W - \mu)^{2} + \Delta^{2}} + \sqrt{\mu^{2} + \Delta^{2}}
= 2 \epsilon_{F},        \label{number.2D.band}
\end{equation}
where we have used the simplest quadratic
dispersion law $\epsilon_{F} = \pi n_{f}/m$ appropriate to 2D metals.

          Equations (\ref{gap.2D.band}) and
(\ref{number.2D.band}), which were obtained in the mean field
approximation (as is well adequate for $T=0$)\footnote{Of course,
there are also residual interactions related
to quantum fluctuations, etc.}, form a complete set
for finding the quantities $\Delta$ and $\mu$ as functions of $W$ and
$\epsilon_{F}$ (or $n_{f}$). It differs from the similar set in
\cite{Randeria.2D} by the explicit dependence on $W$ that, in
principle, can be important for the case of narrow or multi-band
systems which we are going to discuss
in Sec.~\ref{sec:band}.

          It should also be noted that the need to use
the system of equations to find $\Delta$ and $\mu$ self-consistently
has been known for a long time (see \cite{Schrieffer}).
However, in ordinary 3D metals the number density is practically unchanged,
so that, as a rule, the equation for $\mu$ is
trivialised to the equation $\mu = \epsilon_{F}$ and only the
value of $\Delta$ is regarded as unknown. The importance of
the second equation for small particle densities was first pointed
out in the papers \cite{Eagles.1969,Leggett}.

Equations (\ref{gap.2D.band}) and
(\ref{number.2D.band}) allow, along with the trivial solution
($\Delta=0$, $\mu= \epsilon_{F}$) the nontrivial one:
\begin{eqnarray}
&& \Delta^{2} =
\frac{\epsilon_{F}(W - \epsilon_{F})}{\sinh^{2}(2 \pi/ m U)};
\nonumber                         \\
&& \mu = \epsilon_{F} \coth \frac{2 \pi}{m U} - \frac{W}{2}
\left( \coth \frac{2 \pi}{m U} - 1 \right),
\label{solution.2D.band}
\end{eqnarray}
which is valid for any physically reasonable values of the relevant
parameters. It is clear that this solution is meaningful physically
when $U >0$ only. A solution of precisely this form was originally obtained by
\cite{Sofo} (see Eq.~(\ref{solution.2D.discrete})
in Sec.~\ref{sec:discrete.diagram}).
\begin{figure}[h]
\centering
\resizebox{0.9\textwidth}{!}{%
  \includegraphics[bb=110 280 460 495]{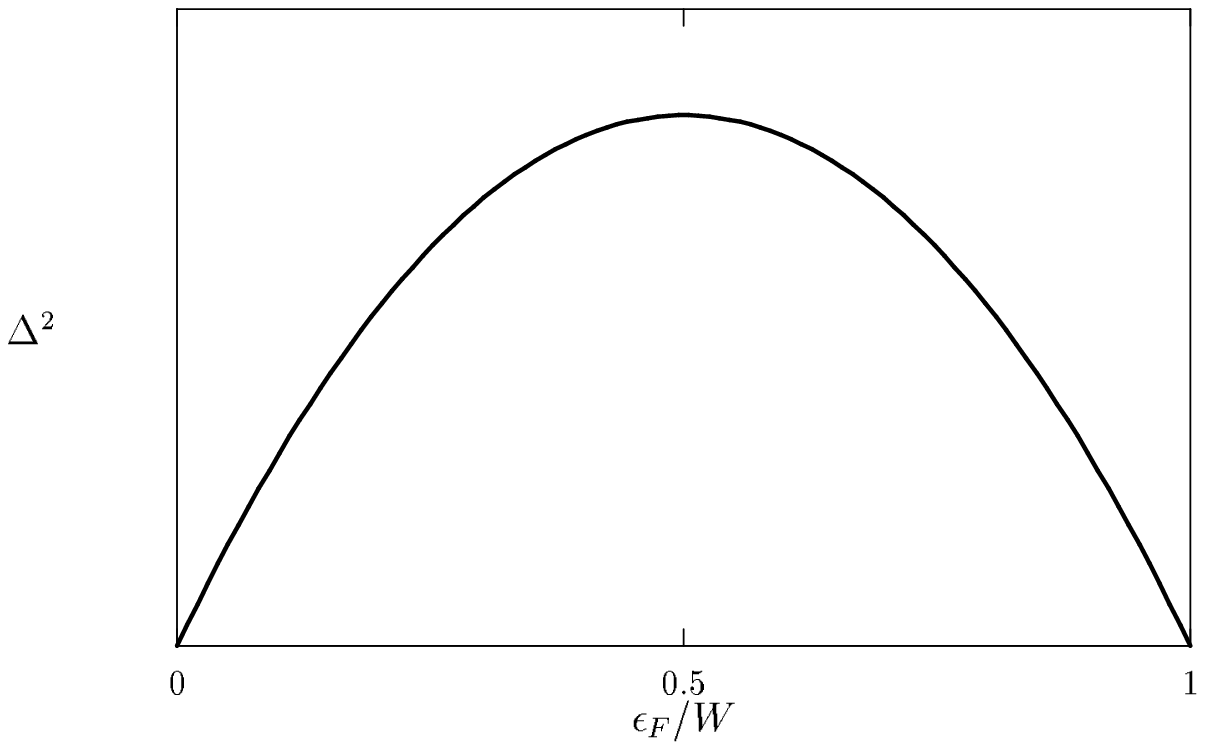}
  \includegraphics[bb=110 280 460 495]{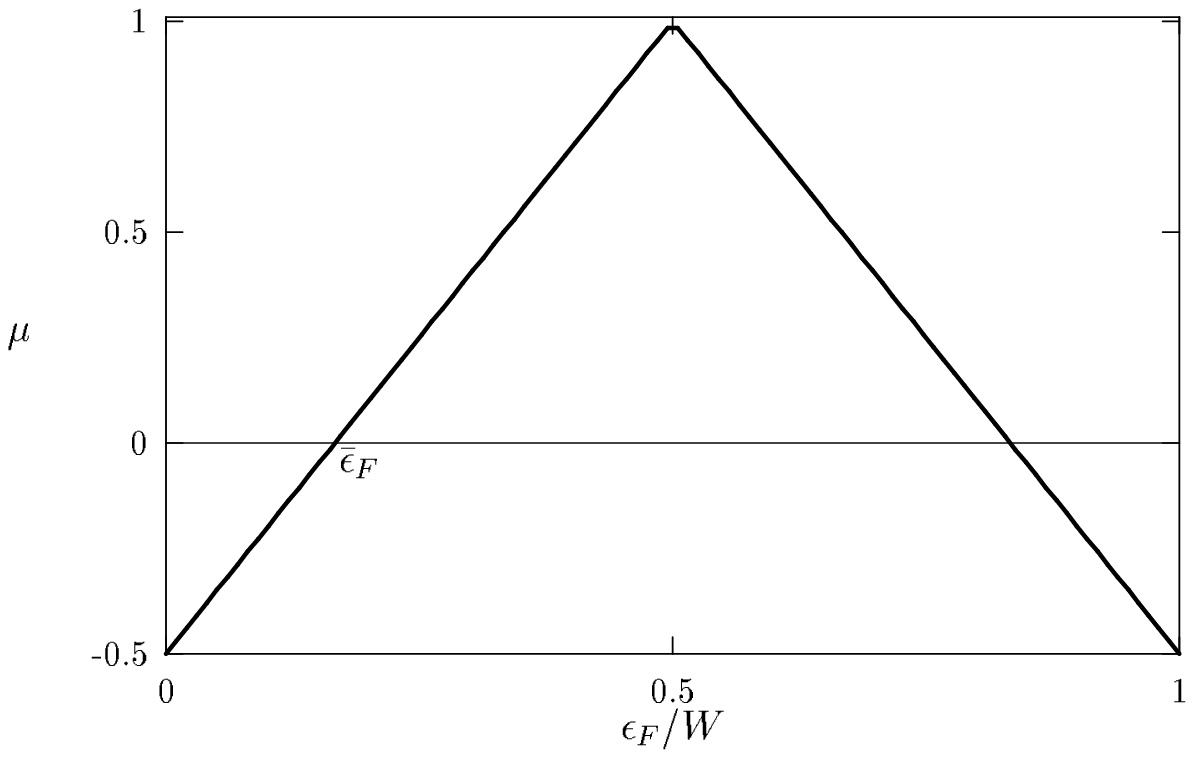}
}
\caption{The functions $\Delta^{2}(\epsilon_{F})$ and $\mu(\epsilon_{F})$
that follow from the solution (\ref{solution.2D.band}) as taken from
\protect{\cite{GGL}}.}
\label{fig:05}
\end{figure}
The functions $\Delta^{2}(\epsilon_{F})$ and $\mu(\epsilon_{F})$
are shown in Fig.~\ref{fig:05}, where the gap value attains
its maximum at $\epsilon_{F} = W/2$, here
$\Delta_{max} = (W/2) \sinh^{-1}(2 \pi/Um)$. At the same
point the value $\mu = W/2 >0$. The symmetry
of the functions $\Delta^{2}(\epsilon_{F})$ and $\mu(\epsilon_{F})$
about the line $\epsilon_F = W/2$ reflects the particle--hole
symmetry about half-filling. One should consider the
hole or anti--particle picture in the region $W/2 < \epsilon_{F} < W$ and
thus change the sign of the mass $m \to -m$.  It is also very important
(see below) that for small $\epsilon_{F}$, there is a region where
$\mu < 0$, and that the sign changeover occurs at a definite
point ${\bar \epsilon_{F}} = W/2 [1 - \tanh(2 \pi/m U)]$.

The expressions that are found in \cite{Miyake,Randeria.2D}
follow directly from (\ref{solution.2D.band}) if, treating $W$ as
large and the attraction $U$ as small, we introduce the 2D two-body
binding energy
\begin{equation}
\varepsilon_{b} = -2W \exp \left(-\frac{4 \pi}{m U} \right),
                \label{bound.energy}
\end{equation}
which does not include any many-particle effects.
The introduction of the expression (\ref{bound.energy}) enables one to
take the limit $W \gg \epsilon_{F}$ and thus justifies to a certain
degree the use of the parabolic dispersion law. In addition the fitting
parameter $\varepsilon_{b}$ is more physically relevant. For example
it is well-defined even for potentials with repulsion. We stress that the
introduction of $\varepsilon_{b}$ instead of $U$ also permits
the regularization of the ultraviolet divergence, which is in fact present
in the gap equation (\ref{gap.2D.band}).

It should be mentioned that, in a dilute gas model,
the existence of
a two-body bound state in vacuum is a {\em necessary\/} (and
sufficient) condition for a Cooper instability
\cite{Randeria.book,Miyake,Randeria.2D}. This statement becomes
nontrivial if one considers two-body potentials $U(r)$ with
short-range repulsion (e.g., hard-core plus long-range attraction),
so that one has to cross a finite (but really very weak \cite{Landau})
threshold in the attraction before a bound state forms in vacuum.

   Making use of (\ref{bound.energy}), it is easy to
simplify (\ref{solution.2D.band})
\cite{Randeria.book,Miyake,Randeria.2D} to
\begin{equation}
\Delta = \sqrt{2 |\varepsilon_{b}| \epsilon_{F}} \,, \qquad
\mu = - \frac{|\varepsilon_{b}|}{2} + \epsilon_{F},
               \label{solution.2D}
\end{equation}
where the chemical potential now changes sign at
${\bar \epsilon_{F}} = |\varepsilon_{b}|/2$.

    To understand the physical significance of these
remarkably simple results we look at the two limits of this
solution. For very weak attraction (or high density) the two-particle
binding energy is extremely small, i.e. $|\varepsilon_{b}| \ll
\epsilon_{F}$, and it is seen that we recover the well-known BCS
results with strongly overlapping (in ${\bf r}$-space) Cooper pairs.
The chemical potential $\mu \simeq \epsilon_{F}$, and the gap function
$\Delta \ll \epsilon_{F}$.

         In the opposite limit of very low particle density
(or a very strong attraction) we have a deep two-body bound state
$|\varepsilon_{b}| \gg \epsilon_{F}$, and find that we are in a
regime in which there is BEC of composite bosons, or "diatomic
molecules". The chemical potential here
$\mu \simeq -|\varepsilon_{b}|/2$, which is one half the energy of
pair dissociation for tightly bound (local) pairs.

      It should also be kept in mind that in the local pair
regime ($\mu < 0$) the gap $E_{gap}$ in the quasi-particle
excitation spectrum equals not $\Delta$ (as in the case $\mu > 0$)
but rather $\sqrt{\mu^{2} + \Delta^{2}}$ (see \cite{Leggett},
the review \cite{Randeria.book} and the discussion of the
Bethe-Salpeter equation in Sec.~\ref{sec:Salpeter}).

   Leaving aside the analysis of $\Omega_{pot}$ for arbitrary
values of the parameters (see \cite{GGL,LSh.review}), let us consider
the most interesting case of $W \to \infty$, $U \to 0$ with finite
$\varepsilon_{b}$. Then finding from (\ref{bound.energy}) the
expression for $4 \pi/m U$ and substituting it into
(\ref{Omega.Potential.2D.expr}) one obtains
\begin{equation}
\Omega_{pot}(v, \mu, \Phi,\Phi^{\ast}) = v
\frac{m}{4\pi} |\Phi|^{2} \left( \ln \frac{\sqrt{\mu^{2} +
|\Phi|^{2}} - \mu}{|\varepsilon_{b}|} - \frac{\mu}{\sqrt{\mu^{2} +
|\Phi|^{2}} - \mu} - \frac{1}{2} \right) .
\label{Omega.Potential.2D.eps}
\end{equation}
Using the potential (\ref{Omega.Potential.2D.expr}) or
(\ref{Omega.Potential.2D.eps}) one can demonstrate that
the superconducting state is indeed energetically favoured.

The equations (\ref{gap.2D.band})
and (\ref{number.2D.band}) can also be written as
\begin{equation}
\sqrt{\mu^{2} + \Delta^{2}} - \mu = |\varepsilon_{b}|; \qquad
\sqrt{\mu^{2} + \Delta^{2}} + \mu = 2\epsilon_{F},
\label{gap+number.2D}
\end{equation}
respectively; their solution, (\ref{solution.2D}), is quoted above.

\subsubsection{Pairs in the Fermi sea}
\label{sec:Salpeter}

To study the possibility of the existence of the bound states in the presence
of a collection of fermions (Fermi sea) the Bethe-Salpeter equation was
analyzed in \cite{GGL}. Here we shall only present the results of
this analysis in the particular limit $W \to \infty$, $U \to 0$ with finite
$\varepsilon_{b}$.\footnote{It is important
to stress the difference between the
bound states in the vacuum which are always present in a 2D system
with attraction and the possibility of existence of such states
in the presence of other particles. To analyse the
conditions for the existence of bound states in the latter case,
the Bethe-Salpeter equation
(see, for example, \cite{Abrikosov.book}) must be studied.}
The analysis in \cite{GGL} showed that if the charge symmetry
is unbroken, i.e. $\Phi = \Phi^{\ast} = 0$, the Bethe-Salpeter
energies are given by
\begin{equation}
E_{\pm} = - \frac{|\epsilon_b|}{2} \left\{ 1 \pm \sqrt{1 - \frac{8
\epsilon_F}{|\epsilon_b|}} \right\}
\label{Saltpeter}
\end{equation}
Thus the system contains non-decaying bound states, i.e. the energies are real,
only if $\epsilon_{F} < \epsilon_{F}^\mathrm{cr} = |\varepsilon_{b}|/8$.
When $\epsilon_{F} > \epsilon_{F}^\mathrm{cr}$ the solutions
acquire an imaginary part which, according to \cite{Schrieffer},
suggests the development of a Cooper instability and vacuum
rearrangement.

For the state with the rearranged vacuum so that
$\Phi = \Phi^{\ast} = \Delta$ the analysis of the Bethe-Salpeter
equation firstly proved the existence of a gapless Goldstone
mode. The second solution of the equation which is only real
for $\mu < 0$ is in agreement with the quasi-particle excitation energy
$E_{gap}$ described above.

\subsubsection{The gradient terms of the effective action
and the correlation length {\em versus\/} doping}
\label{sec:derivative}

        Now we calculate the terms $\Omega_{kin}$ which contain the
derivatives in the expansion (\ref{kinetic+potential}). As before, we
shall assume the inhomogeneities of $\Phi$ and $\Phi^{\ast}$ to be
small, restricting ourselves to terms with only the lowest order
derivatives. For simplicity we shall also consider the stationary
case and calculate only the terms with second-order spatial
derivatives. These terms make it possible to determine the coherence
length $\xi_{\mathrm{coh}}$ and its doping dependence.

        With these restrictions, and taking into account the
invariance of $\Omega(v, \mu, \Phi, \Phi^{\ast})$ (see equation
(\ref{Effective.Action})) with respect to the phase transformations
(\ref{transformation}), one can write the most general form for the
kinetic part of the action as
\begin{eqnarray}
&& \Omega_{kin}(v, \mu, T, \Phi(x),\Phi^{\ast}(x)) =
T \int_{0}^{\beta} d \tau \int d^2 r \,
T_{kin}(\Phi, \Phi^{\ast}, \nabla \Phi, \nabla \Phi^{\ast}) =
\nonumber                       \\
&& \qquad
T \int_{0}^{\beta} d \tau \int d^2 r
\left[ T_{1}(|\Phi|^{2}) |\nabla \Phi|^{2} +
\frac{1}{2} T_{2}(|\Phi|^{2}) (\nabla |\Phi|^{2})^{2} \right].
            \label{Omega.Kinetic}
\end{eqnarray}
Here there are no items with a total derivative since
boundary effects are regarded as unessential, and the
coefficients $T_{1,2}(|\Phi|^{2})$ are assumed to be unknown
quantities. It follows from (\ref{Omega.Kinetic}) that, if one
includes derivatives up to second order, variations in both
the direction (phase) of the field $\Phi$ and its absolute
value are included.

    The calculation for the coefficients $T_{1,2}(|\Phi|^{2})$
is straightforward but rather lengthy. It has been carried out in
\cite{GGL} (see also \cite{LSh.review}) using a general method for
derivative expansion developed in
\cite{Gusynin.1992}. The paper \cite{GGL} also
contain the expressions for the case of a finite band-width $W$.

Knowing $T_{1}(|\Phi|^{2})$ and $T_{2}(|\Phi|^{2})$ one
can find the values for different observables. For practical
purposes one can restrict oneself to considering the
coefficients obtained at the point of minimal effective potential
$|\Phi| = \Delta$. Instead of $T_{1}(|\Phi|^{2})$ and
$T_{2}(|\Phi|^{2})$
it proves to be more convenient to introduce the combination
${\tilde T}_{2}(|\Phi|^{2}) \equiv T_{1}(|\Phi|^{2}) +
2 \Delta^{2} T_{2}(|\Phi|^{2})$ which determines the change in the
$|\Phi|$ value only and arises as the coefficient at $(\nabla|\Phi|)^{2}$.
Furthermore, using (\ref{bound.energy}) and (\ref{solution.2D}), one can
derive $T_{1}(\Delta^{2})$ and ${\tilde T}_{2}(\Delta^{2})$ in the very
simple form:
\begin{equation}
T_{1}(\Delta^{2}) = \frac{1}{16 \pi \hbar^2 |\varepsilon_{b}|};
\qquad
{\tilde T}_{2}(\Delta^{2}) = \frac{1}{24 \pi \hbar^{2}}
\frac{(2 \epsilon_{F} - |\varepsilon_{b}|)^{2}}
{(2 \epsilon_{F} + |\varepsilon_{b}|)^{3}},
           \label{coefficients.energy}
\end{equation}
where for completeness we have restored Planck's constant.

  The explicit forms of $T_{1}(\Delta^{2})$ and
$a \equiv v^{-1} \partial^{2} \Omega_{pot}(\Phi,\Phi^{\ast})/
\partial \Phi \partial \Phi^{\ast}|_{|\Phi|^{2} = \Delta^{2}}$
permit one to calculate the coherence length and the penetration depth. From
Eq.~(\ref{Omega.Potential.2D.eps}) one has
\begin{equation}
a = \frac{m}{2 \pi \hbar^{2}}
\frac{\epsilon_{F}}{2 \epsilon_{F} + |\varepsilon_{b}|}.
\label{a}
\end{equation}
Then, according to general theory of fluctuation phenomena
\cite{Lifshitz}, one obtains
\begin{equation}
\xi_{\mathrm{coh}} = \hbar \left[\frac{T_{1}(\Delta^{2})}{a} \right]^{1/2} =
\hbar \left(\frac{2 \epsilon_{F} + |\varepsilon_{b}|}
{8 m \epsilon_{F} |\varepsilon_{b}|} \right)^{1/2}.
\label{coherence}
\end{equation}
This formula shows the dependence of $\xi_{\mathrm{coh}}$ on $\epsilon_{F}$
(or $n_{f}$). The zero temperature coherence length $\xi_{\mathrm{coh}}$ was
also studied in \cite{Pistolesi.1996} for 2D and 3D cases, where it is
referred as the {\em phase\/} coherence length $\xi_{\mathrm{phase}}$.

It is very interesting and useful to compare
(\ref{coherence}) with the definition of the pair size
\cite{Quick,Carter} (also sometimes incorrectly referred to as the
coherence length (see \cite{Randeria.2D,Pistolesi.1994})), namely
\begin{equation}
\xi_{\mathrm{pair}}^{2} = \frac{\int d^2 r g({\bf r}) {r}^{2}}
{\ds \int d^2 r g({\bf r})} \,,
           \label{size.general}
\end{equation}
where
\begin{equation}
g({\bf r}) = \frac{1}{n_{f}^{2}} \left| \langle
\Psi_{\mathrm{BCS}} | \psi^{\dagger}_{\uparrow} ({\bf r})
\psi^{\dagger}_{\downarrow} (0) | \Psi_{\mathrm{BCS}} \rangle  \right|^2
\end{equation}
is the pair-correlation-function for opposite spins and
$| \Psi_{\mathrm{BCS}} \rangle$ is the usual BCS trial function.
For the 2D model under consideration the general
expression (\ref{size.general}) gives \cite{Randeria.2D,Quick}
\begin{equation}
\xi_{\mathrm{pair}}^{2} = \frac{\hbar^{2}}{4 m} \frac{1}{\Delta}
\left[ \frac{\mu}{\Delta} +
\frac{\mu^{2} + 2 \Delta^{2}}{\mu^{2} + \Delta^{2}}
\left(\frac{\pi}{2} + \tan^{-1} \frac{\mu}{\Delta} \right)^{-1} \right],
  \label{size}
\end{equation}
where $\Delta$ and $\mu$ are given by (\ref{solution.2D}).
So, we are ready now to compare (\ref{coherence}) and (\ref{size}).

   Again to understand the underlying physics it is worthwhile to
look at the two extremes of (\ref{coherence}) and (\ref{size}).
For high carrier densities, $\epsilon_{F} \gg |\varepsilon_{b}|$,
one finds that
$\xi_{\mathrm{coh}} \sim \xi_{\mathrm{pair}} \sim  \hbar v_{F}/\Delta$, i.e.
the well-known Pippard's result is reproduced correctly.
Moreover, if we introduce the pair size
$\xi_{b} = \hbar (m |\varepsilon_{b}|)^{-1/2}$ in vacuum \cite{Miyake},
it is clear that in the high density limit
($|\varepsilon_{b}| \lesssim \epsilon_{F}$ and $\mu \approx \epsilon_{F}$),
$\xi_{\mathrm{coh}} \sim \xi_{\mathrm{pair}} \lesssim \xi_{b}$.
Thus both $\xi_{\mathrm{coh}}$ and $\xi_{\mathrm{pair}}$ prove to be
of the order of the pair size, $\xi_{b}$, which is much larger than the
interparticle spacing.  The latter statement follows from the value of the
dimensionless parameter $\xi_{\mathrm{coh}} k_{\mathrm{F}} \sim
\xi_{\mathrm{pair}}  k_{\mathrm{F}} \sim \epsilon_{F}/\Delta \gg 1$.

        In the opposite limit of very low density
($|\varepsilon_{b}| \gg \epsilon_{F}$ and $\mu<0$)
one can see that $\xi_{b} \ll \xi_{\mathrm{coh}} \sim k_{\mathrm{F}}^{-1}$
while $\xi_{\mathrm{pair}} \sim \xi_{b}$. Consequently the correct
interpretation of $\xi_{\mathrm{pair}}$ is the pair size (in presence
of the Fermi sea (\ref{size.general})) rather than the coherence
length.  The former in the extreme Bose regime is much smaller than
the mean interparticle spacing, i.e. $\xi_{\mathrm{pair}} k_{\mathrm{F}} \ll 1$,
while $\xi_{\mathrm{coh}} k_{\mathrm{F}} \sim 1$.
The meaning of $\xi_{\mathrm{coh}}$ is the coherence length
because it remains finite and comparable with the mean
interparticle spacing even in the limit of infinite binding i.e. when
$|\varepsilon_{b}|$ goes to infinity.
This situation is consistent with the case of $^{4}$He where the
coherence length is nonzero and comparable with the mean
inter-atomic distance although $|\varepsilon_{b}|$
(or energy of nucleon-nucleon binding) is extremely large.

Pistolesi and Strinati in \cite{Pistolesi.1996} obtained the same results,
namely the coincidence of the pair size, $\xi_{\mathrm{pair}}$
and the coherence length $\xi_{\mathrm{coh}}$ in the weak coupling
limit $\xi_{\mathrm{pair}} k_{\mathrm{F}} \gg 1$ and the inequality
$\xi_{\mathrm{coh}} \gg \xi_{\mathrm{pair}}$ in the strong coupling
or Bose limit  $\xi_{\mathrm{pair}} k_{\mathrm{F}} \ll 1$. They established
that $\xi_{\mathrm{coh}}$ coincides with $\xi_{\mathrm{pair}}$ down to
$\xi_{\mathrm{pair}} k_{\mathrm{F}} \simeq 10$.

The above discussion is a clear example of how the two
different energetic scales discussed in \cite{Deutscher} behave
in the low- and high-density limits. Indeed, the single-particle
excitation energy, $\Delta_{\mathrm{pair}} = \Delta$ is obviously
related to the individual pair size, $\xi_{\mathrm{pair}}$, while the energy
coherence range is defined by the coherence length, $\xi_{\mathrm{coh}}$. It
follows from the consideration of $\xi_{\mathrm{pair}}$ and $\xi_{\mathrm{coh}}$
above that the energy scales $\Delta_{\mathrm{pair}}$ and
$\Delta_{\mathrm{coh}}$ are the same in the high-density limit. However they
diverge in the low density limit where $\Delta_{\mathrm{pair}}$ is larger than
$\Delta_{\mathrm{coh}}$.

It is interesting to note that the femtosecond time-domain
spectroscopy \cite{Demsar.1999} (see also the previous papers on
this subject \cite{Kabanov} and the papers discussed above
\cite{Mihailovic,Demsar.1998}) shows the existence of two distinct gaps in the
entire overdoped region of Y$_{1-x}$Ca$_x$Ba$_2$Cu$_3$O$_{7-\delta}$, where
the doping here refers to the chemical fraction of Ca included. It is claimed
in \cite{Demsar.1999} that one of them is a temperature-independent
``pseudogap'' $\Delta_{\mathrm{pair}}$ and the other is a $T$-dependent
collective gap $\Delta_{\mathrm{coh}}(T)$. They suggest that the presence
of two gaps is due to a spatially inhomogeneous picture, supported
experimentally by the observation of an inhomogeneous charge distribution
\cite{Mihailovic.book}. In low carrier density regions $\Delta_{\mathrm{pair}}$ is
simply the energy scale for single pair formation. In high-density regions
one has strong collective effects and a collective temperature dependent gap
$\Delta_{\mathrm{coh}}(T)$ which shows BCS-type closure at $T_c$. If such
a two-phase picture is indeed correct, then one may have a Fermi surface
in that the chemical potential remains positive yet at the same time
in local low carrier density regions one can reach the
Bose limit -- limit of pre-formed pairs.

Another important comment which we have to make here is that the
analysis of the experimental data performed in \cite{Quick}
shows that the optimally doped cuprates are definitely
on the BCS side of the BEC-BCS crossover in the sense that the chemical
potential is close to $\epsilon_F$ if one assumes a spatially homogeneous
system. To make such a conclusion the ratio $\mu/\epsilon_{F}$ was extracted in
\cite{Quick} from the available experimental data for the coherence length.
In particular, it was crudely estimated that for YBaCuO with $T_{c} = 93$K the ratio
$\mu/\epsilon_{F} \simeq 0.9998-0.9995$; for BiSrCaCuO with
$T_{c} = 100$K --- $\mu/\epsilon_{F} \simeq 0.9991-0.9978$ and
for TlBaCaCuO with $T_{c} = 125$K ---
$\mu/\epsilon_{F} \simeq 0.9986-0.9965$. It is clear that for
a spatially homogeneous model that closeness to the Bose
limit implies negative ratio $\mu/\epsilon_{F}$ or at least
$\mu < \epsilon_{F}$ which is apparently not the case for
optimally doped HTSC.

\begin{figure}[h]
\centering \resizebox{0.5\textwidth}{!}{%
  \includegraphics{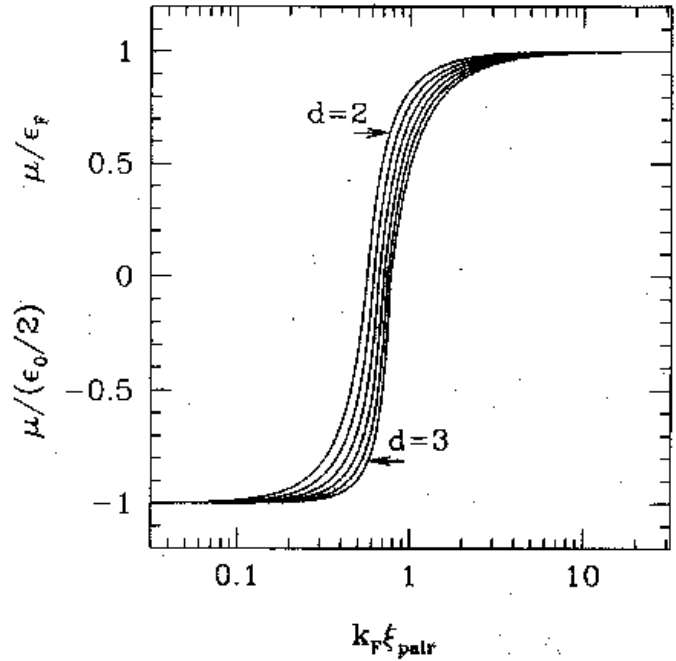}
} \caption{Chemical potential $\mu$ vs $k_{\mathrm{F}} \xi_{\mathrm{pair}}$
(at zero temperature) for ``contact'' potential and the dimensionality of space
$2 \leq d \leq 3$. Different curves are labeled by the values of $d$ (in steps
of 0.2). Positive values of $\mu$ are normalized by the Fermi energy
$\epsilon_{F} = k_{\mathrm{F}}^{2}/2m$, while negative values of $\mu$ are
normalized by half the magnitude $\epsilon_{0}$ of the eigenvalue of the
two-body problem in $d$ dimensions. This plot was taken from
\protect{\cite{Pistolesi.1996}}.} \label{fig:06}
\end{figure}

Note also that, since
$\xi_{b} k_{\mathrm{F}} \sim \xi_{\mathrm{pair}} k_{\mathrm{F}}$ and
$\xi_{b} k_{\mathrm{F}}$ is directly related to the
dimensionless ratio $\epsilon_{F}/|\varepsilon_{b}|$, it can be
inferred that $\xi_{\mathrm{pair}} k_{\mathrm{F}}$ is another physical
parameter which can correctly determine the type of pairing.
There is a very remarkable plot from \cite{Pistolesi.1996}, shown in
Fig.~\ref{fig:06}, which examines the behaviour of the dimensionless chemical
potential versus $k_{\mathrm{F}} \xi_{\mathrm{pair}}$
($|\varepsilon_{b}| \equiv \epsilon_{0}$ in the notations of
\cite{Pistolesi.1996}). This plot
appears to be quite ``universal'', in the sense that it is
remarkably independent of the specific model Hamiltonian and
of the dimensionality (at least on
the mean-field level).\footnote{Notwithstanding these similarities,
one should be careful
with the Bose limit for the discrete Hubbard model. This limit,
as was firstly pointed out in \cite{Nozieres}, is quite different
from that of continuum model (see Sec.~\ref{sec:s.lattice}).}
The Fig.~\ref{fig:06} also shows that the crossover between
BCS and BEC regimes occurs in a rather {\em narrow\/} range
of the parameter $k_{\mathrm{F}} \xi_{\mathrm{pair}}$.

Finally, note that the concentration dependence of
the penetration depth was also studied in \cite{GGL} and
that the 2D crossover model (\ref{Hamiltonian.2D}) in the presence
of a magnetic field was investigated in \cite{Shovkovy}.
In particular the concentration dependence of the derivative
$\partial H_{c2}/\partial T$ was studied and it was shown that
\cite{Shovkovy} this derivative is substantially less in the Bose
than in the BCS limit.

\subsection{BCS-Bose crossover in the multi-band model:
the coexistence of local and Cooper pairs}
\label{sec:band}

Clearly the one band model as considered in Sec.~\ref{sec:one-band}
only describes the part of the phase diagram presented in Fig.~\ref{fig:01}
corresponding to the superconducting state. Since we have considered
zero-temperature we have in fact only described the single left-hand point
with zero $T_{c}$ in Fig.~\ref{fig:01}. At this point $T_c=0$ and it thus
corresponds to the point $\Delta = \epsilon_F=0$ in Fig.~\ref{fig:05}.
We will return to possible theoretical explanations
for the experimental phase diagram Fig.~\ref{fig:01}
when we consider the generalization
of the one-band model to finite temperature.

There is, however, an important question related to the
peculiarities of the crystalline structure of the cuprates.
It is known that the La$_{2-x}$Sr$_x$CuO$_4$ system discovered by
Bednorz and M\"uller \cite{Bednorz} has $T_{c} \approx 40$K
at optimal doping and contains a single CuO$_2$ layer
in a cell. On the other hand the later HTSC compound
YBa$_2$Cu$_3$O$_{6+\delta}$ \cite{Wu} has $T_{c} \approx 90$K
and two cuprate layers per unit cell. The entire homologic family
of cuprates has the composition A-M-Ca-Cu-O, where A $\equiv$ Bi, Tl, Hg;
M $\equiv$ Ba, Sr, and the number of CuO$_2$ layers per unit cell,
$1 <\leq N_{pl} < 6$.
Since these cuprate layers are situated in the same unit cell the
arguments from Sec.~\ref{sec:relevance} about the independence of
the different CuO$_2$ layers are not applicable in this case
(see, for example, a recent paper \cite{Leggett.1999}).
Thus the possibility of coherent tunnelling and pairing between such
layers must be taken into account. Taken together with the specific features
of the BCS-Bose crossover problem discussed in
Sec.~\ref{sec:one-band}, the effect of tunnelling and pairing
between adjacent layers in the same unit cell produces
rather interesting physics which we will
briefly discuss in this section.

Many-band models of superconductivity have been extensively
studied in the context of HTSC \cite{Kochorbe} (see also
the review \cite{Konsin}). Here we consider one such model
\cite{GLSh.FNT.1995} which is directly related to the geometrical
structure of cuprates, and with particular emphasis
placed on the BCS-Bose crossover.

\subsubsection{Model Hamiltonian}

Of course, a model which completely reflects the peculiarities of the
crystalline and electronic structure of such HTSC materials would be
too complicated for consistent calculations. For this reason the
following simplifying assumptions were made in \cite{GLSh.FNT.1995}:
the HTSC material under investigation consists of identical
metal blocks separated by spacers; superconductivity emerges in each
$N_{pl}$-layered metal independently so that all the actual
interactions are confined to a single block; inside this block, the
coupling differs from zero only for the nearest adjacent layers.

Ultimately, we arrive at the Hamiltonian of a multilayered
conductor possessing all the above mentioned characteristics (see
also Fig.~\ref{fig:07}):
\begin{figure}[h]
\centering
\resizebox{0.50\textwidth}{!}{%
  \includegraphics{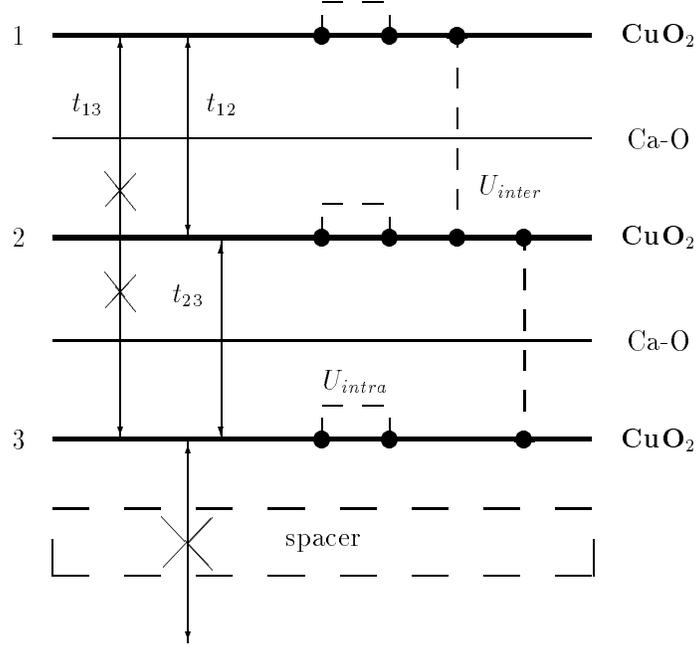}
} \caption{The schematic representation of the interactions included in the
Hamiltonian (\ref{H.plates}).} \label{fig:07}
\end{figure}
\begin{equation}
H = H_{0} + H_{int},                              \label{H.plates}
\end{equation}
where
\begin{eqnarray}
H_{0} & = & \sum_{j=1}^{N_{pl}} \sum_{\sigma} \left\{
\int d^2  r \psi_{j \sigma}^{\dagger}({\bf r},t) \left(
- \frac{\nabla^{2}}{2 m} - \mu
\right) \psi_{j \sigma}({\bf r},t) + \right.
\nonumber                                    \\
& - & \left. \int \! \! \! \int d^2  r_{1} d^2 r_{2}
[\psi_{j \sigma}^{\dagger}({\bf r}_{1},t)
t_{j j+1}({\bf r}_{1} - {\bf r}_{2})
\psi_{j+1 \sigma}({\bf r}_{2},t) + h.c.]
\right\}                                           \label{H0.plates}
\end{eqnarray}
is the Hamiltonian of free charge carriers and
\begin{eqnarray}
H_{int} = - \frac{1}{2} \sum_{j_{1} j_{2}} \sum_{\sigma}
\int \! \! \! \int d^2 r_{1} d^2  r_{2}
\psi_{j_{1} \sigma}^{\dagger}({\bf r}_{1},t)
\psi_{j_{2} \bar \sigma}^{\dagger}({\bf r}_{2},t) \times
\nonumber                                          \\
\times U_{j_{1} j_{2}}({\bf r}_{1} - {\bf r}_{2})
\psi_{j_{2} \bar \sigma}({\bf r}_{2},t)
\psi_{j_{1} \sigma}({\bf r}_{1},t)
                                                   \label{Hint.plates}
\end{eqnarray}
is their interaction.

In Eqs.~(\ref{H0.plates}), (\ref{Hint.plates}), the following notation
is used:
$\psi^{\dagger}_{j \sigma}({\bf r}, t)$,
$\psi_{j \sigma}({\bf r}, t)$ are the Fermi operators of a
particle with an effective mass $m$, spin $\sigma (\equiv - \bar \sigma)$
and ${\bf r} = (x, y)$ from the $j$-th $(1 \leq j \leq N_{pl})$
layer satisfying the boundary condition
$\psi_{0 \sigma}({\bf r}, t) = \psi_{N_{pl}+1, \sigma}({\bf r}, t)$;
$\mu$ is the chemical potential fixing the carrier density
$N_{f} (\equiv N_{pl} n_{f})$;
$t_{j j+1}({\bf r}_{1} - {\bf r}_{2})$ is a $j$-independent
one-particle interlayer tunnelling;
$U_{j_{1} j_{2}}({\bf r}_{1} - {\bf r}_{2})$
are the parameters of the intraplanar $(j_{1} = j_{2})$ and
interplanar $(j_{1} \neq j_{2})$ where the interaction depends only
on $|j_2-j_1|$. Here the sign of the interactions has been
chosen such that $U_{j_{1} j_{2}}({\bf r}_{1} - {\bf r}_{2}) > 0$
corresponds to attraction between carriers with opposite spins
and we used real time $t$.

The component of the Hamiltonian (\ref{H.plates}) corresponding
to the free charge carriers permits diagonalization if we introduce
new Fermi fields in accordance with the representation
\begin{eqnarray}
\psi_{j \sigma}({\bf r}, t) = \sum_{\nu = 1}^{N_{pl}}
u_{j \nu} \varphi_{\nu \sigma}({\bf r}, t) &;& \qquad
u_{j \nu} = \left( \frac{2}{N_{pl} + 1} \right)^{1/2}
\sin{\frac{\pi j \nu}{N_{pl} + 1}};
\nonumber                                      \\
\sum_{j = 1}^{N_{pl}} u_{j \nu_{1}} u_{j \nu_{2}} =
\delta_{\nu_{1} \nu_{2}} & ; & \qquad
\sum_{\nu = 1}^{N_{pl}} u_{j_{1} \nu} u_{j_{2} \nu} =
\delta_{j_{1} j_{2}}.
                                                   \label{replace}
\end{eqnarray}
As a result, the free Hamiltonian is transformed to the following
diagonal form:
\begin{equation}
H_{0}  =  \sum_{\nu = 1}^{N_{pl}} \sum_{\sigma}
\int d^2 r
\varphi_{\nu \sigma}^{\dagger}({\bf r},t)
\left(- \frac{\nabla^{2}}{2 m_{\nu}} - \mu_{\nu} \right)
\varphi_{\nu \sigma}({\bf r},t),                \label{H0.bands}
\end{equation}
which describes an $N_{pl}$-band metal in the effective mass
approximation. In this case, a carrier belonging to the $\nu$-th
band is characterized by the effective mass
$
m_{\nu} = \left( m^{-1} + \Delta m^{-1}
\cos{\pi \nu/(N_{pl} + 1)} \right)^{-1}
$
and by the chemical potential
\begin{equation}
\mu_{\nu} = \mu + t_{\mathrm{eff}} \cos{\frac{\pi \nu}{N_{pl} + 1}}.
\label{mu.band}
\end{equation}
The renormalization of mass and the position of each band on the
energy scale are determined by the constant $t_{j j+1}({\bf r})$
of interlayer hopping since $t_{\mathrm{eff}} \equiv 2 t_{j j+1}(0)$,
and $\Delta m^{-1} \equiv
a^{2}\nabla^{2} t_{j j+1}({\bf r}) \mid _{{\bf r} = 0}$
(a is a parameter having the dimensions of length and associated with the
bandwidths in this approximation through the relation
$W_{\nu} = 2 \pi^{2} / m_{\nu} a^{2}$).
The chemical potential (\ref{mu.band})
allows us to judge (see Sec.~\ref{sec:solution}) what type of
pairing (i.e. local or Cooper) occurs in the $\nu$-th band simply
by looking at the sign of the corresponding $\mu_{\nu}$.

In the new variables (\ref{replace}) and for local interactions
$U_{j_1j_2}({\bf r}_1-{\bf r}_2) = U_{j_1j_2}(0) \delta({\bf r}_1-{\bf r}_2)$,
the interaction (\ref{Hint.plates}) assumes the form
\begin{equation}
H_{int} = -\frac{1}{2} \sum_{\nu, \lambda = 1}^{N_{pl}} \sum_{\sigma}
\int d^2  r
\varphi_{\lambda \sigma}^{\dagger}({\bf r},t)
\varphi_{\lambda \bar \sigma}^{\dagger}({\bf r},t)
U_{\lambda \nu}
\varphi_{\nu \bar \sigma}({\bf r},t)
\varphi_{\nu \sigma}({\bf r},t),
                                             \label{Hint.bands}
\end{equation}
where the matrix element $U_{\lambda \nu}$
can be expressed directly in terms of the initial parameters
$U_{j_{1} j_{2}}(0)$.
For the sake of simplicity we will retain
only the following two values (see above):
$U_{intra} \equiv U_{j j}(0)$ and  $U_{inter} \equiv U_{j j+1}(0)$.
As a result, expressions (\ref{H0.bands}) and (\ref{Hint.bands})
can be regarded as a generalization of the well-known two-band
model of superconductivity \cite{Moskalenko.1959,Suhl} to the case of an
arbitrary number of bands. The symmetry
of the chosen model (the equivalence of all planes) and the nature
of intra- and interlayer interactions are evident. As an example
Eq.~(\ref{Hint.bands}) contains no terms corresponding
to interband pairing which generally appear in phenomenological
models \cite{Kochorbe,Konsin}. Naturally, this does not mean
that each band in the model (\ref{H0.bands}) and (\ref{Hint.bands})
behaves independently because the rearrangement of the vacuum
(the emergence of anomalous mean values or the condensate)
for charge carriers for one of the bands immediately leads
to the same rearrangement for charge carriers from the other bands.

As noted previously many-band models of superconductivity have
been studied in the context of HTSC \cite{Kochorbe} (see also
the review \cite{Konsin}). The new idea that is brought to the many-band
model by using the BCS-Bose crossover formalism is in
establishing the correspondence between the sign of the chemical
potential $\mu_{\nu}$ of the $\nu$-th band and the nature of
pairs.

The further analysis of the many-band Hamiltonian
(\ref{H0.bands}), (\ref{Hint.bands}) is very similar to that
discussed in Sec.~\ref{sec:one-band} and we need not
repeat it. Omitting the general equations derived for the many-band
case \cite{GLSh.FNT.1995} we show only the equations for the
two-band model and their numerical solution.

\subsubsection{Bilayered cuprates}

\begin{figure}[h]
\centering \resizebox{1.0\textwidth}{!}{%
  \includegraphics[bb=110 280 460 495]{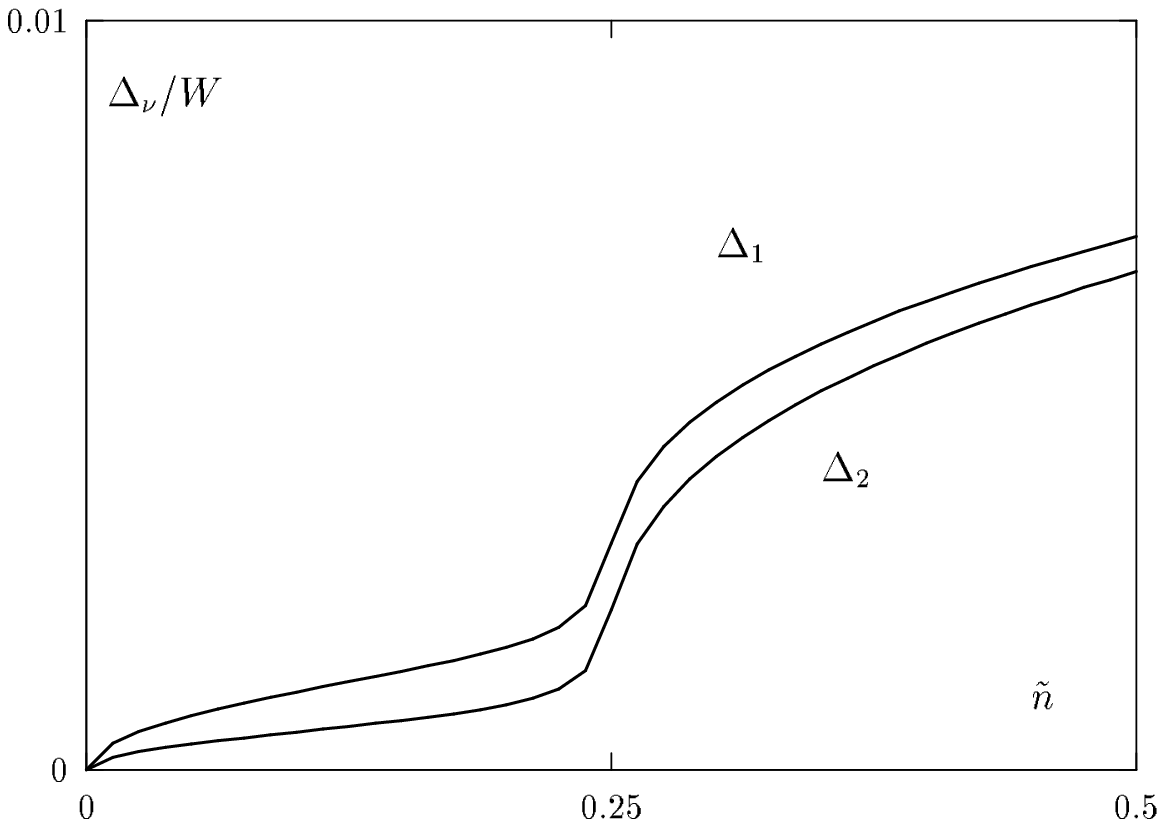}
  \includegraphics[bb=110 280 460 495]{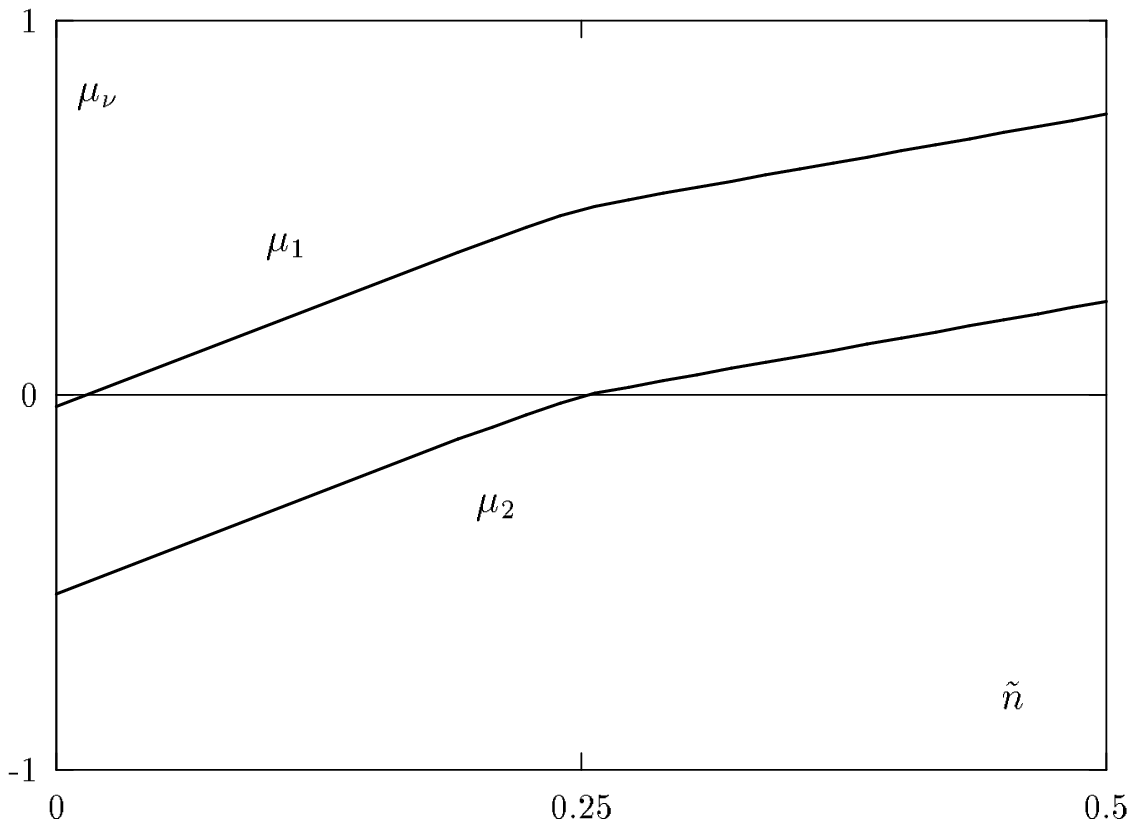}
}
\caption{Dependencies $\Delta_{\nu}(n_{f})$ and $\mu_{\nu}(n_{f})$,
taken from \protect{\cite{GLSh.FNT.1995}}, for $m_{\nu} = m$,
$\tilde n = 2 \pi n_{f}/m$; $W = 8 \epsilon_{0}$
at $4 \pi/(mU_{intra}) =9$; $4 \pi/(mU_{inter}) =21$.}
\label{fig:08}
\end{figure}

It was mentioned above that bilayered materials include Y-Ba-Cu-O
compound. The chemical potentials for the two bands are given by
$\mu_{\nu} = \mu - (-1)^{\nu} t_{\mathrm{eff}}/2$.
It proves convenient to introduce the dimensionless constants
\begin{eqnarray}
g_{\nu}^{-1} = \frac{2 \pi}{m_{\nu}}
\frac{1}{2} (U^{-1}_{intra} + U^{-1}_{inter});
\nonumber                                     \\
g_{1 2}^{-1} = \frac{2 \pi }{\sqrt{m_{1} m_{2}}}
\frac{1}{2} (U^{-1}_{intra} - U^{-1}_{inter}),      \label{V.2band}
\end{eqnarray}
giving the couplings of the superconducting order parameters from the same
and different bands respectively. One may then obtain
(see \cite{GLSh.FNT.1995}) the following system of equations
for the superconducting gaps $\Delta_{1}$, $\Delta_{2}$
and for the chemical potential $\mu$:
\begin{eqnarray}
\Delta_{1} \left[ g_{1}^{-1} -
\ln{ \frac
{\sqrt{(W_{1} - \mu_{1})^{2} + \Delta_{1}^{2}} + W_{1} - \mu_{1}}
{\sqrt{\mu_{1}^{2} + \Delta_{1}^{2}} - \mu_{1}} } \right] +
\Delta_{2} \sqrt{\frac{m_{2}}{m_{1}}} g_{1 2}^{-1} = 0;
\nonumber                                \\
\Delta_{1} \sqrt{\frac{m_{1}}{m_{2}}} g_{1 2}^{-1} +
\Delta_{2} \left[ g_{2}^{-1} -
\ln{ \frac
{\sqrt{(W_{2} - \mu_{2})^{2} + \Delta_{2}^{2}} + W_{2} - \mu_{2}}
{\sqrt{\mu_{2}^{2} + \Delta_{2}^{2}} - \mu_{2}} } \right] = 0;
\nonumber                                \\
\sum_{\lambda = 1,2} m_{\lambda} \left[ W_{\lambda} -
\sqrt{(W_{\lambda} - \mu_{\lambda})^{2} + \Delta_{\lambda}^{2}} +
\sqrt{\mu_{\lambda}^{2} + \Delta_{\lambda}^{2}} \right]
 =  4 \pi n_{f}.                       \label{T=0.System.2bands}
\end{eqnarray}
If one assumes that intra- and interplanar interactions are the
same, i.e. $U_{intra} = U_{inter}$ the system
(\ref{T=0.System.2bands}) can even be investigated analytically
\cite{GLSh.FNT.1995} leading to a better understanding of the underlying
physics. Here however we present only its numerical solution
shown in Fig.~\ref{fig:08}.
To avoid confusion with the two gaps $\Delta_{\mathrm{pair}}$ and
$\Delta_{\mathrm{coh}}$ discussed above, we would like to stress
that both the gaps $\Delta_{1}$ and $\Delta_{2}$ in Fig.~\ref{fig:08}
are single-particle excitation gaps and have the same
origin as $\Delta_{\mathrm{pair}}$. The observation of different gaps
in a similar context has been discussed in the literature
\cite{ManyGaps1,ManyGaps2}. It would be interesting to analyze
the modern experimental data, for example,
\cite{Mihailovic,Demsar.1998,Demsar.1999,Kabanov} bearing in mind
that Y-Ba-Cu-O cuprate may have in fact two single-particle
excitation gaps.

The behaviour of the chemical potentials $\mu_{1,2}$ shown in
Fig.~\ref{fig:08} demonstrates a very interesting feature of
the many-band model. While $\mu_{1} >0$ for almost the whole range
of carrier densities which correspond to the Cooper pair regime in
this band, one finds that $\mu_{2} < 0$ for a reasonably
large range of carrier densities. This means that the pairs in
this band are local and one has a coexistence of local and Cooper
pairs in the same system. Of course, for higher carrier densities
both $\mu_{1}$ and $\mu_{2}$ become positive and the BCS regime
described by the two-band model \cite{Moskalenko.1959,Suhl}
is restored.

In conclusion it is notable that the idea of mixed BCS vs BEC
behaviour in a two gap model of superconductivity was recently
readdressed in \cite{Perali}. In this paper fermions from different
parts of the Fermi surface with small and large Fermi velocities
respectively were treated as belonging to two different bands.

\subsection{Peculiarities of the $s$-wave crossover on the lattice}
\label{sec:s.lattice}

The physics of the BEC -- BCS crossover can be well
understood on the basis of continuum models. Nonetheless real
superconductors are crystals and, if the pair size is not much larger than the
lattice spacing (as in the case of HTSC), lattice effects are important and
should be considered.

Let us review the crossover at $T =0$ for the lattice version of the
continuum model (\ref{Hamiltonian.2D}), given by
\begin{equation}
H =-t\sum_{\langle {\bf n } {\bf m} \rangle, \sigma}
( c^\dagger_{{\bf n} \sigma} c_{{\bf m} \sigma} + {\rm {h.c.}} )
- U \sum_{\bf n} n_{{\bf n}\uparrow} n_{{\bf n} \downarrow} -
\mu_0 \sum_{{\bf n}, \sigma} n_{{\bf n} \sigma} \,,
\label{Hubbard}
\end{equation}
where $t$ is the transfer integral between neighbouring lattice sites ${\bf n}
= (n_x, n_y)$, ${\bf m} = (m_x, m_y)$; $c^\dagger_{{\bf n} \sigma}, c_{{\bf n}
\sigma}$ are electronic creation, annihilation operators for the site ${\bf
n}$ and spin $\sigma$, respectively, $U>0$ is the strength of the on-site
attractive interaction between two electrons occupying the same lattice site,
$n_{{\bf n} \sigma} = c^\dagger_{{\bf n}\sigma} c_{{\bf n}\sigma}$, and
$\mu_0$ is the chemical potential. This Hamiltonian is also called the
attractive or {\em negative-$U$ Hubbard model\/}. The word ``negative''
indicates that there is a negative sign before $U$ in contrast to the positive
sign in the original Hubbard model.

Amongst others, this lattice  model defined by (\ref{Hubbard})
or (\ref{Hubbard.momentum}) has been studied in
\cite{Kagan,Micnas,Gooding,Ichinomiya,Yanase,Sofo} for $T \neq 0$ and
in the many other papers mentioned in Sec.~\ref{sec:survey}.

The number of charge carriers per lattice site reads
\begin{equation}
n \equiv n(\mu, T) = \frac{1}{N} \sum_{{\bf n}, \sigma}
\langle n_{{\bf n} \sigma} \rangle \,
\label{number.definition}
\end{equation}
where $n \in [0,2]$.
In momentum representation the Hamiltonian (\ref{Hubbard})
can be written as follows
\begin{equation}
H = \sum_{{\bf k} \sigma} (\varepsilon_{{\bf k}} - \mu_0)
c^{\dagger}_{{\bf k} \sigma} c_{{\bf k} \sigma}
- \frac{U}{N} \sum_{{\bf k}, {\bf p}, {\bf q}}
c^{\dagger}_{{\bf k} \uparrow} c_{{\bf k} + {\bf q} \uparrow}
 c^{\dagger}_{{\bf p} \downarrow} c_{{\bf p} - {\bf q} \downarrow} \,,
\label{Hubbard.momentum}
\end{equation}
Here we have used the notation ${\bf k}$ as an index,
$\varepsilon_{{\bf k}} = -2t (\cos k_x  + \cos k_y )$
is the simplest band energy in the nearest neighbour approximation,
we have set the lattice constant $a = 1$,
$N$ is the number of lattice sites, and
$c^{\dagger}_{{\bf k} \sigma} (c_{{\bf k} \sigma})$
is the creation (annihilation) operator for momentum ${\bf k}$
and spin $\sigma$.

We shall predominantly follow the work in \cite{Andrenacci} and write down
the standard coupled equations
for the gap $\Delta$ and the effective chemical
potential $\mu$\footnote{Eqs.~(\ref{gap.2D.band}) and
(\ref{number.2D.band}) give these equations for the
case of a quadratic dispersion law $\varepsilon({\bf k})$
(see (\ref{Green.fermion.momentum}))}
\begin{equation}
1 = \frac{U}{N} \sum_{{\bf k}} \frac{\Delta}{2 E_{{\bf k}}}
\label{gap.lattice}
\end{equation}
\begin{equation}
n =  \frac{1}{N}
\sum_{\bf k} \left( 1 - \frac{\xi_{{\bf k}}}{E_{{\bf k}}} \right)
     \label{number.lattice}
\end{equation}
where here (compare with (\ref{Green.fermion.momentum}))
$E_{\bf k} = \sqrt{\xi_{\bf k}^2 + \Delta^2}$ and
$\xi_{\bf k} = \varepsilon_{\bf k} - \mu$, where the effective chemical
potential is $\mu = \mu_0 + Un/2$ i.e.
it contains the Hartree-Fock shift $-nU/2$. This shift is essential to ensure
that the true chemical potential $\mu_0$ coincides with $-|\varepsilon_b|/2$
in the Bose limit.

\begin{figure}[h]
\centering
\resizebox{0.50\textwidth}{!}{%
  \includegraphics{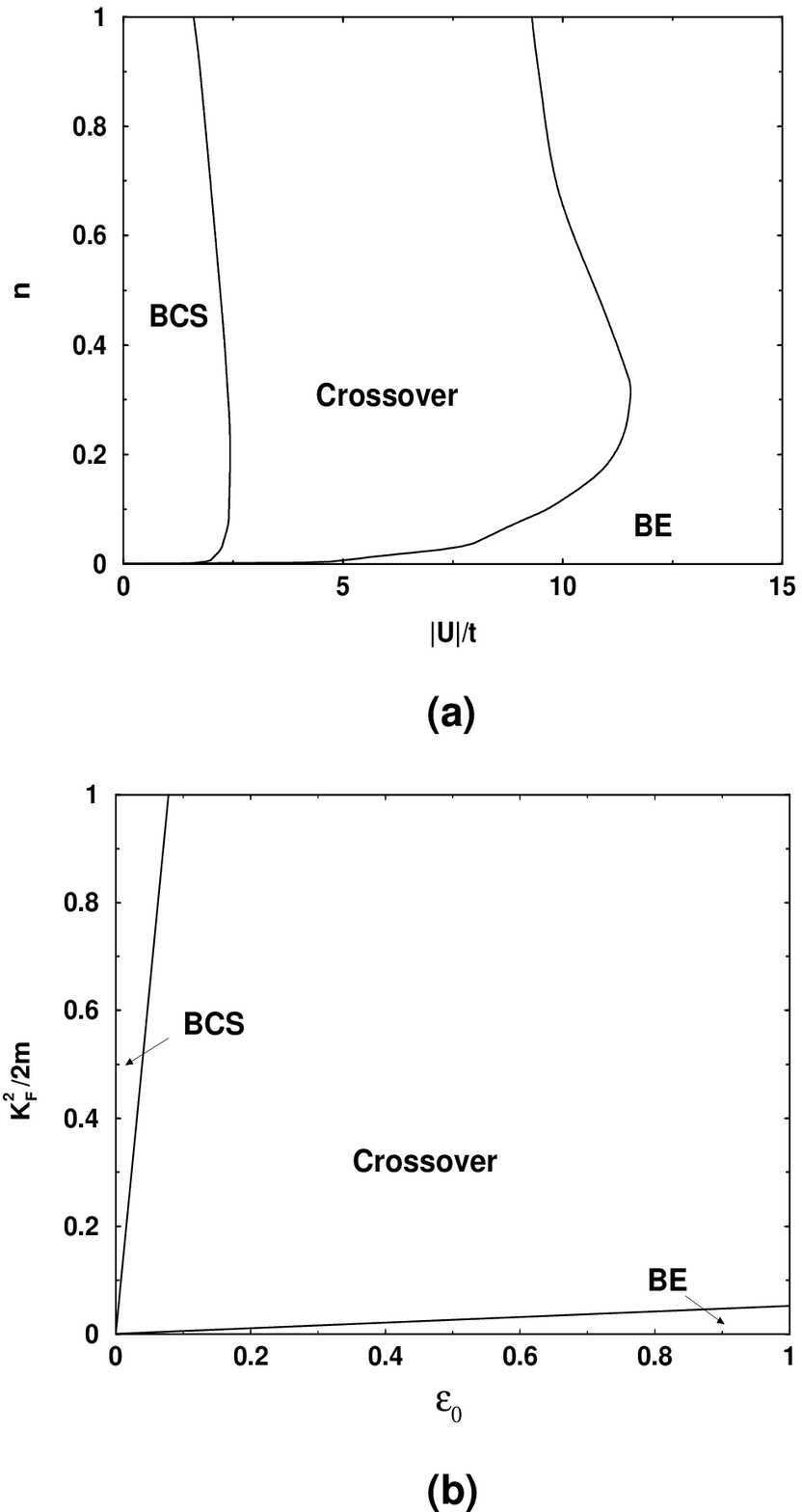}
}
\caption{(a) -- Phase diagram (U/t,n) for the $s$-wave solution of the
negative-$U$ Hubbard model; (b) -- Phase diagram
$(\epsilon_0 \equiv |\varepsilon_b|, k_F^2/2m)$ for the contact potential.
These diagrams were taken from \protect{\cite{Andrenacci}}.}
\label{fig:09}
\end{figure}

The crossover point itself is simply the point where $\mu = 0$.
However, as motivated at the end of Sec.~\ref{sec:one-band},
it proves physically meaningful to consider the Bose and BCS regions as
defined in terms of the pair-size $\xi_{\mathrm{pair}}$
(see Eq.~(\ref{size.general})) for opposite spin-fermions:
the condition $k_{\mathrm{F}} \xi_{\mathrm{pair}} \le 1/\pi$
identifies the Bose region and $k_{\mathrm{F}} \xi_{\mathrm{pair}} \ge 2 \pi$
the BCS region. From the solution of the equations one can then construct
the ``phase diagram'' \cite{Andrenacci} (the bound state energy
there is denoted as $\epsilon_{0}$)
shown in Fig.~\ref{fig:09}~(a) which should be compared
to that for the continuum contact potential shown in Fig.~\ref{fig:09}~(b)
considered in Sec.~\ref{sec:one-band}.

We note that the ``phase diagram'' for the lattice case is symmetric
about half filling $n=1$. From the diagram it is clear that to obtain
a crossover from the BE to BCS regions as one changes the density (at fixed
interaction strength $U/t$) one needs $U/t \lesssim 2.4$. In this case the BE
and crossover regions occur only at extremely low densities. This is in
contrast to the continuum case where one can obtain a density-induced
crossover for all coupling strengths. The reason for this difference
is the reentrant shape of the curves in the lattice case resulting
from the van Hove singularity in the density of states as one approaches
half-filling. Note that in contrast for the continuum case the density of
states is constant.

One should also note that, in contrast to the continuum case, one
cannot always reach the ``dilute'' boson limit on the lattice at high
densities even in the limit of very strong coupling. This effect is
not the result of the finite size of the composite bosons which may
be regarded as point-like in the infinite coupling limit. Instead
it is due to the ``overlap'' of the centers of mass of the composite
bosons since near half-filling the distance between any two bosons
is a minimum of a single lattice spacing. Thus one expects the fermionic
degrees of freedom to again  predominate near half-filling on the lattice
\cite{Nozieres}.

This argument may be quantified \cite{Andrenacci}
by considering the commutator of the following boson-like operator
\begin{equation}
b^\dagger = \sum_{\bf k} g_{\bf k} c^\dagger_{{\bf k} \uparrow}
c^\dagger_{-{\bf k} \downarrow} \,,
\end{equation}
where $g_{\bf k}$ represents the pair wave function.
The commutator $[b,b^\dagger]$ may be regarded as a $c$-number provided
that the occupation number of each relevant state
$\langle n_{{\bf k }\sigma} \rangle < 1$.
This can be achieved at high density (i.e. a large total number of particles)
only if there is an infinite number of ${\bf k}$ states available. This
condition is clearly not satisfied on the
lattice where only a finite number of ${\bf k}$ states exist. Thus, in the
lattice case, one must satisfy both the condition
$k_{\mathrm{F}} \xi_{\mathrm{pair}} < 1/\pi$ and the condition
$n << 1$  for the system to be regarded as a composite Bose gas.

\subsection{Crossover in the models with $d$-wave pairing}

In this Section we consider the problem of $d-$wave pairing in 2D at
zero temperature both in lattice
\cite{Andrenacci,LGN.1994,Ferrer,denHertog,Annett}
and in continuum models \cite{Zwerger,Borkowski}. The motivation
for considering this type of pairing is the experimental observation in the
ARPES and other measurements
\cite{Timusk,Campuzano,Wollman,Hardy,Scalapino.review}
of a $d_{x^2-y^2}$ symmetry in HTSC.
We shall analyse the crossover behaviour both as a function of density
at fixed interaction strength and as a function of interaction strength
at fixed density. The interesting question as to which type of pairing
symmetry is actually present as one changes the interaction parameters
and the density has been considered in \cite{LGN.1994,Ferrer,Annett}
but will not be addressed here.

Firstly one should note that pairs of $d$-wave symmetry cannot contract to
point like bosons. This automatically implies that the interaction between
composite bosons which results from the Pauli exclusion principle is of finite
range. This in turn leads to a larger range of correlations between these
bosons which has a dramatic impact on the crossover. In particular at
moderately high densities and large couplings these correlations
suppress the Bose degrees of freedom and give rise to a larger (fermionic)
BCS region \cite{denHertog}. Secondly for the simplest model which permits
$d$-wave pairing there exists a critical interaction strength for the
formation of a two--body bound state on the empty lattice while this threshold
is zero for $s$-wave pairing. This eliminates the possibility for a density
induced BCS-BE crossover in this model. Thirdly the $d$-wave symmetry has
important implications for both the excitation spectrum and the momentum
distribution since at least for the chemical potential greater than the
critical (or crossover) value the excitation spectrum is gapless in certain
directions.

The lattice models that have been considered are usually 2D Hubbard
models. To obtain a $d$-wave solution the fermionic potential must
contain an inter-site term of strength $V$ in addition to the on-site
term of strength $U$ considered in the previous section. The Hamiltonian
is then given by
\begin{equation}
H =-t\sum_{\langle {\bf n} {\bf m}\rangle, \sigma}
( c^\dagger_{{\bf n} \sigma} c_{{\bf m} \sigma} + {\rm {h.c.}} )  -
U \sum_{\bf n} n_{{\bf n}\uparrow} n_{{\bf n} \downarrow}
- \mu_0 \sum_{{\bf n}, \sigma} n_{{\bf n} \sigma}  -
V \sum_{\langle {\bf n} {\bf m}\rangle}
c^\dagger_{{\bf n} \uparrow} c^\dagger_{{\bf m} \downarrow} c_{{\bf m} \downarrow}
c_{{\bf n} \uparrow}\,,
                      \label{ext.Hubbard}
\end{equation}
where $V >0$ corresponds to attraction and the notation $\langle {\bf n} {\bf
m} \rangle$ denotes as above nearest neighbour pairs. The case of $V=0$ and $U
>0$ simply gives the Hubbard Hamiltonian (\ref{Hubbard}) which only has an
$s$-wave solution. When $V >0$ (inter-site attraction) and $U<0$ (on-site
repulsion) one can also find a $d-$wave solution of the type
\begin{equation}
\Delta({\bf k}) = \Delta_{d} (\cos k_x - \cos k_y )\,.
\end{equation}

We first review the results for this model, the so-called $t-V$ model,
with only nearest neighbour hopping and for which the dispersion
relation takes the form
\begin{equation}
\xi_{\bf k} = -2t (\cos k_x + \cos k_y ) -  \mu
\end{equation}
where $\mu$ is the effective chemical potential
\begin{equation}
\mu = \mu_0 + n(U+4V)/2
\label{chemsh}
\end{equation}
which, as in the case of $s$-wave pairing, incorporates the Hartree shift.

In this case there is a critical coupling strength
$V_{\mathrm{cr}}/4t = 1.83$ below which no two-body ground state on
the empty lattice exists and which gives the start of the
BE region at zero density. This is in contrast to the $s$-wave case
which has no threshold for pair formation. For a many-body $s$-wave
Cooper instability to occur in 2D one needs the existence of a $s$-wave
bound state for the two body problem on the empty lattice
\cite{Randeria.book}.\footnote{We note that the two-body problem on
the empty square lattice was solved many years ago without reference to
superconductivity \cite{Gaididei} and more fully investigated for the $t-J$
model by Kagan and Rice \cite{Kagan.1994}.}
However no such condition applies to higher angular momenta and one can
therefore obtain a BCS ground state for couplings less than
$V_{\mathrm{cr}}$. There is however some controversy as to the minimal
coupling strength actually required for such a paired ground state.
In \cite{Andrenacci} it is claimed that pairing always occurs, or
equivalently one always has a non-zero value of $\Delta_{d}$. As
stated in this work \cite{Andrenacci} this is in direct contradiction
to the earlier work of \cite{denHertog} which finds an unpaired region
for very weak coupling.
In our opinion, the result of \cite{Andrenacci} is more plausible
since the presence of the Fermi surface reduces the dimensionality of
the problem \cite{Polchinski}, and
therefore favours the formation of a paired
many-body ground state.
\begin{figure}[h]
\centering
\resizebox{0.50\textwidth}{!}{%
  \includegraphics{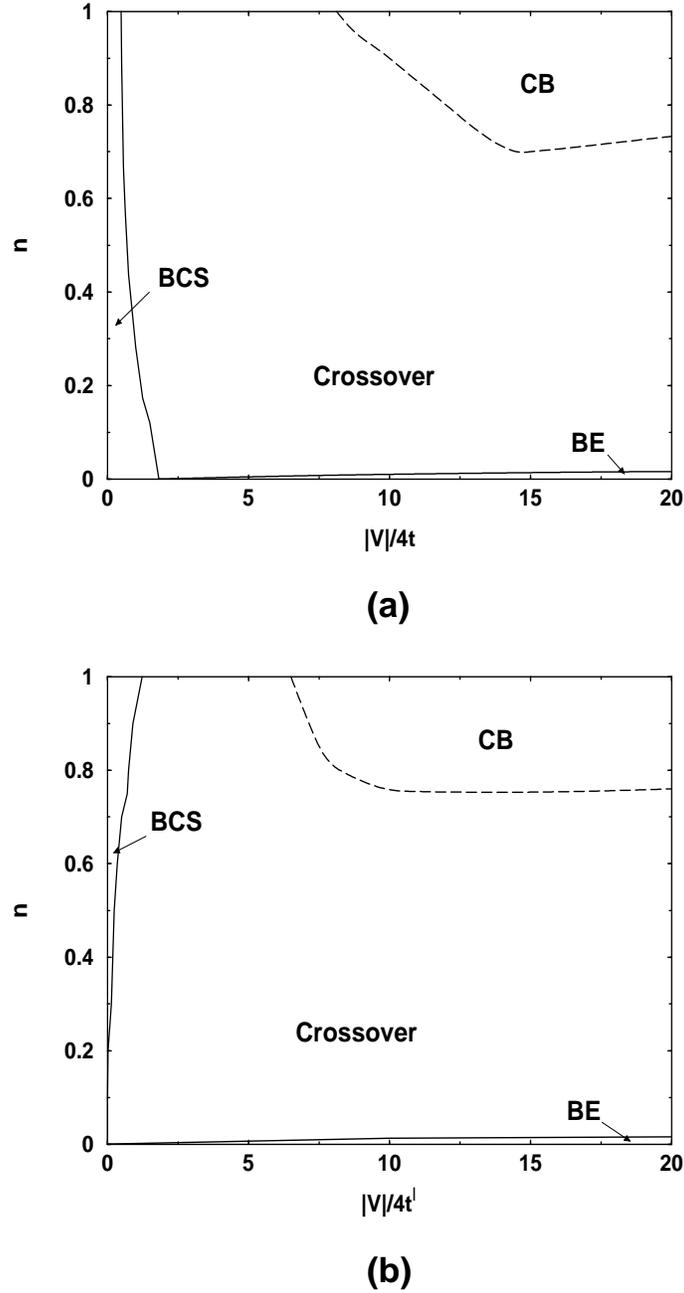}
} \caption{Phase diagram ($V/4t \equiv |V|/4t,n$) for the $d$-wave solution of
the extended Hubbard model with attraction between nearest-neighbour sites in
two dimensions, considering (a) -- nearest-neighbour or (b) -- second- and
third-neighbour hopping. These diagrams were taken from
\protect{\cite{Andrenacci}}.} \label{fig:10}
\end{figure}

For coupling strengths less than $V_{\mathrm{cr}}$ no density induced
BCS-Bose crossover is possible.
For larger couplings strengths, $V > V_{\mathrm{cr}}$, it is clear from the
phase diagram in \cite{denHertog} that one may achieve the crossover point
for which $\mu$ crosses the bottom of the band by increasing the density.
One cannot however reach a true BCS phase for which
$k_{\mathrm{F}} \xi_{\mathrm{pair}} > 2 \pi$ by varying density
as can be seen in Fig.~\ref{fig:10} (a).
Thus independent of the coupling strength and in contrast to the
$s$-wave case, a density induced crossover is not possible.
This is a result of the presence of a pairing threshold
\cite{Andrenacci} as evidenced by considering the
$t^\prime-t^{\prime \prime}-V$ model, or extended attractive
Hubbard model \cite{Nazarenko}, with second and third
neighbour hopping.
For this model the dispersion relation is
\begin{equation}
\xi_{\bf k} = 4 t^\prime \cos k_x \cos k_y + 2 t^{\prime \prime}
\left( \cos 2 k_x + \cos 2k_y \right) - \mu
\end{equation}
with $t^\prime > 0$ and $t^{\prime \prime} > 0$ and where $\mu$ is again
the effective chemical potential given by (\ref{chemsh}).\footnote{This
dispersion law captures the small-arc features of the Fermi surface
detected in underdoped cuprates.}
In this case the critical value $V_{\mathrm{cr}}$ vanishes for
$t^{\prime \prime} < 0.5 t^\prime$ and a density induced crossover
is indeed possible as seen in Fig.~\ref{fig:10} (b).

For moderate densities, one becomes aware not only of the finite lattice
spacing, but also of the intrinsic finite size of $d$-wave bosons. As stated
above this finite size enhances the interactions between composite bosons
leading to an enhanced fermionic region. For an $s$-wave system one can
obtain a crossover from fermionic superconductivity to bosonic degrees
of freedom with increasing interaction strength for any choice of the
density. For a $d$-wave system such a crossover is only possible in
the dilute regime.

One can describe the crossover in terms of the effective chemical potential which is
plotted for the $t-V$ model in \cite{denHertog} as a function of the coupling
strength for fixed density. Here one confirms that at large doping the
chemical potential differs little from its normal state value for a large
range of the coupling strength in contrast to both the low-density and the
$s$-wave results. The limiting case here is half-filling $n=1$ for which the
effective chemical potential is zero, and thus the
system is always a fermionic superconductor. It is clear from this work
\cite{denHertog} that one can access the crossover point at which the
effective chemical potential reaches the bottom of the tight-binding band,
albeit at ever increasing coupling strengths with increasing densities, for all values
of the density below half-filling. One can never however reach the Bose-limit
for higher densities ($n > 0.016$) for both this model and the $t^\prime-
t^{\prime \prime}-V$ model \cite{Andrenacci} since for these densities
$k_F \xi_{\mathrm{pair}} > 1/\pi$ even for infinite coupling. This is apparent
from the line labelled BE in the phase diagrams shown in Fig.~\ref{fig:10}.
This should be contrasted to the corresponding line for the $s$-wave case.
It is also confirmed for the $t-V$ model by a plot of the pair-size in the
infinite coupling limit as a function of density \cite{denHertog}. From
such a plot one can see that the pair-size in the infinite coupling
limit even diverges as one approaches half-filling.

The chemical potential also highlights the interactions between
the composite bosons in the BE region $(n < 0.016)$.
In the $s$-wave case the chemical potential approaches minus one
half the two-body binding energy in the limit of infinite coupling
corresponding to non-interacting bosons. This is only true
for $d$-wave symmetry if the density is zero
\cite{denHertog,Andrenacci}. Defining the boson
chemical potential by $\mu_B = 2 \mu_0 + |\varepsilon_b|$ where $\varepsilon_b$
is the two-body binding energy its behaviour in the BE region is given by
\begin{equation}
\mu_B \rightarrow - (U+2V) n
\end{equation}
One thus sees that one has $\mu_B < 0$ for $0 < -U < 2V$ corresponding to
an average inter-boson attraction and $\mu_B > 0$ for $0 < 2V < -U$
corresponding to an inter-boson repulsion which increases with increasing
interaction strength $V$ (and $-U$).

Returning to the pair-size it can be shown that it does not decrease
monotonically with increasing $V$ but converges asymptotically to a value
larger then the lattice spacing for all $n < 1$. The asymptotic value
increases with $n$ and diverges as one approaches half-filling. This
unusual behaviour is absent in the $s$-wave case and is related to
quasi-long-range-order correlations among the bosons which reside individually
on nearest neighbour sites. The pair-size in this case measures the
correlation between opposite spin fermions belonging to {\it different}
composite bosons. This can most clearly be seen in the lattice Fourier
transform of the pair-wavefunction $\Delta_{\bf k} /E_{\bf k}$
at half-filling. In the very large coupling limit $V/t >> 1$ and at
half-filling
\begin{equation}
\frac{\Delta_{\bf k}}{E_{\bf k}} = \frac{1}{2} \: \mbox{sgn} \: \left(
\cos k_x - \cos k_y \right)
\end{equation}
with lattice Fourier transform
\begin{equation}
\phi({\bf R}_n) = \frac{1}{\pi^2} \frac{\left[ 1 - (-1)^{n_x+n_y}
\right]}{n_x^2 - n_y^2}
\end{equation}
where ${\bf R}_n = (n_x,n_y)$. It decays as a power law with increasing
density when $n_x+n_y$ is an odd integer and vanishes when $n_x + n_y$ is an
even integer (including the case $n_x=n_y=0$ for same-site correlation). In
contrast in the $s$-wave case the pairing wavefunction tends to a constant
with lattice Fourier transform $\Delta_{{\bf R}_n,{\bf 0}}$. The divergence
of $\xi_{\mathrm{pair}}$ is related to the strong interaction $V/4t
\rightarrow \infty$ between fermions of opposite spin but on neighbouring
lattice sites.  It is not related to the divergence of the size of the
composite bosons which is related to correlations between fermions of
opposite spins but on the same site. For this reason the additional area in
the top left of the phase diagrams (Fig.~\ref{fig:10}) refers to ``correlated
bosons'' (CB) rather than a BCS state even though $k_{\mathrm{F}}
\xi_{\mathrm{pair}} > 2 \pi$.

It is interesting to note that the phase diagram shown in
Fig.~\ref{fig:10} resembles the phase diagram obtained in
\cite{Beck.1995} (see also the review \cite{Micnas.review})
for the same model, yet taking into account antiferromagnetic
ordering competing with superconducting ordering.

$d$-wave pairing has also been considered in the context of a continuum
model \cite{Borkowski} where the system must be regarded as dilute
for the approximations used. For low densities the results are
similar to those for the $t-V$ model \cite{denHertog} including
the existence of a coupling threshold $V_{\mathrm{cr}}$.

Lastly the $d$-wave symmetry has an important effect on the excitation
spectrum and the momentum distribution \cite{Borkowski,denHertog,Quader}.
The gap parameter is zero at azimuthal angles $\phi = \pm \pi/4, \pm
3 \pi/4$. Thus for the continuum model with $\mu>0$, or the
lattice model with $\mu$ above the bottom of the band, the quasi-particle
spectrum is gapless at the four Dirac points given by
$\phi = \pm \pi/4, \pm 3 \pi /4$ and $k=k_\mu$
defined by $\xi_{{\bf k}_\mu} = 0$. Near these points the excitation spectrum
is linear. Now consider the direction $\phi = 0$ (or the equivalent directions
$\phi = \pm \pi/2, \pi$) and let $E_g(k_\mu)$ denote the energy gap at
the point $k = k_\mu$ in this direction. The $E_g(k_\mu)$ is a non-monotonic
function of $\mu$ reaching a maximum at intermediate values of $\mu$ above the
crossover point. As $\mu$ reaches the crossover value, $k_\mu$ becomes zero.
The minimal value of the energy gap is thus at the single point
${\bf k} = 0$ and is given by $E_g(0) = |\Delta({\bf 0})| = 0$
implying that the spectrum disperses quadratically for small momenta
for all $\phi$. As soon as $\mu < 0$ for the continuum model, or $\mu$
(which is physically the same) is below
the band edge for the lattice model, the spectrum becomes fully gapped but the
minimal gap remains at ${\bf k} = {\bf 0}$ and takes the value of $|\mu|$.
This behaviour is strikingly different to that for the $s$-wave superconductor
which is always gapped.

Furthermore this behaviour of the excitation spectrum implies that the single
particle distribution strength $n_{{\bf k},\sigma}$ changes dramatically as
one passes through the crossover value for $\mu$ for $d$-wave pairing
\cite{Borkowski,denHertog,Quader}. This discontinuity can be clearly seen in
the momentum distribution at ${\bf k} = 0$ which drops to zero discontinuously
as the chemical potential crosses the critical value \cite{Borkowski}.
No such discontinuity is seen in the $s$-wave case.
It is also illustrated by the contour plots in \cite{denHertog}). For weak
coupling, the chemical potential is essentially equal to its normal state
value, and one has a slightly smeared Fermi-surface. The momentum distribution
is thus maximal at the centre of the Brillouin zone $(0,0)$ and falls off
least rapidly in the anti-nodal directions $\phi = 0,\pm \frac{\pi}{2}, \pi$.
For strong coupling, when the chemical potential has crossed the bottom of the
band, the momentum distribution is zero at the centre of the Brillouin zone
and has a lobe-like structure centered around $(\pm \pi,0)$ and $(0,\pm \pi)$.
These directions with high pair occupation probability may well be the preferred
direction for the opening of the pseudogap \cite{denHertog}.

This qualitative changes in the excitation spectrum and momentum distribution
substantially affect the quasi-particle density of states \cite{Andrenacci}.
For low frequencies this is always zero in the $s$-wave case because the
system is gapped. For the $d$-wave case it changes discontinuously with
$\mu$ from linear in $\omega$ above the crossover value, to a constant at the
crossover value, to zero below the crossover value where the spectrum is again
gapped.

\subsection{Peculiarities of the crossover in models with retarded attraction}
\label{sec:indirect}

As seen in Sec.~\ref{sec:relevance} Uemura suggested
\cite{Uemura} (see Fig.~\ref{fig:04}) that
the attractive interaction between carriers leading to superconductivity
is non-retarded on the Bose side of the BCS-Bose crossover,
while on the BCS side the interaction becomes retarded.
We could not consider this feature of the Bose-BCS
crossover in the previous sections since all the models discussed
were based on postulated predominantly local
(as in Secs.~\ref{sec:one-band} and \ref{sec:band}) direct
non-retarded attraction. It should be admitted
that this type of interaction is not realistic and must be
regarded as a consequence of a more fundamental interaction
whose nature has not yet been established for HTSC. It cannot be
ruled out that, as in traditional superconductors, this is the
electron-phonon interaction \cite{Ginzburg} or the electron-magnon
(or electron spin-fluctuation)
interaction \cite{Pines.review} in view of the peculiar magnetic properties of
HTSC.  In any case in any model describing HTSC correctly, the attraction
between charge carriers must be derived from the exchange
of excitations of a bosonic field. In such a model, we can not only
obtain the attraction between charge carriers, but also estimate
the role of such an important factor as the retardation of
the interaction. The corresponding generalization introduces one
more parameter into the system, viz., the characteristic frequency
$\omega_{0}$ of elementary bosonic excitations. In low-temperature
superconductors, $\omega_{0}$ is the Debye frequency for which the
inequality $\omega_{0} \ll \epsilon_{F}$ is always satisfied.
However, in the case of HTSC with a low number density of charge
carriers, the situation with $\omega_{0} < \epsilon_{F}$ is more
likely if, for example, optical phonons
are responsible for the interaction. Alternatively we can assume that
$\omega_{0} \approx \epsilon_{F}$ if the interaction is realized through
magnons; it might even be the case that $\omega_{0} > \epsilon_{F}$ (see e.g.
\cite{Loktev.review}).

The inclusion of non-retarded interactions into the Eliashberg equations
\cite{Eliashberg}
(see also the textbooks \cite{Schrieffer,Vonsovsky,Rickayzen.book} and the
review \cite{Carbotte}) has been extensively
studied and we will give the references shortly. However here we focus on
the simplest model with an electron-phonon interaction\footnote{It was
noted above that elementary Bose excitations of
different kinds may play the role of an intermediate boson,
but we will simply refer to them as ``phonons''.}
but with a variable carrier density which makes possible to obtain
a BCS-Bose crossover as considered in \cite{LSh.FNT.1996}
(see also the more formal treatment in \cite{LTSh.TMF.1998}). This is to the
best of our knowledge the only attempt
to consider the BCS-Bose crossover itself using a model with indirect
interaction and we will briefly present the model and discuss
the main results. Another related and very interesting question
which we discuss here is a possible violation of
{\em Migdal's theorem\/} \cite{Migdal} in HTSC.

\subsubsection{Model and basic equations}

It is known that (see \cite{Abrikosov.book,Vonsovsky}) the Hamiltonian
density describing a system with electron-phonon interaction can
be written in the form
\begin{equation}
\mathcal{H} = - \psi^{\dagger}_{\sigma}({\bf r})
\left( \frac{\nabla^{2}}{2 m} + \mu \right) \psi_{\sigma}({\bf r})
+ \mathcal{H}_{ph}(\varphi({\bf r})) +
  g_{\mathrm{el-ph}} \psi^{\dagger}_{\sigma}({\bf r})\psi_{\sigma}({\bf r})
      \varphi({\bf r}),
\label{Hamiltonian.phonon}
\end{equation}
where $\mathcal{H}_{ph}$ is the operator describing phonons with
the dispersion $\omega({\bf k})$; ${\bf r}$ is a 2D vector;
$\psi_{\sigma}({\bf r})$, $\varphi({\bf r})$
are the Fermi and Bose fields; $\sigma$ is the spin;
$g_{\mathrm{el-ph}}$ is the constant of electron-phonon interaction.

In order to clarify the influence of retardation effects on the
electron spectrum as a function of $n_{f}$, one must analyze a
system of two equations. The first equation is the Eliashberg
equation \cite{Eliashberg} (see also the textbooks
\cite{Schrieffer,Vonsovsky,Rickayzen.book} and the review \cite{Carbotte})
for the self-energy of the electron, $\Sigma(p)$ written in
Nambu formalism \cite{Nambu} as
\begin{equation}
\Sigma(p) = i g_{\mathrm{el-ph}} \int \frac{d^{3} p'}{(2 \pi)^{3}}
\tau_{3}G(p') \tau_{3} \Gamma_{\mathrm{el-ph}}(p', p; p - p')
D_{\mathrm{ph}}(p - p'),
                                    \label{Eliashberg}
\end{equation}
where $p = (p_0,{\bf p})$ (${\bf p}$ being a two-dimensional vector) and
one integrates over both the frequency and the momentum.
$\Gamma_{\mathrm{el-ph}}(p', p; p - p')$ is the
vertex function and $D_{\mathrm{ph}}(p - p')$ is the ``dressed''
phonon propagator.

The second equation is the well-known condition imposed on the Green's
electron function $G(p)$\footnote{Note
that Eq.~(\ref{number.Green}) is exactly equivalent to
Eq.~(\ref{number}) which was written in terms of the effective potential.}
relating the chemical potential $\mu$ to
$n_{f} = \epsilon_{F} m / \pi$:
\begin{equation}
- i \int \exp(i \delta p_0 \tau_{3})
  \mbox{tr}[\tau_{3} G(p_0, {\bf p})]
  \frac{d^3 p}{(2 \pi)^{3}} = n_{f}, \qquad
  \delta \to 0^{+} \,.
\label{number.Green}
\end{equation}

Obviously, it is impossible to solve the system of equations
(\ref{Eliashberg}), (\ref{number.Green}) in such a general form, and
one must make certain simplifying assumptions.

One such assumption, which is often made in an analysis of
the Eliashberg equation, is to disregard effects that are
not directly connected to the emergence of anomalous mean values.

The second assumption follows from {\em  Migdal's theorem\/}
\cite{Migdal} (see also the textbooks
\cite{Abrikosov.book,Fetter,Rickayzen.book}) stating that for
$\omega_{0} \ll \epsilon_{F}$, the electron-phonon vertex
function can be replaced by its ``bare'' value $g_{\mathrm{el-ph}}$.
Note that one cannot use directly the result obtained by Migdal
since the standard inequality $\omega_{0} \ll \epsilon_{F}$
can be violated in HTSC as mentioned above. However,
Migdal's theorem was recently generalized to the case when
$\omega_{0} \gg \epsilon_{F}$ \cite{Ikeda}. Hence the vertex function
$\Gamma$ in (\ref{Eliashberg}) can also be replaced
by $g_{\mathrm{el-ph}}$ in the case of sufficiently low densities (Bose limit)
that $\omega_{0} \gg \epsilon_{F}$. Since the results will focus on the Bose
and BCS limits, where Migdal's theorem appears to be valid, one assumes
that the ``bare'' vertex can be used throughout.

It is known \cite{Pietronero.1992,Pietronero.1995}, however,
that the complete vertex should be taken into account
for $\omega_{0} \sim \epsilon_{F}$ and this consideration
leads to a significant increase in $T_{c}$. The problem of
solving the Eliashberg equation with a vertex correction \cite{Pietronero.1995}
is very complicated and in fact deserves its own review.
Nevertheless, for the limiting cases $\omega_{0} \ll \epsilon_{F}$
and $\omega_{0} \gg \epsilon_{F}$ considered here, one does not
need to include the vertex correction. Nonetheless the final result
obtained shows that, if one wants to explain why the BCS
regime is not the most optimal for high $T_{c}$, one may well need
vertex corrections. The theory of the Eliashberg equation with vertex
corrections is currently a very rapidly developing branch of superconductivity
and we simply mention here some of the latest papers
\cite{Ummarino,Grimaldi,Cappelluti,Bang} where the relevance of
this theory to HTSC and new theoretical developments were discussed.

Furthermore, there is an interesting correlation between
the need to include the vertex corrections into the Eliashberg
theory and obtaining the correct limiting behaviour in 2D in the study of
superconducting fluctuations. For example, it is known that the
$T$-matrix approximation which is used extensively
(see Secs.~\ref{sec:survey}, \ref{sec:limitations}) fails to
describe BKT physics \cite{Micnas,Tremblay.1997,Kagan,Pieri}.
At precisely the same time vertex corrections are also crucial.

Finally, the third assumption is associated with the choice
of the form of the ``dressed'' boson propagator. As usual
(see \cite{Schrieffer}), it is replaced by the ``bare'' propagator
\cite{Abrikosov.book},
\begin{equation}
D_{\mathrm{ph}}(k) = \frac{\omega^{2}({\bf k})}
{k_{0}^{2} - \omega^{2}({\bf k}) + i \delta},
\qquad k \equiv k_{0}, {\bf k}, \qquad \delta \to 0^{+} \,.
                                       \label{D}
\end{equation}

In the subsequent analysis, the Einstein model has been used for the
phonon dispersion $\omega({\bf k})$, i.e.
$\omega({\bf k}) = \omega_{0}$. Using this approximation and
taking the function $\Sigma(p)$ to be proportional to the
order parameter $\Phi(p)$, one can transform Eq.~(\ref{Eliashberg})
into an equation for the order parameter $\Phi$ (see the details
in \cite{LTSh.TMF.1998}), which depends on the frequency $p_{0}$:
\begin{equation}
\Phi(p_{0}) = - \imath g_{\mathrm{el-ph}}^{2} \int \frac{d^{3} p'}{(2 \pi)^{3}}
\frac{\Phi(p_{0}')}
{(p_{0}')^{2} - \xi({\bf p}')^{2} - |\Phi(p_{0}')|^{2} + i \delta}
\frac{\omega_{0}^{2}}
{(p_{0} - p_{0}')^{2} - \omega_{0}^{2} + i \delta} \,,
                                           \label{Elias.simple1}
\end{equation}
where again $p \equiv \{p_{0}, {\bf p}\}$. The problem is thus reduced to an
analysis of the system of equations (\ref{Elias.simple1}) and
(\ref{number.Green}) as a function of the carrier concentration. It should be
emphasized that the Eliashberg equation \ref{Elias.simple1} had been
investigated  and solved several times  previously (see e.g.
Refs.~\cite{Schrieffer,Abrikosov.book,Vonsovsky,Eliashberg,Morel}) but only
when the value of $\mu$ coincided with $\epsilon_{F}$ (i.e. one did not
consider the BCS-Bose crossover problem). In this latter case the expansion
$\xi({\bf p}) \approx {\bf p}_{F} ({\bf p} - {\bf p}_{F}) / m$ (${\bf p}_{F}$
is the Fermi momentum) was then used for the normal state spectrum.

\subsubsection{Electron spectrum}

The desired concentration dependence of the electron spectrum
can be characterized by the quantity
$\Delta(n_{f}) \equiv |\Phi(p_0 =0, n_{f})|$ which defines the energy gap.
The system of equations (\ref{Elias.simple1}),
(\ref{number.Green}) was studied in \cite{LSh.FNT.1996} using
the rather crude approximation that $\Phi(p_0) =$ const. It was also
studied in \cite{LTSh.TMF.1998} by applying a method (see Refs. in
\cite{LTSh.TMF.1998}) which allows one to reduce the integral
Schwinger-Dyson equation (\ref{Elias.simple1}) to a differential
one, which can then be solved relatively easily. This more
accurate treatment showed good agreement with \cite{LSh.FNT.1996}.
Thus here we simply present the main classes of solutions of
the system derived in \cite{LSh.FNT.1996}. We assume everywhere
that the conduction bandwidth $W \gg \omega_{0}, \epsilon_{F}, \Delta$
as is appropriate for HTSC.

\paragraph{BCS or high density limit}

In the BCS limit, the inequality $\epsilon_{F} \gg \omega_{0}, \Delta$
holds. In this case, it follows from (\ref{number.Green})
that $\mu \approx \epsilon_{F}$.

a) If we assume that additionally $\Delta \ll \omega_{0}$,
we reproduce by solving (\ref{Elias.simple1}) the following
formula from BCS theory \cite{Abrikosov.book}:
\begin{equation}
\Delta = \Delta_{\mathrm{BCS}} =
2 \omega_{0} \exp{\left( - \frac{2 \pi}{m g_{\mathrm{el-ph}}^{2}}\right)},
                                      \label{deltaBCS}
\end{equation}
Thus the above inequality is, as expected, satisfied if the coupling is weak
($g_{\mathrm{el-ph}}^{2} m / 2 \pi \ll 1$). Note that Eq.~(\ref{deltaBCS})
corresponds to the result of the classical Morel-Anderson \cite{Morel} model
(if we adopt the notation $m g_{\mathrm{el-ph}}^{2} / 2 \pi \equiv \lambda$).
The quantity $\Delta \equiv \Phi(0)$ was obtained in \cite{Morel} by
substitution of a suitably constructed trial function $\Phi(p_0) \neq const$
into an equation similar to (\ref{Elias.simple1}) and this represents the
first known way of approximating the analytical solution of the Eliashberg
equation.

b) If one assumes the opposite limit, $\Delta \gg \omega_{0}$, we obtain
\begin{equation}
\Delta = \omega_{0} \sqrt{1 + \left( \frac{g_{\mathrm{el-ph}}^{2} m}{16} \right)^{2}}
\simeq \omega_{0} \frac{g_{\mathrm{el-ph}}^{2} m}{16}.        \label{deltaBCS.strong}
\end{equation}
This result is valid if we assume that the coupling is strong
($g_{\mathrm{el-ph}}^{2} m / 16 \gg 1$). If $\Delta$ is much greater than
$\omega_{0}$, it can be verified that the formula
(\ref{deltaBCS.strong}) corresponds to ultrastrong coupling
\cite{Allen} (see also \cite{Vonsovsky}). Note that to be consistent
in this limit the equation for the renormalization factor $Z(p)$ must
also be studied.

\paragraph{Low-density limit}

Small values of $n_{f}$ correspond to the condition
$\epsilon_{F},$ $\Delta \ll \omega_{0}$. Since the value
$\Delta$ in this case can be of the order of $\epsilon_{F}$,
Eq.~(\ref{number.Green}) becomes nontrivial (see the discussion
in Sec.~\ref{sec:solution}) and does not lead to $\mu = \epsilon_{F}$.

a) If one also assumes that $|\mu| \ll \omega_{0}$, one arrives at
the following solution
\begin{equation}
\Delta = 2 \sqrt{\omega_{0} \epsilon_{F}}
\exp{\left( - \frac{2 \pi}{m g_{\mathrm{el-ph}}^{2}} \right)},
\qquad \mu = - \omega_{0} \exp{\left(- \frac{4 \pi}{m g_{\mathrm{el-ph}}^{2}}\right)}
+ \epsilon_{F}.                               \label{sol.weak}
\end{equation}
It can be seen that the condition $|\mu| \ll \omega_{0}$
holds for $g_{\mathrm{el-ph}}^{2} m / 4 \pi \ll 1$ i.e. weak coupling. Comparing
(\ref{bound.energy}) and (\ref{solution.2D}) with (\ref{sol.weak}),
one can easily see that the latter equation can be written
in the form
\begin{equation}
\Delta =
\sqrt{2 |\varepsilon_{b}^{\mathrm{weak}}| \epsilon_{F}},
\qquad \mu = - \frac{|\varepsilon_{b}^{\mathrm{weak}}|}{2}  + \epsilon_{F},
                                     \label{sol.weak.Eb}
\end{equation}
where
\begin{equation}
\varepsilon_{b}^{\mathrm{weak}} \equiv -2 \omega_{0}
\exp{\left(-\frac{4 \pi}{m g_{\mathrm{el-ph}}^{2}}\right)}.       \label{Eb.weak}
\end{equation}

Since the expressions (\ref{Eb.weak}) and (\ref{bound.energy})
are similar, it is natural to assume that
$\varepsilon_{b}^{\mathrm{weak}}$
has the same meaning as $\varepsilon_{b}$ defined by
(\ref{bound.energy}), i.e. is the energy of the bound state
which emerges in this case due to the electron-phonon interaction.
For this reason, expression (\ref{Eb.weak}) contains the quantity
$\omega_{0}$ characterizing the energy of phonons instead of
the bandwidth $W$.

At the same time, solution (\ref{sol.weak.Eb}) differs significantly
from (\ref{solution.2D}). The latter expression is valid for all
values of $\epsilon_{F}$ and all interaction strengths. The former applies
only for weak coupling and low densities $\epsilon_{F} \ll \omega_{0}$.
If the inequality $\epsilon_{F} \gg \omega_{0}$ holds, expression
(\ref{deltaBCS}) must be used. For $\epsilon_{F} \ll \omega_{0}$ the
charge carriers form an ``adiabatically'' slow subsystem,
and the retardation of indirect interaction can be disregarded
as is shown in Fig.~\ref{fig:04}. However, as the value
$\epsilon_{F}$ increases (and especially in the region
$\epsilon_{F} \gg \omega_{0}$), the role of retardation becomes
decisive as is evident in formula (\ref{deltaBCS}).

\begin{figure}[h]
\centering \resizebox{0.6\textwidth}{!}{%
  \includegraphics[bb=110 280 460 495]{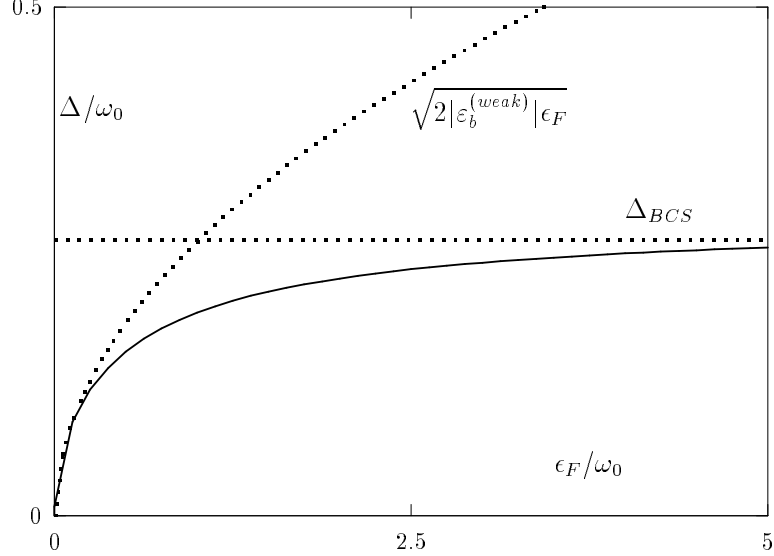}
} \caption{$\Delta(\epsilon_{F})$ for $m g_{\mathrm{el-ph}}^2/4 \pi =0.25$
(solid curve) as taken from \protect{\cite{LTSh.TMF.1998}}. Dashed lines
correspond to solution of (\ref{sol.weak.Eb}) for $\Delta$ and the value of
$\Delta_{\mathrm{BCS}}$ determined from (\ref{deltaBCS}).}
                     \label{fig:11}
\end{figure}

The numerical solution of (\ref{Elias.simple1}) and (\ref{number.Green})
is shown in Fig.~\ref{fig:11}.
It can be seen that the asymptotic expression (\ref{sol.weak.Eb}) for $\Delta$
coincides with the numerical solution for $\epsilon_{F} \ll \omega_{0}$,
while for $\epsilon_{F} \approx \omega_{0}$ and $\epsilon_{F} \gg \omega_{0}$
it lies above the corresponding numerical curve
(Expression (\ref{sol.weak}) can increase without bound, while if one incorporates
the band structure then $\Delta$ achieves its maximal value
at half-filling, see (\ref{solution.2D.band}) and Fig.~\ref{fig:05}).
The maximal value attained is however much lower in the system with
an electron-phonon interaction. The gap cannot increase
without limit as suggested by Eq. (\ref{sol.weak}) but attains a
saturation value equal to $\Delta_{\mathrm{BCS}}$ in the limit
$\epsilon_F >> \omega_0$.

The chemical potential $\mu$ (which is not shown on the figure)
depends linearly on $\epsilon_{F}$ and is negative for very low
concentrations of the charge carriers.
It can also be seen from (\ref{sol.weak.Eb}) that for
$\epsilon_{F} < |\varepsilon_{b}^{\mathrm{weak}}| / 2$, the chemical
potential $\mu < 0$, and hence the local pair mode
is realized (see Sec.~\ref{sec:solution}). It is important to
note that although the quantity $\varepsilon_{b}$ was an independent
parameter in Sec.~\ref{sec:one-band} , in the case
under investigation $\varepsilon_{b}^{\mathrm{weak}}$ is expressed
in terms of electron-phonon interaction parameters.
It can be seen from (\ref{deltaBCS}) and (\ref{Eb.weak}) that
$\Delta_{\mathrm{BCS}}$ and $\varepsilon_{b}^{\mathrm{weak}}$ are connected
by the relation $|\varepsilon_{b}^{\mathrm{weak}}| =
\Delta_{\mathrm{BCS}} \exp(- 4 \pi / m g_{\mathrm{el-ph}}^{2})$, i.e.
$\varepsilon_{b}^{\mathrm{weak}}$ is exponentially smaller than
$\Delta_{\mathrm{BCS}}$.  Thus in the case of
weak coupling, the region of existence of local pairs is very
small and is only of theoretical interest.

b) Let us now assume that although $\epsilon_{F}, \Delta \ll \omega_{0}$,
the absolute value of $|\mu| \gg \omega_{0}$ ($\mu < 0$).
In this case one can obtain the following solution of the system
(\ref{Elias.simple1}) and (\ref{number.Green}):
\begin{eqnarray}
\Delta &=& 2 \sqrt{\omega_{0}
\coth{\left(\frac{4 \pi}{m g_{\mathrm{el-ph}}^{2}}\right)}\epsilon_{F}},
\nonumber               \\
\mu &=& - \omega_{0} \coth{\left(\frac{4 \pi}{m g_{\mathrm{el-ph}}^{2}}\right)}
+ 2 \epsilon_{F} \coth^{2}{\left(\frac{4 \pi}{m g_{\mathrm{el-ph}}^{2}}\right)}.
                 \label{sol.strong}
\end{eqnarray}
It can be seen that the condition $|\mu| \gg \omega_{0}$
holds in the strong coupling limit ($g_{\mathrm{el-ph}}^{2} m / 4 \pi \gg 1$).
In this case, we can again introduce the energy
\begin{equation}
\varepsilon_{b}^{\mathrm{strong}} \equiv -2 \omega_{0}
\coth{\left(\frac{4 \pi}{m g_{\mathrm{el-ph}}^{2}}\right)} \approx
-\omega_{0} \frac{g_{\mathrm{el-ph}}^{2} m}{2 \pi}          \label{Eb.strong}
\end{equation}
and write the expressions ($\ref{sol.strong}$) in the form
\begin{equation}
\Delta =
\sqrt{2 |\varepsilon_{b}^{\mathrm{strong}}| \epsilon_{F}},
\qquad \mu = - \frac{|\varepsilon_{b}^{\mathrm{stong}}|}{2}  +
2 \epsilon_{F} \coth^{2}{\left(\frac{4 \pi}{m g_{\mathrm{el-ph}}^{2}}\right)}.
                                \label{sol.strong.Eb}
\end{equation}

Virtually all that has been said about the solution for
(\ref{sol.weak.Eb}) remains in force for (\ref{sol.strong.Eb}).
However, some differences do exist. In particular it can be seen
from (\ref{Eb.strong}) and (\ref{sol.strong.Eb}) that the region of
existence of local pairs is substantially larger in the case of strong
coupling.

There is also some evidence that the strong coupling case may
be relevant to HTSC due to the presence of an extended van Hove
singularity.

\subsubsection{Discussion}
\label{sec:discussion.indirect}

The results presented here correspond to an oversimplified model. Nonetheless
they are a good illustration of the importance of the position of the Fermi
level $\epsilon_{F}$ relative to the limiting boson frequency $\omega_{0}$,
which is so transparently shown on Uemura's version of the HTSC
phase diagram (see Fig.~\ref{fig:04} where an arbitrary boson frequency
$\omega_B$ replaces $\omega_0$).  The relationship between
carrier density and retardation is clear from the forms of the solutions
both in the high density (see Eq.~(\ref{deltaBCS})) and low density
(see Eqs.~(\ref{sol.weak.Eb}) and (\ref{sol.strong.Eb})) limits :
\begin{equation}
\Delta(\epsilon_{F}) = \left\{
\begin{array}{c c c}
\sqrt{2 |\varepsilon_{b}^{\mathrm{ph}}| \epsilon_{F}},
& \epsilon_{F} \ll \omega_0, & \mbox{(non-retarded)}, \\
\sqrt{2 \omega_0 |\varepsilon_{b}^{\mathrm{ph}}|}, &
\epsilon_{F} \gg \omega_0 , & \;\; \mbox{(retarded)}
\end{array}    \right.
\label{solution.phonon.general}
\end{equation}
where $\varepsilon_{b}^{\mathrm{ph}}$ is given by
$\varepsilon_{b}^{\mathrm{weak}}$ from Eq.~(\ref{Eb.weak})
or, for the non-retarded case only, by
$\varepsilon_{b}^{\mathrm{strong}}$ from Eq.~(\ref{Eb.strong})
respectively.

However, the approximations above have raised certain questions.
In particular the quantity $\Delta$
and hence $T_{c}$ attains its maximal value for
$\epsilon_{F} \gg \omega_{0}$. The following
question thus arises: why do HTSC then attain maximal
values of $T_{c}$ at relatively small values of $\epsilon_{F}$?
Indeed if one assumes spatially homogeneous systems
the presence of real local pairs as suggested by
\cite{Mihailovic,Demsar.1998} even demands
negative $\mu$. This argument does not apply
to spatially inhomogeneous systems. Moreover
if one suggests that the optimally doped
cuprates correspond to the BCS regime
(see Sec.~\ref{sec:derivative}), the model above proposes an
explanation as to why this limit gives the highest $T_{c}$ values.

There also exists a possibility that one can exceed the value
of $\Delta_{\mathrm{BCS}}$ by taking into account the vertex
correction for the region $\epsilon_{F} \sim \omega_{0}$
\cite{Pietronero.1992,Pietronero.1995,Ummarino,Grimaldi,Cappelluti,Bang}.

We mention also other recent models \cite{Onoda,Kobayashi} with an indirect
(through spin fluctuations) interaction between carriers which have been
studied recently and where the pseudogap behaviour is the result of a complex
interplay between antiferromagnetic and $d$-wave pairing
fluctuations. It is very likely that the ultimate explanation
of the pseudogap behaviour is only possible within such models.
The models with direct interactions between carriers, which we
have predominantly discussed here, are only the very first step in
understanding the unusual properties of HTSC. However, as we will see in the
next chapter there are many unanswered questions even within
the relatively simple attractive Hubbard model so often
used for the study of crossover and pseudogap behaviour.

\section{Self-consistent $T$-matrix approximation and
    its limitations in the 2D limit}
\setcounter{equation}{0}
\label{sec:T.matrix}

In the previous chapter we considered a number of different models in the
mean-field approximation which is of course justified at $T=0$ when continuous
symmetry breaking is possible even in 2D \cite{Coleman}. Any generalization
of these models to $T \neq 0$ necessarily requires going beyond simple
mean-field theory.
Here we will describe the different forms of the $T$-matrix approximation
widely used to study the Hamiltonian (\ref{Hubbard}) or its continuum version
(\ref{Hamiltonian.2D}) at nonzero temperatures following mostly the articles
\cite{Levin,Serene,Micnas,Gooding}. Unfortunately, as we will see in
Sec.~\ref{sec:limitations}, even quite sophisticated approximations
may not be sufficient for 2D systems.

\subsection{Self-consistent, conserving approximation}
\label{sec:self}

\begin{figure}[h]
\centering \resizebox{0.6\textwidth}{!}{%
  \includegraphics[bb=50 420 580 590]{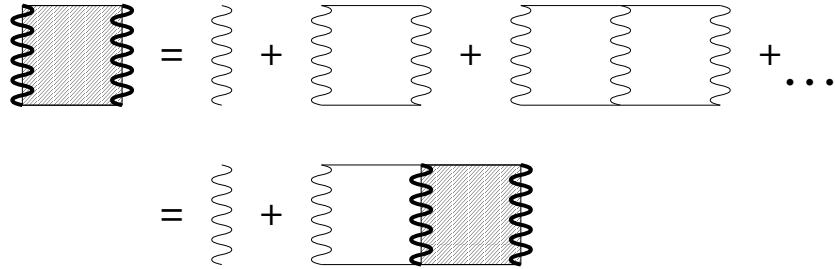}
}
\caption{Diagrammatic representation of Eq.~(\ref{T.matrix})
taken from \protect{\cite{Gooding}}.
The $T$-matrix contains the repeated
scattering of two-particles.}
\label{fig:12}
\end{figure}

We first discuss the $T$-matrix approach to the Hubbard model
in the absence of a superconducting condensate i.e
in the normal state. The $T$ matrix consists of the sum of
all particle-particle ladder diagrams (see Fig.~\ref{fig:12})
It is known that the $T$-matrix is the ladder approximation
to the Bethe-Salpeter equation \cite{Fetter}
\begin{equation}
T (i \Omega_m, {\bf q}) =  \frac{-U}
{1 -  U \chi(i \Omega_m,{\bf q})} \,,
\label{T.matrix}
\end{equation}
where $\chi(i \Omega_m, {\bf q})$ is the independent pairing
susceptibility given by
\begin{equation}
\chi(i \Omega_m, {\bf q}) = \frac{1}{\beta N} \sum_{n, {\bf k}}
G(i \Omega_m - i \omega _n, {\bf q}-{\bf k}) G(i\omega_n, {\bf k}) \,.
\label{chi}
\end{equation}
Recall that $\omega_{n}$, and $\Omega_m$ are the fermionic
(odd) and bosonic (even) Matsubara frequencies, respectively.
This approximation is expected to be accurate in the
low density limit when one can neglect particle-hole scattering
and is valid when the product of the Fermi momentum, ${\bf k}_{\mathrm{F}}$ and
the scattering length, $a$ is small.
Note that the function $\chi(i \Omega_m, {\bf K})$
is chosen to have different signs by different authors (compare, e.g.
\cite{Micnas} and \cite{Gooding}), and the reader should take
note of the consequence of this sign choice in the subsequent equations.

\begin{figure}[h]
\centering \resizebox{0.6\textwidth}{!}{%
  \includegraphics[bb=35 400 580 530]{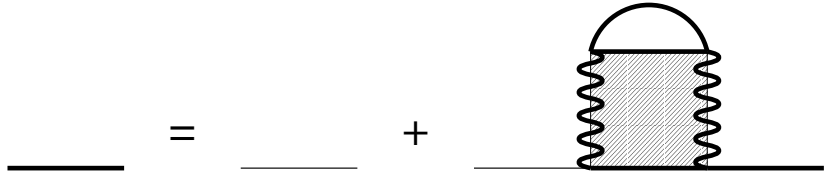}
} \caption{Diagram for the single-particle Green's function (solid line) in
the self-consistent, conserving approximation given by Eqs.~(\ref{Dyson}) --
(\ref{self.energy}). The thin solid lines represent the non-interacting Green's
function $G_{0}$.}
\label{fig:13}
\end{figure}

The corresponding one-particle Green's function in (\ref{chi})
satisfies the Dyson equation (see the diagram in Fig.~\ref{fig:13})
\begin{equation}
G(i\omega_n, {\bf k}) =
( G_0(i\omega_n, {\bf k})^{-1}- \Sigma(i\omega_n, {\bf k}))^{-1} \,,
                     \label{Dyson}
\end{equation}
with
\begin{equation}
G_0(i\omega_n, {\bf k}) =
\frac{1}{i\omega_n - \varepsilon_{\bf k}} \,,
                     \label{Green.0}
\end{equation}
where $\Sigma(i\omega_n, {\bf k})$ is the electron self-energy.

In the self-consistent $T$-matrix approach the self-energy is given by
\begin{equation}
\Sigma(i \omega _n, {\bf k}) =  \frac{1}{\beta N} \sum_{m,{\bf q}}
T (i \Omega_m, {\bf q })
G(i\Omega_m - i\omega_n, {\bf q} - {\bf k})
\label{self.energy}
\end{equation}
which closes the system of self-consistent equations.

In most cases the results for real frequencies are
obtained from a numerical analytic continuation
$i\omega_n \to \omega +i0$  \cite{Vidberg} (see also the review
\cite{Carbotte}). Since this continuation
is inaccurate (see an example in Sec.~\ref{sec:continuation}
and recent comprehensive analysis of the method in \cite{Beach})
a real-time technique is also used to avoid this problem
\cite{Gooding,Kornilovitch,Dagotto}.

The above system of equations, valid in the normal state,
is conserving, i.e. satisfying Kadanoff-Baym criteria
\cite{Kadanoff} and ``$\Phi$ derivable'' in the sense
of Baym \cite{Baym} and known to include Gaussian fluctuations
of the pairing field in a self-consistent manner.
``$\Phi$ derivable'' approximation means that the corresponding
self-consistent approximation for the free energy (per lattice site)
is given by the functional
\begin{eqnarray}
\Omega[\Sigma,G] & = &  -\frac{2}{\beta N} \sum_{n,{\bf k}}
\exp(i\omega_n \delta)
\left\{\Sigma(i \omega _n, {\bf k}) G(i \omega _n, {\bf k}) \right.
\nonumber                    \\
& + & \left. \ln [-G_{0}(i \omega _n, {\bf k})^{-1} +
\Sigma(i \omega _n, {\bf k})] \right\}
+ \Phi[G], \qquad \delta \to 0^{+} \,,
                         \label{Omega}
\end{eqnarray}
with
\begin{equation}
\Phi[G] = \frac{1}{\beta} \sum_{m,{\bf q}} \left\{
\ln[1 - U \chi(i \Omega_m, {\bf q})] + U \chi(i \Omega_m, {\bf q})
\right\}
      \label{Phi}
\end{equation}
and at the stationary point $G$, $\Sigma$ and $\Phi$ are
related by Dyson's equation (\ref{Dyson}) and
\begin{equation}
\Sigma(i \omega _n, {\bf k}) = \frac{1}{2}
\frac{\delta \Phi[G]}{\delta G(i \omega _n, {\bf k})} \,
\label{stationary}
\end{equation}
which reproduces Eq. (\ref{self.energy}).
Using short-hand notation one may rewrite (\ref{Omega})
as (compare with Eq.~(\ref{Effective.Action}))
\begin{equation}
\Omega[T,\mu, \Sigma,G] = -
2\mbox{Tr}[\Sigma G + \mbox{Ln}(-G_0 + \Sigma)] + \Phi[G] \,,
               \label{Omega.short}
\end{equation}
where $\mbox{Tr} A = 1/(N \beta) \sum_{n,{\bf k}} A(i \omega_{n}, {\bf k})$.

The term ``conserving approximation''implies that the approximation conserves
the number of particles, the energy, the momentum, and the angular momentum.
As proven by Baym \cite{Baym} ``$\Phi$ derivable'' theory is
conserving \cite{Kadanoff}, although this is not a
{\em necessary \/} condition for the theory to be conserving.

In fact the functional (\ref{Omega}) (or its short-hand form
(\ref{Omega.short})) is closely related to the Cornwall-Jackiw-Tomboulis
effective potential \cite{Cornwall} (see also the textbook
\cite{Miransky}) which is widely used in quantum field
theory for the description of systems of
equations similar to those given here. For example, this method
easily permits one to generalize the system
(\ref{T.matrix}) - (\ref{self.energy}) to a state with a
broken continuous symmetry e.g the superconducting state in 3D.
Note that such equations were obtained in \cite{Pedersen}
(see also \cite{Haussmann.1993,Schafroth.1997,Keller} where these equations
were also discussed and solved by different methods) for a 2D model
using the functional derivative technique of \cite{Baym,Kadanoff} and
solved approximately using a momentum approach.

The electron density as a function of $T$ and $\mu$ is given in
terms of the thermodynamical potential by
\begin{equation}
n(\mu,T) = - \frac{\partial \Omega}{\partial \mu}
\label{number.stationary}
\end{equation}
If one calculates $n(T,\mu)$ from $\Omega[\Sigma,G]$
and exploits the stationary properties of this functional,
one obtains immediately the result,
\begin{equation}
n(\mu,T) =
\frac{2}{\beta N} \sum_{n,{\bf k}} G(i\omega_n, {\bf k})
\exp (i \omega_n \delta), \qquad \delta \to 0^{+} \,,
\label{number.Green.T}
\end{equation}
which is equivalent to that given by (\ref{number.definition})
as expected.

The superconducting instability of the normal state at $T_{c}$
is signalled by the divergence of the $T$-matrix
at zero energy (at chemical potential) and zero momentum
(the well-known Thouless criterion):
\begin{equation}
1 -  U  \chi({\bf q = 0}, \omega = 0) = 0 \,.
               \label{Thouless}
\end{equation}
This equation, and the associated normal state properties, can
be examined in a variety of approximations. Approximations are
often used to simplify the fully self-consistent, conserving
theory described by Eqs.~(\ref{T.matrix}) -- (\ref{self.energy}),
(\ref{number.Green.T}) since they cannot be studied analytically without
additional simplification. However they can be and have been
studied numerically \cite{Haussmann.1994,Gooding,Keller}.

Many of the discussions related to the different
predictions and scenarios of these theories, which are all based on
superconducting precursor fluctuations, are related to the fact
that these additional approximations lead to quite
different results (see, for example, the different opinions about
the most adequate approach expressed in \cite{Levin,Levin.1999}
and \cite{Ichinomiya,Yanase}). It is thus very difficult to say, at this
stage, whether a particular theory really predicts different
physics or whether the differences are simply an artifact of the
approximation used to solve the $T$-matrix equations.\footnote{There
are probably two criteria one can use as a guide to decide which
of the existing approaches is the best. Firstly, a good approach
should violate as few exact results as possible. Secondly,
a good approach should reproduce the exact or numerically accurate results
whenever these are known. Based on these criteria the best approach
of the $T$-matrix type for the repulsive Hubbard model in the weak and
intermediate coupling regimes was probably presented in
\cite{Tremblay.1997}. We refer the reader to this paper where
the generalization of these results to the negative-$U$
Hubbard model is also discussed in detail.}
Our point of view (see Sec.~\ref{sec:limitations})
is, however, that even in its best form presented in this
section, $T$-matrix approximation
may not be sufficient for the description of vortex
phase fluctuations which are important in the 2D limit at finite
temperatures and are the main subject of this review.

\subsection{Non self-consistent, non-conserving approximation}
\label{sec:non}

The self energy in this case is given by
(compare with (\ref{self.energy}))
\begin{equation}
\Sigma_{0}(i \omega _n, {\bf k}) = \frac{1}{\beta N} \sum_{m,{\bf q}}
T_{0} (i \Omega_m, {\bf q })
G_{0}(i\Omega_m - i\omega_n, {\bf q} - {\bf k}) \,.
              \label{self.energy.non}
\end{equation}
The index ``0'' indicates the use of free Green's functions
so that the $T$-matrix (\ref{T.matrix}) is also calculated
in terms of the free Green's functions
\begin{eqnarray}
T_0^{-1}(i \Omega_m, {\bf q}) &=& -\frac{1}{U} + \chi_0(i \Omega_m,{\bf q})
\nonumber \\
&=& - \frac{1}{U} + \frac{1}{\beta N} \sum_{n, {\bf k}}
G_0(i \Omega_m - i \omega _n, {\bf q}-{\bf k})
G_0(i\omega_n, {\bf k}) \,.
\label{T.matrix.non}
\end{eqnarray}
The summation over Matsubara frequencies for this lowest order
approximation can be easily performed and
the $T$-matrix (\ref{T.matrix.non}) is given in terms of the corresponding
pair susceptibility (\ref{chi}) which acquires a rather simple form
\begin{equation}
\chi_0(\Omega, {\bf q}) =  \frac{1}{N} \sum_{{\bf k}}
\frac{1 - f(\xi_{{\bf k}+{\bf q}/2}) -f(\xi_{-{\bf k}+{\bf q}/2})}
{\xi_{{\bf k}+{\bf q}/2} + \xi_{-{\bf k}+{\bf q}/2} - \Omega}
                          \label{chi.non}
\end{equation}
with $\xi_{\bf k} = \varepsilon_{\bf k} - \mu$ and
$f(x) = (e^{\beta x} +1)^{-1}$ is the Fermi function.

The full Green's function in this
non self-consistent and non-conserving approximation is
the expansion to the first order in the self-energy of
(\ref{Dyson}):
\begin{equation}
G(i \omega _n, {\bf k}) = G_0(i \omega _n, {\bf k})
+ G_0(i \omega _n, {\bf k}) \Sigma_0 (i \omega _n, {\bf k})
G_0(i \omega _n, {\bf k}) \,,    \label{Green.non}
\end{equation}
which means that $G$ in (\ref{Green.non}) and $\Sigma$
(\ref{self.energy.non}) are not calculated self-consistently.

This system corresponds to the approximation used in the
earlier work of Nozi\`eres and Schmitt-Rink \cite{Nozieres}
(see also the reviews \cite{Randeria.book,LSh.review})
to analyze the BCS-Bose crossover in 3D theory, by solving
the system of equation comprising the Thouless
criterion (\ref{Thouless})
\begin{equation}
1 - U \chi_0(0,0) = 0
\label{Thouless.non}
\end{equation}
in terms of $\chi_{0}$ given by Eq. (\ref{chi.non}),
and the number equation (\ref{number.Green.T}) (see below its final
form given by Eq.~(\ref{number.non})).
In particular, to calculate the fluctuation correction in the number
equation (\ref{number.Green.T}) they replaced $\Omega[\Sigma,G]$ by
$\Omega[0,G_{0}]$ before differentiating. In practice, Nozi\`eres and
Schmitt-Rink replace the frequency sum in $\Omega$ by the standard
contour integral representation \cite{Schrieffer,Rickayzen.book,Fetter},
and deform the contour to lie along the real axis before
differentiating with respect to $\mu$. An alternative (but formally
equivalent) procedure is to retain the explicit frequency sum,
and to evaluate the derivative $- \partial \Phi[G_0]/ \partial \mu$
by first differentiating $\Phi$ with respect to $G_0$ using
Eq.~(\ref{stationary}), and then differentiating $G_0$ with respect
to $\mu$. The result is
\begin{equation}
n(\mu,T) = n_0 +
\frac{2}{\beta N} \sum_{n,{\bf k}} G_0(i \omega _n, {\bf k})
\Sigma_0 (i \omega _n, {\bf k}) G_0(i \omega _n, {\bf k}) \,,
\label{number.non}
\end{equation}
which exactly corresponds to the use of the Green's
function (\ref{Green.non}) defined above.
In (\ref{number.non}) the density $n_0$ is expressed via non-interacting
Green's function $G_0$.

Note also that the thermodynamical potential
\begin{equation}
\Omega_{L} \equiv \Phi[G_{0}] =
\frac{1}{\beta} \sum_{m,{\bf q}} \left\{
\ln[1 - U \chi_{0}(i \Omega_m, {\bf q})] +
U \chi_0(i \Omega_m, {\bf q}) \right\}
        \label{Thermopotential.Thouless}
\end{equation}
which includes the contribution from the sum of all particle-particle
ladder diagrams evaluated with non-interacting Green's functions
was firstly studied by Thouless for separable
electron-electron interaction \cite{Thouless.review}.
The equivalent representation of this contribution to the free
energy is given by Gaussian fluctuations of the pair field
in a functional integral representation of the partition function,
as was pointed out by Langer \cite{Langer} (see also the more modern
consideration in \cite{Kleinert}).

An extension of the Nozi\`eres and Schmitt-Rink
approach \cite{Varma} to the case of 2D continuum gas
in the same approximation as described here
suggested the breakdown of the Fermi-liquid picture for
any attraction. Moreover, for any electron density
the chemical potential goes to the energy of two particle bound
states as $T \to 0$ in evident contradiction with the results
of the mean field theory presented in Sec.~\ref{sec:one-band}.
This specific behaviour of $\mu$ has been associated
in \cite{Varma} with the existence of large $\bf q$ bound
states in the 2D system mediated by $s$-wave attraction.
This was, possibly, one of the earliest papers where the
composite hard core bosons were replaced by less stable objects
in an attempt to explain the observed properties of cuprate
superconductors.

However, the approximation scheme applied in \cite{Varma} is not
sufficient for 2D systems because it is not conserving
\cite{Serene,Sofo}, and neglects the interaction between
fluctuations (see also Sec.~\ref{sec:limitations}).
In particular, it was shown in \cite{Sofo} that the values
of entropy, $S$ and specific heat, $C$ become negative in
some temperature interval if the thermodynamical potential
$\Omega[0,G_{0}]$ which leads to the non self-consistent Green's
function (\ref{Green.non}) is used to calculate them.
The Green's function (\ref{Green.non}) even leads to a negative
spectral weight, $A(\omega, {\bf k}) = - \mbox{Im} G(\omega+i0,{\bf k})$.
Indeed, taking $G_{0}(i\omega_n, {\bf k})$ given by (\ref{Green.0})
and $\Sigma_{0}(i\omega_n, {\bf k}) = \mbox{const}$ or simply a
function that does not have a zero at
$i \omega_n = \varepsilon({\bf k})$, one can see that one of the
contributions to $A(\omega, {\bf k})$ coming from the last term
of Eq.~(\ref{Green.non}) is
\begin{equation}
- \mbox{Im} \left( \frac{1}{\omega + i 0 - \varepsilon(\bf k)}\right)^2 =
\mbox{Im} \frac{\partial}{\partial \omega}
\left( \frac{1}{\omega + i 0 - \varepsilon(\bf k)} \right) =
\pi \frac{\partial}{\partial \omega}
\delta(\omega - \varepsilon({\bf k}) )\,.
\end{equation}
The derivative of the delta function is negative in certain regions,
as one can see by taking one of the many representations
of the delta function as the limit of a regular function.
In other words, all double poles in $(G_{0}(i\omega_n, {\bf k}))^2$
result in a negative contribution to the spectral weight. That is
one of the reasons why Dyson's equation (\ref{Dyson}) is necessary:
to sum high-order poles to infinite order and transform them
into an eigenvalue shift.

All of this represents a serious {\em gaffe\/} of
non-conserving, non self-consistent approximation.

\subsection{Non self-consistent, conserving approximation}
\label{sec:Serene}

\begin{figure}[h]
\centering \resizebox{0.6\textwidth}{!}{%
  \includegraphics[bb=35 415 585 530]{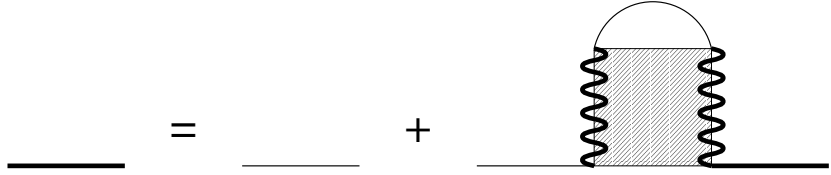}
}
\caption{Diagram for the single-particle Green's function
(solid line) in the non self-consistent, conserving approximation
given by Eqs.~(\ref{Dyson}) and (\ref{self.energy.non}).
The thin solid lines represent the non-interacting Green's
function $G_{0}$. This diagram has been taken from \protect{\cite{Gooding}}.}
\label{fig:14}
\end{figure}

The conserving character of the approximation described in
Sec.~\ref{sec:non} can be, however, restored if we do not
expand the Green's function (\ref{Dyson}) in
terms of $\Sigma$ as we did with Eq.~(\ref{Green.non}). At the same time
one retains the same approximations for the T matrix (\ref{T.matrix.non})
and self-energy (\ref{self.energy.non}) so that the Thouless criterion
is unchanged and given by Eq. (\ref{Thouless.non}).
This conserving but still non self-consistent approximation
 was proposed by Serene \cite{Serene} and
the Green's function in this case is shown diagrammatically
in Fig.~\ref{fig:14}.
As discussed in \cite{Serene}, such an approach entails
taking the sum of all repeated scatterings of an electron by
independent pair fluctuations, but omitting the interaction
between fluctuations and vertex corrections. There is a debate
(see, e.g. \cite{Serene,Sofo,Gooding}) as to whether the restoration
of the conserving character of approximation will preserve
the physics suggested in \cite{Varma}.
Finally we write the number equation (\ref{number.Green.T})
for the non self-consistent, conserving approximation
\begin{equation}
n(\mu,T) =
\frac{2}{\beta N} \sum_{n,{\bf k}} \exp (i \omega_n \delta)
[G_0^{-1}(i\omega_n, {\bf k}) - \Sigma_0(i\omega_n, {\bf k})]^{-1} \,,
\quad \delta \to 0^{+} \,.
             \label{number.Green.cons}
\end{equation}
It is clear that Eqs.~(\ref{number.non}) and
(\ref{number.Green.cons}) differ because the first of them
is obtained in non-conserving approximation. The two approximations
are equivalent when the self-energy corrections are small, but
if the corrections are not small, Eq.~(\ref{number.Green.cons})
might yield a sensible result when the non-conserving
approximation does not.

\subsection{Pairing approximation}

An alternative approximation to that discussed in the previous
section has been used in the set of papers by Levin et al.
\cite{Levin,Levin.quasi2D,Levin.1999}.

The feedback effect of the self-energy $\Sigma$ on the $T$-matrix
has been included into the theory by using the following equation for
the T matrix (compare with Eqs.~(\ref{T.matrix}) and (\ref{T.matrix.non})):
\begin{equation}
T^{-1}(i \Omega_m, {\bf q}) = - \frac{1}{U} +
\frac{1}{\beta N} \sum_{n, {\bf k}}
G_0(i \Omega_m - i \omega _n, {\bf q}-{\bf k})
G(i\omega_n, {\bf k}) \,,
\label{T.pairing}
\end{equation}
where  the full Green's function $G$ is given by the Dyson equation
(\ref{Dyson}) and the self-energy is calculated using $T$-matrix
(\ref{T.pairing}) and the bare Green's function $G_{0}$
\begin{equation}
\Sigma(i \omega _n, {\bf k}) = \frac{1}{\beta N} \sum_{m,{\bf q}}
T (i \Omega_m, {\bf q })
G_{0}(i\Omega_m - i\omega_n, {\bf q} - {\bf k}) \,.
              \label{self.energy.pairing}
\end{equation}
Thus one can see that the feedback effect is introduced into
the theory via the presence of only one full Green's function
in Eq.~(\ref{T.pairing}). And as usually this system should be
completed by the equation for the chemical potential
(\ref{number.Green.T}).

It is stated in \cite{Levin} that the system (\ref{T.pairing}),
(\ref{Dyson}), (\ref{self.energy.pairing}) and
(\ref{number.Green.T}) is conserving because it satisfies
Baym-Kadanoff \cite{Kadanoff} criteria. It is clear, however,
that the theory is not ``$\Phi$-derivable'' and non
self-consistent. Based on the early experimental data which
showed the absence of the pseudogap in the optimally doped
cuprates it was assumed in \cite{Levin} that the correct theory
should be consistent with BCS theory.
Claiming that the fully self-consistent theory  discussed
Sec.~\ref{sec:self} does not reproduce the BCS limit in the weak
coupling regime (see, e.g. \cite{Haussmann.1993,Haussmann.1994})
it was argued in \cite{Levin} that the mode
coupling approach is more appropriate, since it more closely
reproduces the results of BCS theory when the coupling
is small. More recently however it has become clear that
pseudogap phenomena are also observable in the optimally doped region
\cite{Renner}, but over a far smaller range of temperatures than for the
underdoped cuprates. This implies that one needs
reproduce the BCS limit only asymptotically
(see Sec.~\ref{sec:results}), so that the original objection against
the fully self-consistent theory \cite{Levin} no longer seems valid.

It was also shown in \cite{Levin} that the mode coupling approximation
leads to the desired gap-like self-energy
$\Sigma^{R}(\omega, {\bf k}) \sim
1/(\omega+ \epsilon_{\bf k} + i \Gamma_0)$, where $\Gamma_0$ is pair-breaking
scattering (see Sec.~\ref{sec:Franz}).
However it is stated in
\cite{Yanase} that this result is in contradiction with the
fully self-consistent theory. This clearly shows that more
studies are necessary to determine which approximations
are sufficiently good that the results obtained
are not simply an artifact of the approximation.

\subsection{Limitations of the $T$-matrix approximation}
\label{sec:limitations}

In the previous Section we have discussed the discrepancies
between different forms of the $T$-matrix approximation
when it is applied to predominantly 3D models. We note also that
this approximation neglects the particle-hole channel important
for high densities. In 2D there is, however,
a more fundamental problem with the $T$-matrix approximation
used in any, including the fully self-consistent, form.
The essence of the problem is rather well-known: since
the superconducting fluctuations are treated in an RPA-like
or Gaussian approximation the BKT transition is not recovered
by a $T$-matrix approach \cite{Micnas,Kagan,Tremblay.1997}
(see also \cite{Tokumitu,Traven}, where the effect of interaction
between fluctuations was considered).
One may hope however to derive BKT physics, or to be more exact
the properties of the system for $T < T_{\mathrm{BKT}}$\footnote{The
region $T > T_{\mathrm{BKT}}$ is evidently more complicated due to
the presence of vortices which are singular objects.}
only going beyond the $T$-matrix approximation, e.g. including
the equation for the vertex. The $T$-matrix approximation in fact represents
only the lowest order approximation to the infinite chain of Schwinger-Dyson
equations for the fermion and boson Green's functions where these have been
truncated by neglecting the equation for the vertex function and
using the bare vertex.
It is known that in the $T$-matrix approximation in the form
discussed here there is no BKT solution below the mean-field
critical temperature $T_{c}{^\mathrm{MF}}$, i.e. there is
no solution for the boson propagator behaving like $q^{-2 \alpha}$
with $\alpha < 2$ (see Sec.~\ref{sec:phase.mode}).\footnote{Note
that studying the Schwinger-Dyson equations one can speak about
the BKT phase transition only in terms of the propagators.}
As was shown in \cite{Gusynin.PRD} for the 4D gauged
Nambu-Jona-Lasinio model solving the equation for vertex leads to the
desired behaviour of the boson propagator, but the solution
obtained is valid only on the critical line and cannot
be analytically continued into a broken phase (which exists in this model).
This clearly shows that
a more elaborate scheme for the $T$-matrix approximation should
be investigated. Note also, in this connection, that the importance
of the consistency between the one- and two-particle properties was
stressed in \cite{Tremblay.1997}.

We mention also the recent paper \cite{Keller} where it is
stated that $T$-matrix approximation does not yield the pseudogap
behaviour found in quantum Monte-Carlo simulations
\cite{Moreo,Trivedi,Singer,Tremblay.1998,Tremblay.1999,Tremblay.1999a}
and always leads to a Fermi-liquid normal state.

At the same time there is some evidence for pseudogap behaviour
even for the repulsive 2D Hubbard model, but within a more refined
form of the $T$-matrix approximation \cite{Tremblay.1997}.
If the coupling constant is not too large, one can convince
oneself that the spectral weight reveals a two-peak structure near the Fermi
level below a {\em crossover temperature\/}, $T_{X} < T_{c}^{\mathrm{MF}}$.
According to this paper (see also the more recent Ref. \cite{Tremblay.1999a})
the pseudogap starts to appear (in $d=2$, but not in $d =3$)
when the two-particle correlation length becomes
larger than the single-particle de Broglie wave length and the system
enters into a so-called renormalized classical regime, as
confirmed both analytically and numerically. A similar picture
applies to pairing fluctuations slightly away from half-filling
in the attractive Hubbard model where the pseudogap opens,
initially in the region of $k$-space where the Fermi velocity is
smallest.

The Gaussian fluctuations destroy long-range order in 2D \cite{Rice},
and if solving equation (\ref{Thouless}) one searches for the
$T_{c}$ at which long-range order sets in, one obtains zero in
accordance with the CMWH theorem \cite{Coleman,Su}. In some sense
this is an excellent check whether one has a good solution to
the $T$-matrix equations:
the correct solution should give $T_{c} =0$ in 2D.

The most unambiguous acknowledgement of the problem with $T_{c}$
is evident in \cite{Yanase}, where the authors
phenomenologically introduce three-dimensionality and define the
superconducting critical temperature to correspond to a factor of 100
enhancement of the superconducting susceptibility, i.e. the value
of $T_{c}$ follows from the equation (compare with Eq.~(\ref{Thouless}))
\begin{equation}
1 -  U  \chi({\bf q = 0}, \omega = 0) = 0.01 \,.
\label{Thouless.spoiled}
\end{equation}
One can see that this choice of the definition of $T_{c}$ is
entirely arbitrary and with the same success one could
propose an enhancement of $\chi$ by a factor of 1000.
In fact to avoid problems with the correct definition of $T_{c}$ most
$T$-matrix 2D theories are restricted to the discussion
of the phase above $T_{c}$ and one cannot then investigate the
evolution of the pseudogap into the superconducting gap.

There is no doubt that the quasi-2D character of cuprates
must be taken into account and many attempts have
been made to consider quasi-2D models (see, for example,
\cite{Carlson,Levin.quasi2D,Sudbo,Li,our.quasi2D,Roddick,Preosti}).
However for a pure 2D theory one cannot use the $T$-matrix
approach to find the superconducting temperature
and we outline an alternative approach in the next chapter.

\section{The superconducting transition as a BKT transition and the
    temperature scale for the opening of the pseudogap}
\setcounter{equation}{0}
\label{chap:diagram}

\subsection{Phase diagram based on the classical phase
fluctuations in the absence of Coulomb repulsion}
\label{sec:our.diagram}

As discussed in the previous chapter most previous analyses of
the behaviour of 2D systems at $T \neq 0$ have been based on
different extensions of the Nozi\`eres and Schmitt-Rink
\cite{Nozieres} approach (see, e.g. \cite{Varma,Serene,Tokumitu,Traven}).
This approach, and its $T$-matrix extensions as discussed in
Sec.~\ref{sec:limitations}, do not go beyond the Gaussian approximation
for the pairing fluctuations which may explain the abovementioned
difficulties faced in these calculations \cite{Randeria.book}.
It is an interesting problem in the theory of superconductivity to
develop a finite temperature crossover approach in 2D but based on
the BKT transition. Such an analytical treatment
did not however prove easy to develop \cite{Randeria.book}.

Nonetheless, there has been some progress using approaches
which are specifically designed to deal with the phase degree
of freedom. The initial development of the crossover from superconductivity
to superfluidity in 2D as a function of the carrier density $n_f$ was carried
out as long ago as 1992 by Drechsler and Zwerger \cite{Drechsler}
(see also Ref. \cite{Zwerger}, where the $d$-wave case was studied).
By means of a Hubbard-Stratonovich transformation a statistical
Ginzburg-Landau theory was derived. By applying the correct normalization to
make the order parameter a Bose field the following free energy functional
results
\begin{eqnarray}
\Omega(\Phi,{\bf A}) &=& \frac{1}{\beta}
\int d^2r \int_0^{\beta} d \tau \left( - \mu^* |\Phi|^2
+ \frac{1}{2m^*} \left| \left( \frac{1}{i} {\bf \nabla} +
\frac{2e}{c} {\bf A} \right) \Phi \right|^2 \right. \nonumber \\
&+& \left. \Phi^{\ast} \partial_\tau \Phi
+ \frac{g^*}{2} |\Phi|^4 \right) + \frac{{\bf h}^2}{8 \pi}
\end{eqnarray}
where ${\bf h} = {\bf \nabla} \times {\bf A}$ is the magnetic field, $c$
is the speed of light and the effective chemical potential $\mu^*$, effective
mass $m^*$ and effective coupling strength $g^*$ are all determined by
the parameters of the original Ginzburg-Landau theory.
The critical temperature must still however be determined by the introduction
of a third dimension to avoid divergences in the number equation
\cite{Drechsler,Shovkovy,Zwerger}. It can then be shown that one obtains the
correct limiting behaviour for the effective parameters e.g.
the chemical potential at the critical temperature
$\mu_c \equiv \mu^*(T_c)$ attains the value
$-|\varepsilon_b|/2$ in the Bose limit and $\epsilon_F$ in the Fermi limit.
Remarkably one also recovers certain features of BKT physics e.g. in the Bose
limit the critical temperature is close to the BKT critical temperature
for a dilute hard core Bose gas of boson mass $2m$ and density $n_f/2$ on
a lattice
\begin{equation}
T_c \simeq T_{\mathrm{BKT}} = 0.89 \frac{\pi n_f}{4 m}
\end{equation}
We defer the discussion of the Nelson-Kosterlitz jump in the full
(renormalised) superfluid density to Sec.~\ref{sec:impure}. As noticed in
\cite{Alvarez}, the method of \cite{Drechsler} is fully justified only close to
$T_{c}^{\mathrm{MF}}$ where the linearized Ginzburg-Landau theory is valid.
Nor does this method satisfy the important requirement of self-consistency
\cite{Sofo}.

A crossover theory based directly on the BKT transition which avoids
these problems has however been studied in relativistic $2+1$-theory
\cite{MacKenzie}. We now present this approach which has recently been
applied to the theory of superconductivity in a continuum
model \cite{Gusynin.JETP} (see also the generalization to the phonon
model (\ref{Hamiltonian.phonon}) in \cite{Turkowski}). Analogous results,
which will also be discussed, have previously been obtained for the discrete
Hubbard model \cite{Alvarez,Beck.1997,Ariosa.1998}.

For a 2D system one should rewrite the
complex ordering field $\Phi(x)$\footnote{Because
an order parameter is absent in 2D
systems we will use this notation in what follows.}
in terms of its modulus $\rho(x)$ and phase $\theta(x)$ i.e.
$\Phi(x)=\rho(x)\exp[i \theta(x)]$, as
was discussed by Witten in the context of 2D quantum field theory
\cite{Witten} (see also the review \cite{Rosenstein}), and used in the theory
of the superconducting properties of quasi-1D systems \cite{Efetov}
and thin films \cite{Ramakrishnan}.
It is impossible to obtain $\Phi \equiv\langle \Phi(x) \rangle \ne 0$
at finite $T$ since this would correspond to the formation of symmetry
breaking homogeneous long-range order which is forbidden by the CMWH
theorem \cite{Coleman}.  However, it is possible to obtain
$\rho\equiv \langle\rho(x) \rangle \ne 0$ but at the same time
$\Phi = \rho \langle \exp[i\theta(x)] \rangle = 0$ due to random
fluctuations of the phase $\theta(x)$ (i.e. due to transverse
fluctuations of the order field proceeding from the modulus
conservation principle \cite{Patashinskii}). We stress that
$\rho\ne 0$ does not imply any long-range superconducting order
(which is destroyed by the phase fluctuations) and, therefore, is
not in contradiction with the abovementioned theorem. At the same time
one must take into account that $\rho$ is a collective variable which
corresponds to some sort of collective behaviour in the 2D system.

The basic concepts that originate from the modulus-phase representation can be
summarized as follows. Firstly, as in the paper of \cite{Drechsler},
one can develop a low-energy theory for the phase fluctuations in terms of an
effective $XY$ model with a superconducting temperature that is given by the BKT
temperature \cite{Alvarez,Beck.1997,Ariosa.1998,Gusynin.JETP}
(see also \cite{Chakraverty.1997}).
Secondly the amplitude (or modulus) of pairing fluctuations can be non-zero
\cite{Abrikosov.1997,MacKenzie,Gusynin.JETP,Beck.1997,Ariosa.1998,Chakraverty.1997}
even above the BKT transition temperature and first becomes non-zero at roughly
the pairing temperature. We now illustrate these concepts using the
simple continuum model of \cite{Gusynin.JETP} for which one can do much of the
calculation analytically and then discuss the effect of the lattice
\cite{Alvarez} and of quantum phase fluctuations and Coulomb repulsion
\cite{Beck.1997,Ariosa.1998,Chakraverty.1997}.

We stress that the two concepts above are not specific to the pairing
Hamiltonian and/or the BCS-Bose crossover but are related to the
dimensionality of the system. Any 2D theory which is characterized by
a single complex order parameter has a corresponding effective low-energy
theory in terms of the phase of this parameter. The resulting Hamiltonian
is often of $XY$ type with a corresponding BKT superconducting transition
temperature. Furthermore the modulus or amplitude of the order parameter
can be non-zero above the superconducting transition temperature since
this non-zero value does not correspond to long-range order. Thus these
ideas were first proposed in a phenomenological model by Emery and
Kivelson \cite{Emery.1995a,Emery.1995b}
in which there are two temperature scales as discussed above.

In the continuum model (and in fact in general) three regions can be
identified in the 2D phase diagram. The first one is the superconducting
(here BKT) phase with
$\rho \ne 0$ at $T < T_{\mathrm{BKT}}$, where $T_{\mathrm{BKT}}$ is the
temperature of the BKT transition, which plays the role of $T_{c}$ in pure
2D superconducting systems.  In this region there is algebraic order,
or a power law decay of the $\langle\Phi^{\ast}\Phi\rangle$
correlations.  The second region corresponds to 
a phase with non-zero $\rho$ ($T_{\mathrm{BKT}} < T < T_{\rho} $),
where $T_\rho$ is the temperature at which the homogeneous (by definition)
$\rho$ becomes zero.
In this phase although $\rho$ is still non-zero the correlations
mentioned decay exponentially.  The third is a normal
(Fermi-liquid) phase at $T > T_{\rho}$ where $\rho= 0$.\footnote{It
might be the case that
$\rho(x)$ never becomes strictly zero. In this case one would have
crossover behaviour instead of a phase transition, see, for example,
P.~Curty, H.~Beck, Phys.~Rev.~Lett. 85 (2000) 796.}
Note that $\Phi = 0$ everywhere as is the case for all other
correlators which violate the
symmetry e.g. $\langle\Phi(x)\Phi(0)\rangle = 0$. Note that while this
phase diagram was derived for the idealized 2D model, there are
indications that even in layered systems as complicated as HTSC the
critical temperature may nonetheless be well estimated using
$T_{\mathrm{BKT}}$ \cite{Abrikosov.1997,our.quasi2D}, even though the
transition undoubtedly belongs to the 3D $XY$ class (see also
\cite{Babaev}). The identification of the intermediate phase with
the pseudogap phase is somewhat controversial and we will return to
this point in some detail. We note however that it has been
pointed out by Abrikosov \cite{Abrikosov.1997} that a nonzero gap
in the one-particle excitation spectrum can persist even without
long-range order, being a local characteristic (see the discussion
in Sec.~\ref{sec:form}).

The proposed description of the phase fluctuations and the BKT
transition is of course very similar to that proposed earlier by
Emery and Kivelson \cite{Emery.1995a,Emery.1995b}. However in their
phenomenological approach the field $\rho(x)$ does not appear explicitly
while in the present microscopic approach it occurs naturally. We mention
here also the application of similar ideas to the 3D case \cite{Babaev},
where instead of the 2D temperature $T_{\mathrm{BKT}}$
one has the temperature of the phase transition in the 3D $XY$-model,
$T_{c}^{XY}$.

            There is no need to write down the model Hamiltonian
which is studied here, since it is identical with that described
in Sec.~\ref{sec:one-band}.

            The desired phase diagram consisting of normal,
anomalous and superconducting phases was calculated in
\cite{Gusynin.JETP} employing the Hubbard-Stratonovich method (see
Sec.~\ref{sec:formalism}, the equations (\ref{partition.Hubbard}) --
(\ref{Effective.Action})). In the 2D case at nonzero $T$, however,
instead of using the accepted method for the calculation of the partition
function $Z(v, \mu, T,)$ (see (\ref{partition.Hubbard})), one must
perform the calculation in modulus-phase variables. This allows one
to avoid any subsequent treatment of the phase fluctuations at Gaussian
level only.  Thus, one is able to take into account the phase
degree of freedom with the  needed accuracy.

         The modulus-phase variables were introduced in accordance
with \cite{Thouless}, where the parameterization
\begin{equation}
\Phi(x) = \rho(x) e^{i \theta(x)}, \qquad
\Phi^{\ast}(x) = \rho(x) e^{-i \theta(x)},
                      \label{parametrization.Hubbard}
\end{equation}
was used.
At the same time as this replacement (\ref{parametrization.Hubbard})
is implemented, one makes the gauge transformation
\begin{equation}
\psi_{\sigma}(x) = \chi_{\sigma}(x) e^{i \theta(x)/2}, \qquad
\psi_{\sigma}^{\dagger}(x) = \chi_{\sigma}^{\dagger}(x)
e^{-i \theta(x)/2} \,.
                        \label{parametrization.fermi}
\end{equation}
\cite{Witten} (see also Refs. \cite{Rosenstein,Ramakrishnan,Thouless}).
Physically, this amounts to replacing the charged fermion
$\psi_\sigma(x)$ with a neutral fermion $\chi_\sigma(x)$ and a spinless
charged boson $e^{i\theta(x)/2}$.
These parameterizations, (\ref{parametrization.Hubbard}) and
(\ref{parametrization.fermi}), prove more appropriate for presenting the
corresponding functional integral
in two dimensions \cite{Witten}.
In Nambu variables (\ref{Nambu.variables}) the transformation
(\ref{parametrization.fermi}) takes the following form
\begin{equation}
\Psi(x) = e^{i \tau_{3} \theta(x)/2} \Upsilon(x), \qquad
\Psi^{\dagger}(x) = \Upsilon^{\dagger}(x) e^{-i \tau_{3} \theta(x)/2}.
                          \label{parametrization.Nambu}
\end{equation}
Note that the neutral fermion variables are still Grassmann
fields since they satisfy the appropriate anti-commutation algebra.

Making the corresponding substitutions (\ref{parametrization.Nambu})
into the representation (\ref{partition.Hubbard}) and integrating over
the fermi-fields $\Upsilon$ and $\Upsilon^{\dagger}$ we arrive at the
expression (compare with (\ref{partition.effective}) and
(\ref{Effective.Action}))
\begin{equation}
Z(v, \mu, T) = \int \rho \mathcal{D} \rho \mathcal{D} \theta
\exp{[-\beta \Omega (v, \mu, T, \rho(x), \partial \theta (x))]},
                                \label{partition.2D.effective}
\end{equation}
where
\begin{equation}
\beta \Omega (v, \mu, T, \rho (x), \partial \theta (x)) =
\frac{1}{U} \int_{0}^{\beta}
d \tau \int d^2 r \rho^{2}(x) -
\mbox{Tr Ln} G^{-1} + \mbox{Tr Ln} G_{0}^{-1}
                        \label{Effective.Action.2D}
\end{equation}
is, as (\ref{Effective.Action}), the one-loop effective action, which,
however, depends on the modulus-phase variables. The action
(\ref{Effective.Action.2D}) is expressed through the Green's function of
the initial (charged) fermions that has in the new variables the
following operator form\footnote{It may
be obtained as a solution of a differential
equation with anti-periodic boundary conditions
(see (\ref{Green.fermion}) and (\ref{boundary})).}
\begin{equation}
G^{-1}  =  -\hat{I} \partial_{\tau} +
\tau_{3} \left(\frac{\nabla^{2}}{2 m} + \mu \right) +
\tau_{1} \rho(\tau,  {\bf r})
 -  \Sigma(\partial \theta)
                          \label{Green.fermion.phase}
\end{equation}
with
\begin{equation}
\Sigma(\partial \theta) \equiv
\tau_{3} \left[\frac{i \partial_{\tau} \theta}{2} +
\frac{(\nabla \theta)^{2}}{8 m} \right] -
\hat{I} \left[\frac{i \nabla^{2} \theta}{4 m} +
\frac{i \nabla \theta(\tau,  {\bf r}) \nabla}{2 m} \right].
                          \label{Sigma}
\end{equation}

The free fermion Green's function $G_{0} = G|_{\mu, \rho, \theta=0}$
provides a convenient regularization in the process of calculation.
It is important that neither the smallness nor slowness of the variation
of the phase of the order parameter is assumed in obtaining expression
(\ref{Effective.Action.2D}). In other words, it is formally exact.

Since the low-energy dynamics of phases for which
$\rho \neq 0$ is governed mainly by long-wavelength fluctuations
of $\theta(x)$, only the lowest-order derivatives of the phase need be
retained in the expansion of
$\Omega(v, \mu, T, \rho(x), \partial \theta(x))$:
\begin{equation}
\Omega (v, \mu, \rho(x), \partial \theta(x))  \simeq
\Omega _{\rm kin} (v, \mu, T, \rho, \partial \theta(x)) +
\Omega _{\rm pot} (v, \mu, T, \rho),
\label{kinetic.phase+potential}
\end{equation}
where
\begin{equation}
\Omega _{\rm kin} (v, \mu, T, \rho, \partial \theta(x))
 = T\, \mbox{Tr} \sum_{n=1}^{\infty}
\left. \frac{1}{n} (\mathcal{G} \Sigma)^{n} \right|_{\rho = const}
\label{Omega.Kinetic.phase}
\end{equation}
and
\begin{equation}
\Omega _{\rm pot} (v, \mu, T, \rho)  =
\left. \left(\frac{1}{U} \int d^2  r \rho^{2} -
 T\, \mbox{Tr}\ln\mathcal{G}^{-1} +
 T\, \mbox{Tr}\ln G_{0}^{-1} \right) \right|_{\rho = const}.
\label{Omega.Potential.modulus}
\end{equation}
The kinetic $\Omega_{\mathrm{kin}}$ and potential
$\Omega_{\mathrm{pot}}$ parts can be expressed in terms of the Green's
function of the neutral fermions, which satisfies the equation
\begin{equation}
\left[-\hat I \partial_{\tau} + \tau_{3}
\left(\frac{\nabla^{2}}{2 m} + \mu \right) + \tau_{1} \rho \right]
\mathcal{G}(\tau,  {\bf r}) = \delta(\tau) \delta( {\bf r})
\label{Green.fermion.modulus}
\end{equation}
and the operator (\ref{Sigma}).

The representation (\ref{kinetic.phase+potential}) enables
one to obtain the full set of equations necessary to find
$T_{\mathrm{BKT}}$, $\rho(T_{\mathrm{BKT}})$, and $\mu(T_{\mathrm{BKT}})$ at
given $\epsilon_{F}$ (or, for example, $\rho(T)$ and $\mu(T)$ at given $T$
and $\epsilon_{F}$).  While the equation for $T_{\mathrm{BKT}}$ will be
written using the kinetic part (\ref{Omega.Kinetic.phase}) of the effective
action, the equations for $\rho(T_{\mathrm{BKT}})$ and
$\mu(T_{\mathrm{BKT}})$ (or $\rho(T)$ and $\mu(T)$) can be obtained using the
mean field potential (\ref{Omega.Potential.modulus}). It turns out that in a
phase for which $\rho \neq 0$, the mean-field approximation for the modulus
variable describes the system quite well. This is mainly related to the
non-perturbative character of the Hubbard-Stratonovich method, i.e., most
of the pairing effects are included in the nonzero value of $\rho$.

It is clear that the CMWH theorem \cite{Coleman} does not preclude
nonzero $\langle\rho\rangle$ and, as a consequence,  an energy gap
for fermion $\chi$, since no continuous symmetry is broken when such a
gap appears. Despite  strong phase fluctuations in the two-dimensional
case, the energy gap in the spectrum of the neutral fermion $\chi$
still persist in the spectrum of the charged fermion $\psi$
(see Chap.~\ref{chap:spectral} and \cite{Gusynin.new}), even well
above the critical temperature.\footnote{We note that the specific
heat experiments \cite{Timusk,Randeria.review} demonstrated a loss
of entropy at temperatures much higher than $T_{c}$. This
can be considered indicative of a degenerate normal state, consistent
with the existence of a nonzero order parameter $\langle\rho(x) \rangle$.}

Thus, within the proposed scenario, the pseudogap properties might
conceivably be attributable to the energy gap of a neutral fermion introduced
in the way described above, so that the pseudogap itself can be considered
as a remnant of the superconducting gap.  The condensate of neutral
fermions is completely unrelated to the superconducting transition; the
latter is only possible when the superfluid density of bosons becomes
large enough to stiffen the phase $\theta(x)$. The temperature $T_\rho$
at which nonzero $\langle\rho(x)\rangle$ develops should be identified in
this approach with the pseudogap onset temperature $T^{\ast}$ (see
Sec.~\ref{sec:results} item {\rm a} and the discussion about heavily
underdoped region in Sec.~\ref{sec:pairing.temperature}).

The strategy of treating charge and spin degrees
of freedom as independent seems to be a useful, and at the same
time general feature of two-dimensional systems \cite{Witten,Ramakrishnan} and
low-dimensional systems in general
(see also the review \cite{Wiegmann}).

\subsubsection{Derivation of self-consistent  equations for
$T_{\mathrm{BKT}}$, neutral order parameter, and chemical potential}
\label{sec:derivation}

When the model under consideration is reduced to  some known model
describing the BKT phase transition, one can easily write down an
equation for $T_{\mathrm{BKT}}$, which in the present approach can be
identified with the superconducting transition temperature $T_{c}$.
Indeed, in the lowest orders the kinetic term (\ref{Omega.Kinetic.phase})
coincides with the classical spin $XY$-model
\cite{Minnhagen,Pierson,Izyumov,Plischke,Huang}, which has the continuum
Hamiltonian
\begin{equation}
H_{\mathrm{XY}} = \frac{J}{2} \int d^2 r\,
[\nabla \theta ({\bf r})]^{2} .      \label{XY.Hamiltonian}
\end{equation}
Here $J$ is the some coefficient (in the original classical discrete
$XY$-model it is the stiffness of the relatively small spin rotations) and
$\theta$ is the angle (phase) of the two-component vector in the plane.

The temperature of the BKT transition is, in fact, known
for this model:
\begin{equation}
T_{\rm BKT} = \frac{\pi}{2} J.
                        \label{BKT.temperature}
\end{equation}
Despite the very simple form of Eq.~(\ref{BKT.temperature}), it was
derived (see, e.g., Refs. \cite{Minnhagen,Pierson,Izyumov,Plischke,Huang})
using the renormalization group technique, which takes into account the
non-single-valuedness of the phase $\theta$.
Thus, fluctuations of the phase are taken into account in a higher
approximation than Gaussian and this approximation may well be better
than the $T$-matrix even in its fully self-consistent conserving form.
However, the disadvantage of such an effective theory of the phase
fluctuations is that one loses contact with the underlying fermions
(they are essentially integrated over to obtain the effective
theory\footnote{This contact, however, would not be lost if one
retained the source terms for the fermi-fields in the corresponding
functional integral. See also the approach of \cite{Dorsey} where the
low energy fermions are not integrated out.}). From this point of
view, these two approaches are essentially complementary.

The $XY$-model was assumed to be
adequate for a qualitative phenomenological
description of the underdoped cuprates
\cite{Emery.1995a,Emery.1995b,Carlson} (see also
Refs.~\cite{Doniach,Beck.1997,Sachdev}), and the relevance of the BKT
transition to Bose- and BCS-like superconductors was recently
discussed in Ref.~\cite{Zwerger} (see also \cite{Halperin}).
The most important difference between superfluidity in $^4$He films,
which is adequately described by the $XY$-model \cite{Minnhagen}, and
superconductivity is that the superconducting liquid is charged.
This results in coupling to the magnetic field, which inevitably
accompanies any current flow in the superconductor, e.g. in the presence of
vortices. However, as was shown in \cite{Halperin,Zwerger} the effective
penetration depth for the magnetic field in thin films is of the order
of 1~cm, and thus for sample sizes smaller than this, magnetic screening
becomes irrelevant. In such a situation the
difference between charged and neutral superfluids becomes irrelevant
and superconducting films may undergo the BKT transition.
Note also that this difference leads also to electric
coupling between Cooper pairs, in particular those situated in
different superconducting layers. The effect of this coupling
is considered below in Sec.~\ref{sec:Coulomb}.

To expand $\Omega _{\rm kin}$ up to $\sim (\nabla \theta)^{2}$, it is
sufficient to restrict  ourselves to terms with $n=1,2$ in the expansion
(\ref{Omega.Kinetic.phase}). The calculation is similar to that employed in
\cite{Thouless}, where only high densities $n_{f}$ were considered at $T=0$.
(We mention also the derivation of the effective action at $T =0$ in
\cite{Palo.1999} where the fluctuations in the density (conjugated to the
phase $\theta$) were also treated.) Thus, to obtain the kinetic part, one
should directly calculate the first two terms of the series
(\ref{Omega.Kinetic.phase}), which can be formally written
$\Omega_{\mathrm{kin}}^{(1)} = T \mbox{Tr} (\mathcal{G} \Sigma)$ and
$\Omega_{\mathrm{kin}}^{(2)} = \frac{1}{2}T \mbox{Tr} (\mathcal{ G} \Sigma
\mathcal{ G} \Sigma)$. We note that $\Sigma$ has the structure $\Sigma =
\tau_{3} O_{1} + \hat{I} O_{2}$, where $O_{1}$ and $O_{2}$ are differential
operators (see (\ref{Sigma})).  One can see, however, that the part of
$\Sigma$ proportional to the unit matrix $\hat{I}$ does not contribute to
$\Omega_{\rm kin}^{(1)}$.  Hence,
\begin{equation}
\Omega_{\rm kin}^{(1)} = T \int_{0}^{\beta} d \tau \int d^2 r\,
\frac{T}{(2 \pi)^{2}} \sum_{n = - \infty}^{\infty}
\int d^2 k\, \mbox{Tr} [\mathcal{ G}(i \omega_{n},  {\bf k})
\tau_{3}] \left( \frac{i \partial_{\tau} \theta}{2} + \frac{(\nabla
\theta)^{2}}{8 m}\right),         \label{K1}
\end{equation}
where (compare with (\ref{Green.fermion.momentum}))
\begin{equation}
\mathcal{ G}(i \omega_{n},  {\bf k}) = - \frac{
i \omega_{n} \hat{I} + \tau_{3} \xi( {\bf k}) - \tau_{1} \rho}
{\omega_{n}^{2} + \xi^{2}( {\bf k}) + \rho^{2}}     \label{K2}
\end{equation}
is the Green's function of neutral fermions in the frequency-momentum
representation, with
$\xi( {\bf k})$ and $\varepsilon ( {\bf k})$ defined after
(\ref{Green.fermion.momentum}).
The summation over the fermionic Matsubara
frequencies $\omega_{n} = \pi (2n + 1) T$ and integration over
$ {\bf k}$ in (\ref{K1}) can be easily performed
using the sum (\ref{A8});  one thus obtains
\begin{equation}
\Omega_{\rm kin}^{(1)} =
T \int_{0}^{\beta} d \tau \int d^2  r\,
n_{F}(\mu, T, \rho) \left(
\frac{i \partial_{\tau} \theta}{2} +
\frac{(\nabla \theta)^{2}}{8 m}
\right),                \label{K3}
\end{equation} where
\begin{equation}
n_{F}(\mu, T, \rho(\mu, T)) = \frac{m}{2 \pi} \left\{
\sqrt{\mu^{2} + \rho^{2}} + \mu +
2 T \ln{\left[
1 + \exp{\left(-\frac{\sqrt{\mu^{2} + \rho^{2}}}{T}\right)}
\right]}\right\}.
                 \label{K4}
\end{equation}
This has the form of a Fermi quasi-particle density
(for $\rho = 0$ the expression (\ref{K4}) is
simply the density of free fermions given by the first term $n_0$ in
(\ref{number.non})).

For the case $T = 0$ \cite{Thouless}, in which the real time $t$
replaces the imaginary time $\tau$, one can argue from Galilean
invariance that the coefficient of $\partial_{t} \theta$
is rigorously related to the coefficient of $(\nabla \theta)^{2}$.
It therefore does not appear in $\Omega_{\rm kin}^{(2)}$.
We wish, however, to stress that these arguments cannot be used
to eliminate the term $(\nabla \theta)^{2}$
from $\Omega_{\rm kin}^{(2)}$ when $T \neq 0$, so that we must calculate
it explicitly.

The $O_{1}$ term in $\Sigma$ yields
\begin{eqnarray}
\Omega_{\rm kin}^{(2)} (O_{1}) & = & - \frac{T}{2}\int_{0}^{\beta} d \tau
\int d^2 r\,\frac{T}{(2 \pi)^{2}} \sum_{n = - \infty}^{\infty}
\int d^2  k\, {Tr} [\mathcal{ G}(i \omega_{n},  {\bf k})
\tau_{3}\mathcal{ G}(i \omega_{n},  {\bf k}) \tau_{3}]
\nonumber                    \\
&& \times \left( \frac{i
\partial_{\tau} \theta}{2} + \frac{(\nabla \theta)^{2}}{8 m}\right)^{2}.
\label{K5}
\end{eqnarray}
Using (\ref{A12}) to calculate the sum over the Matsubara
frequencies, we find that
\begin{equation}
\Omega_{\rm kin}^{(2)} (O_{1}) = - \frac{T}{2}
\int_{0}^{\beta} d \tau \int d^2  r\, K(\mu, T, \rho)
\left( i \partial_{\tau} \theta +
\frac{(\nabla \theta)^{2}}{4 m}\right)^{2},
                                 \label{K6}
\end{equation}
where
\begin{equation}
K(\mu, T, \rho(\mu, T)) =
\frac{m}{8 \pi} \left(1 + \frac{\mu}{\sqrt{\mu^2 + \rho^2}}
\tanh{\frac{\sqrt{\mu^2 + \rho^2}}{2T}} \right).
                                 \label{K7}
\end{equation}

Obviously, the $O_{1}$ term does not affect the coefficient
of $(\nabla \theta)^{2}$.
Further, it is easy to make sure that the cross term involving
$O_{1}$ and $O_{2}$ in $\Omega_{\rm kin}^{(2)}$ is absent.
Finally, calculations of the $O_{2}$ contribution
to $\Omega_{\rm kin}^{(2)}$ yield\footnote{
Derivatives higher than $(\nabla \theta)^{2}$
were not found here.}
\begin{equation}
\Omega_{\rm kin}^{(2)} (O_{2}) =  T
\int_{0}^{\beta} d \tau \int d^2 r\,
\frac{T}{(2 \pi)^{2}} \sum_{n = - \infty}^{\infty}
\int d^2 k\,  {\bf k}^{2}
\mbox{Tr} [\mathcal{ G}(i \omega_{n},  {\bf k}) \hat{I}
           \mathcal{ G}(i \omega_{n},  {\bf k}) \hat{I}]
\frac{(\nabla \theta)^{2}}{16 m^{2}}.                 \label{K8}
\end{equation}
Thus, summing over the Matsubara frequencies (see Eq.~(\ref{A13})),
one obtains
\begin{equation}
\Omega_{\rm kin}^{(2)} (O_{2}) = - \int_{0}^{\beta} d \tau \int d^2  r\,
\frac{1}{128 \pi^{2} m^{2}} \int d^2  k\,\frac{ {\bf k}^{2}}{\cosh^{2}
\frac{\ds \sqrt{ \xi^{2}( {\bf k}) + \rho^{2}}}{\ds 2T}}(\nabla \theta)^{2}.
                                                      \label{K9}
\end{equation}
As expected, this term vanishes when $T \to 0$,
but at finite $T$ it is comparable with (\ref{K3}).

Combining (\ref{K3}), (\ref{K9}), and (\ref{K6}) we finally obtain
\begin{equation}
\Omega _{\rm kin} =
\frac{T}{2}
\int_{0}^{\beta} d \tau \int d^2  r
\left[n_{F}(\mu, T, \rho) i \partial_{\tau} \theta +
J(\mu, T, \rho) (\nabla \theta)^{2} +
K(\mu, T, \rho) (\partial_{\tau} \theta)^{2}
\right],
                 \label{Omega.Kinetic.phase.final}
\end{equation}
where
\begin{equation}
J(\mu, T, \rho(\mu, T)) = \frac{1}{4 m} n_{F}(\mu, T, \rho) -
\frac{T}{4 \pi} \int_{-\mu/2T}^{\infty} dx\,
\frac{x + \mu/2T}{\cosh^{2} \sqrt{x^{2} +
\frac{\ds \rho^{2}}{\ds 4 T^{2}}}}
                              \label{K10}
\end{equation}
characterizes the phase stiffness and governs the spatial variation of the
phase $\theta( {\bf r})$. One can see that the value of the phase stiffness
$J(T=0)$ coincides with the nonrenormalized stiffness used in Ref.
\cite{Emery.1995a}.

During the derivation of the effective action for the phase
field we have neglected the so-called {\em Landau terms\/}
(see \cite{Aitchison} and Refs. therein) which
have a non-local form and thus cannot be presented in terms
of the derivative expansion. These terms are important
since they make the $\theta$-mode decaying for any $T \neq 0$.
However, as recently shown in \cite{Aitchison}, for $s$-wave
superconductors their contribution is negligible at least for
$T \lesssim 0.6 T_{\rho}$. This is not the case for $d$-wave
superconductors where these terms are expected to be important
even in the low temperature region.

The quantity $J(\mu,T, \rho)$ vanishes at $\rho=0$, which means that
above $T_{\rho}$ the modulus-phase variables are meaningless; to
study the model in this region one must return to the old variables
$\Phi$ and $\Phi^{\ast}$. Near  $T_\rho$ one can
obtain from (\ref{K10}) in the high-density limit (see below)
\begin{equation}
J(\mu\simeq\epsilon_{F}, T \to T_{\rho}^{-}, \rho \to 0) =
\frac{7 \zeta(3)}{16 \pi^{3}} \frac{\rho^{2}}{T_{\rho}^{2}} \epsilon_{F}
\simeq 0.016 \frac{\rho^{2}}{T_{\rho}^{2}} \epsilon_{F} \,,
\label{J.bcs}
\end{equation}
where $\zeta$ is the zeta function.

Direct comparison of (\ref{Omega.Kinetic.phase.final}) with the Hamiltonian
of the $XY$-model
(\ref{XY.Hamiltonian}) makes it possible to write
Eq.~(\ref{BKT.temperature}) for $T_{\mathrm{BKT}}$ directly:
\begin{equation}
T_{\mathrm{BKT}} =
\frac{\pi}{2} J(\mu, T_{\mathrm{BKT}}, \rho(\mu, T_{\mathrm{BKT}})) .
\label{BKT.equation}
\end{equation}
Although mathematically this  reduces to a well-known problem,
the analogy is incomplete. Indeed, in the standard $XY$-model
(as well as the nonlinear $\sigma$-model) the vector (spin)
subject to ordering is assumed to be a unit vector with no dependence
on $T$.
In this case this is definitely not the case, and a
self-consistent calculation of $T_{\mathrm{BKT}}$ as a function of $n_{f}$
requires additional equations for $\rho$ and $\mu$, which together
with (\ref{BKT.equation}) form a complete set.
We note, however, that the dependence of $J$ on $T$ was neglected in
\cite{Emery.1995a}, where the nonrenormalized phase stiffness $J(T=0)$ was
used to write Eq.~(\ref{BKT.equation}). Using this nonrenormalized phase
stiffness one would obtain that $T_{\mathrm{BKT}} \sim \epsilon_{F}$ which
is, as we will see, apparently wrong for high carrier densities since
one does not recover the BCS limit.

Using the definition (\ref{Omega.Potential.modulus}), one
can derive the effective potential
$\Omega _{\mathrm{pot}}(v, \mu, T, \rho)$
(see Appendix~\ref{sec:A}).
Then the desired missing equations are the condition \\
$\partial \Omega_{\mathrm{pot}}(\rho)/ \partial \rho = 0$
that the potential (\ref{A11}) be minimized, and the equality
$v^{-1} \partial \Omega_{\mathrm{pot}} /\partial \mu = -n_{f}$, which
fixes $n_{f}$.
These are, respectively
(compare with Eqs.~(\ref{minima}) -- (\ref{number.2D.band}))
\begin{equation}
\frac{1}{U} = \int \frac{d^2  k}{(2 \pi)^{2}}
\frac{1}{2\sqrt{\xi^{2}( {\bf k}) + \rho^{2}}}
\tanh{\frac{\sqrt{\xi^{2}( {\bf k}) +
\rho^{2}}}{2 T}},
\label{rho}
\end{equation}
\begin{equation}
n_{F}(\mu, T, \rho) = n_{f},
\label{number.rho}
\end{equation}
where $n_{F}(\mu, T, \rho)$ is defined by (\ref{K4}).

Equations (\ref{rho}) and (\ref{number.rho})
comprise a self-consistent system for determining the modulus $\rho$
of the order parameter and the chemical potential $\mu$ in the
mean-field approximation for fixed $T$ and $n_{f}$.

It is very interesting to note that to resolve the problems which
are present in the non-conserving, non self-consistent
approach (see Sec.~\ref{sec:non}), the following form of the phenomenological
pair susceptibility (compare with Eqs.~(\ref{chi}) and (\ref{chi.non}))
was suggested in \cite{Sofo}
\begin{equation}
\chi(\Omega, {\bf q}) =  \frac{1}{N} \sum_{{\bf k}}
\frac{1 - \langle n_{{\bf k}+{\bf q}/2} \rangle
- \langle n_{-{\bf k}+{\bf q}/2} \rangle }
{\xi_{{\bf k}+{\bf q}/2} + \xi_{-{\bf k}+{\bf q}/2} - \Omega} \,;
                          \label{chi.phen}
\end{equation}
where $ \langle n_{\bf k} \rangle$ is a function with a free
parameter $\varphi$, and has the form (compare with (\ref{K4}))
\begin{equation}
\langle n_{\bf k} \rangle = \frac{1}{2} \left[1 -
\frac{\xi_{\bf k}}{\sqrt{\xi_{\bf k}^2 + \varphi^{2}}}
\tanh \frac{\sqrt{\xi_{\bf k}^2 + \varphi^{2}}}{2T} \right].
\label{n}
\end{equation}
It was stressed in \cite{Sofo} that $\varphi$  is not
an off diagonal long-range order parameter but is related to
pair binding. The phenomenological susceptibility
(\ref{chi.phen}) was then used in \cite{Sofo} to write down the
thermodynamical potential $\Omega[0,G]$ (see \ref{Omega}) where
$\Phi[G]$ (see \ref{Phi}) is given in terms of (\ref{chi.phen}).
This thermodynamical potential was then used to
derive equations for the chemical potential
(which coincides with (\ref{number.rho})), entropy and specific
heat. An additional equation for $\varphi$ was derived in
\cite{Sofo} by minimizing this potential with respect to $\varphi$ subject to
the number constraint $2 \sum_{\bf k} \langle n_{\bf k} \rangle = n_f$.
It was shown in \cite{Sofo} that the presence of $\varphi$ avoids problems with
negative entropy and specific heat which are present in the non-conserving, non
self-consistent approximation discussed in Sec.~\ref{sec:non}.

Eqs.~(\ref{rho}) and (\ref{number.rho}) seem to yield a
reasonable approximation at high densities $n_{f}$, since they include
the condensed boson pairs in a non-perturbative way via nonzero $\rho$.
Nonetheless they must certainly be corrected in the strong coupling regime (low
densities $n_{f}$) to take into account the contribution of noncondensed
bosons (this also appears to be important for Eq.~(\ref{BKT.equation}),
which determines $T_{\mathrm{BKT}}$).  The extent to which this alters
the present results is not completely clear. Previously, the best way
to incorporate noncondensed pairs appears to have been the self-consistent
$T$-matrix approximation (see e.g.
\cite{Levin,Tchernyshyov,Serene,Micnas} and other references in
Secs.~\ref{sec:survey} and \ref{sec:T.matrix}), which allows one to
account for the feedback of pairs on the self-energy of fermions.
However as we discussed above, the $T$-matrix approach, at least in
its standard form presented in Sec.~\ref{sec:T.matrix}, fails to
describe the BKT phase transition, for which one must consider the
equation for the vertex. On the other hand, in this approach the BKT
phase transition is realized by the condition (\ref{BKT.temperature}),
while an analog of the $T$-matrix approximation in terms of propagators
of the $\rho$-particle and the neutral fermion $\chi$ has yet to be
developed.

Again the energy of two-particle bound states in a vacuum
defined by (\ref{bound.energy})
is more convenient to use than the four-fermion constant $U$.
For example, one can easily take the limits $W \to \infty$ and
$U \to 0$ in Eq.~(\ref{rho}), which after this renormalization
becomes
\begin{equation}
\ln{\frac{|\varepsilon_{b}|}{\sqrt{\mu^{2} +
\rho^{2}} - \mu }} = 2 \int_{-\mu/T}^{\infty} d u\,
\frac{1}{\ds \sqrt{u^2 + \left( \frac{\rho}{ T} \right)^2}
\left[\exp{\sqrt{u^2 + \left( \frac{\rho}{T} \right)^2}} + 1 \right]} .
\label{rho.bound}
\end{equation}
Thus, in practice, we solve Eqs.~(\ref{BKT.equation}),
(\ref{number.rho}), and (\ref{rho.bound}) analytically and numerically
to study $T_{\rm BKT}$ as function of $n_{f}$
(or equivalently, of the Fermi energy $\epsilon_{F} = \pi n_{f}/m$).
It is easy to see that at $T = 0$, the system (\ref{number.rho}),
(\ref{rho.bound})  transforms into a previously studied system
(\ref{gap+number.2D}).
Setting $\rho = 0$ in Eqs.~(\ref{rho}) and
(\ref{number.rho}), we obtain (in the same approximation) the
equations for the critical temperature $T_{\rho}$:
\begin{equation}
\ln{\frac{|\varepsilon_{b}|}{T_{\rho}}\frac{\gamma}{\pi}} =
- \int_{0}^{\mu/2T_{\rho}} d u\, \frac{\tanh{u}}{u} \,,
\label{temperature.rho}
\end{equation}
where $\gamma \simeq 1.781$ ($ C = \ln \gamma \simeq 0.577$ is the
Euler's -- Mascheroni's constant) and the corresponding value of $\mu$:
\begin{equation}
T_{\rho} \ln{\left[1 +
\exp{ \left( \frac{\mu}{T_{\rho}} \right)}\right]} = \epsilon_{F}\,,
\label{number.temperature.rho}
\end{equation}

Note that these equations coincide with the system that determines
the mean-field temperature $T_{c}^{(2D) \mathrm{MF}}\,(=T_\rho)$ and
$\mu(T_{c}^{(2D) \mathrm{MF}})$, evidently as a result of the
mean-field approximation for the variable $\rho$ used here.
Thus in this approximation the average amplitude $\rho$ of the pairing
fluctuations is simply given by the mean-field gap $\Delta^{\mathrm{MF}}$ which is
non-zero below  $T_c^{(2D) \mathrm{MF}}$ ($= T_\rho$). This non-zero amplitude then
plays the role of the ``pseudogap''. Note that the paramagnetic susceptibility
(see \cite{Gusynin.JETP}), specific heat in the mean-field approximation
exhibit kink-like behaviour at $T_{\rho}$ while at the superconducting
transition temperature $T_{\mathrm{BKT}}$ they do not show any singularities.
The picture presented here is thus to all practical purposes identical to that
given in \cite{Beck.1997,Ariosa.1998}.

There is an important difference between the
temperatures $T_{c}^{2D}$ and $T_\rho$.  Specifically, if one takes
fluctuations into account, $T_{c}^{2D}$ goes to zero, while the value of
$T_{\rho}$ remains finite. The crucial point is that the perturbation theory
in the variables $\rho$ and $\theta$ does not contain any infrared singularities
\cite{Witten,Ichinose}, in contrast to the perturbation theory in
$\Phi,\Phi^{\ast}$; thus the fluctuations do not reduce $T_\rho$ to zero.
This is why the temperature $T_{\rho}$ can have a physical meaning:
incoherent (local or Cooper) pairs begin to form (at least at high enough
$n_{f}$, see Secs.~\ref{sec:results} item {\rm c}
and \ref{sec:pairing.temperature}) just below $T_{\rho}$.  At higher
temperatures, only pair fluctuations exist (see, e.g.
\cite{LSh.FNT.1997}).

\subsubsection{Main results}
\label{sec:results}

The interested reader may find a more detailed discussion of the results
for the continuum model in \cite{Gusynin.JETP}. Here we present only the
issues directly  related to the phase diagram of the model
(\ref{Hamiltonian.2D}).
A numerical and analytical investigation of the systems
(\ref{BKT.equation}), (\ref{number.rho}), (\ref{rho.bound}), and
(\ref{temperature.rho}), (\ref{number.temperature.rho}) yields the
following results, which are displayed graphically as the phase
diagram of the system.

\begin{figure}[h]
\centering \resizebox{0.6\textwidth}{!}{%
  \includegraphics{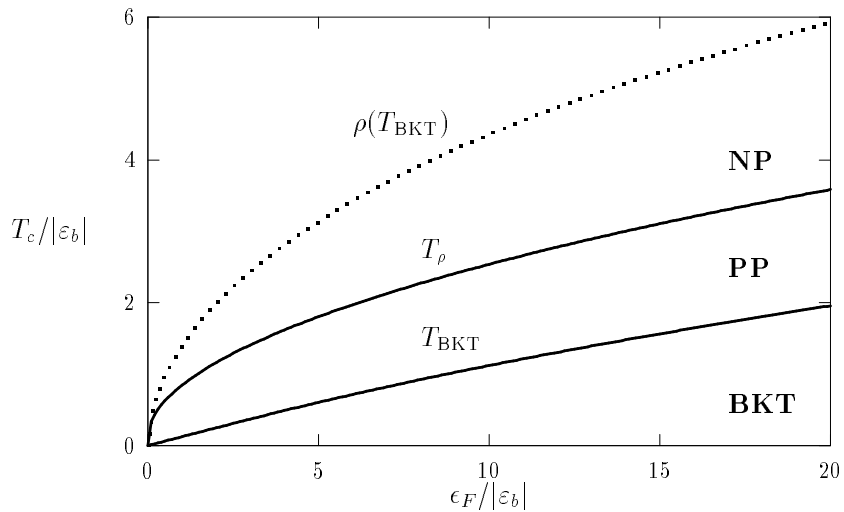}}
\caption{$T_{\mathrm{BKT}}$ and $T_{\rho}$ versus the noninteracting fermion
density (taken from \protect{\cite{Gusynin.JETP}}). Dots represent the function
$\rho(\epsilon_{F})$ at $T = T_{\mathrm{BKT}}$.  The regions of normal phase
(NP), pseudogap phase (PP), and BKT phase are indicated.} \label{fig:15}
\end{figure}

{\rm a)} For rather low carrier densities, the pseudogap phase area
(see Fig.~\ref{fig:15}) is comparable with the BKT area.

For high carrier densities
($\epsilon_{F} \gtrsim 10^2 \, - \, 10^3 |\varepsilon_{b}|$),
inserting into (\ref{J.bcs}) the well-known dependence of $\rho(T)$
(see, for example \cite{Schrieffer})
\begin{equation}
\rho^{2}(T \to T_{\rho}^{-}) =
\frac{8 \pi^{2}}{7 \zeta(3)} T_{\rho}^{2} \left(1 - \frac{T}{T_{\rho}}
\right)
\label{rho.dependence}
\end{equation}
and then substituting (\ref{J.bcs}) into
(\ref{BKT.equation}) one obtains the following
asymptotic expression for the BKT temperature (also given in
\cite{Halperin,Randeria.review,Babaev} and \cite{Turkowski},
where the corresponding asymptotical expression for the boson-exchange
model (\ref{Hamiltonian.phonon}) was obtained)
\begin{equation}
T_{\mathrm{BKT}} = T_{\rho}
\left(1 - \frac{4T_{\rho}}{\epsilon_{F}} \right),
\qquad T_{\mathrm{BKT}} \lesssim T_{\rho} \,,
\label{asymp}
\end{equation}
which shows that the pseudogap region shrinks asymptotically.
This behaviour qualitatively restores the BCS
limit.  This is a good place to consider the difference between
the two high-density limits shown in Fig.~\ref{fig:02}-A and -B
respectively. As one can easily see, Eq.~(\ref{asymp}) implies
that the present results are in favour of the picture presented in
Fig.~\ref{fig:02}-B.\footnote{Of course, $T_{\mathrm{BKT}}$ does
not reach zero at high doping due to the oversimplified
character of the model. For example, we saw in Sec.~\ref{sec:one-band}
that $T_{\mathrm{BKT}}$  may decrease due to band
filling or at the very least stabilizes as in the model with phonon
interaction from Sec.~\ref{sec:indirect}. Nonetheless one clearly
sees that $T_{\rho}$ and $T_{\mathrm{BKT}}$ are merging.}
The merger of $T^{\ast}$ and $T_{\mathrm{BKT}}$ in the approach
based on the superconducting fluctuations is probably not that
surprising since both these temperatures have the same superconducting
origin.
However, as one can see from Fig.~\ref{fig:02}-A this merger
would be rather unusual if one accepts the antiferromagnetic
pseudogap scenario and considers $T^{\ast}$ to have a
magnetic nature, different from that of $T_{c}$.
Thus, as was suggested in \cite{Zachar}, additional studies on the
overdoped cuprates in a magnetic field may also help to clarify the
pseudogap origin.

It has been shown that for optimal doping, the dimensionless ratio can be
estimated as $\epsilon_{F}/|\varepsilon_{b}| \sim 3 \cdot10^{2}\,-\, 10^{3}$
\cite{Carter}.  This implies that the size of the pseudogap phase predicted by
(\ref{asymp}) to be of order $T_{\mathrm{BKT}}^{2}/\epsilon_{F}$
\cite{Halperin,Randeria.review,Babaev} is (even in the underdoped region) much
smaller than the difference $T^{\ast} - T_{c}$ observed in cuprates. One might
however be able to explain this discrepancy either by the presence of 
impurities as done in Sec.~\ref{sec:impure} or by the formation of spatial
inhomogeneities e.g. stripes.

{\rm b)} For $\epsilon_{F} \leq (10\,-\,15) |\varepsilon_{b}|$, the function
$T_{\rm BKT}(\epsilon_{F})$ is linear, as is confirmed by the analytic
solution of the system (\ref{BKT.equation}), (\ref{number.rho}), and
(\ref{rho.bound}), which yields $T_{\rm BKT} = \epsilon_{F}/8$. As we
discussed in Sec.~\ref{sec:relevance} (see also the Uemura plot in
Fig.~\ref{fig:03}) such behaviour of $T_{c}(\epsilon_{F})$ is observed for all
families of HTSC cuprates in their underdoped region though with a smaller
coefficient of proportionality ($0.01\,-\, 0.1$). Again this discrepancy can
be in principle explained by the presence of non-magnetic impurities (see
Sec.~\ref{sec:impure}). Other reasons such as, for example, a contribution due
to noncondensed pairs in Eq.~(\ref{BKT.equation}) which defines
$T_{\mathrm{BKT}}$, could also be important.

\begin{figure}[h]
\centering \resizebox{0.6\textwidth}{!}{%
  \includegraphics{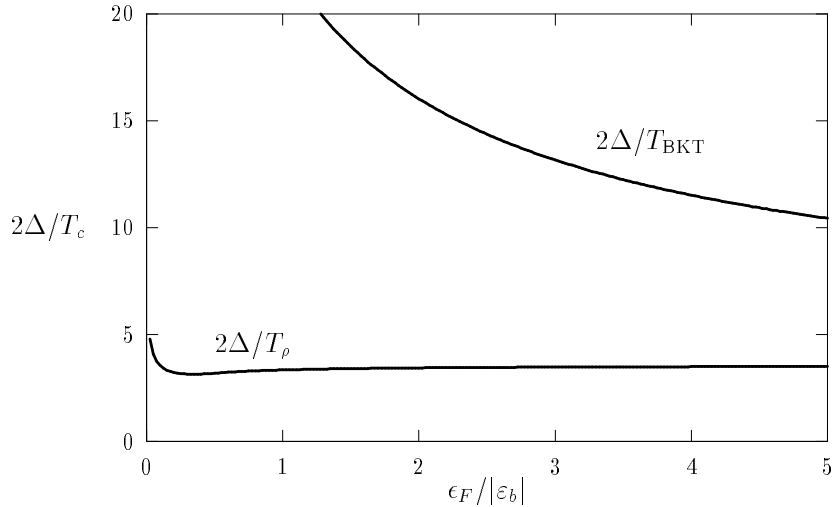}
} \caption{$2 \Delta/T_{\mathrm{BKT}}$ and $2 \Delta/T_{\rho}$ versus the
non-interacting fermion density (taken from \protect{\cite{Gusynin.JETP}}).}
\label{fig:16}
\end{figure}

{\rm c)} Finally the calculations showed (see Fig.~\ref{fig:16})
that the ratio $2 \Delta/T_{\mathrm{BKT}}$ for the carrier densities
shown in the figure is greater than 4.7 and grows as the doping decreases.
This result is also confirmed by the more general consideration
of Abrikosov \cite{Abrikosov.1997} which is also based on phase
fluctuations of the order parameter but takes into account
interlayer tunnelling.
This concentration behaviour is consistent with numerous measurements
of the ratio $2 \Delta/T_{c}$ in HTSC \cite{Timusk}. The value
$2 \Delta/T_{\rho} (= 2\Delta/T_{c}^{\mathrm{MF}})$ is, however,
somewhat lower and reaches the BCS theory limit of 3.52 only for
$\epsilon_{F} \gtrsim  |\varepsilon_{b}|$. It is very interesting
and important that the measurements of \cite{Oda} show that
for all doping levels
$2 \Delta/T^{\ast} \sim 4 \, - \, 5$ which is very close to
the BCS value, $2 \Delta_{d}/T_{c}^{\mathrm{MF}} \simeq 4.3$
but for a $d$-wave superconductor.
This presents a strong argument in a favour of the theories
which relate the pseudogap formation at $T^{\ast}$ to
the precursor superconducting fluctuations. In this case the scale
for $T^{\ast}$ is the BCS temperature $T_{c}^{\mathrm{MF}}$, or as in
\cite{Gusynin.JETP} $T_{\rho}$ which, as we showed, has a quite rigorous
definition in the modulus-phase approach. Another important conclusion which
one can draw from the experimental values of  $2 \Delta/T^{\ast}$
\cite{Oda} is that the simple Bose-BCS crossover theories with
pre-formed local pairs existing in the region with
$\mu/\epsilon_{F} < 1$ (or even $\mu < 0$) with
$\epsilon_{F}/|\varepsilon_{b}| \lesssim 1$ are unable
to give $2 \Delta/ T_{\rho} < 4$ since Fig.~\ref{fig:16}
clearly shows that the ratio grows for very low carrier densities.
As we discussed in Sec.~\ref{sec:derivative} one may still have
local Bose pairs in a more sophisticated picture where the system
is inhomogeneous
(see also the discussion in Sec.~\ref{sec:impure}
about the differences between the approaches of Uemura \cite{Uemura} and
Emery-Kivelson \cite{Emery.1995a,Emery.1995b}).

Note also that the divergence of $2 \Delta/T_{\mathrm{BKT}}$ and
$2 \Delta/T_{\rho}$ at $\epsilon_{F} \to 0$ is directly related to
the definition of $\Delta$ at $\mu < 0$.

\subsubsection{Pairing temperature $T_{\mathrm{pair}}$ versus carrier density.
More about the meaning of $T_{\rho}$}
\label{sec:pairing.temperature}

There is no disagreement concerning the asymptotic behaviour of
$T_{\mathrm{BKT}}$ (or $T_{c}$ shown in Fig.~\ref{fig:01})
proportional to $\epsilon_{F}$ in the region of very low carrier densities.
In contrast, the behaviour of the temperature $T^{\ast}$, below which we assumed
pairs are formed, cannot be considered to be generally accepted. It is
an open question as to how it behaves in the heavily underdoped cuprates:
namely whether it increases being related to the N\'eel temperature
$T_{\mathrm{N}}$ or decreases. The opinion of theoreticians on this subject
is also divided.

For example, on the phase diagram suggested by Uemura in
Fig.~\ref{fig:04}, the temperature $T^{\ast}$ is taken to be the
temperature $T_{\mathrm{pair}}$ of local uncorrelated pairing which increases
with decreasing $n_{f}$. Randeria (see Ref. \cite{Randeria.book} and
references therein), to define the pairing temperature $T_{\mathrm{pair}}$,
uses the system of equations for the mean-field transition temperature
and the corresponding chemical potential, which is essentially
identical to the system (\ref{temperature.rho}),
(\ref{number.temperature.rho}).  Thus his
$T_{\mathrm{pair}} \to 0$ as $n_{f} \to 0$. This is accordance to the
 behaviour of $T^{\ast}$ shown in the theoretical phase
diagram from Fig.~\ref{fig:01} suggested by Zachar \cite{Zachar}.

It is also well known \cite{Randeria.book,Haussmann.1993,Varma} that
in the low-density limit, it is vital to include quantum fluctuations,
at least in the number equation \cite{Nozieres}, in the calculation of
the critical temperature at which long-range order forms in 3D.

Certainly quantum fluctuations are also important in
the calculation of $T_\rho$ in the limit $n_{f} \to 0$ and, in
particular, in the number equation.  However, as already stressed in
Sec.~\ref{sec:derivation}, these corrections are quite different from
that obtained using the variables $\Phi,\Phi^{\ast}$, since perturbation
theory in the variables $\rho$ and $\theta$ does not contain any
infrared singularities \cite{Witten,Ichinose}, and the fluctuations do
not yield $T_\rho \equiv 0$.

In our opinion, the temperature $T_{\rho}$ has its own physical
interpretation: this is the temperature of a smooth transition to the
state in which the neutral order parameter $\rho \neq 0$, and below
which one can observe pseudogap manifestations.
In this respect the temperature $T_{\rho}$ can be compared with the
crossover temperature $T_{X}$ introduced for the repulsive and
attractive Hubbard models in \cite{Tremblay.1997} (see also the
explanation in Sec.~\ref{sec:limitations}). At the same time note that both
these temperatures should not be strictly identical to the mean-field
temperature $T_{c}^{\mathrm{MF}}$
(see the end of Sec.~\ref{sec:derivation}).

There is also a very interesting and important question about the
character of the transition (see, for example, \cite{Chakraverty.1997}).
Certainly in the simplest Landau theory one appears to have a second-order phase
transition, since $\rho$ takes a nonzero value only below $T_{\rho}$
\cite{MacKenzie} and since
$\rho(x)$ has only been treated in the mean-field approximation i.e.
one has neglected the fluctuations in both $\rho(x)$ and $\theta(x)$,
a second-order phase transition was obtained at $T_{\rho}$.
However, as it was stressed in Introduction, experimentally the
formation of the pseudogap phase does not display any sharp transition
and the temperature $T^{\ast}$ observed in various experiments is to be
considered as a characteristic energy scale and not as a temperature
where the pseudogap is reduced to zero \cite{Timusk,Renner}.
We believe that taking into account the fluctuations of $\rho(x)$ may
resolve the discrepancy between the experimental behaviour of
$T^{\ast}$ and the temperature $T_{\rho}$ introduced in the theory.

To define the temperature $T_{\mathrm{pair}}$ properly, one should study
the spectrum of bound states either by solving the Bethe-Salpeter equation
\cite{GGL} (see also Sec.~\ref{sec:Salpeter}) or by analyzing the
corresponding Green's functions as is done here.  It turns out that there is no
difference between $T_{\mathrm{pair}}$ and $T_{\rho}$ in the Cooper pair regime
($\mu > 0$), while in the local pair region ($\mu < 0$) these temperatures
exhibit different behaviour.

Indeed, let us study the spectrum of bound states in both the normal
($\rho = 0$) and pseudogap ($\rho \neq 0$) phases. We are especially
interested in determining the conditions under which  real bound states
(with zero total momentum $ {\bf K} = 0$) become unstable.  For this
purpose one can look at the propagator of the $\rho$-particle in the
pseudogap phase:
\begin{equation}
T_{\rho}^{-1}(\tau,  {\bf r}) =  \left. \frac{1}{2}
\frac{\beta \delta^{2} \Omega(v, \mu,T, \rho(\tau,  {\bf r}),
\partial \theta(\tau,  {\bf r}))}
{\delta \rho(\tau,  {\bf r}) \delta \rho(0, 0)}
\right|_{\rho = \rho_{\rm min} = const},
\label{Gamma.rho}
\end{equation}
where $\rho_{\mathrm{min}}$ is defined by the minimum condition
(\ref{rho}) (or (\ref{rho.bound})) of the potential part (\ref{A11}) of the
effective action (\ref{Effective.Action.2D}).
In the momentum representation, the spectrum of bound states is
usually determined by the condition
\begin{equation}
\frac{1}{T_{\rho}^{R} (\omega,  {\bf K})} = 0,
\label{spectrum}
\end{equation}
where $T_{\rho}^{R} (\omega,  {\bf K})$ is the retarded Green's
function obtained directly from the finite temperature Green's function
(\ref{Gamma.rho}) in frequency-momentum representation,
$T_{\rho}(i\Omega_{n},  {\bf K})$ using analytical continuation
$i\Omega_{n} \to \omega + i0$.
Recall that such analytical continuation must be performed after
evaluating the sum over the Matsubara frequencies. In the case of
vanishing total momentum $ {\bf K} = 0$, one arrives at the energy
spectrum equation
\begin{equation}
\frac{1}{T_{\rho}^{R} (\omega, 0)} = \frac{1}{U} + 2 \int
\frac{d^2  k}{(2 \pi)^{2}} \frac{\xi^{2}( {\bf
k})}{\sqrt{\xi^{2}( {\bf k}) + \rho^{2}}} \frac{\tanh \sqrt{
\xi^{2}( {\bf k}) + \rho^{2}}/2T} {\omega^{2} - 4 [\xi^{2}
( {\bf k}) + \rho^{2}]} = 0.
\label{spectrum.rho}
\end{equation}
From the explicit
expression (\ref{spectrum.rho}) for $T_{\rho}^{R} (\omega, 0)$,
this function
obviously has a branch cut at frequencies
\begin{equation} |\omega| \geq 2\,
\mbox{min} \sqrt{ \xi^{2}( {\bf k}) + \rho^{2}} =
\left\{\begin{array}{cc} 2 \rho,  \qquad \quad & \mu \geq 0 \\ 2
\sqrt{\mu^{2} + \rho^{2}}, & \mu < 0.  \end{array} \right.
\label{cut}
\end{equation}
Thus, bound states can exist only below this cut.

\begin{figure}[h]
\centering \resizebox{0.6\textwidth}{!}{%
  \includegraphics{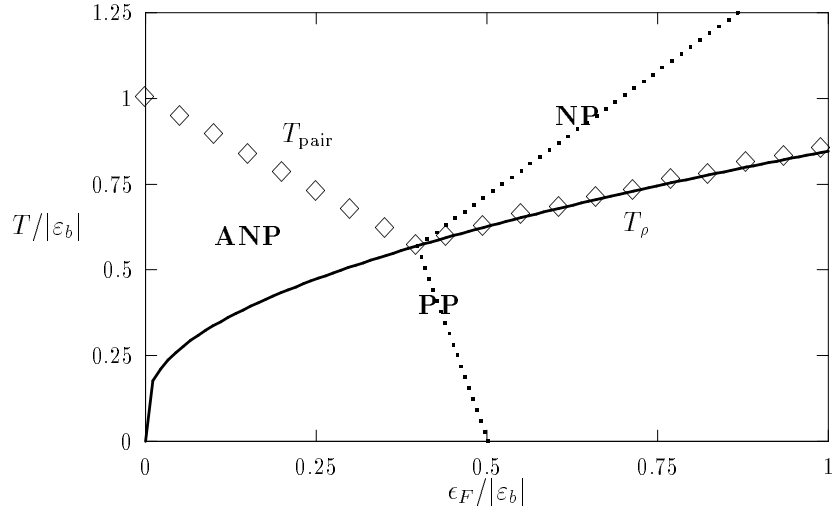}
}
\caption{Phase diagram of the 2D metal at low concentrations (taken from
\protect{\cite{Gusynin.JETP}}). The dotted line corresponds to $\mu =0$, and
the pairing temperature $T_{\mathrm{pair}}$ (squares) separates the abnormal
normal phase (ANP) from the normal phase (NP). The region below the solid line
is the pseudogap phase (PP). The critical temperature $T_{\mathrm{BKT}}$ is
not shown.} \label{fig:17}
\end{figure}

Real bound states decay into two-fermion states
when the energy of the former reaches the branch point
$2\,\mbox{min} \sqrt{ \xi^{2}( {\bf k}) + \rho^{2}}$. Since
$1/T_{\rho}^{R}(\omega,0)$ is a monotonically decreasing function of
$\omega^{2}$, the solution for the bound-state energy must be
unique and is given by $\omega \equiv -\varepsilon_{b}(T) = 2 \rho(T)$.
At this point Eq.~(\ref{spectrum.rho}) coincides exactly with the mean-field
equation (\ref{rho}) for $\rho(T)$ so that $\rho(T)$ takes on the mean-field
value. It is also clear that for $\mu < 0$ we have real bound states with
energy $\varepsilon_{b}(T) = - 2 \rho$ below the two-particle scattering
continuum at $\omega = 2\sqrt{\mu^{2} + \rho^{2}}$, while at $\mu \geq 0$ there
are no stable bound states. The line $\mu(T, \epsilon_{F}) =0$
in the $T$--$\epsilon_{F}$ plane at $\rho \neq 0$ separates the
negative $\mu$ region, where local pairs exist, from that
in which only Cooper pairs exist (positive $\mu$).
This line (see Fig.~\ref{fig:17}) begins at the point
$T = (e^{\gamma}/\pi) |\varepsilon_{b}| \approx 0.6 |\varepsilon_{b}|$,
$\epsilon_{F} \approx 0.39 |\varepsilon_{b}|$ and ends at $T=0$,
$\epsilon_{F} = |\varepsilon_{b}|/2$. (The latter follows directly
from the solution at $T=0$ given by (\ref{solution.2D}).)

To find the equivalent line in the normal phase with $\rho=0$ i.e. above
$T_\rho$, we consider the corresponding equation for the bound states.
The propagator of these states (in imaginary time formalism) is defined to be
\begin{equation}
T^{-1}(\tau,  {\bf r}) =  \left.
\frac{\beta \delta^{2} \Omega(v, \mu,T, \Phi(\tau,  {\bf r}),
\Phi^{\ast}(\tau,  {\bf r}))}
{\delta \Phi^{\ast}(\tau,  {\bf r}) \delta \Phi(0, 0)}
\right|_{\Phi = \Phi^{\ast} = 0}.
\label{Gamma.Phi.def}
\end{equation}
(In the normal phase, where $\rho =0$, we must again use  the initial
auxiliary fields $\Phi$ and $\Phi^{\ast}$.)
Then in the momentum representation (after summing  over the Matsubara
frequencies) we have
\begin{eqnarray}
T^{-1}(i\Omega_{n}, {\bf K}) = &&
\frac{1}{U} - \frac{1}{2} \int \frac{d^2  k}{(2 \pi)^{2}}
\frac{\tanh \xi_{+}( {\bf k}, {\bf K})/2T +
      \tanh \xi_{-}( {\bf k}, {\bf K})/2T}
{\xi_{+}( {\bf k}, {\bf K}) + \xi_{-}( {\bf k}, {\bf K})
- i \Omega_{n}},
\nonumber                                    \\
&& \xi_{\pm}( {\bf k}, {\bf K}) \equiv \frac{1}{2m}
\left( {\bf k} \pm \frac{ {\bf K}}{2} \right)^2 - \mu,
                      \label{Gamma.Phi}
\end{eqnarray}
where $ {\bf k}$ is the relative momentum of the pair.
It is seen that Green's function (\ref{Gamma.Phi}) coincides with
the $T$-matrix $T_{0}(i\Omega_{n}, {\bf K})$
(compare with Eqs.~(\ref{T.matrix.non}) and (\ref{chi.non}))
in the non self-consistent, non-conserving approximation, up to an overall
minus sign.
The spectrum of bound states is given again by Eq.~(\ref{spectrum}).
Using the energy $\varepsilon_{b}$ (see Eq.~(\ref{bound.energy})) of
the bound state at $T = 0$, for $ {\bf K} = 0$ we obtain the
following equation for the energies of these states in the normal phase:
\begin{equation}
\int_{0}^{\infty} dx \left[\frac{1}{x + |\varepsilon_{b}|/2} -
\frac{\tanh(x - \mu)/2T}{x - \mu - \omega/2}\right] = 0.
\label{spectrum.Phi}
\end{equation}
Such states can exist provided
$-2 \mu - |\varepsilon_{b}| < \omega < - 2 \mu$. The left-hand side
of Eq.~(\ref{spectrum.Phi}) is positive at
$\omega = -2\mu - |\varepsilon_{b}|$ and tends to $+ \infty$
($\mu > 0$) or $- \infty$ ($\mu < 0$)  when $\omega \to - 2 \mu$.
This equation thus always has a solution between $-2\mu-|\varepsilon_b|$
and $-2 \mu$ for $\mu < 0$, i.e. bound states with zero total momentum exist
for negative $\mu$.

For $\mu > 0$, the analytic analysis becomes more complicated, and requires
numerical study.  One can easily find from (\ref{spectrum.Phi}) that for
the special case $T=0$, which is not in fact in the normal phase but for
which one can use the variables $\Phi$ and $\Phi^\ast$, that such stable
bound states exist up to $\mu < |\varepsilon_{b}|/8$
(see also proof in Sec.~\ref{sec:Salpeter}). In fact, numerical study for
$T \geq T_{\rho}$ shows that the trajectory $\mu (T,\epsilon_{F}) = 0$
(or $T = \epsilon_{F}/ \ln2$, see (\ref{number.temperature.rho}))
approximately divides the normal phase into two qualitatively different
regions -- with ($\mu < 0$) and without ($\mu >
0$) stable (long-lived) pairs. This result is the same as that in the
phases with $\rho \ne 0$, so that the line $\mu(T,
\epsilon_{F})=0$ (Fig.~\ref{fig:17}) separates the regions with
stable and unstable pairs for all non-zero temperatures. However this line
(squares) does not correspond to any phase transition.

Given the two-particle binding energy as a function of temperature, it is
natural to define the pairing temperature $T_{\mathrm{pair}}$ by
$T_{\mathrm{pair}} \approx |\varepsilon_{b}(T_{\mathrm{pair}},
\mu(T_{\mathrm{pair}},\epsilon_{F}))|$. This equation  can be easily analyzed
in the region $\epsilon_{F} \ll |\varepsilon_{b}|$, for which we directly
obtain $T_{\mathrm{pair}} \approx |\varepsilon_{b}|$, which clearly coincides
with the standard estimate \cite{Uemura,Micnas}. This means in turn that the
curve $T_{\mathrm{pair}}(\epsilon_{F})$ starting at $T_{\mathrm{pair}}(0)
\approx |\varepsilon_{b}|$ will decrease until the point
$T_{\mathrm{pair}}(0.39 \epsilon_{F}) \approx 0.6 |\varepsilon_{b}|$, which
lies on the line $T_{\rho}(\epsilon_{F})$ (see Fig.~\ref{fig:17}). It is
important that this line is not a phase transition curve; it merely divides
the fermion system diagram into temperature regions where one has either a
prevailing mean number of local pairs ($T \lesssim T_{\mathrm{pair}}$) or
unbound carriers ($T \gtrsim T_{\mathrm{pair}}$).  This is the region of the
abnormal normal phase where one has some density of pre-formed boson pairs. It
is widely accepted, however, that this case is only of theoretical interest,
since there is no Fermi surface ($\mu<0$) in this phase (see nevertheless the
discussion in Secs.~\ref{sec:precursor}, \ref{sec:discussion.indirect} and
Refs. \cite{Mihailovic,Demsar.1998,Alexandrov,Mott}). The phase area or the
difference $T_P(\epsilon_F)- T_{\rho}(\epsilon_F)$ is an increasing function
as $\epsilon_{F} \to 0$, which corresponds to the behaviour usually assumed
\cite{Uemura,Emery.1995a}.

When $\mu>0$ there are no stable bound states
($\varepsilon_{b}(T) = 2 \rho(T) =0$) in the normal phase.
Formally, using $\rho(T)=0$ in Eq.~(\ref{spectrum.rho}),
we immediately obtain (\ref{temperature.rho}) or, in other words, here
$T_{\mathrm{pair}} = T_{\rho}$. Such a conclusion is in accordance with the
generally accepted definition of $T_{\mathrm{pair}}$ in the BCS case
\cite{Micnas}.

Thus the phase diagram of a 2D metal above $T_{c}$ acquires the form
shown in Fig.~\ref{fig:17}. It is interesting that if there are no stable
pairs i.e. $T_{\mathrm{pair}}(\epsilon_{F})$ is not well defined, then the
temperature $T_{\rho}(\epsilon_{F})$ is the line below which pairs reveal
some signs of collective behaviour. Moreover, for $T < T_{\rho}$ one can
speak of a real pseudogap in the one-particle
spectrum, while in the region $T_{\rho} < T < T_{\mathrm{pair}}$ only strongly
developed pair fluctuations (a finite number of pairs) exist, although they
probably suffice to reduce the  spectral quasi-particle weight, and to produce
other observed manifestations that mask pseudogap (spin gap) formation.

\subsection{The peculiarities of the phase diagram for the
lattice model}
\label{sec:discrete.diagram}

The discrete version of the attractive Hubbard Hamiltonian
(\ref{Hubbard}) was also mapped \cite{Alvarez} into the classical
$XY$-model. Although we disagree on some of the fine points in the
derivation we agree entirely with the results.
It is instructive to consider these fine points since they
yield a better understanding of the modulus-phase formalism described
in Sec.~\ref{sec:our.diagram}. The statistical sum
(\ref{partition.effective}) is written in \cite{Alvarez} for
(\ref{Hubbard}) in the momentum representation. Then the
saddle-point approximation described by (\ref{Omega.Potential})
is considered. Of course, this approximation leads to nothing
but the finite temperature generalization of the zero
temperature equations for $\Delta$ (\ref{minima}) and $\mu$
(\ref{number}) considered in Sec.~\ref{sec:solution}. The
value $\Delta$ in these equations is called  in \cite{Alvarez}
``order parameter'' in accordance with the saddle-point character
of the derived approximation for the complex Hubbard-Stratonovich
fields $\Phi$ and $\Phi^{\ast}$. These equations
are, of course, formally identical to Eqs.~(\ref{rho}) and
(\ref{number.rho}) derived in Sec.~\ref{sec:derivation},
but as we have discussed the meaning of $\rho$
is completely different allowing nonzero $\rho$ but keeping
$\langle \Phi \rangle = \rho \langle \exp(i \theta(x)) \rangle = 0$.

To go beyond the saddle-point approximation, it is proposed
in \cite{Alvarez} to consider a variation of the
Hubbard-Stratonovich fields around their saddle-point values.
In doing this, one looks for the low energy excitations.
It is claimed in \cite{Alvarez} that choosing the order
parameter to be real (or taking it any specific value),
one breaks its continuous symmetry, provided
$\Delta \ne 0$. According to Goldstone's theorem, this
symmetry breaking implies that a soft mode exists as
${\bf q} \to 0$. These modes are associated in \cite{Alvarez}
with the phase fluctuations of the order parameter.
It is clear, however, that this breaking of the continuous
symmetry and appearance of the Goldstone mode in 2D at $T \neq 0$
are strictly forbidden by the CMWH theorem \cite{Coleman}.
In fact the above conclusions of \cite{Alvarez} are obviously
related to the use of the saddle-point approximation
in the form described in Sec.~\ref{sec:formalism}.
To be consistent with the CMWH theorem one should (following
by Witten \cite{Witten}) use the set of transformations
(\ref{parametrization.Hubbard}) and (\ref{parametrization.fermi})
for any 2D theory. As we have explained in
Sec.~\ref{sec:our.diagram} these transformations make
the perturbation theory in the modulus-phase variables free
from symmetry violating terms. Furthermore, as we will see in
Sec.~\ref{sec:phase.mode} in 2D the propagator associated with the
phase fluctuations does not have the canonical behaviour
$\sim 1/{\bf q}^2$ and becomes softer.

All these comments do not however change the results
of the calculations in \cite{Alvarez} but rather reflect subtle
points in their interpretation. Thus we present the
effective action for the phase mode derived in \cite{Alvarez}\footnote{Note
that the phase was assumed to be time-independent in \cite{Alvarez}.}
\begin{equation}
\Omega_{\mathrm{eff}}[\theta] =
\sum_{\bf q} J_{\mathrm{eff}} ({\bf q})
\theta_{\bf q} \theta_{-{\bf q}}
                        \label{Effective.phase}
\end{equation}
with
\begin{eqnarray}
J_{\mathrm{eff}}({\bf q})  =  \frac{\Delta^{2}}{U} \left\{ 1-
U \sum_{\bf k} \right.
&& \left[ (u_{+} u_{-} + v_{+} v_{-})^{2}
\frac{1 - f_{+} - f_{-}}{E_{+} + E_{-}}  \right.
\nonumber            \\
 - && \left. \left.
(u_{+} v_{-} - u_{-} v_{+})^{2} \frac{f_{+} - f_{-}}{E_{+} - E_{-}}
\right] \right\} \,,
                       \label{J.effective}
\end{eqnarray}
where $\Delta$ is the solution of the mean-field gap
equation (see e.g. Eq.~(\ref{rho})), the subscript $\pm$ indicates
${\bf k} \pm {\bf q}/2$,
$u_{\pm}^{2} = \frac{1}{2}(1 + \xi_{\pm}/E_{\pm})$,
$v_{\pm}^{2} = \frac{1}{2}(1 -\xi_{\pm}/E_{\pm})$, and
$f_{\pm} = f(E_{\pm})$ is the usual Fermi functions with
$E_{\bf k} = \sqrt{\xi_{\bf k}^{2} + \Delta^{2}}$.
As in Sec.~\ref{sec:derivation} one may check that
$J_{\mathrm{eff}}$ is nonzero only below the mean-field
(BCS) critical temperature where $\Delta \neq 0$.

To provide a link with the $XY$-model, one expands $J_{\mathrm{eff}}$
up to second order in ${\bf q}$ to obtain the low energy
behaviour
\begin{equation}
H = J \sum_{\bf q} {\bf q}^{2} \theta_{\bf q} \theta_{- {\bf q}}
= J \int d {\bf r} [\nabla \theta({\bf r})]^{2} \,.
\end{equation}
This expansion of $J_{\mathrm{eff}}$ in momentum space
is an equivalent way to understand the approximations used in
the derivative expansion in Sec.~\ref{sec:derivation}.
Note that the definition of the phase stiffness $J$ used here
and in Sec.~\ref{sec:our.diagram} differ by a factor of 2.
The effective model is then discretized in \cite{Alvarez} by
introducing a lattice parameter $a$:
\begin{equation}
H = J a^{d-2} \sum_{\langle i j \rangle}
(\theta_{i} - \theta_{j})^{2} \,, \qquad
|\theta_{i} - \theta_{j}| \ll 1\,,
\label{XY.discrete}
\end{equation}
where $d=2$ is the dimensionality of the system.

One can also derive an effective $XY$-model starting from
the Ginzburg-Landau theory \cite{Li,Drechsler,Zwerger}. As
noticed in \cite{Alvarez}, this procedure which allows
one to write parameters of the effective $XY$-model
in terms of the Ginzburg-Landau parameters, is only fully justified
close to $T_{c}^{\mathrm{MF}}$ where the linearized Ginzburg-Landau
theory is valid. However, as we saw in Sec.~\ref{sec:results}
the most interesting region i.e. that with a relatively large pseudogap
 $(T^{\ast} - T_{\mathrm{BKT}})/T^{\ast} \sim 1$ corresponds
to $\rho/T \sim 1$ (or $\Delta/T \sim 1$) which is clearly
beyond the limits of validity of the Ginzburg-Landau theory.

\begin{figure}[h]
\centering \resizebox{0.6\textwidth}{!}{%
  \includegraphics{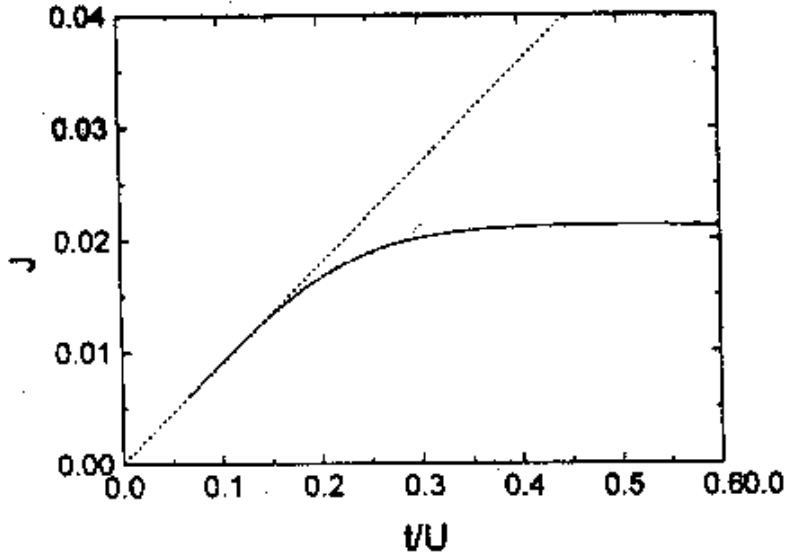}
}
\caption{The low temperature coupling constant $J$ vs the
inverse of $U$, calculated for a square lattice with
electron density $n=0.2$. This figure has been taken from
\protect{\cite{Alvarez}}.}
\label{fig:18}
\end{figure}

Arguing that the phase stiffness $J$ is only weakly
temperature dependent up to a temperature of the order of
the BCS temperature $T_{c}^{\mathrm{MF}}$ it was suggested in \cite{Alvarez}
to use
\begin{equation}
J(T =0) = \frac{\Delta^{2}}{2} \sum_{\bf k}
\frac{2 \xi_{\bf k}^{3} \xi_{\bf k}^{''} + \Delta^{2}
(3 (\xi_{\bf k}^{'})^{2} + 2 \xi_{\bf k} \xi_{\bf k}^{''})}
{E_{\bf k}} \,,
                    \label{J.Alvarez.original}
\end{equation}
where $\xi_{\bf k}^{'} = \frac{1}{2} \partial \xi_{\bf k}/\partial k_x$
and
$\xi_{\bf k}^{''} = \frac{1}{8} \partial^2 \xi_{\bf k}/\partial^2 k_x$.
The low temperature value of $J$ as a function of $1/U$ is shown
in Fig.~\ref{fig:18}. One can see that for large $U$, $J$ goes as
$1/U$.

Assuming a rectangular density of states of width $2W$
(i.e. a square band extending from $-W$ to $W$)
and an electron density $n$ consistent with its definition given by
Eq.~(\ref{number.definition}), one obtains \cite{Sofo} the following
zero temperature solution for the system of Eqs.~(\ref{gap.lattice})
and (\ref{number.lattice})
(compare with Eq.~(\ref{solution.2D.band}) taking into
account the difference in the definitions of the bandwidth in
Sec.~\ref{sec:solution} and here)
\begin{equation}
\Delta = W \frac{\sqrt{n(2-n)}}{\sinh \frac{2W}{U}} \,, \qquad
 \mu = W \frac{n-1}{\tanh \frac{2W}{U}}\,,
                      \label{solution.2D.discrete}
\end{equation}
where $n$ is the carrier density per lattice site
(see Eq.~(\ref{number.definition})).
Note that the chemical potential $\mu$ includes the normal
Hartree-Fock energy shift. This solution is in fact equivalent
to (\ref{solution.2D.band}), but we present it here for
convenience, because it is expressed in terms of often used
density $n$ per lattice site for the lattice model (\ref{Hubbard}).
Then, if in contrast to Eq.~(\ref{bound.energy}),
one considers the strong coupling limit $W \ll U$, the solution
(\ref{solution.2D.discrete}) can be written as
$\mu = -(1-n)U/2$ and $\Delta = \sqrt{n(2-n)}U/2$.
In the same limit $J$ has been obtained in \cite{Alvarez} as
\begin{equation}
J = \frac{3}{2}\frac{n(2-n)}{U} \,.
\label{J.Alvarez}
\end{equation}
which implies that $J \ll \Delta$ for strong coupling.

The equation for $T_{\mathrm{BKT}}$ was written in
\cite{Alvarez} for the discrete version of $XY$-model
(\ref{XY.discrete}):
\begin{equation}
T_{\mathrm{BKT}} = A J\,,
                  \label{BKT.discrete}
\end{equation}
where $A$ is a dimensionless number of the order of 1 which
depends on the details of the short distance physics \cite{Emery.1995a}.
For the model on the square lattice considered here $A \simeq 0.89$.
Note that using the definition of $J$ \cite{Alvarez} in this
section the value of $A$ for the continuum model is $\pi/4 \simeq 0.79$.

\begin{figure}[h]
\centering \resizebox{0.6\textwidth}{!}{%
  \includegraphics{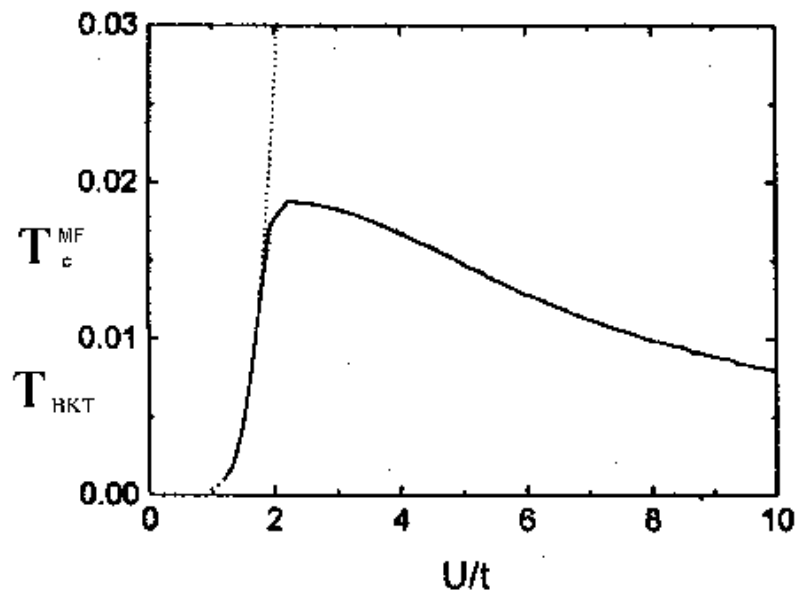}
}
\caption{The critical temperature, $T_{c}^{\mathrm{MF}}$
of the mean-field (dashed line) and BKT critical temperature,
$T_{\mathrm{BKT}}$ (full line) vs $U$, calculated for a square lattice
with electron density $n=0.2$. This figure has been taken from
\protect{\cite{Alvarez}}.}
\label{fig:19}
\end{figure}

The authors of \cite{Alvarez} classified weak
and strong coupling regimes by the ratio between $J$
and $\Delta$. In the weak coupling limit at low temperatures,
$J \gg \Delta$ and pair breaking, BCS-type excitations
dominate the low temperature thermodynamical properties.
In this regime, the phases exhibit quasi-long-range (algebraic)
order up to the temperatures of the order of the BCS critical
temperature $T_{c}^{\mathrm{MF}}$. It is only very close to this
temperature, when $J$ rapidly decreases to zero, that the phase
becomes really disordered. In this regime, the BKT temperature
$T_{\mathrm{BKT}}$ lies near $T_{c}^{\mathrm{MF}}$, and the
transition to the superfluid phase occurs concurrently with the appearance
of Cooper pairs. In the strong coupling limit, the low
temperature value of the phase stiffness is such that
$J \ll \Delta$ and the thermodynamic properties at low
temperatures are dominated by phase fluctuations with
$T_{\mathrm{BKT}} \ll T_{c}^{\mathrm{MF}}$. In this regime, when
the temperature is decreased, first Cooper pairs without off-diagonal
long-range order are formed and then, below $T_{\mathrm{BKT}}$,
the system becomes superfluid. In Fig.~\ref{fig:19},
$T_{\mathrm{BKT}}$ is shown as a function of $U$, and for comparison
$T_{c}^{\mathrm{MF}}$ is shown on the same graph.

Thus, apart from some details related to the formalism
and the behaviour of the system at very low carrier densities $\mu <0$
the phase diagram built in Sec.~\ref{sec:our.diagram} and its physical
interpretation for the continuum model (\ref{Hamiltonian.2D}) in fact coincides
with that for the discrete model \cite{Alvarez}.

\subsection{The effects of Coulomb repulsion and quantum
phase fluctuations}
\label{sec:Coulomb}

In this section we will discuss the influence of Coulomb
repulsion and quantum phase fluctuations on the phase
diagram built in the previous section. We will follow the papers
of Ariosa {\em et al.}
\cite{Ariosa.1991,Ariosa.1992,Beck.1997,Ariosa.1998}. Since the
discussion of the pairing mechanism  is beyond the scope of the
current review, we will not consider any effects directly related
to the pairing mechanism suggested in \cite{Ariosa.1998}.

\subsubsection{Derivation of the quantum $XY$-model}

Our starting point is the Hamiltonian \cite{Beck.1997} which
describes charge carriers on a lattice, subject to an on-site
attraction and a long-range Coulomb repulsion,
$e^2 V_{\mathrm{C}}$, acting between particles on different sites:
\begin{eqnarray}
H & = & H_{0} + H_{U} + H_{\mathrm{C}} \,,
\nonumber                \\
H_{0} & = & - \sum_{\langle {\bf n}, {\bf m} \rangle, \sigma}
t c^{\dagger}_{{\bf n} \sigma} c_{{\bf m} \sigma} - \mu \sum_{{\bf n}} n_{{\bf n}} \,,
\nonumber                \\
H_{U} & = & -\frac{U}{2} \sum_{{\bf n}, \sigma}
c^{\dagger}_{{\bf n} \sigma} c_{{\bf n} \sigma}
c^{\dagger}_{{\bf n} -\sigma} c_{{\bf n} -\sigma} \,,
\nonumber                 \\
H_{\mathrm{C}} & = & \frac{1}{2} \sum_{{\bf n} \neq {\bf m}}
(n_{{\bf n}} - n) e^2 V_{\mathrm{C}}({\bf n}, {\bf m}) (n_{{\bf m}} -n) \,,
                  \label{Hamiltonian.Coulomb}
\end{eqnarray}
where $U >0$. Here, as before, (see Hamiltonian (\ref{Hubbard}))
$c^{\dagger}_{{\bf n} \sigma}$ ($c_{{\bf n} \sigma}$) are creation
(annihilation) operators, $\langle {\bf n}, {\bf m} \rangle$ denotes denotes
pairs of nearest-neighbour sites, $n$ is the neutralizing background charge
which is of course equal to the carrier density per lattice site, and $n_{{\bf
n}} = \sum_{\sigma} c^{\dagger}_{{\bf n} \sigma} c_{{\bf n} \sigma}$.

The partition function can be written as a functional integral by
means of two successive Stratonovich-Hubbard transformations
\cite{Umezawa} (see also \cite{Kleinert,Capezzali.thesis}
and compare with Sec.~\ref{sec:formalism}), decoupling the two interaction
terms in $H$ with the help of a complex field $\Phi({\bf n},\tau)$ and a
real field $V({\bf n},\tau)$ both of which depend on the site
${\bf n}$ and the imaginary time $\tau$:
\begin{eqnarray}
Z & = & \mbox{Tr} e^{-\beta H}
\nonumber            \\
& = & \int \mathcal{D} \Phi \mathcal{D} \Phi^{\ast}
\int \mathcal{D} V \left\{ \mbox{Tr}  \left[
e^{-\beta H_{0}} T_{\tau} e^{-i \int_{0}^{-i\beta} d \tau
( \tilde{H}(\Phi,V,\tau)+\epsilon (\Phi,V,\tau) )} \right] \right\} \,,
\end{eqnarray}
with a part linear in the Hubbard-Stratonovich fields
\begin{equation}
\tilde{H}(\Phi,V,\tau)= \sum_{{\bf n}}^{}
{\left[ \Phi^{\ast}({\bf n},\tau)c^{}_{{\bf n}\uparrow}c^{}_{{\bf n} \downarrow}
+\Phi({\bf n},\tau)c^{\dagger}_{{\bf n} \downarrow}c^{\dagger}_{{\bf n}\uparrow}
+iV({\bf n},\tau) \left(n_{{\bf n}}(\tau)-n \right) \right]}  \, ,
\end{equation}
and a part containing quadratic terms
\begin{equation}
\epsilon(\Phi,V,\tau)={\frac{1}{U}}\sum_{{\bf n}}^{}{
|\Phi({\bf n},\tau)|^{2}}+{\frac{1}{2e^{2}}} \sum_{{\bf n}\neq {\bf m}}^{}
{V({\bf n},\tau)V_{\mathrm{C}}^{-1}({\bf n},{\bf m})V({\bf m},\tau)} \,.
\end{equation}
As in the earlier sections $\Phi({\bf n},\tau)$ denotes a complex pairing field
which couples to the local pair operator $c^\dagger_{{\bf n} \downarrow}
c^\dagger_{{\bf n} \uparrow}$ while the second real field represents a local
electric potential which acts on the local charge density fluctuations
$n_{\bf n}(\tau)-n$. One evaluates the trace over the electronic degrees of
freedom, with the definition :
\begin{equation}
\mbox{Tr} \left\{
e^{-\beta H_{0}}T_{\tau}e^{-i\int_{0}^{-i\beta}{
d\tau\left[ \tilde{H}(\Phi,V,\tau)+\epsilon(\Phi,V,\tau) \right]}}
\right\}= e^{-i\int_{0}^{-i\beta}{d\tau \Omega(\Phi,V,\tau)}}
\end{equation}
Then expanding the ``free energy'' $\Omega$ to fourth order in $\Phi$,
to second order in $V$ and  to leading terms in the space
and time gradients of the two fields one arrives at:
\begin{equation}
\Omega =\Omega_{0} + \Omega(\Phi,V, \tau) \,,
\end{equation}
where $\Omega_0$ gives the result for the free electron contribution and
\begin{eqnarray}
& &\Omega(\Phi,V,\tau) \nonumber \\
& & =\sum_{{\bf n}}^{}{
\left[ a|\Phi({\bf n},\tau)|^{2} - \frac{d}{2} \left\{ i
\Phi^{\ast}({\bf n},\tau) \left(\frac{\partial}{\partial \tau}-
2V({\bf n},\tau) \right)\Phi({\bf n},\tau) + h.c. \right\} \right]}
\nonumber             \\
& & + c\sum_{\left \langle {\bf n}\neq {\bf m}\right\rangle}^{}{
|\Phi({\bf n},\tau)-\Phi({\bf m},\tau)|^{2}}+i\sum_{{\bf n}}^{}{ V({\bf n},\tau)
\left( \left\langle n_{{\bf n}} \right\rangle-n \right)}
\nonumber             \\
& &+ {\frac{1}{2e^{2}}}\sum_{{\bf n},{\bf m}}^{}{V({\bf n},\tau)
V_{\mathrm{SC}}^{-1}({\bf n},{\bf m})V({\bf m},\tau)}+b\sum_{{\bf
n}}^{}{|\Phi({\bf n},\tau)|^{4}} \,.        \label{Omega.Delta.V}
\end{eqnarray}
Here the Ginzburg-Landau coefficients $a$, $b$, $c$ and $d$ are related
to the free electron particle-particle propagator
\cite{Randeria.book,Kleinert,Drechsler,Zwerger,LSh.FNT.1997,Umezawa}
and are thus temperature dependent in  general;
$V_{\mathrm{SC}}^{-1}({\bf n},{\bf m})=V_{\mathrm{C}}^{-1}({\bf n},{\bf m})+ \chi_{0}({\bf n},{\bf
m})$
is the screened Coulomb potential (approximated by its static limit)
with $\chi_{0}({\bf n},{\bf m})$ being the electronic polarizability.

The terms containing the Ginzburg-Landau coefficients $a$, $b$,
$c$ and $d$ give the usual time-dependent Ginzburg-Landau
free energy\footnote{Note that
the coefficient $c$ is related to the kinetic
term which is not obvious from the discrete form
(\ref{Omega.Delta.V}).},
which is usually supposed to be valid only very close to the
transition temperature equal to $T_{c}^{\mathrm{MF}}$ or $T_{\rho}$
in our case. Thus these coefficients are derived in the same
approximation as the phase stiffness (\ref{J.bcs}) for the BCS
limit. This, of course, does not allow one to investigate
the equation equivalent to the self-consistent equation
(\ref{BKT.equation}) in the regime $\rho/T \sim 1$ (see
the discussion in Sec.~\ref{sec:discrete.diagram}). However,
the advantage of the representation (\ref{Omega.Delta.V})
is its relative simplicity which allows one to integrate
out the dependence on the electric potential $V$.
Commenting further on Eq.~(\ref{Omega.Delta.V}) note that
its third term describes the coupling between the two
space- and time-fluctuating Hubbard-Stratonovich fields;
the fifth term in the above equation takes into account
the coupling between the local electric potential and the
number density fluctuations of the charge carriers,
while the last double-sum is the contribution of the screened
Coulomb potential to the thermodynamical potential.

Integrating over the electric potential $V$ yields
\begin{eqnarray}
\Omega(\Phi, \tau) && = \sum_{{\bf n}}^{}\left[
a |\Phi({\bf n},\tau)|^{2} + b |\Phi({\bf n},\tau)|^{4} \right]
\nonumber                      \\
&& - \frac{d}{2} \left[
i \Phi^{\ast}({\bf n},\tau)
{\frac{\partial {\Phi({\bf n},\tau)} }{ \partial \tau}}  -
i \Phi({\bf n},\tau)
{\frac{\partial {\Phi^{\ast}({\bf n},\tau)} }{ \partial \tau}}
\right]
\nonumber                       \\
&& + c \sum_{\left\langle {\bf n},{\bf m} \right\rangle}^{}{|\Phi({\bf n},\tau)
- \Phi({\bf m},\tau)|^{2}} \nonumber \\
&& + {\frac{1}{2e^{2}}}\sum_{{\bf n},{\bf m}}^{}
{\rho({\bf n},\tau)V_{\mathrm{SC}}({\bf n},{\bf m})\rho({\bf m},\tau)}
\end{eqnarray}
where
$\rho({\bf n},\tau)=2d|\Phi({\bf n},\tau)|^{2}+\left\langle n_{{\bf n}} \right\rangle-n$
represents the number density fluctuations in the presence of the
fluctuating pairing field at site ${\bf n}$. Roughly speaking, if one
interprets $\sqrt{d} \Phi({\bf n},\tau)$ as the wave-function of a
local pair at site ${\bf n}$ and, thus, its squared modulus as a
``pair-density'', we note that $\rho({\bf n},\tau)$ is composed of two
separate contributions, namely one from the ``unpaired'' charge
carriers (``normal'' channel) and one from the ``paired'' fermions
(``superfluid'' channel), respectively. Obviously, the charge
neutrality constraint imposes $\sum_{{\bf n}} \rho({\bf n},\tau) =0$.

Next the amplitude and phase of $\Phi$ was introduced in
\cite{Beck.1997} in a way similar to that discussed in the
previous section.
In the following discussion, relatively strong coupling was assumed
\cite{Beck.1997} so that the fermions bind into
on-site singlet pairs at a temperature of the order of the mean
field transition temperature $T_{c}^{\mathrm{MF}}$, well above the
superconducting phase transition, the latter being finally triggered
by the onset of phase order.
Below $T_{c}^{\mathrm{MF}}$, $a<0$, so that the average amplitude
has a non-vanishing mean value $\Delta = \langle \Phi({\bf n}, \tau) \rangle$
given by $a+2b\Delta^{2} = 0$.
Charge neutrality implies
$ 2d\Delta^{2}+\left\langle n_{{\bf n}} \right\rangle -n=0$ and,
for $t\ll U$
\cite{Randeria.book,Kleinert,Drechsler,Zwerger,LSh.FNT.1997,Umezawa}:
\begin{equation}
c= \frac{2t^{2}}{U^{3}}, \qquad
d= \frac{1}{U^{2}}, \qquad
\Delta^{2}= \frac{1}{4} n(2-n) U^{2} \approx
\frac{1}{2} n U^{2} \quad \mbox{for} \quad n \ll 1.
\end{equation}
Note that, as explained above, the expressions  for $c$, $d$
and the relationship between $a$, $b$ and $\Delta$
are, strictly speaking, valid in the vicinity of $T_{c}^{\mathrm{MF}}$
while one needs to use them well below this temperature.
This is the reason why in \cite{Beck.1997} the proposal was made to use
these expressions together with the {\em zero\/} temperature expression
for $\Delta$ taken from \cite{Alvarez}
(see Eq.~(\ref{solution.2D.discrete}) and the strong-coupling limit below).
Splitting the number of pairs at a given site into
$|\Phi({\bf n},\tau)|^{2}=\Delta^{2}+\frac{1}{2}U^{2} \delta
n_{p}({\bf n},\tau)$, the integration over $\delta n_{p}({\bf n},\tau)$
was performed in \cite{Beck.1997} which (neglecting gradient terms)
yields a free energy functional for phase fluctuations only:
\begin{eqnarray}
\Omega (\theta, \tau) & = &  J \sum_{\left\langle {\bf n},{\bf m} \right\rangle}^{}
{\left[ 1-\cos {\left( \theta({\bf n},\tau) - \theta({\bf m},\tau) \right)} \right]}
+ \frac{n}{2} \sum_{{\bf n}}^{}
\frac{\partial {\theta({\bf n},\tau)} }{ \partial \tau }
\nonumber                         \\
&& - \frac{1}{2} \sum_{{\bf n},{\bf m}}^{}
\frac{ \partial {\theta({\bf n},\tau)} }{ \partial \tau} \frac{1}{(2e)^{2}}
W^{-1}({\bf n},{\bf m}) \frac{\partial {\theta({\bf m},\tau)}} { \partial \tau}
\label{final.phase.Coulomb}
\end{eqnarray}
with the phase stiffness \cite{Umezawa,Alvarez}
$J=2c\Delta^{2}=2n t^{2}/ U$.\footnote{It may also be called
 Josephson coupling due to the close analogy to the quantum
analysis of the Josephson junction, see the review \cite{Schon}
and Refs. therein. This analogy is widely used in
\cite{Ariosa.1991,Ariosa.1992,Beck.1997,Capezzali.thesis} for
physical interpretation.}
Note that since the value of $J$ was expressed via the zero
temperature gap $\Delta$ and neglects the gradient terms for
$\delta n_{p}(l,\tau)$, the treatment of the classical part of the
phase fluctuations becomes equivalent to that in the previous
section. The interaction matrix $W({\bf n},{\bf m})$ in
(\ref{final.phase.Coulomb}) is given by
\begin{eqnarray}
V({\bf n},{\bf m})= \left\{
\begin{array}{cc}
V_{\mathrm{SC}}({\bf n},{\bf m}), &  {\bf n}\neq {\bf m} \\
(2e)^{2} \frac{\ds b}{\ds d^{2}}, &  {\bf n}={\bf m} \,.
\end{array}  \right.
\end{eqnarray}
The expansion of $\Omega(\Phi,V,\tau)$ in powers of
$\Phi({\bf n}, \tau)$ has yielded a spurious
on-site repulsion $(2e)^{2}{b / d^{2}}$ between pairs. However,
due to the exclusion principle, two pairs cannot sit on the
same lattice site.  Thus, in the following,  ${\bf n} = {\bf m}$
is excluded in
the last term of (\ref{final.phase.Coulomb}); this is equivalent
to the usual introduction of hardcore bosons.

The final step in the derivation in
\cite{Beck.1997,Capezzali.thesis} is a standard Legendre
transformation from the "phase velocities"
$\partial {\theta}/ \partial \tau$ to the conjugate momenta
$p({\bf n},\tau)$, that are interpreted as the number density fluctuations
of charge carriers, at site ${\bf n}$ and imaginary time $\tau$.
The partition function that emerges from the latter
transformation is equal to the one that would have been obtained
from the following Hamiltonian:
\begin{eqnarray}
H & = & \frac{1}{2}\sum_{{\bf n},{\bf m}}^{}{
\left( p({\bf n})- \frac{n}{2} \right)
\left[ (2e)^{2}V({\bf n},{\bf m}) \right]
\left( p({\bf m})- \frac{n}{2} \right)}
\nonumber                  \\
&& + J\sum_{\left\langle {\bf n},{\bf m} \right\rangle}^{}{
\left[ 1-\cos{\left( \theta({\bf n})-\theta({\bf m}) \right)} \right]}.
\label{Hamiltonian.final.phase.Coulomb}
\end{eqnarray}

The Eq.~(\ref{Hamiltonian.final.phase.Coulomb}) is the central
result of the present section: the partition function of the Hubbard
model with local attraction and long-range Coulomb repulsion
has been written as a functional integral with a generalized
action (or thermodynamical potential) involving a fluctuating pairing
field $\Phi({\bf n}, \tau)$ and a local electric potential $V({\bf n},\tau)$.
After functional integration over the latter Hubbard-Stratonovich
field and over fluctuations in $|\Phi({\bf n}, \tau)|^2$, the final form
of the generalized action involves:

\noindent
$(i)$ an effective phase coupling
term between the local spatial variations of the phases of
$\Phi({\bf n}, \tau)$, which presents the same analytical structure
as the Josephson coupling term appearing in the Hamiltonian of the
classical $XY$-model;

\noindent
$(ii)$ a ``kinetic energy'' term, representing
the screened Coulomb interaction between local fluctuations of
the number density of charge carriers. The Hamiltonian
(\ref{Hamiltonian.final.phase.Coulomb}) represents a non-local
(extended) version of so-called {\em 2D quantum $XY$-model\/}
which, in addition to the usual Josephson coupling term, involves
a non-local kinetic energy term; in the case of underdamped
Josephson junction arrays, the latter describes capacitive
(quantum) effects which may arise between the superconducting
islands themselves but not between the superconducting islands
and the normal substrate.

\subsubsection{Mapping to the quantum $XY$-model with local
kinetic energy. Phase diagram}
\label{sec:mapping}

Here, following \cite{Beck.1997,Ariosa.1998,Capezzali.thesis}
we discuss the application of the Hamiltonian
(\ref{Hamiltonian.final.phase.Coulomb}) for calculating the transition
temperature of strongly anisotropic superconductors, such as
superlattices and bulk systems in the underdoped regime. To do this
the following approximations are made:

\noindent
$(i)$ $H$ is restricted to one superconducting layer;

\noindent
$(ii)$ the screened Coulomb interaction, which takes into
account the electric coupling between layers, is modelled by an
analytical form, which is similar to the usual
Yukawa potential:
\begin{equation}
V_{\mathrm{SC}}(r) = \frac{e^2}{\varepsilon a^2}
\frac{\exp(-r/ \lambda_{\mathrm{TF}})}{r} \,,
               \label{Yukawa}
\end{equation}
where $\varepsilon$ is the dielectric constant of the interlayer
material, $a$ is the lattice constant and
$\lambda_{\mathrm{TF}}$ is Thomas-Fermi screening length
depending on the density $n$ of charge carriers according to
the well-known formula
\begin{equation}
\lambda_{\mathrm{TF}}^{-1} =
\frac{2.95}{2 \pi} \sqrt{\frac{a_{\mathrm{B}}}{r_s}} \, [\mbox{A}^{-1}] \,.
\end{equation}
In the last expression $a_{\mathrm{B}}$ is the Bohr radius and
$r_{s}=\left( {3/ 4\pi n} \right)^{{1/3}}$ is a sphere diameter,
related to the density of charge carriers $n$.

\noindent
$(iii)$ considering only $n\ll 1$,
the "background shift", $-n/ 2$, in the first term of
(\ref{Hamiltonian.final.phase.Coulomb}) is neglected.

At this point one can make a connection to another, simpler
model \cite{Ariosa.1991,Ariosa.1992} (which is rather similar
to that proposed by Doniach and Inui \cite{Doniach}), based on
the quantum phase fluctuation of the $XY$-model, in which the
BKT transition temperature, $T_{\mathrm{BKT}}$ is identified with
the superconducting critical temperature in the absence of charging
effects and in which $T_{\mathrm{BKT}}$ is depressed from its
classical value by charging (quantum) effects \cite{Makivic}
(see also \cite{Sachdev}).
The basic Hamiltonian for this system is the quantum $XY$-model with
{\em local\/} kinetic energy:
\begin{equation}
H_{\mathrm{loc}} = J \sum_{\left\langle {\bf n},{\bf m} \right\rangle}^{}{
\left[ 1-\cos{\left( \theta({\bf n})-\theta({\bf m}) \right)} \right]} +
\frac{(2e)^2}{2C_{\mathrm{cap}}} \sum_{{\bf n}} N_{{\bf n}}^{2} \,,
               \label{local.quantum.XY}
\end{equation}
where $N_{\bf n}$ is equivalent to the number of ``2e-charges'' which are
present at site ${\bf n}$, $C_{\mathrm{cap}}$ is the self-capacitance of the
site.  A fundamental property of the quantum $XY$-model with a local
capacitive term, is provided by the fact that $n_{\bf n}$ and $\theta({\bf
n})$ are canonically conjugate variables and therefore satisfy the following
commutation relation (we restored temporarily the Planck constant $\hbar$ in
this relation):
\begin{equation}
[N_{{\bf n}}, \theta({\bf m})] = - i \hbar \delta_{{\bf n},{\bf m}} \,.
         \label{commutator}
\end{equation}
Thus, the fluctuations of these two are intimately related. In
the approach that has been used in \cite{Ariosa.1991,Ariosa.1992,Doniach},
the relation between $T_{\mathrm{BKT}}$ and real superconducting
temperature, which we denote as $T_{c}$, is given by
(again we restore the Boltzmann constant $k_{\mathrm{B}}$):
\begin{equation}
T_{c} = T_{\mathrm{BKT}} -
\frac{e^2 \pi}{12 k_{B} \langle C_{\mathrm{cap}} \rangle}\,, \qquad
\alpha \equiv \frac{2 e^2}{J \langle C_{\mathrm{cap}} \rangle} \ll 1 \,,
                       \label{Tc.quantum}
\end{equation}
where $\langle C_{\mathrm{cap}} \rangle$ is the average effective self-capacitance
and $\alpha$ is the ratio between charging and Josephson
energy.
Thus, within the approach \cite{Ariosa.1991,Ariosa.1992}, one only
has to evaluate the appropriate $\langle C_{\mathrm{cap}} \rangle$. It is
obvious that the quantum phase fluctuations decrease $T_{c}$
with respect to $T_{\mathrm{BKT}}$ evaluated in the presence
of the classical phase fluctuations only : $T_{\mathrm{BKT}}$
is an upper bound to the actual transition temperature
\cite{Emery.1995a}.

It is also important to mention the results of \cite{Roddick}
where for the quasi-2D version of the Hamiltonian
(\ref{local.quantum.XY}) it was shown that phase fluctuations
produce a linear increase in the penetration depth $\lambda(T)$
with increasing $T$ at low temperatures. This in turn implies
a linear decrease of the full superfluid density
(see the discussion in Sec.~\ref{sec:full.superfluid}).

In order to apply the framework sketched above one must
map the Hamiltonian (\ref{Hamiltonian.final.phase.Coulomb})
onto a ``local capacitance'' Hamiltonian (\ref{local.quantum.XY})
using the following approximation:
\begin{equation}
\frac{1}{2} \sum_{{\bf n},{\bf m}}^{}{p(l) \left[ (2e)^{2}V({\bf n},{\bf m}) \right] p({\bf m})}
\approx \frac{(2e)^{2}}{2C_{\mathrm{cap}}}\sum_{{\bf n}}^{}{p({\bf n})^{2}} \,,
\end{equation}
where
\begin{equation}
\frac{1}{2C_{\mathrm{cap}}}=
\frac{1}{e^{2}} \sum_{{\bf n}}^{}{V({\bf n},0)}=
\frac{2\pi \lambda_{\mathrm{TF}}}{ \varepsilon a^{2}}
e^{-a / \lambda_{\mathrm{TF}}} \,.
\label{capacitance}
\end{equation}
In Ref.~\cite{Ariosa.1991} the critical temperature $T_c$ for
the Hamiltonian density (\ref{local.quantum.XY}) has been evaluated in the
``self-consistent harmonic approximation'' (SCHA) which gives results for
$\alpha \ll 1$ in good agreement with Monte Carlo simulations.
A good overall fit of the numerical SCHA result is
(see Refs. in \cite{Beck.1997})
\begin{equation}
T_{c}(\alpha) \approx
T_{BKT} \sqrt{1- \frac{\alpha}{\alpha_{c} }} \,.
   \label{Tc.quantum.all}
\end{equation}
When $\alpha$ approaches $\alpha_{c}$=6.2, $T_{c}$ goes to zero.

\begin{figure}[h]
\centering \resizebox{0.7\textwidth}{!}{%
  \includegraphics{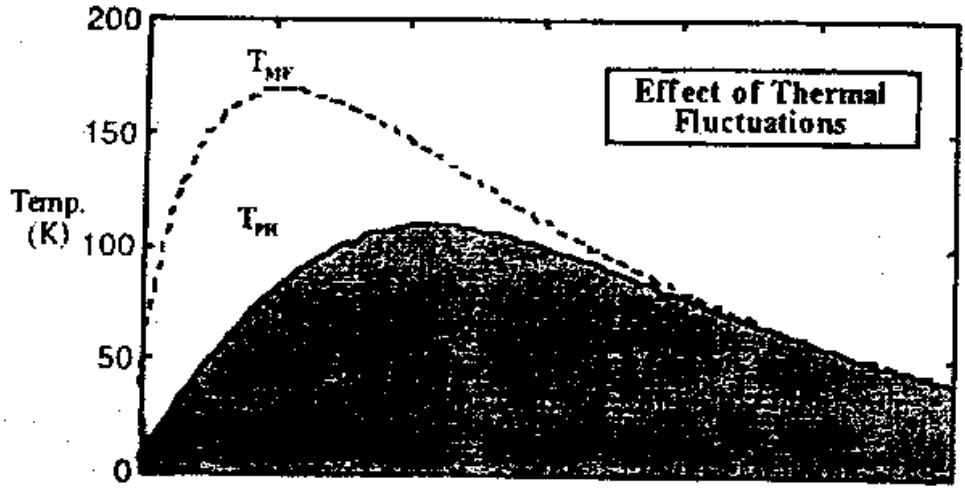}
}
\caption{The phase diagram based on the contribution
of the classical phase fluctuations. The $x$-axis shows the number of
carriers per Cu atom, and here $T_{\mathrm{BKT}}$ has been denoted by
$T_{\mathrm{PH}}$. The non-monotonic behaviour of the BCS temperature
$T_{\mathrm{MF}}$ is related to the peculiarities of the pairing mechanism.
This diagram is taken from \protect{\cite{Ariosa.1998}}.}
\label{fig:20}
\end{figure}

One has now derived an effective Hamiltonian of the form
(\ref{local.quantum.XY}), where the capacitance $C_{\mathrm{cap}}$ is given by
Eq.~(\ref{capacitance}), and $J$ is the phase stiffness, which is in fact
temperature dependent. As seen earlier, this temperature dependence is
particularly important in the high-density limit where $T_{BKT} \simeq
T_{c}^{\mathrm{MF}}$ and thus $J(T_{BKT}) \simeq J(T_{c}^{\mathrm{MF}}) = 0
\ll J(T=0)$. In \cite{Ariosa.1998} the temperature dependence is approximated
by a simple linear form chosen precisely such that $J(T_{c}^{\mathrm{MF}}) =
0$ i.e.
\begin{equation}
J(T) = J(T=0) \left( 1 - \frac{T}{T_{c}^{\mathrm{MF}}} \right),
\end{equation}
where $J(T=0)$ is the zero-temperature phase stiffness. Given this fit to the
phase stiffness, one can then calculate the critical temperature for the
classical phase fluctuations as a function of doping. This is nothing but the
BKT transition temperature, $T_{BKT}$, given by Eq.~(\ref{BKT.discrete}) with
the value of $A = 0.9$ taken from Monte-Carlo simulations. The critical
temperature for the quantum fluctuations (i.e. including charging effects) can
then also be calculated as a function of doping using
Eq.~(\ref{Tc.quantum.all}). Note that, it follows from the form of the
equations, that both $T_c$ and $T_{BKT}$ are bounded above by
$T_{c}^{\mathrm{MF}}$. The phase diagram based on only the classical phase
fluctuations is shown in Fig.~\ref{fig:20}.
This version of the phase diagram is in fact very similar to
the phase diagram presented in Fig.~\ref{fig:15} except
for details related to the pairing mechanism in
Ref.~\cite{Ariosa.1998} which results in a non-monotonic
behaviour for $T_{c}^{\mathrm{MF}}(n_{f})$.
The idea that, in the above context, the so-called pseudogap be
identified with the standard mean field superconducting gap
is also proposed \cite{Ariosa.1998}. Nonetheless, these gaps
should not be identical due to the quantum corrections, as is the case
for the temperatures $T_{c}^{\mathrm{MF}}$ and $T_{\rho}$
or $T_{X}$ from \cite{Tremblay.1997} (see the discussion in
Sec.~\ref{sec:pairing.temperature}).

\begin{figure}[h]
\centering \resizebox{0.7\textwidth}{!}{%
  \includegraphics{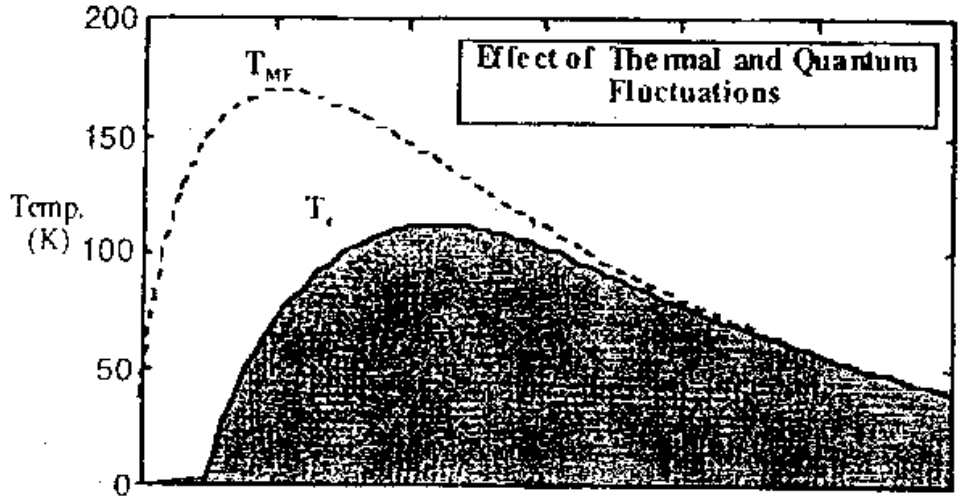}
}
\caption{The phase diagram from
\protect{\cite{Ariosa.1998}} based on both classical and quantum
phase fluctuations. Again the $x$-axis shows the number of
carriers per Cu atom.}
\label{fig:21}
\end{figure}

The phase diagram where the effect of the quantum phase
fluctuations is taken into account is shown in Fig.~\ref{fig:21}.
The most important effect of these fluctuations is that
they push the critical temperature to zero in the underdoped
region. Thus one can see that the quantum fluctuations lead
to a greater overlap between the experimental phase
diagram in Fig.~\ref{fig:01} and its theoretical version in
Fig.~\ref{fig:21}.

Furthermore, a possible mechanism which also pushes $T_c$ to zero for
the overdoped region was suggested in \cite{Ariosa.1998}. Its discussion
is however beyond the phase fluctuation scenario of this review.
Finally, to make a connection with the next section, we note that
another reason why $T_c$ is zero for overdoped systems
is the presence of impurities which can also destroy
superconductivity.

\subsection{The effect of non-magnetic impurities}
\label{sec:impure}

Up till this point we have considered models for the crossover and
the pseudogap behaviour which do not take into account the effect of
impurities. However, a discussion of the effect of impurities seems to be
crucial for any realistic model of the HTSC. Indeed, it is known, that the
itinerant holes in HTSC are created by doping which in turn
introduces a considerable disorder into the system, for instance,
from the random fields of chaotically distributed charged impurities
\cite{Pogorelov}.  Thus the purpose of this section is to study
the influence of non-magnetic impurities on superconducting properties
including the superfluid density.

In the theory of ``common metals'' the Fermi energy $\epsilon_{F}$
and the mean transport quasi-particle time $\tau_{\mathrm{tr}}$ are
independent quantities which are always assumed to satisfy the
criterion  $\epsilon_{F} \tau_{\mathrm{tr}} \gg 1$. In HTSC which are
``bad metals'' \cite{Emery.1995b} both $\epsilon_{F}$ and
$\tau_{\mathrm{tr}}$ are dependent on the doping and the abovementioned
criterion may fail \cite{Pogorelov}. In particular it has been shown
\cite{Pogorelov}, that for strongly disordered metallic systems,
superconductivity is absent if the scattering-to-pairing ratio
exceeds a critical value. Secondly, for such
systems \cite{Pogorelov}, even if this ratio is not exceeded,
superconductivity only exists for a finite range of doping.

We shall study two cases here. The first is the case of weak disorder
originally studied in the papers of Anderson \cite{Anderson.theorem} and
Abrikosov-Gor'kov (AG) \cite{Abrikosov} (see also \cite{Abrikosov.book}).
when the superconducting order is preexisting and the criterion
$\epsilon_{F} \tau_{\mathrm{tr}} \gg 1$ is satisfied.
In this case superconductivity in disordered thin films was first
considered by Ramakrishnan \cite{Ramakrishnan} using an approach
very similar to the method presented in the review.
The second case is that of strong disorder where this criterion is not
satisfied, and superconductivity is destroyed by sufficient disorder.

We concentrate on several important aspects. The first is
the influence of impurities on the phase diagram particularly
at high-densities where the Fermi surface is well-developed.
The second is the behaviour of superfluid density.
Its temperature dependence at low ($T \ll T_{c}$) and high
($T_{c} < T < T^{\ast}$) temperatures is not yet understood and
presents a challenge for competing theories.
One of the consequences of the presence of
impurities is that the value of the zero temperature superfluid density
can be less than the total density of carriers, so that their presence
may help to explain the experimental results for these quantities,
\cite{Emery.1995a,Emery.1995b}. The third is the creation of
inhomogeneities and the ultimate destruction of superconductivity by
strong disorder.

It is interesting to note an alternative explanation for the lowered
zero temperature superfluid density as a result of quantum phase
fluctuations has been recently presented \cite{Chakraverty} (see also
\cite{Corson.low}).
In the previous section the importance of the contribution from
the quantum phase fluctuations is clearly evident. It follows that
the lowering of the superfluid density is the result of at least two
factors -- namely impurities and quantum phase fluctuations.

\subsubsection{Weak disorder limit}
\label{sec:weak.disorder}

Anderson's theorem \cite{Anderson.theorem}
(see also \cite{Rickayzen.book}) states that in 3D the BCS
critical temperature is unchanged in the presence of weak disorder
in the form of non-magnetic impurities. As discussed in
Sec.~\ref{sec:our.diagram}, the BCS critical temperature (in 2D) is the
temperature $T^{\ast}$ at which the pseudogap opens and is unchanged
by the presence of impurities. On the other hand
the superconducting transition temperature is the temperature
$T_{\mathrm{BKT}}$ of the BKT transition. In contrast to the former, the
latter (as seen in Sec.~\ref{sec:our.diagram}) is defined in terms of the bare
superfluid density or the phase stiffness $J$ (given by the delocalized
carriers) which decreases with an increasing concentration of
impurities. Thus in 2D the temperature of the superconducting transition,
$T_{\mathrm{BKT}}$ decreases with increasing impurity concentration.

Thus, in the model with impurities, the relative size of the
pseudogap phase, $(T^{\ast} - T_{\mathrm{BKT}})/T^{\ast}$ could be larger
in the presence of impurities \cite{Loktev.dirty}
than in the clean limit \cite{Gusynin.JETP}
and might therefore be observed over a wider range of densities.

\paragraph{Model and main equations}
Our starting point is, as in Secs.~\ref{sec:one-band} and
\ref{sec:our.diagram}, a continuum version of the 2D attractive Hubbard model
defined by the
simplest Hamiltonian (\ref{Hamiltonian.2D}) in the presence of impurities
\cite{Abrikosov.book}
\begin{eqnarray}
H  = \int d^2 r \left[
\psi_{\sigma}^{\dagger}(x)
\left(- \frac{\nabla^{2}}{2 m} - \mu \right) \psi_{\sigma}(x) \right.
& - & U \psi_{\uparrow}^{\dagger}(x) \psi_{\downarrow}^{\dagger}(x)
     \psi_{\downarrow}(x) \psi_{\uparrow}(x)
\nonumber                   \\
& + & \left. V_{imp} (\mbox{\bf r})
\psi_{\sigma}^{\dagger}(x) \psi_{\sigma}(x) \right] \,,
                     \label{Hamilton}
\end{eqnarray}
where as before $x= \mbox{\bf r}, \tau$ denotes the space and
imaginary time variables, $\psi_{\sigma}(x)$ is a fermion field
with spin $\sigma =\uparrow,\downarrow$, $m$ is the effective
fermion mass, $\mu$ is the chemical potential, $U$ is an effective
local attraction constant, and $V_{imp}({\bf r})$ is the static
potential of randomly distributed impurities.

From the Hamiltonian (\ref{Hamilton}), following the derivation
given in Sec.~\ref{sec:derivation},
one can again derive the effective Hamiltonian of
the classical $XY$-model (\ref{XY.Hamiltonian})
with the bare (i.e. unrenormalized by the phase fluctuations, but
including pair breaking thermal fluctuations) superfluid stiffness:
\begin{eqnarray}
J(\mu, T, \rho) = && \frac{T}{16 m \pi^2} \sum_{n = -\infty}^{\infty}
\int d^2 k \mbox{tr} [\tau_3
\langle \mathcal{G}(i \omega_{n}, {\bf k}) \rangle]
\nonumber                      \\
&& + \frac{T}{32 m^2 \pi^2} \sum_{n = -\infty}^{\infty}
\int d^2 k k^2 \mbox{tr}
[ \langle \mathcal{G}(i \omega_{n}, {\bf k}) \rangle
  \langle \mathcal{G}(i \omega_{n}, {\bf k}) \rangle ] \,.
\label{J}
\end{eqnarray}
Here
\begin{equation}
\langle \mathcal{G}(i \omega_{n}, {\bf k}) \rangle = -
\frac{ (i \omega_{n} \hat{I} - \tau_{1} \rho ) \eta_{n}
+ \tau_{3} \xi({\bf k})}
{(\omega_{n}^{2} + \rho^{2}) \eta_{n}^{2} + \xi^{2}(\mbox{\bf k})},
                       \label{Green.impurity}
\end{equation}
with
\begin{equation}
\eta_{n} = 1 +
\frac{1}{2 \tau_{\mathrm{tr}} \sqrt{\omega_{n}^{2} + \rho^{2}}},
\end{equation}
is the AG \cite{Abrikosov.book} Green's function of neutral fermions
averaged over a random distribution of impurities and written in the
Nambu representation \cite{Evans,Rickayzen.book}.
In writing (\ref{J}) we assumed that
$ \langle \mathcal{G}(i \omega_{n}, {\bf k})
\mathcal{G}(i \omega_{n}, {\bf k}) \rangle \simeq
\langle \mathcal{G}(i \omega_{n}, {\bf k}) \rangle
\langle \mathcal{G}(i \omega_{n}, {\bf k}) \rangle$.
This approximation, as shown by AG \cite{Abrikosov.book},
does not change the final result for $J$. Note also that the Green's
function (\ref{Green.impurity}) is valid only when
$\epsilon_{F} \tau_{\mathrm{tr}} \gg 1$ which demands the presence of
a well developed Fermi surface, which in turn implies that
$\mu \simeq \epsilon_{F}$. Thus one cannot use the expression
(\ref{Green.impurity}) in the Bose limit with
$\mu < 0$.

Substituting (\ref{Green.impurity}) into (\ref{J}), and
using the inequalities $\mu \gg T, \rho$ to extend the limits of
integration to infinity, one arrives at
\begin{equation}
J = \frac{\mu}{4 \pi} + \frac{T \mu }{4 \pi}
\sum_{n=-\infty}^{\infty} \int_{-\infty}^{\infty}
dx \left( \frac{1}{x^2 + (\omega_{n}^{2} + \rho^{2}) \eta_{n}^{2}}
- \frac{2 \omega_{n}^{2} \eta_{n}^{2}}
{[x^2 + (\omega_{n}^{2} + \rho^{2}) \eta_{n}^{2}]^2} \right) .
\label{J.intermediate}
\end{equation}
Eq.~(\ref{J.intermediate}) is formally divergent and demands special care due
to the fact that one, in contrast to Sec.~\ref{sec:derivation}, has to perform
the integration over $x$ before the summation \cite{Abrikosov.book}. Finally
one can formally cancel the divergence \cite{Abrikosov.book} to obtain
\begin{equation}
J = \frac{\mu \rho^{2} T}{4} \sum_{n = -\infty}^{\infty}
\frac{1}{(\omega_{n}^{2} + \rho^{2})
\left[ \sqrt{\omega_{n}^{2} + \rho^{2}} +
\frac{\ds 1}{\ds 2 \tau_{\mathrm{tr}}} \right]}.
\label{J.final}
\end{equation}

The temperature of the BKT transition for the Hamiltonian
(\ref{XY.Hamiltonian}) is determined by the self-consistent
equation (\ref{BKT.equation}).

The self-consistent calculation of $T_{\mathrm{BKT}}$ as a function
of the carrier density $n_f = m \epsilon_{F}/\pi$ as before requires additional
equations for $\rho$ and $\mu$, which together with
(\ref{BKT.equation}) form a complete set
(see Sec.~\ref{sec:derivation}).

When the modulus of the order parameter $\rho(x)$ is treated
in the mean field approximation, the equation for $\rho$
(compare with (\ref{rho})) takes in the presence of
impurities the following form
\begin{equation}
\frac{2 \rho}{U} = \sum_{n = -\infty}^{\infty}
\int \frac{d^2 k}{(2 \pi)^{2}} \mbox{tr}
[\tau_1 \langle \mathcal{G}(i \omega_{n}, \mbox{\bf k}) \rangle] ,
\label{rho.gap}
\end{equation}
which formally coincides with the gap equation of the BCS theory.
This coincidence confirms Anderson's theorem
\cite{Anderson.theorem} which states that in this limit the dependence of
$\rho(T)$ is the same as that for the clean superconductor and is not affected
by the presence of non-magnetic impurities.

We recall, however, that there is both physical and mathematical
differences between the gap in the BCS theory and $\rho$
(see Sec.~\ref{sec:our.diagram} above).
The main point, which we would like to stress here, is that due
to the Anderson theorem \cite{Anderson.theorem} the value of $T_{\rho}$
does not depend on the presence of impurities at least when
$\epsilon_F \tau_{\mathrm{tr}} \gg 1$, while the
temperature $T_{\mathrm{BKT}}$, as we will see below, is really
lowered.

The chemical potential $\mu$ is defined by the number equation
written for the Green's function (\ref{Green.impurity}) in
the presence of impurities:
\begin{equation}
\sum_{n = -\infty}^{\infty} \int \frac{d^2 k}{(2 \pi)^{2}} \mbox{tr}
[\tau_3 \langle \mathcal{G}(i \omega_{n}, \mbox{\bf k}) \rangle]
\exp (i \delta \omega_n \tau_{3})= n_f \,.
\label{number.impure}
\end{equation}
Since we are interested in the high carrier density region
the solution of (\ref{number}) is $\mu \simeq \epsilon_{F}$, so
that in Eqs.~(\ref{J.final}) -- (\ref{rho.gap}) one can replace
$\mu$ by $\epsilon_{F}$.

\paragraph{The effect of weak disorder on the pseudogap region}
In the clean limit the transport time $\tau_{\mathrm{tr}}$ is infinite,
so that
\begin{equation}
J(\epsilon_{F}, T, \rho(\epsilon_{F},T)) =
\frac{\epsilon_{F} \rho^{2} T}{4} \sum_{n = -\infty}^{\infty}
\frac{1}{(\omega_{n}^{2} + \rho^{2})^{3/2}} .
\label{J.clean}
\end{equation}
Near $T_{\rho}$ one can obtain from (\ref{J.clean}) its asymptotical
form (\ref{J.bcs}) which has been re-derived here
using the reverse order for the summation and integration.

As has been explained in Sec.~\ref{sec:results},
if one inserts the dependence of $\rho(T)$ given by (\ref{rho.dependence})
into (\ref{J.bcs}) and then substitutes (\ref{J.bcs}) into (\ref{BKT.equation})
one obtains the asymptotic expression (\ref{asymp}) for the BKT
temperature in the clean limit for high carrier densities.

In the high density limit one can also use the following
solution of Eq.~(\ref{temperature.rho})
for $T_{\rho}$
\begin{equation}
T_{\rho} = \frac{\gamma}{\pi} \sqrt{2 |\varepsilon_{b}| \epsilon_{F}}\,.
\label{rho.temperature}
\end{equation}
(Compare Eq.~(\ref{rho.temperature}) with (\ref{solution.2D}) which
reproduce together the standard BCS ratio
$2 \Delta/T_{\rho} = 2 \pi/\gamma \simeq 3.5$.)

It is obvious from (\ref{asymp}) and (\ref{rho.temperature})
that for a clean system the pseudogap region shrinks rapidly for high carrier
densities. It is therefore natural to ask (see, for example,
\cite{Randeria.review,Halperin} and the discussion in
Sec.~\ref{sec:results}) whether precursor superconducting phase fluctuations
can explain the pseudogap anomalies which are observed over a wide range of
temperatures and carrier densities, since in the clean limit the relative size
of the pseudogap region $(T_{\rho} - T_{\mathrm{BKT}})/T_{\rho}$ is, for
instance, less than $1/2$ when the dimensionless ratio
$\epsilon_{F}/|\varepsilon_{b}| \lesssim 128 \gamma^{2}/\pi^2 \simeq 41$.
If one makes a crude estimate for the dimensionless ratio for optimally doped
cuprates one obtains $\epsilon_{F}/|\varepsilon_{b}| \sim 3 \cdot 10^2$ - $10^3$
\cite{Carter} which indicates that in the clean superconductor
the pseudogap region produced by the phase fluctuations is too small.
Of course, all these estimations are rather qualitative due to the
simplicity of the model.

We now turn to the dirty limit where the transport time
$\tau_{\mathrm{tr}}$ is small ($\tau_{\mathrm{tr}} \ll \rho^{-1}(T=0)$).
This inequality is not in contradiction with the weak disorder limit
but merely implies high density i.e. that $\sqrt{\epsilon_F} >>
\sqrt{|\epsilon_b|}$. In this limit one can neglect the radical in the
bracket of (\ref{J.final}) \cite{Abrikosov.book}. The remaining series is
easily summed and one obtains the following expression for the bare
superfluid stiffness,
\begin{equation}
J(\epsilon_{F}, T, \rho(\epsilon_{F},T), \tau_{\mathrm{tr}}) =
\frac{\epsilon_{F} \tau_{\mathrm{tr}} \rho}{4} \tanh \frac{\rho}{2T} .
\label{J.dirty}
\end{equation}

As in the clean limit, due to Anderson's theorem, the expressions
(\ref{rho.dependence}) for $\rho$ and (\ref{rho.temperature}) for
$T_{\rho}$ remain unchanged in the presence of
impurities. Again substituting (\ref{rho.dependence}) into
(\ref{J.dirty}) one obtains
\begin{equation}
T_{\mathrm{BKT}} = T_{\rho}
\left( 1 - \frac{14 \zeta(3)}{\pi^3}
\frac{1}{\epsilon_{F} \tau_{\mathrm{tr}}}  \right), \qquad
T_{\mathrm{BKT}} \lesssim T_{\rho}.
\label{BKT.dirty}
\end{equation}
One can see that the size of the pseudogap region is now
controlled by the new parameter $\tau_{\mathrm{tr}}$ which is an
unknown function of $\epsilon_{F}$ for HTSC. If one assumes that
one indeed has weak disorder i.e. $\epsilon_F \tau_{\mathrm{tr}} >>1$
clearly the pseudogap region remains small. If one however moves away
from the weak disorder lint and uses values for $\tau_{\mathrm{tr}}$
extracted from experiment \cite{Timusk,Puchkov}, for which
$\tau_{\mathrm{tr}}^{-1} \gg \rho(T=0) \sim T_{\rho}$ then
Eq.~(\ref{BKT.dirty}) one does obtain an increase in the pseudogap
region with impurities. Such an estimate is of necessity only qualitative.

Certainly impurities which are always present in HTSC do increase the
size of the pseudogap region in the weak disorder limit. The effect
however appears to be small in this limit and it seems likely that the
effect of quantum phase fluctuations \cite{Chakraverty} is more important
in allowing one to reproduce the experimentally observed extent of the
pseudogap region. It would however be interesting to readdress this
problem in the case of strong disorder.

The difference in the temperature scales for pairing and phase coherence
for higher densities where $\mu > 0$ is vital to the debate, previously
mentioned in Sec.~\ref{sec:relevance}, as to how the approaches of Uemura
\cite{Uemura} and  Emery-Kivelson \cite{Emery.1995a,Emery.1995b}
(see in particular \cite{Emery.conference}) are related. While
Uemura \cite{Uemura} considers that the approach based on the
phase fluctuations as ``essentially identical to the BEC-BCS crossover
picture'', Emery and Kivelson \cite{Emery.conference} insist
that the separation of the temperature scales
for pairing and phase coherence is a consequence of the fact
that HTSC are doped insulators. Here we would like to present a point
of view which is closer to that expressed in \cite{Emery.conference}.
Indeed, as we saw in the previous chapters the indication that one
is on the BEC side of the BEC-BCS is a negative sign for the chemical
potential $\mu$. In Sec.~\ref{sec:results} we saw that in the
region with a large pseudogap $\mu$ was indeed mostly negative
phase, so that the statement of Uemura \cite{Uemura} applies.
However, as we discussed optimally doped cuprates correspond
to higher carrier densities where $\mu$ is already positive
and $\mu \simeq \epsilon_{F}$. The presence of quantum phase fluctuations,
and to a lesser extent impurities, however permits one to have a relatively big
pseudogap region even on the BCS side of the crossover.
Another argument in favour of the picture presented by
Emery and Kivelson is that the value
$2 \Delta_{\mathrm{pair}} /T^{\ast} \sim 4\, - \,5$ \cite{Oda} is
almost independent of doping while it increases on the BEC side of
crossover, see Sec.~\ref{sec:results}.
Thus the BEC-BCS crossover and the present approach based on the phase
fluctuations need not be related.

We stress however that this argument about the sign of the chemical
potential is not valid in the presence of spatial inhomogeneities (see
discussion in Sec.~\ref{sec:derivative}) as for example seen in the stripe
picture of Emery, Kivelson and Zachar \cite{Zachar}. This implies that
BEC-BCS crossover and phase fluctuation picture
are not mutually exclusive since for a spatially
inhomogeneous system it is possible to have local regions with so small
a carrier density that pre-formed Bose pairs do exist there, while
in average the whole system is still on the BCS side of the crossover.
This is, in some sense, similar to the many-band model
considered in Sec.~\ref{sec:band} where coexisting local and
Cooper pairs were separated not in real, but in momentum space.

\paragraph{The superfluid density}
\label{sec:superfluid}

In this section we contrast the experimental results for the bare and full
superfluid densities with those for the clean superconductor and indicate
the likely role played by impurities.

The bare superfluid density is straightforwardly
expressed via the bare phase stiffness, $n_s(T) = 4m J(T)$.
In particular in the clean limit, it follows from
(\ref{J.clean}) (or (\ref{K10}) with (\ref{number.rho}))
that $n_s(T=0) = n_f$.
This is not surprising since $n_s(T=0)$ must be equal to the
full density $n_f$ for any superfluid ground state in a
translationally invariant system \cite{Leggett.theorem} and the clean
system is translationally invariant. We note, however, as stated above
that in HTSC $n_{s}(T=0) \ll n_f$ \cite{Emery.1995b}.
Substituting (\ref{rho.dependence}) into (\ref{J.bcs}) one
obtains for $T$ close to $T_{\rho}$ the bare superfluid
density
\begin{equation}
n_{s}(T \to T_{\rho}^{-}) = 2 n_{f} (1 - T/T_{\rho}).
\label{eq:barebcs}
\end{equation}
This behaviour of the bare superfluid density is formally the same
as the behaviour of the full superfluid density in the BCS theory.
Nevertheless it is important to remember that the full superfluid
density in the present model undergoes the Nelson-Kosterlitz
jump at $T_{\mathrm{BKT}}$ and is zero for $T > T_{\mathrm{BKT}}$.

We note that one can experimentally probe both the bare superfluid
density in high-frequency measurements \cite{Corson} and the full
superfluid density in low-frequency measurements
\cite{Panagopoulos}.
The former describes mainly the local condensate density distribution
which is, of course, nonzero above $T_c$ in the precursor
superconductivity theories, while the latter corresponds to an
average condensate density and is zero above $T_c$.

\begin{figure}[h]
\centering \resizebox{0.5\textwidth}{!}{%
  \includegraphics{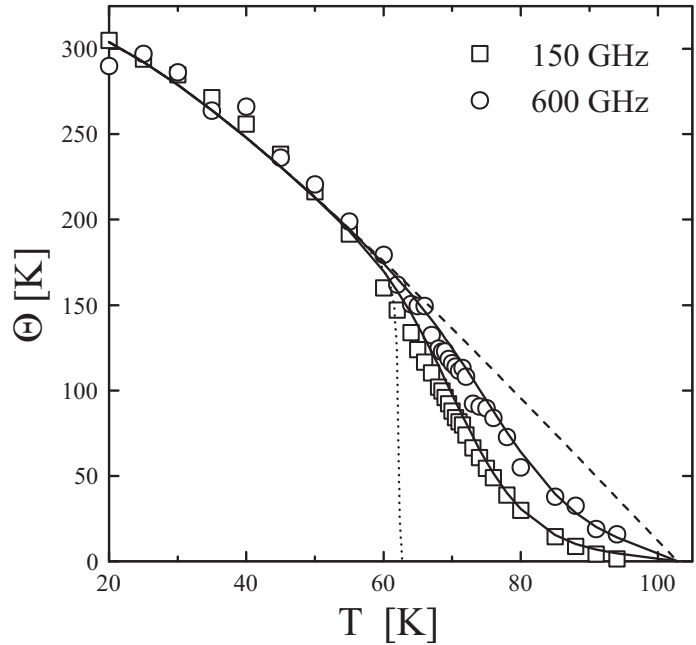}
}
\caption{The phase stiffness, $\Theta(T)$ of the
Bi$_2$Sr$_2$CaCu$_2$O$_{8+\delta}$ sample with $T_c =71$K
plotted at 140 GHz, open squares, and 600 GHz, open circles.
The long-dashed line is the bare phase-stiffness temperature.
The dotted line is the prediction for $\Theta$ in the low-frequency
limit, illustrating the universal superfluid jump. This figure
is taken from \protect{\cite{Corson}}.}
\label{fig:22}
\end{figure}

A direct comparison with experiment reveals further
discrepancies between experiment and the predictions given above for
the clean limit.
The experimental shape of the bare superfluid density
(see Fig.~\ref{fig:22} taken from \cite{Corson}) disagrees with that
given by Eq.~(\ref{eq:barebcs}).
Moreover, one of the most striking features is that although
$n_{s}(T)$ remains nonzero well above $T_c$, it does not persist
throughout the pseudogap temperature regime. Instead it falls
linearly to zero at a temperature $T_{n_{s}}$ well below $T_c$.
Even though the presence of impurities leads one closer to
the experimentally observed behaviour of
$n_{s}(T)$ in the region $T_{c} < T < T_{\rho}$
(see \cite{Loktev.dirty}), the superfluid density still becomes zero
only at $T = T_{\rho}$.
Thus the observed behaviour of the bare superfluid density
at finite temperature has not yet been fully explained in this
approach and presents an open
question to the current theories of phase fluctuations
(for another problem related to the superfluid density
in the theories based on phase fluctuations
see also the end of Sec.~\ref{sec:full.superfluid}).

On the other hand,
the zero value of $n_{s}(T)$ above $T_{n_{s}} < T^{\ast}$
may be well understood within other scenarios which claim that
pseudogap is not the result of the superconducting fluctuations
but is related, for example, to charge- and/or spin-density
waves as discussed in \cite{Klemm.1998,Krasnov}.
In such a case there is still a superconducting contribution
to the pseudogap even in the region $T_{c} < T < T_{n_{s}}$.
Since $T_{n_{s}}$ is less than $T^{\ast}$, one can speculate that
it could be much easier to explain the observed size of the
superconducting pseudogap region even within a simple model from
Sec.~\ref{sec:our.diagram}.
Further experimental studies of the doping dependence of
$T_{n_{s}}$ and its clear resolution from $T_{c}$ and $T^{\ast}$
are necessary.

The experimental data of \cite{Panagopoulos}
also shows that the full superfluid density, $\mathcal{N}_{s}(T)$,
does not display the Nelson-Kosterlitz jump. This is probably related
to the influence of the interlayer coupling as discussed in
\cite{Carlson}.

One can however directly obtain a reduced value
of the zero temperature bare superfluid density in the dirty limit.
In this limit the value of the zero temperature superfluid density is
given by $n_{s}(T=0) = \pi n_{f} \tau_{\mathrm{tr}} \rho \ll n_{f}$
since $\tau_{\mathrm{tr}} \rho \ll 1$. This does not contradict the
results of \cite{Leggett.theorem} because the system is not
translationally invariant in the presence of impurities.

We conclude that the low value of the superfluid
density in HTSC \cite{Emery.1995b} may at least in part
(for other mechanisms, for example, based on the quantum phase
fluctuations, see \cite{Chakraverty,Corson.low}) be straightforwardly
related to the impurities which are inevitably present in HTSC. On the
other hand its temperature dependence remains an important unsolved
problem.

\subsubsection{Strong disorder}

As seen in Sec.~(\ref{sec:weak.disorder}) impurities are vital for an
understanding of the superconducting properties of HTSC.
We now turn to a discussion of these properties in the complementary
case of strong disorder. In the presence of strong
disorder Anderson's theorem no longer holds, and the pairing
amplitude $\rho$ need no longer be constant in space
but is in fact spatially dependent i.e. $\rho \equiv \rho({\bf r})$. An
analysis of this effect on the superconducting properties of $s$-wave
superconductors has been presented in \cite{Trivedi.1998} and we simply
represent the main results here.

\paragraph{Model and main equations}

The model is the $2D$ attractive Hubbard model (see Eq.~(\ref{Hubbard})) but
with an additional site dependent disorder potential $V^{imp}_{\bf n}$ with
values chosen randomly from the uniform distribution $[-V,V]$ where $V$
controls the strength of the disorder
\begin{equation}
H =-t\sum_{\langle {\bf n } {\bf m} \rangle, \sigma}
( c^\dagger_{{\bf n} \sigma} c_{{\bf m} \sigma} + {\rm {h.c.}} )
- U \sum_{\bf n} n_{{\bf n}\uparrow} n_{{\bf n} \downarrow} +
\sum_{{\bf n}, \sigma} (V^{imp}_{\bf n} - \mu) n_{{\bf n} \sigma} \,,
\label{Hubbard.imp}
\end{equation}
Firstly one treats the spatial fluctuations of the modulus of the order
parameter in the mean field approach, using the standard
Bogolyubov-de Gennes (BdG)
equations \cite{deGennes.book}
\begin{equation}
\left( \begin{array}{cc} \hat{\xi} & \rho({\bf r}_{\bf n}) \\
\rho^*({\bf r}_{\bf n}) & - \hat{\xi}^*  \end{array} \right)
\left( \begin{array} {c} u_{\bf i}({\bf r}_{\bf n}) \\
v_{\bf i}({\bf r}_{\bf n}) \end{array} \right)
= E_{\bf i} \left( \begin{array} {c} u_{\bf i}({\bf r}_{\bf n})
\\ v_{\bf i}({\bf r}_{\bf n})
\end{array} \right)
                                 \label{BdG}
\end{equation}
where the site-dependent modulus is given by
\begin{equation}
\rho({\bf r}_{\bf n}) = U \sum_{\bf i} u_{\bf i}({\bf r}_{\bf n})
v_{\bf i}^*({\bf r}_{\bf n}),
\end{equation}
$\hat{\xi} u_{\bf i}({\bf r}_{\bf n}) = - t \sum_{\hat{\delta}}
u_{\bf i}({\bf r}_{\bf n}+\hat{\delta})+ (V^{imp}_{\bf n}-\tilde{\mu})
u_{\bf i}({\bf r}_{\bf n})$ and similarly for $v_{\bf i}({\bf r}_{\bf n})$.
Here $\hat{\delta} = \pm \hat{\bf x}, \pm \hat{\bf y}$ as appropriate to a
square lattice and $\tilde{\mu}_{\bf n} = \mu + U n_{\bf n}/2$ incorporates a
site-dependent Hartree shift dependent on the local number density,
\[ n_{\bf n} = \sum_{\bf i} |v_{\bf i}({\bf r}_{\bf n})|^2 \]
The chemical potential, $\mu$, is simply given by fixing the average number
density, $n$.
\begin{equation}
\frac{1}{N} \sum n_{\bf n}  = n
\end{equation}
where $N$ is the number of sites. These equations are non-linear and have to
be solved self-consistently
for the BdG eigenvalues $E_{\bf i}$ and the eigenvectors
$[u_{\bf i}({\bf r}_{\bf n}),v_{\bf i}({\bf r}_{\bf n})]$.
In practice they are solved numerically on a finite lattice with
periodic boundary conditions.

\paragraph{Single-particle spectral gap and bare superfluid stiffness}

Several random realizations for the disorder
are considered. For each realization one solves the BdG equations to find
$\rho({\bf r}_{\bf n})$ and then averages over the disorder to obtain a
probability distribution $P(\rho)$ for the local pairing amplitudes. For weak
disorder, $V < 0.25 t$, $P(\rho)$ is strongly
peaked at its mean-field (BCS) value and one thus recovers a spatially uniform
$\rho$. The distribution broadens significantly for $V \simeq t$ and is rather
skewed for $V \simeq 2t$ with a large weight near $\rho = 0$.

One can then also construct the disorder averaged one-particle density
of states from the BdG eigenvalues
\begin{equation}
N(\omega) = \frac{1}{N} \sum_{\bf i} \delta(\omega-E_{\bf i})
\end{equation}
and this has been plotted in \cite{Trivedi.1998} for three different
disorder strengths $V$. The most important feature of these plots is
the fact that there is a finite spectral gap even at high disorder.
For each individual representation of the disorder, the lowest eigenvalue
$E_{\bf i}$, remains non-zero and of the order of the zero disorder BCS gap
and this is the origin of the spectral gap. This occurs despite the fact that,
for high disorder, a large fraction of the sites have near zero pairing
amplitudes.

There is however a particularly simple explanation for this
behaviour in the high disorder limit as follows. Plotting
$\rho({\bf r}_{\bf n})$ one observes superconducting ``islands'',
(i.e. clusters of sites where $\rho({\bf r}_{\bf n})$ is large)
which correspond to regions where $|V^{imp}_{\bf n}|$ is small.
These are surrounded by a larger ``sea'' of sites with
$\rho \simeq 0$ and where $|V^{imp}_{\bf n}|$ is large. The lower lying BdG
eigenvalues correspond to the superconducting islands and so have finite gap.
The BdG eigenvalues corresponding to the ``sea'' have even higher energies
because the corresponding disorder potential $|V^{imp}_{\bf n}|$ is large.

One notes that in this procedure \cite{Trivedi.1998} one calculates the
spectrum $E_{\bf i}$ and $N(\omega)$ for each individual disorder
representation and then disorder averages. This will clearly give different
results from the procedure \cite{Xiang.1995} of constructing a disorder
averaged $\rho({\bf r}_i)$ and then looking at the energy gap. In the
latter case one smoothes out the amplitude fluctuations and, in contrast to the
picture presented here, one obtains a gap which closes with increasing
disorder \cite{Xiang.1995}. Note that it is not possible to prove
that the single-particle strength  at $\omega = 0$ is exactly zero
(particularly for weak coupling $U/t << 1$) since the numerical calculations are
by necessity in a finite system. Furthermore, note that the gap will be filled
by the presence of phase fluctuations, as in the case of zero disorder
which will be discussed in Chap.~\ref{chap:spectral}.

\begin{figure}[h]
\centering \resizebox{0.50\textwidth}{!}{%
  \includegraphics{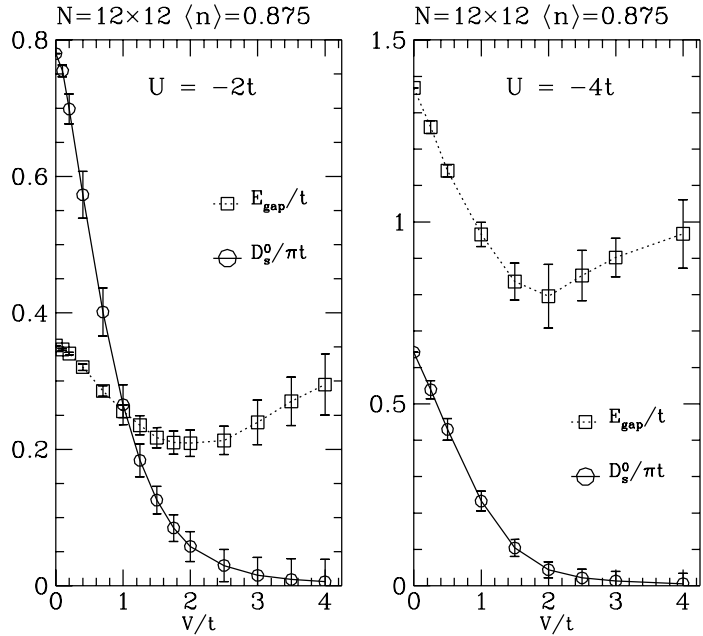}
}
\caption{The $T=0$ BdG bare superfluid stiffness $D_s^0$ and the spectral gap
$E_{\mathrm gap}$ plotted as a function of disorder strength for two
different values $U (>0)$ of attraction. Note that the gap persists
with increasing disorder while stiffness decreases. This figure has
been taken from \protect{\cite{Trivedi.1998}}.}
\label{fig:23}
\end{figure}

One can also calculate the bare superfluid stiffness, $D_s^0$, given by
\cite{Scalapino.1993}
\begin{equation}
\frac{D_s^0}{\pi} = \langle -k_x \rangle  - \Lambda_{xx}(q_x=0,q_y \rightarrow
0,\omega=0).
\end{equation}
The diamagnetic term is one-half (in 2D) the kinetic energy $\langle -
\mathcal{K} \rangle$ and the paramagnetic term is the (disorder averaged)
transverse current-current correlation function. From Fig. (\ref{fig:23})
one can see a reduction in $D_s^0$ of two orders of magnitude with increasing
disorder. It never however becomes zero showing that superconductivity is not
destroyed by amplitude fluctuations.

In the strong disorder limit $D_s^0 << E_{gap}$ regardless of the
dimensionless coupling $U/t$. It is thus vital to consider phase
fluctuations and preliminary calculations in \cite{Trivedi.1998}
show that it is these fluctuations which are responsible for the
destruction of superconductivity at high disorder.

\subsubsection{Discussion}
\label{sec:full.superfluid}
Since in HTSC the pairing scale $T^\ast$ is different from the
superconducting transition temperature the role of non-magnetic
impurities is not traditional and they in fact define the
superconducting properties of a ``bad metal''.
In particular even for weak disorder we have seen that these impurities
strongly depress the superfluid density at zero temperature bringing
it closer to that experimentally observed.
In the case of strong disorder, impurities lead to large
inhomogeneities in the system with the existence of isolated
superconducting islands. This effect then dramatically changes the
superconducting properties, as evidenced by a greatly reduced bare
superfluid stiffness, and ultimately leads to the destruction of
superconductivity.

Of course, the results presented are only qualitative since only
the simplest model with non-retarded $s$-wave attraction and an
isotropic fermion spectrum was considered.
However, there are some indications that for
$d$-wave pairing, the properties obtained will persist
\cite{Franz.dirty}.\footnote{Note that $d$-wave superconductivity
appears to be completely destroyed by relatively weak disorder if
one describes the system within the AG theory used in
Sec.~\ref{sec:weak.disorder}. However, it is shown in \cite{Franz.dirty}
using the finite temperature version of the BdG equations (\ref{BdG})
for $d$-pairing, that $d$-wave superconductivity may well
survive even for strong disorder. This difference results from the more
adequate treatment of the spatial fluctuations of the complex ordering field
modulus $\rho(\mbox{\bf r})$ in \cite{Franz.dirty}.
One also notes that $d$-wave superconductivity
can be stable in the strong coupling limit, or
on the BEC side of the BCS-BEC crossover \cite{Sadovskii}.}

Further studies of the superfluid density are
necessary. For example, it is important to explain the concentration
dependence of the full superfluid slope, $d n_{s} (T)/dT$ at $T=0$
\cite{Panagopoulos,Mesot}.\footnote{It appears to be
more difficult to understand why $n_{s} \to 0$ as $n_{f} \to 0$
in the weak coupling model. However, as  mentioned in the beginning
of Sec.~\ref{sec:one-band} it is the doped carriers that are weakly coupled
and in fact a half-filled band in the strong coupling model corresponds
here to the empty conduction band.}
As already mentioned in Sec.~\ref{sec:mapping} the classical phase
fluctuations\footnote{While the quantum $XY$-model was considered
in \cite{Roddick} for more exact quantitative estimates, it was shown
in this paper (see also more recent publication \cite{Carlson}) that
the linear slope of  the low temperature full superfluid density is
produced by the classical phase fluctuations.}
result in the linear slope of the full superfluid density.
Nevertheless, as shown in \cite{Durst} (see also \cite{Lee})
the linear dependence can be also produced by the nodal (Dirac)
quasiparticles with a relativistic dispersion law which are
present in a d-wave superconductor. It is now widely discussed whether
the main contribution to the observed slope of the low temperature
superfluid density is due to these quasiparticles (see e.g.,
\cite{Chiao,Randeria.quasiparticles}) or phase fluctuations
\cite{Carlson,Cochec}.
It is likely that the dominant contribution to the slope
varies with doping as suggested in \cite{Cochec}. While
in the optimally- and over-doped samples it is produced
by the nodal quasiparticles, in the underdoped samples it can
be compatible with the model of classical phase fluctuations.
We stress only at the end that the problem of the superfluid slope
in HTSC should be considered also in the presence of defects
of different origin (including dopants).

\section{The Green's function in modulus-phase representation and
    non-Fermi liquid behaviour}
\setcounter{equation}{0}
\label{chap:our.Green}

The main quantity of interest in the present chapter is the
one-fermion Green's function $G(\omega, {\bf k})$.
The associated spectral function
$A(\omega,{\bf k})=-(1/\pi){\rm Im}G(\omega+i0,{\bf k})$,
being proportional to the intensity of the
ARPES \cite{Dessau,Campuzano},
encodes information about the pseudogap and quasi-particles.
Following \cite{Gusynin.new}
the Green's function for the charged (physical) fermions is given
by the convolution (in momentum space) of the propagator for
neutral fermions which has a gap $\rho\neq0$ and the Fourier
transform of the phase correlator
$\langle\exp(i\tau_3\theta(x)/2)\exp(-i\tau_3\theta(0)/2)\rangle$.

Thus the approximation employed here assumes the absence of coupling between
the spin and charge degrees of freedom. This can however be taken into account
at the next stage of approximation. Nonetheless, as we shall see, the phase
fluctuations result in {\em non-Fermi liquid\/} behaviour of the system both
below and above the superconducting transition temperature, $T_{\mathrm{BKT}}$.

\subsection{The modulus-phase representation for
         the fermion Green function}
\label{sec:form}

The starting point is a continuum version of the two-dimensional
attractive Hubbard model (\ref{Hamiltonian.2D}).
As discussed in Sec.~\ref{sec:formalism} the model with the
Hamiltonian density (\ref{Hamiltonian.2D}) is equivalent to a model
with an auxiliary BCS-like pairing field. This equivalent model has a
Hamiltonian density, which is given in terms of the Nambu variables
$\Psi(x)$ and $\Psi^{\dagger}(x)$
(see Eq.~(\ref{Nambu.variables})) and the auxiliary complex
Hubbard-Stratonovich fields $\Phi(x)= U \Psi^\dagger(x)\tau_{-}\Psi(x)=
U \psi_\downarrow \psi_\uparrow$
and
$\Phi^{\ast}(x)= U \Psi^\dagger(x)\tau_{+}\Psi(x)=
U \psi^{\dagger}_{\uparrow} \psi^{\dagger}_{\downarrow}$,
as
\begin{equation}
\mathcal{H} =
\Psi^{\dagger}(x)\left[\tau_3 \left(- \frac{\nabla^{2}}{2 m} - \mu \right)
 - \tau_{+}\Phi(x) - \tau_{-}\Phi^\ast(x)\right]\Psi(x)+
 \frac{|\Phi(x)|^2}{U} \,.
\end{equation}

Let us consider the full fermion Green's function in the
Matsubara finite temperature formalism
\begin{equation}
G(x) = \langle \Psi(x) \Psi^{\dagger}(0) \rangle,
\qquad x \equiv {\bf r}, \tau \,.
\label{Green.def}
\end{equation}
The frequency-momentum representation for (\ref{Green.def}) in
the case of the BCS theory in 3D (i.e. in the mean field approximation)
is given by (\ref{Green.fermion.momentum}).

However, a problem arises when one tries to apply
Eq.~(\ref{Green.fermion.momentum}) directly to 2D systems at
$T \neq 0$, since a non-zero value for $\Phi \equiv \langle \Phi(x) \rangle$
is forbidden by the CMWH theorem \cite{Coleman}
(see also the discussion in Secs.~\ref{sec:survey}).
Nevertheless as was discussed in
Sec.~\ref{sec:our.diagram} one can assume that the modulus of the
order parameter $\rho = |\Phi|$ has a nonzero value, while its phase
$\theta(x)$  defined by (\ref{parametrization.Hubbard}) is a random
quantity. To be consistent with (\ref{parametrization.Hubbard}) in
Sec.~\ref{sec:our.diagram} the spin-charge variables
(\ref{parametrization.fermi}), which separate the Nambu spinors
into their modulus and phase, were simultaneously introduced.

Applying (\ref{parametrization.fermi}) the Green's function
(\ref{Green.fermion.momentum})
can be split into spin and charge parts
\begin{equation}
G_{\alpha \beta}(x) = \sum_{\alpha^\prime, \beta^{\prime}}
\mathcal{G}_{\alpha^\prime\beta^\prime}(x)
\langle (e^{i\tau_3\theta(x)/2})_{\alpha\alpha^\prime}
(e^{-i\tau_3\theta(0)/2})_{\beta^\prime\beta} \rangle,
\label{Green.splited}
\end{equation}
where
\begin{equation}
\mathcal{G}_{\alpha\beta}(x)=\langle \Upsilon_{\alpha}(x)
\Upsilon^{\dagger}_{\beta}(0) \rangle
\end{equation}
is the Green's function for neutral fermions. We note here that
this equation implies that there is no interaction between the spin and
charge degrees of freedom. Introducing the
projectors $P_{\pm}= \frac{1}{2}(\hat I\pm \tau_3)$ one gets
\begin{equation}
e^{i\tau_3\theta/2} = P_{+}e^{i\theta/2}+P_{-}e^{-i\theta/2},
\quad
e^{-i\tau_3\theta/2} = P_{-}e^{i\theta/2}+P_{+}e^{-i\theta/2},
\end{equation}
so that Eq.~(\ref{Green.splited}) may be rewritten as
\begin{equation}
G(x) = \sum_{\alpha,\beta=\pm} P_\alpha \mathcal{G}(x) P_\beta
\langle \exp [i\alpha\theta(x)/2 ] \exp [- i\beta\theta(0)/2] \rangle,
\label{Green.projectors}
\end{equation}
where $\alpha=\beta$ and $\alpha=-\beta$ correspond to  the diagonal and
non-diagonal parts of the Green's function respectively.

For the frequency-momentum representation of
(\ref{Green.projectors}) one has
\begin{equation}
G(i \omega_{n}, {\bf k}) = T \sum_{m = - \infty}^{\infty} \int
\frac{d^2 p}{(2 \pi)^{2}}
\sum_{\alpha,\beta=\pm}
P_\alpha \mathcal{G}(i \omega_{m}, {\bf p}) P_\beta D_{\alpha
\beta} (i \omega_{n} - i \omega_{m}, {\bf k} - {\bf p}),
\label{Green.projectors.momentum}
\end{equation}
where
\begin{equation}
\mathcal{ G}(i \omega_{m}, {\bf k}) =
\int_{0}^{1 / T} d \tau \int d^2 r
\exp [i \omega_{m} \tau - i {\bf k} {\bf r}]
\mathcal{ G}(\tau, {\bf r})
\label{Green.neutral.Fourier}
\end{equation}
 and
\begin{equation}
D_{\alpha \beta}(i \Omega_{n}, {\bf q}) =
\int_{0}^{1 /T} d \tau \int d^2 r
\exp \left(i \Omega_{n} \tau - i {\bf q} {\bf r}\right)
\langle \exp [ i\alpha \theta(\tau, {\bf r}) / 2 ]
 \exp [ - i\beta \theta(0) / 2 ] \rangle  \,.
\label{Green.phase.Fourier}
\end{equation}

There is good reason to believe (see Sec.~\ref{sec:our.diagram})
that, for $T$ close to $T_{\mathrm{BKT}}$, fluctuations of the modulus
$\rho$ of the order parameter\footnote{These
are so-called longitudinal fluctuations
which correspond to fluctuations of the carrier density and
undoubtedly must be taken into account in the very underdoped
region.} are irrelevant. Thus one may safely use for the Green's
function (\ref{Green.neutral.Fourier}) of the neutral fermions the
mean-field approximation given by Eq.~(\ref{K2})
(compare with (\ref{Green.fermion.momentum})).

As we shall see, fluctuations of the $\theta$-phase transform the
pole in the Green's function for the neutral fermions (\ref{K2})
into a branch cut in the Green's function for the charged particles
(\ref{Green.def}) for $T < T_{\rho}$.  The CMWH theorem \cite{Coleman}
concerning the absence of spontaneous breaking of a continuous symmetry
in 2D means that symmetry-violating Green's functions which
are described by terms with nonzero $\Phi$ and $\Phi^{\ast}$
must vanish.  However, it says nothing about the gap in the spectrum of
excitations, as is sometimes incorrectly stated.
The correct explanation is that if the symmetry is unbroken,
{\em and\/} the fermion excitation appears as a pole in the $\Psi$
two-point function (\ref{Green.def}), then the fermion must be gapless.
If the fermion does not have the same quantum numbers as $\Psi$
(e.g. the neutral fermion $\Upsilon$) and so does not appear in the $\Psi$
two-point function (\ref{Green.def}) as a one-particle state, then
symmetry does not tell us whether the fermion ($\Upsilon$) will be gapless
or not. Thus there is no contradiction
between the gapped form (\ref{K2}) and the CMWH theorem.
We note that this formal consideration \cite{Witten} closely resembles
the more physical arguments presented by Abrikosov \cite{Abrikosov.1997}
about the possible existence of a gap in the one-particle excitation spectrum
without the presence of long-range order.

\subsection{The correlation function for the phase fluctuations}
\label{sec:phase.green}

As stated above, phase fluctuations are expected to be
responsible for the differences between the properties of the charged
(physical) and the neutral fermions defined above. The latter are described
by the Green's function (\ref{K2}) which coincides with the BCS Green's
function (\ref{Green.fermion.momentum})) only if one assumes that
the phase $\theta$ of the order parameter $\Phi=\rho \exp(i\theta)$ is a
constant and can be chosen to be zero. This is not the case in the 2D
model, where there is a decay of phase correlations and the
Green's functions of charged and neutral fermions are non-trivially
related via Eq.~(\ref{Green.projectors.momentum}). To establish
their relationship one needs to know the correlator for phase
fluctuations. Its calculation is quite straightforward for $T <
T_{\mathrm{BKT}}$, while for $T > T_{\mathrm{BKT}}$ one may apply
the results of the theory of the BKT transition (see e.g.
\cite{Plischke} or the other textbooks mentioned).

\subsubsection{The correlator for $T < T_{\mathrm{BKT}}$}
\label{subsec:phase.green}

In the superconducting phase free vortex excitations are absent
and the exponent correlator is easily expressed in terms of the
Green's function
\begin{equation}
D_{\theta} (x) =
\langle \theta(x)\theta(0) \rangle
\end{equation}
(here as above $x \equiv \tau,{\bf r}$) via the Gaussian
functional integral
\begin{eqnarray}
&& D_{\alpha \beta}(x)  =
                           \\
&& \int \mathcal{D}\theta(x)
\exp \left\{ -\int_{0}^{1/T} \! \! d \tau_{1}
\int  \! \! d^{2}  r_{1}\left[ \frac{1}{2} \theta(x_1)
D_{\theta}^{-1} (x_1)\theta(x_1) + I(x_1)
\theta(x_1) \right] \right\} =
\nonumber                  \\
&& \exp \left[- \frac{1}{2}
\int_{0}^{1/T} \! \! d \tau_{1} \int_{0}^{1/T} \! \! d \tau_{2}
\int  \! \! d^{2} r_{1} \int  \! \! d^{2} r_{2}
I(\tau_1,{\bf r}_1) D_\theta (\tau_{1} - \tau_{2}, {\bf r}_{1}
-{\bf r}_{2}) I(\tau_2, {\bf r}_2 )\right],
\nonumber
\label{Gaussian}
\end{eqnarray}
where the source is given by
\begin{equation}
I(x_1) = -i \frac{\alpha}{2}
\delta(\tau_{1}- \tau) \delta({\bf r}_{1} - {\bf r}) +
i \frac{\beta}{2} \delta(\tau_{1}) \delta({\bf r}_{1}),
\quad (\alpha,\beta=\pm) \,.
\end{equation}
The Green's function
\begin{equation}
D_{\theta}^{-1} (x) =- J(\mu, T, \rho) \nabla^{2}_{r} -
K(\mu, T,\rho) (\partial_{\tau})^{2}
\label{Green.our}
\end{equation}
for the phase fluctuations has already been found in
Sec.~\ref{sec:derivation} and
follows directly from Eq.~(\ref{Omega.Kinetic.phase.final}).
Recall that the superfluid stiffness $J$ (see Eq.~(\ref{K10}))
and the compressibility $K$ (see Eq.~(\ref{K7})) are
both functions of $\mu$, $T$ and $\rho$, and also that the Green's
function (\ref{Green.our}) includes only the lowest order derivatives
of the phase $\theta$.
Higher terms are also present in the
expansion, but will be neglected.

Substituting (\ref{Green.our}) into (\ref{Gaussian}) one gets
\begin{equation}
D_{\alpha \beta}(x)= \exp
\left[ - \frac{T}{4} \sum_{n=-\infty}^\infty
\int \frac{d^2 q}{(2 \pi)^2}
\frac{1- \alpha \beta\cos({\bf q} {\bf r} - \Omega_n \tau)}
{J {\bf q}^2 + K \Omega^2_n} \right].
\label{correl.1}
\end{equation}
It is easy to see that for zero frequency $\Omega_n = 0$ the
integral in Eq.~(\ref{correl.1}) is divergent at ${{\bf q}} = 0$
unless $\alpha =\beta$. Thus in the sums over $\alpha, \beta$  in
Eq.~(\ref{Green.projectors.momentum}) only two terms survive,
namely
\begin{equation}
P_{-} \mathcal{G}(i \omega_{n}, {\bf k})P_{-} +
P_{+} \mathcal{G}(i \omega_{n}, {\bf k})P_{+} =
- \frac{i \omega_{n} \hat I + \tau_{3} \xi({\bf k})}
{\omega_{n}^{2} + \xi^{2}({\bf k}) + \rho^{2}}.
\label{survive}
\end{equation}
It is important that terms like $P_{\pm} G(i \omega_{n},
{\bf k})P_{\mp}$ which are proportional to $\tau_1$ and thus
violate the gauge symmetry, do not contribute in
Eq.~(\ref{Green.projectors.momentum}) due to vanishing of the
$D_{+ -}$ and $D_{- +}$ correlators which follow them. This
explicitly demonstrates that the non-diagonal part of the 2D Green's
function is equal to zero at all finite temperatures\footnote{Strictly
speaking, one can state this, based on Eq.~(\ref{correl.1}), only for
$T < T_{\mathrm{BKT}}$. However, there are relatively simple
arguments \cite{Gusynin.unpublished} why a correlator of the form
$\langle \exp(i\theta(x)\exp(i\theta(0) \rangle$
must be identically zero both below and above $T_{\mathrm{BKT}}$.
Indeed, we know that the correlator
$\langle \exp(i\theta(x)\exp(-i\theta(0) \rangle$
(see Eq.~\ref{correlator.>} below) tends to zero as $r \to \infty$.
On the other hand according to the principle of weakening correlations
our correlator must reduce to the product
$\langle \exp(i\theta(x) \rangle \langle \exp(i\theta(0)\rangle$ at large
$r$, i.e.the vacuum expectation value $\langle\exp(i\theta)\rangle$
must be identically zero.
This of course simply means that one is in the symmetric phase, i.e.
one has an exact symmetry with respect to the transformation
$\theta \to \theta + \mbox{const}$. In turn this means that
all correlators which violate this symmetry are identically zero.}
in accordance with the CMWH theorem.

For the non-zero correlators one has
\begin{eqnarray}
D(x) & \equiv & D_{+ +}(x) = D_{- -}(x)
\nonumber                \\
& = & \exp \left[- \frac{T}{4} \sum_{n=-\infty}^\infty \int
\frac{q d q d \varphi}{(2\pi)^2}
\frac{1- \cos (q  r \cos \varphi) \cos \Omega_n \tau }
{J q^2 + K \Omega^2_n} \right].
\label{correl.2}
\end{eqnarray}

In the following discussion only the static case $\tau=0$
will be considered. The restriction to this case is one of the few main
assumptions used throughout this chapter. Recall also another
essential assumption made in this chapter which is the absence of
coupling between the spin and charge degrees of freedom.

The summation over $n$ and the integration over $\varphi$ in
(\ref{correl.2}) can readily be done yielding the following
expression for the argument of the exponent in (\ref{correl.2})
\begin{equation}
- \frac{1}{16 \pi \sqrt{JK} } \int \limits_0^\infty d q
e^{-q / \Lambda} [1 - J_0(q r)] \tanh \frac{qr_{\mathrm{av}}}{4} ,
\label{exponent}
\end{equation}
where we have introduced the scale
\begin{equation}
r_{\mathrm{av}} =  \frac{2}{T} \sqrt{\frac{J}{K}},
\label{scale}
\end{equation}
which is a function of the variables used (in the simplest case
$r_{\mathrm{av}}\sim\sqrt n_f /T$).

In (\ref{exponent}) we introduced a cutoff $\Lambda$ by hand
with the introduction of the corresponding exponential function. This cutoff
represents the maximal possible momentum in the theory -- the
Brillouin momentum.

One can derive from (\ref{exponent})
(see the details in \cite{Gusynin.new})
the following asymptotic results
\begin{equation}
D(0, {\bf r}) \sim
\left\{\begin{array}{cc}
\left( \frac{\ds r}{\ds r_{\mathrm{av}}} \right)^{-\frac{\ds T}{\ds 8 \pi J}},
&  r \gg r_{\mathrm{av}} \gg \Lambda^{-1} \\
\left(\frac{\ds \Lambda r} {\ds 2} \right)^{-\frac{\ds T}{\ds 8 \pi J}},
&  r \gg \Lambda^{-1} \gg r_{\mathrm{av}}.
\end{array} \right.
\label{asymptotic}
\end{equation}
This long-distance behaviour then governs the $\theta$-fluctuations physics to
be studied below.

Let us discuss the meaning of the value $r_{\mathrm{av}}$. Using again the
phase stiffness $J(T=0)$ and  compressibility $K$  from
Eqs.~(\ref{K10}) and (\ref{K7})
one readily gets that $r_{\mathrm{av}} = 2
\sqrt{\epsilon_{F}/m} /T$ -- the single-particle thermal de Broglie
wavelength ($\epsilon_F=\pi n_f/m$ is the Fermi energy). Then,
considering that $T\sim T_{\mathrm{BKT}}$ and taking $T_{\mathrm{BKT}}$
as given by $T_{\mathrm{BKT}} \simeq\epsilon_F/8$
(see Sec.~\ref{sec:results}), one can estimate
\begin{equation}
r_{\mathrm{av}} \sim \frac{16}{\sqrt{\epsilon_{F} m}} =
\frac{16\sqrt{2}}{k_\mathrm{F}}.
\label{a.estimate.Fermi}
\end{equation}

The value of the Fermi momentum $k_\mathrm{F}$ for cuprates
is less than the Brillouin momentum $\Lambda$. Thus the
first line of (\ref{asymptotic}) appears to be more relevant.

There is a second way to estimate $r_{\mathrm{av}}$. One can rewrite it in
terms of the value of $2 \Delta/ T_{c}$ in the form
\begin{equation}
r_{\mathrm{av}} \sim \sqrt{2}\pi\frac{2\Delta}{T_c} \xi_{\mathrm{pair}},
\label{a.estimate.coherence}
\end{equation}
where $\xi_{\mathrm{pair}} = v_{F}/(\pi \Delta)$ is the BCS coherence
length. (It is assumed, that for the carrier densities
of interest, $\mu \simeq \epsilon_{F}$, so that the coherence length
$\xi_{\mathrm{coh}}$ and the pair size
$\xi_{\mathrm{pair}}$ are the same. See Sec.~\ref{sec:derivative}).
This shows that $r_{\mathrm{av}}$ has the meaning of a coherence length, which
appears to be rather natural since the minimal size of the region
where there is the phase coherence should be of the order of
$\xi_{0}$. Since the coherence length in cuprates is larger than
the lattice space $\Lambda^{-1}$, one again obtains that the first
line of (\ref{asymptotic}) should be applied. Therefore for $T <
T_{\mathrm{BKT}}$ and for static fluctuations we have that
\begin{equation}
D({\bf r}) =
\left( \frac{r}{r_{\mathrm{av}}} \right)^{-\frac{\ds T}{\ds 8 \pi J}},
\label{correlator.<}
\end{equation}
where $r_{\mathrm{av}} = 16/ \sqrt{\epsilon_{F} m}$. Note that all these
estimations are correct for the clean system. For the dirty
system one should replace them by the results from
Sec.~\ref{sec:impure}.

\subsubsection{The correlator for $T > T_{\mathrm{BKT}}$}

For $T > T_{\mathrm{BKT}}$ the expression for the static correlator
(\ref{correlator.<}) can be generalized using well-known
results from the theory of the BKT transition as
\cite{Minnhagen,Pierson,Plischke}:
\begin{equation}
D({\bf r}) = \left( \frac{r}{r_{\mathrm{av}}} \right)^{-\frac{\ds T}{\ds 8 \pi J}}
\exp \left( - \frac{r}{\xi_{+}(T)} \right) \,,
                         \label{correlator.>}
\end{equation}
where
\begin{equation}
\xi_{+}(T) = \xi_{+}(T_{\rho}) \exp
\sqrt{\frac{T_{\rho} - T}{T - T_{\mathrm{BKT}}}}
\label{BKT.length}
\end{equation}
is the BKT coherence length and the value of
$\xi_{+}(T_{\rho})$ to be used in numerical calculations
will be discussed later.

Note that Eq.~(\ref{BKT.length})
represents a very specific temperature dependence of the coherence
length which is completely different from the temperature
dependence of the Ginzburg-Landau coherence length.
The charge and spin susceptibilities for the negative-$U$
Hubbard model (\ref{Hubbard}) were estimated in \cite{Benfatto}
using the BKT and Ginzburg-Landau coherence lengths. It appears that
the spin susceptibility calculated with the BKT coherence length
better fits the Monte Carlo data. The fact that the BKT
coherence length, with its specific temperature dependence, is used
allows one to go beyond the $T$-matrix approximation, and to adequately
describe the physics of the topological BKT superconducting
fluctuations.

There is also an analogy between the renormalized classical regime
discovered in \cite{Tremblay.1997}, when below the temperature
$T_{X}$ (see Sec.~\ref{sec:limitations}) the two-particle
correlation length grows exponentially, and the exponential growth of the
BKT coherence length below $T_{\rho}$.

One may consider
Eq.~(\ref{correlator.>}) as a general representation for
$D({\bf r})$ both for $T > T_{\mathrm{BKT}}$ and
$T < T_{\mathrm{BKT}}$ if the coherence length $\xi_{+}(T)$ is
considered to be infinite for $T < T_{\mathrm{BKT}}$. The
pre-exponential factor in Eq.~(\ref{correlator.>}) is related to the
longitudinal (spin-wave) phase fluctuations, while the exponent is
responsible for the transverse (vortex) excitations, which are
present only above $T_{\mathrm{BKT}}$. The pre-exponential factor
appears to be important for the presence of non-Fermi liquid behaviour
which we discuss below. Note, however, that the longitudinal phase
fluctuations could be suppressed by the Coulomb interaction
\cite{Franz} which is not included in the present simple model. One
further comment is that while the approximation used in \cite{Franz}
to study the vortex fluctuations is good for $T$ well above
$T_{\mathrm{BKT}}$, the form of the correlator $D$ here is appropriate
for $T$ close to $T_{\mathrm{BKT}}$.

The constant $\xi_{+}(T_{\rho})$ may be estimated from the condition that
$\xi_{+}(T)$ cannot be much less than the parameter $r_{\mathrm{av}}$ which
appears in the theory as a natural cutoff, so we will use in our
numerical calculations $\xi_{+}(T_{\rho}) = r_{\mathrm{av}}/4$. In any case,
for $T \gtrsim T_{\mathrm{BKT}}$ where the expression (\ref{BKT.length}) is
valid, the value $\xi_{+}(T)$ is large and not as sensitive to the initial
value of $\xi_{+}(T_{\rho})$.

There is also a dynamical generalization of (\ref{correlator.>})
\begin{equation}
D(t, {\bf r}) = \exp (- \Gamma t)
\left( \frac{r}{r_{\mathrm{av}}} \right)^{-\frac{\ds T}{\ds 8 \pi J}}
\exp \left( - \frac{r}{\xi_{+}(T)} \right) \,,
\label{correlator.dynamic}
\end{equation}
proposed in \cite{Capezzali.1999} from a phenomenological background.
Note that $t$ is the real time and $\Gamma$ is here the decay constant,
so that (\ref{correlator.dynamic}) is the retarded Green's function.
The temperature dependence of $\Gamma(T)$ for the case of a classical
2D planar magnet was studied in \cite{Huber} and showed a
critical slowing down as $T \to T_{\mathrm{BKT}}^{+}$ due to
the disappearance of mobile, free vortices.

The more general case of the dynamical phase
fluctuations (\ref{correlator.dynamic}) was considered
in a related
calculation \cite{Capezzali.1999}
(see its discussion in Sec.~\ref{sec:Capezzali})
where a correlator
$\langle\exp(i\theta(t, {\bf r})\exp(-i\theta(0))\rangle$\footnote{It
differs from the correlator (\ref{correlator.dynamic}) only by a
factor $1/2$ multiplying the phase.},
which includes dynamical
phase fluctuations, has been used in the numerical calculation
of the self-energy of the fermions and the subsequent extraction of
the spectral function from the fermion Green's function.

\subsection{The Fourier transform of $D({\bf r})$}
\label{sec:phase.mode}

For the Fourier transform (\ref{Green.phase.Fourier}) of
(\ref{correlator.>}) one has
\begin{eqnarray}
D(i \Omega_{n}, {\bf q}) & = &
\int_{0}^{1 /T} d \tau \int d^2 r\exp\left(i \Omega_{n} \tau
- i {\bf q} {\bf r}\right)( r / r_{\mathrm{av}} )^{-T/ 8 \pi J}
\exp (- r / \xi_{+}(T) )
\nonumber                         \\
& = & 2\pi \frac{\delta_{n,0}}{T}
r_{\mathrm{av}}^{T / 8 \pi J}\int_0^\infty d r r^{1-
T/ 8\pi J}J_0(qr)\exp(- r / \xi_{+}(T)).
\label{Fourier.D1}
\end{eqnarray}
The integral in (\ref{Fourier.D1}) can be calculated
(see, for example, \cite{Gradshteyn}) leading to
\begin{equation}
D(i \Omega_{n}, {\bf q}) =
\frac{\delta_{n, 0}}{T}
\frac{2\pi r_{\mathrm{av}}^{2(1 - \alpha)} \Gamma(2\alpha)}
{[q^2 + \xi_{+}^{-2}(T)]^{\alpha}} \,\,
{ }_2F_{1} \left(\alpha,
-\alpha+ \frac{1}{2}; 1; \frac{q^2}{q^2 + \xi_{+}^{-2}(T)} \right).
\label{Fourier.D2}
\end{equation}
The hypergeometric function $F(a,b;c;z)$ in (\ref{Fourier.D2}) may
be well approximated by a constant since it is slowly varying at
all values of ${\bf q}$. As such a constant we can take the value
of hypergeometric function at $q=\infty$.  Thus,
\begin{equation}
D(i \Omega_{n}, {\bf q}) = \frac{\delta_{n, 0}}{T}
A [q^2 +  \xi_{+}^{-2}(T)]^{-\alpha},
\label{Fourier.D.final}
\end{equation}
where
\begin{equation}
A\equiv\frac{4 \pi \Gamma(\alpha) }{ \Gamma(1-\alpha) }
\left( \frac{2}{r_{\mathrm{av}}} \right)^{2(\alpha - 1)},
\qquad \alpha\equiv 1 - \frac{T}{16 \pi J}.
\label{A,alpha}
\end{equation}

It should be stressed that for $T > T_{\mathrm{BKT}}$ the parameter
$\alpha$ quickly deviates from unity as $\epsilon_{F}$ decreases;
in other words, the underdoped region will reveal more
non-standard properties than the overdoped one.

Note that for $\xi_{+}^{-1}(T) = 0$ ($T < T_{\mathrm{BKT}}$)
Eq.~(\ref{Fourier.D.final}) is the exact Fourier transform for the correlator
(\ref{correlator.<}).

One should take into account that even for $T < T_{\mathrm{BKT}}$ the
propagator (\ref{Fourier.D.final}) does not have the canonical behaviour $\sim
1/q^2$ which is typical, for example, for the Bogolyubov mode in dimensions $d
> 2$. In 2D, modes with a propagator $\sim 1/q^2$, would lead to severe
infrared singularities \cite{Coleman}, and to avoid this, the modes transform
into the softer ones ($\sim 1/q^{2\alpha}$, $\alpha < 1$).

Finally, substituting (\ref{survive}) and (\ref{Fourier.D.final})
into (\ref{Green.projectors.momentum}) one obtains
\begin{equation}
G(i \omega_{n}, {\bf k}) = - A \int \frac{d^{2} q}{(2 \pi)^2}
\frac{i \omega_{n} + \tau_{3} \xi({\bf q})}
{\omega_{n}^2 + \xi^2({\bf q}) + \rho^{2}}
\frac{1} {[({\bf k} - {\bf q})^{2} +  \xi_{+}^{-2}(T)]^{\alpha}}.
\label{Green.for.derivation}
\end{equation}
The coincidence of the Matsubara frequency in the left
and right sides of Eq.~(\ref{Green.for.derivation})
is straightforwardly related to the static approximation.
It is truly remarkable that the Green's function
(\ref{Green.for.derivation}) may be evaluated exactly,
as may the expressions for the spectral density and for the
density of states. We note, however, that the static approximation
can be justified only {\it a posteriory\/} when the dynamical case
is also considered \cite{GLQSh.new}.

\subsection{The derivation of the fermion Green's function
in Matsubara representation and its analytical continuation}
\label{sec:green.fermion}

The calculation of the fermion Green's function can proceed
analytically along the same lines of calculation as in the relativistic
case of the Gross-Neveu model at $T=0$ \cite{Gusynin.unpublished}
One splits the fermion part of (\ref{Green.for.derivation}) in the
following manner
\begin{equation}
\frac{i\omega_n{\hat I}+\tau_3\xi({\bf k})}
{\omega_n^2+\xi^2(\bf k)+\rho^2}=
\frac{A_1}{\xi({\bf k})+i\sqrt{\omega_n^2+\rho^2}} +
\frac{A_2}{\xi({\bf k})-i\sqrt{\omega_n^2+\rho^2}}\,,
\label{relativistic.spliting}
\end{equation}
where
\begin{equation}
A_{1} = \frac{1}{2} \left(\tau_{3} -
\frac{\omega_{n}}{\sqrt{\omega_{n}^{2} + \rho^{2}}}\right),
\qquad
A_{2} = \frac{1}{2} \left(\tau_{3} +
\frac{\omega_{n}}{\sqrt{\omega_{n}^{2} + \rho^{2}}}\right).
\label{A1,2.Matsubara}
\end{equation}
Then , using the representations
\begin{equation}
\frac{1}{a\pm i b} = \mp i\int\limits_0^\infty ds
\exp{[\pm i s(a \pm i b)]},
\label{representation.Im}
\end{equation}
and
\begin{equation}
\frac{1}{c^\alpha} =
\frac{1}{\Gamma(\alpha)} \int \limits_0^\infty dtt^{\alpha-1}e^{-ct},
\label{representation.Re}
\end{equation}
and taking into account (\ref{relativistic.spliting}), one may
rewrite (\ref{Green.for.derivation}) as
\begin{eqnarray}
G(i \omega_{n}, {\bf k}) & = & \frac{i A}{\Gamma(\alpha)}
\int_{0}^{\infty} ds \int_{0}^{\infty} dt t^{\alpha - 1}
e^{-\xi_{+}^{-2}(T) t - s \sqrt{\omega_{n}^{2} + \rho^{2}}} \times
\nonumber                  \\
\int \frac{d^{2} q}{(2 \pi)^{2}}  \! \! \! \!
&& \left\{ A_{1} \exp \left[ i s
\frac{q^{2}}{2m} - i\mu s - ({\bf k} - {\bf q})^{2} t \right] -
\right.
\nonumber                  \\
&& \left.
A_{2} \exp \left[- i s \frac{q^{2}}{2m} + i\mu s - ({\bf k}
-{\bf q})^{2} t \right] \right\}.
\label{Green.integral}
\end{eqnarray}
Note that the special form of the integral representation
(\ref{representation.Im}) (compare with the representation
(\ref{representation.Re})) guarantees that the Gaussian integral
over $q$ is well-defined independently of the sign of
$\xi({\bf q})={\bf q}^2/2m -\mu$. Now the Gaussian integration
over momenta $q$ in (\ref{Green.integral}) can be done
explicitly.
Subsequently changing the variables $s \to 2m s$ and then
$t \to st$ one can integrate over $s$. Finally, making
the substitution $t \to -iu$ and expanding the quadratic polynomial
in the denominator, gives (see the details in \cite{Gusynin.new}):
\begin{eqnarray}
G(i \omega_{n}, {\bf k})
= -\frac{A m \xi_{+}^{2\alpha}(T)}{2\pi}
&& \! \! \! \!
\left\{
\int_{0}^{i \infty} du \frac{A_1 u^{\alpha - 1} (u + 1)^{\alpha - 1}}
{[(u + u_1) (u + u_2)]^{\alpha}} + \right.
\nonumber               \\
&& \left.
\int_{0}^{-i \infty} du \frac{A_2 u^{\alpha - 1} (u + 1)^{\alpha - 1}}
{[(u + \tilde u_1) (u + \tilde u_2)]^{\alpha}} \right\} \,,
\label{Green.imaginary.limits}
\end{eqnarray}
where
\begin{eqnarray}
u_1 & = & m \xi_{+}^{2}(T) \left(
\frac{k^2\xi_{+}^2(T) + 1}{2m\xi_{+}^2(T)} - \mu + i \sqrt{\omega_{n}^{2} +
\rho^{2}} +\sqrt D \right),
\nonumber             \\
u_2 & = & m \xi_{+}^2(T) \left(
\frac{k^2\xi_{+}^2(T)+1}{2m\xi_{+}^2(T)} - \mu + i \sqrt{\omega_{n}^{2} +
\rho^{2}} -\sqrt D \right)
\label{roots.Matsubara}
\end{eqnarray}
with
\begin{equation}
D\equiv \left( \frac{k^2\xi_{+}^2(T)+1}{2m\xi_{+}^2(T)} - \mu + i
\sqrt{\omega_{n}^{2} +\rho^{2}} \right)^2 + \frac{2}{m \xi_{+}^2(T)}
(\mu - i \sqrt{\omega_{n}^{2} + \rho^{2}})
\label{D.Matsubara}
\end{equation}
and
\begin{equation}
\tilde u_i = u_i(\sqrt{\omega_{n}^{2} + \rho^{2}} \to -
\sqrt{\omega_{n}^{2} + \rho^{2}}) \,.
\end{equation}

One can check from (\ref{roots.Matsubara}) that $\mbox{Re} u_i > 0$
for $\mu < 0$, so that one can rotate the integration contour to the
real axis:
\begin{equation}
G(i \omega_{n}, {\bf k}) =
-\frac{A m \xi_{+}^{2\alpha}(T)}{2\pi}
\left\{
\int_{0}^{\infty} du \frac{A_1 u^{\alpha - 1} (u + 1)^{\alpha - 1}}
{[(u + u_1) (u + u_2)]^{\alpha}} +
(\sqrt{\omega_{n}^{2} +
\rho^{2}} \to -\sqrt{\omega_{n}^{2} + \rho^{2}})
\right\}.
\label{Green.real.limits}
\end{equation}
Then the integral representation (\ref{Green.real.limits}) may be
analytically continued to $\mu > 0$. The change of variable $z
= u/(u+1)$ allows Eq.(\ref{Green.real.limits}) to be expressed in terms
of Appell's function \cite{Bateman}
\begin{equation}
F_{1} (\alpha, \beta, \beta^{\prime}, \gamma; x, y) =
\frac{\Gamma(\gamma)}{\Gamma(\alpha) \Gamma(\gamma - \alpha)}
\int_{0}^{1} \frac{z^{\alpha-1} (1 - z)^{\gamma-\alpha-1}}
{(1 - zx)^{\beta} (1 - zy)^{\beta^{\prime}}} dz,
\label{Appell.definition}
\end{equation}
so that
\begin{eqnarray}
G(i \omega_{n},\mbox {\bf k})=
-\frac{Am\xi_{+}^{2\alpha}(T)}{2\pi\alpha}
&&\left[\frac{A_1 }{(u_1 u_2)^{\alpha}} F_1 \left(\alpha, \alpha, \alpha;
\alpha + 1; \frac{u_1 - 1}{u_1}, \frac{u_2 - 1}{u_2} \right) \right.
\nonumber\\
&&  + \left. (\sqrt{\omega_{n}^{2} +
\rho^{2}} \to -\sqrt{\omega_{n}^{2} + \rho^{2}}) \right].
\label{Green.final.Matsubara}
\end{eqnarray}

For $T < T_{\mathrm{BKT}}$ the BKT coherence length is infinite
($\xi_{+}^{-1}(T) = 0$) so that the first argument of the Appell's
function $(u_1-1)/u_1 = 1$. This allows one to apply the reduction
formula \cite{Bateman}
\begin{equation}
F_1(\alpha, \beta, \beta^{\prime}, \gamma;1,x) =
\frac{\Gamma(\gamma) \Gamma(\gamma - \alpha - \beta)}
{\Gamma(\gamma - \alpha) \Gamma(\gamma - \beta)}
\, \, {}_2F_{1} (\alpha, \beta^\prime; \gamma-\beta;x)
                       \label{AppelltoHyper.in1}
\end{equation}
and express the result via the hypergeometric function
\begin{eqnarray}
G(i \omega_{n}, {\bf k}) && = -
\Gamma^{2}(\alpha) \left(\frac{2}{m r_{\mathrm{av}}^{2}} \right)^{\alpha - 1}
\times
\nonumber                  \\
&&
\left\{\frac{A_1}{[- (\mu - i \sqrt{\omega_{n}^{2} + \rho^{2}})]^\alpha}
\, \, { }_2F_{1} \left(\alpha, \alpha; 1;
\frac{k^2/2m}{\mu - i \sqrt{\omega_{n}^{2} + \rho^{2}})} \right)
\right.
\nonumber                   \\
&& \left. +
\frac{A_2}{[- (\mu + i \sqrt{\omega_{n}^{2} + \rho^{2}})]^\alpha}
\, \, { }_2F_{1} \left(\alpha, \alpha; 1;
\frac{k^2/2m}{\mu + i \sqrt{\omega_{n}^{2} + \rho^{2}})} \right)
\right\} \,,
\label{Green.<.final}
\end{eqnarray}
where the value of $A$ from (\ref{A,alpha}) has been substituted.

\subsubsection{The retarded fermion Green's function}

To obtain the expression for spectral density, one firstly needs to
obtain the retarded real-time Green's function from the temperature
Green's function. One does this by means of the analytical continuation
$i \omega_n \to
\omega + i 0$, so that $\sqrt{\omega_{n}^{2} + \rho^{2}} \to i
\sqrt{\omega^{2} -
\rho^{2}}$. This results in the following rules (compare with
(\ref{A1,2.Matsubara}), (\ref{roots.Matsubara}),
(\ref{D.Matsubara}))
\begin{equation}
A_{1} \to \mathcal{A}_{1} = \frac{1}{2} \left(\tau_{3} +
\frac{\omega}{\sqrt{\omega^{2} - \rho^{2}}}\right),
\qquad
A_{2} \to \mathcal{A}_{2} = \frac{1}{2} \left(\tau_{3} -
\frac{\omega}{\sqrt{\omega^{2}-\rho^{2}}}\right);
\label{A1,2.real}
\end{equation}
\begin{eqnarray}
u_1 \to v_1 & = & m \xi_{+}^{2}(T) \left(
\frac{k^2\xi_{+}^2(T)+1}{2m\xi_{+}^2(T)} - \mu - \sqrt{\omega^{2} -
\rho^{2}} +\sqrt \mathcal{D} \right),
\nonumber             \\
u_2 \to v_2 & = & m \xi_{+}^{2}(T) \left(
\frac{k^2\xi_{+}^2(T)+1}{2m\xi_{+}^2(T)} - \mu - \sqrt{\omega^{2} -
\rho^{2}} -\sqrt \mathcal{D} \right)
\label{roots.real}
\end{eqnarray}
with
\begin{equation}
D \to \mathcal{D} =
\left( \frac{k^2\xi_{+}^2(T)+1}{2m\xi_{+}^2(T)} - \mu - \sqrt{\omega^{2} -
\rho^{2}} \right)^2 + \frac{2}{m \xi_{+}^{2}(T)}
(\mu + \sqrt{\omega^{2} - \rho^{2}})
\label{D.real}
\end{equation}
and
\begin{equation}
\tilde v_i = v_i(\sqrt{\omega^{2} - \rho^{2}} \to -
\sqrt{\omega^{2} - \rho^{2}}).
\end{equation}
Thus for the retarded Green's function one has
\begin{eqnarray}
G(\omega, {\bf k}) = -\frac{A m \xi_{+}^{2\alpha}(T)}{2\pi\alpha}
&&
\left\{ \frac{\mathcal{A}_1 }{(v_1 v_2)^{\alpha}} F_1
\left(\alpha, \alpha, \alpha; \alpha + 1;
\frac{v_1 - 1}{v_1}, \frac{v_2 - 1}{v_2} \right) \right.
\nonumber          \\
&&  + \left. (\sqrt{\omega^{2} - \rho^{2}} \to
-\sqrt{\omega^{2} - \rho^{2}}) \right\}.
\label{Green.final.real}
\end{eqnarray}
It is also easy to see that
\begin{equation}
v_1 v_2 = - 2m \xi_{+}^{2}(T) (\mu + \sqrt{\omega^{2} - \rho^{2}}).
\label{product}
\end{equation}

Let us now discuss the conditions under which the imaginary part of
$G(\omega + i 0, {\bf k})$ is nonzero.

For $|\omega| < \rho$ one can see that, $\tilde v_1 = v_1^\ast$,
$\tilde v_2 = v_2^\ast$ so that $G(\omega, {\bf k})$ is real
and $\mbox{Im} G(\omega+ i 0, {\bf k}) =0$. The case
$|\omega|> \rho$ is more complicated. It follows from the
Appell's function transformation property \cite{Bateman}
\begin{equation}
F_1 (\alpha, \beta, \beta^{\prime}, \gamma; x, y) =
(1 - x)^{-\alpha}
F_1 \left(\alpha, \gamma-\beta-\beta^{\prime}, \beta^{\prime},
\gamma; \frac{x}{x - 1}, \frac{y - x}{1 - x} \right).
\label{Applel.property}
\end{equation}
that for real $x$ and $y$ the function $F_1$ becomes complex if $x
> 1$ or/and $y > 1$. This implies that $G(\omega, {\bf k})$
has an imaginary part if $v_1 < 0$ or/and $v_2 < 0$. Looking at the
expressions (\ref{roots.real}) for $v_1$ and $v_2$ one can see that
$v_1$ is always positive, while $v_2$ may be negative. This means
that $G(\omega, {\bf k})$ has an imaginary part if $v_1 v_2 <
0$.  Now using (\ref{product}) the condition for existence of
nonzero imaginary part of $G(\omega, {\bf k})$ can be written
in the following form $\mu + \sqrt{\omega^{2} - \rho^{2}} > 0$.

\subsection{The branch cut structure of $G(\omega, {\bf k})$ and
non-Fermi liquid behaviour}

Let us consider firstly the retarded fermion Green's function
(\ref{Green.<.final}) for $T < T_{\mathrm{BKT}}$. Applying the rules for
analytical continuation from the previous subsection to
Eq.~(\ref{Green.<.final}) one gets
\begin{eqnarray}
G(\omega, {\bf k}) & = &
-  \Gamma^{2}(\alpha) \left(\frac{2}{m r_{\mathrm{av}}^{2}} \right)^{\alpha - 1}
\times
\nonumber              \\
&& \left[ \frac{ \mathcal{A}_{1}}
{[- (\mu + \sqrt{\omega^{2} - \rho^{2}})]^\alpha} \, \, { }_2F_{1}
\left(\alpha, \alpha; 1;
\frac{k^2/2m}{\mu + \sqrt{\omega^{2} - \rho^{2}})} \right)\right.+
\nonumber              \\
&& \left. \frac{\mathcal{A}_{2}} {[- (\mu - \sqrt{\omega^{2} -
\rho^{2}})]^\alpha} \, \, { }_2F_{1}
\left(\alpha, \alpha; 1;
\frac{k^2/2m}{\mu - \sqrt{\omega^{2} - \rho^{2}})} \right)\right] \,.
\label{Green.<.final.real}
\end{eqnarray}

Near the quasi-particle peaks when $\omega \approx \pm
E({\bf k}) = \pm \sqrt{\xi^{2}({\bf k}) + \rho^{2}}$ the arguments of the
hypergeometric function in (\ref{Green.<.final.real}) are close to 1.
One can consider, for
instance, the first hypergeometric function, so that
\begin{equation}
z_{1} \equiv \frac{k^2/2m}{\mu + \sqrt{\omega^{2} - \rho^{2}}}
\simeq 1.
\end{equation}
Using the following relation between the hypergeometric functions
\cite{Bateman}
\begin{eqnarray}
{}_2F_1(a,b;c;z) = &&
\frac{\Gamma(c) \Gamma(c-a-b)}{\Gamma(c-a)\Gamma(c-b)}
\, \, {}_2F_1(a, b; a+b+1-c; 1-z) +
\nonumber   \\
\frac{\Gamma(c)\Gamma(a+b-c)}{\Gamma(a)\Gamma(b)} && (1-z)^{c-a-b}
\,\, {}_2F_1(c-a,c-b;c+1-a-b;1-z)
\end{eqnarray}
one gets that near $z_1 \simeq 1$
\begin{eqnarray}
&& G(\omega, {\bf k}) \sim
- \Gamma^{2}(\alpha)
\left(\frac{2}{m r_{\mathrm{av}}^{2}}\right)^{\alpha - 1} \times
\nonumber         \\
&& \qquad
\frac{\mathcal{A}_1}{[- (\mu + \sqrt{\omega^{2} - \rho^{2}})]^\alpha}
\left\{  \frac{\Gamma(1-2\alpha)}{\Gamma^{2}(1-\alpha)}
+ \frac{\Gamma(2\alpha-1)}{\Gamma^{2}(\alpha)}\frac{1}{(1-z_1)^{
2\alpha - 1}}\right\}.
\label{cut.<}
\end{eqnarray}
It is evident that the expression obtained for the Green's function is
non-standard. In addition to containing a branch cut, it clearly displays its
non-pole character. The latter in its turn corresponds to the non-Fermi liquid
behaviour of the system as a whole. It must be underlined that the non-Fermi
liquid peculiarities are strictly related to the charged (i.e. observable)
fermions only -- the Green's function (\ref{K2}) of the neutral ones has a
typical (pole type) BCS form. It also follows from (\ref{cut.<}) that the new
properties appear as a consequence of the $\theta$-particle presence (leading
to $\alpha\neq1$), and because the parameter $\alpha$ is a function of $T$
(see (\ref{A,alpha})) non-Fermi liquid behaviour increases with increasing
temperature and is preserved until $\rho$ vanishes.

It is interesting that, in Anderson's theory
\cite{Anderson,Sudbo.non-Fermi} (see also \cite{Anderson.book}),
it was postulated that Fermi liquid theory breaks down in the normal state
as a result of strong correlations. To introduce
this breakdown the following form of the single-particle fermion
propagator for the normal state was {\em assumed \/}
(see e.g. \cite{Sudbo.non-Fermi,Crisan}):
\begin{equation}
G(\omega, {\bf k}) = \frac{g(\alpha)}
{\omega_{c}^{1- \alpha} (\omega - \xi({\bf k}))^{\alpha}} \,,
       \label{Green.non-Fermi}
\end{equation}
where $\omega_c$ is a frequency cutoff which is introduced
to make the Green's function dimensionally correct,
$0 < \alpha < 1$ is independent of $T$ and $g(\alpha)$
is some function of $\delta$ (see more detail in \cite{Crisan}
where this form of Green's function was used as the
starting point for the evaluation of the pair fluctuation
propagator and the electronic self-energy).
It follows from (\ref{Green.non-Fermi}) that the associated spectral
function $A(\omega, {\bf k})$  satisfies close to the Fermi surface
a non-Fermi liquid homogeneity relation:
\begin{equation}
A (\Lambda \omega, \Lambda k) =
\Lambda^{-\alpha} A (\omega, k) \,,
                \label{homogen}
\end{equation}
where the limiting case of a Fermi liquid has $\alpha =1$.

Here we started from Fermi liquid theory and found that it has broken down due
to strong phase fluctuations, so that the final equation (\ref{cut.<}) reveals
the same analytical structure as (\ref{Green.non-Fermi}). The non-Fermi liquid
behaviour may, as suggested in \cite{Anderson,Anderson.book}, lead to the
suppression of coherent tunnelling between layers, that in turn confines
carriers in the layers and leads to strong phase fluctuations. However, in
contrast to \cite{Anderson,Sudbo.non-Fermi,Crisan}, the present model predicts
the restoration of Fermi liquid behaviour as $T$ decreases, since $\alpha \to
1$ when $T \to 0$.

The limit $T=0$ can be also obtained in the following way. Strictly speaking
one cannot estimate the value of $r_{\mathrm{av}}$ in the limit $T \to 0$ in
(\ref{Green.<.final.real}) via Eq.~(\ref{a.estimate.Fermi}) because the
substitution of the relationship $T_{\mathrm{BKT}}\simeq \epsilon_F/8$ into
(\ref{scale}) is not valid in this case. However, this is not essential because
$T/ 8\pi J \to 0$ so that the correlator (\ref{correlator.<}), $D({\bf r}) \to
1$ which clearly establishes long-range order in the system.  Furthermore, the
value of $\alpha$ in (\ref{A,alpha}) goes to 1 for $T \to 0$ so that the
hypergeometric function in (\ref{Green.<.final.real}) directly reduces to the
geometrical series:
\begin{equation}
{ }_2F_{1} (1, 1; 1; z) = \frac{1}{1-z}.
\label{geometrical}
\end{equation}
Therefore, substituting
(\ref{geometrical}) into (\ref{Green.<.final.real}) one gets for
the diagonal component $G_{11}(\omega, {\bf k})$ of the
Nambu-Gor'kov Green's function $G(\omega, {\bf k})$ the
ordinary BCS expression
\begin{equation}
G_{1 1}(\omega, {\bf k}) =
\frac{\omega + \xi({\bf k})}
{\omega^{2} - \xi^{2}({\bf k}) - \rho^{2}}.
\label{BCS.Green11}
\end{equation}
Clearly Eq.~(\ref{BCS.Green11}) results in the standard BCS spectral density
\cite{Schrieffer} with two $\delta$-function peaks
\begin{equation}
A(\omega, {\bf k}) =
\frac{1}{2} \left[1 + \frac{\xi({\bf k})}{E({\bf k})} \right]
\delta(\omega - E({\bf k})) +
\frac{1}{2} \left[1 - \frac{\xi({\bf k})}{E({\bf k})} \right]
\delta(\omega + E({\bf k}))\,.
\label{BCS.spectral.density}
\end{equation}
To recover the non-diagonal components of $G$ one has to restore the
correlators $D_{- +}({\bf r})$ and $D_{+
-}({\bf r})$ that were omitted in Sec.~\ref{subsec:phase.green}.

\section{The spectral function in the modulus-phase representation and
    filling of the gap}
\setcounter{equation}{0}
\label{chap:spectral}

As is well known, \cite{Schrieffer,Rickayzen.book,Abrikosov.book,Fetter},
the spectral features of any system are entirely controlled by
its spectral density. In terms of the matrix Greens' function
(\ref{Green.final.real}) this is given by
\begin{equation}
A(\omega, {\bf k}) = -
\frac{1}{\pi} \mbox{Im} G_{11}(\omega+i 0, {\bf k})
              \label{spectral.definition}
\end{equation}
and, for example, in the cuprates is measured in ARPES experiments
\cite{Timusk,Dessau,Campuzano,Norman.1999} (see also \cite{Chuang},
where a transparent example shows how interaction, finite experimental
resolution and other effects destroy a simple $\delta$ function-like
spectral function).  The spectral function defines the spectrum
anisotropy, the presence of a gap, the DOS, etc. Thus it
is important to discuss its behaviour for the different models based
on phase fluctuations.

\subsection{Absence of gap filling for the Green's
function calculated for the static phase fluctuations in the
absence of spin-charge coupling}

In this section we discuss $A(\omega,{\bf k})$ for the Green's function
obtained above in Sec.~\ref{chap:our.Green}. This example is
rather instructive since it allows one to obtain an analytical
expression for the spectral function \cite{Gusynin.new}, check
the sum rule and explicitly show the limitations of the static,
non-interacting approximation used to calculate the Green's function
and its spectral density.

\subsubsection{An analytical expression for the spectral density}

For $v_1 > 0$ and $v_2 < 0$ the retarded fermion Green's
function (\ref{Green.final.real}) can be rewritten
(see Appendix~\ref{sec:B}) in the following form
\begin{eqnarray}
&& G(\omega, {\bf k})  =
\nonumber       \\
 - && \frac{A m \xi_{+}^{2\alpha}(T)}{2 \pi}
\left\{ \mathcal{A}_1 \left[ \frac{(-1)^{\alpha} \Gamma(\alpha)
\Gamma(1-\alpha)}{ [v_1 (1-v_2)]^{\alpha}}
\, \, {}_2F_{1} \left(\alpha, \alpha; 1;
\frac{v_2 (1-v_1)}{v_1(1-v_2)} \right) +
\frac{1}{|v_2|}
\frac{\Gamma(1-\alpha)}{\Gamma(2-\alpha)} \times
\right. \right.
\nonumber                    \\
&& \left. \left.
F_1 \left(1, \alpha, 1-\alpha; 2-\alpha;
\frac{v_1}{v_2}, \frac{1}{u_2} \right) \right]
+ (\sqrt{\omega^{2} - \rho^{2}} \to
-\sqrt{\omega^{2} - \rho^{2}}) \right\}.
\label{Green.real.spectral}
\end{eqnarray}
Then, according to (\ref{spectral.definition}) the spectral density
for the Green's function (\ref{Green.real.spectral}) has the form:
\begin{eqnarray}
& &A(\omega, {\bf k})=
\frac{A m \xi_{+}^{2\alpha(T)}\sin(\pi\alpha)}{2 \pi^{2}}
\mbox{sgn}\omega\,\theta(\omega^2 - \rho^2)
\left[(\mathcal{A}_1)_{11}  \frac{\Gamma(\alpha) \Gamma(1-\alpha)}
{[v_1(1-v_2)]^{\alpha}} \times \right.
\nonumber       \\
& & \left.
{}_2F_{1}\left(\alpha,\alpha; 1;\frac{v_2(1-v_1)}{v_1(1-v_2)}
\right)\theta(\mu + \sqrt{\omega^{2} - \rho^{2}})
 -  (\sqrt{\omega^{2} - \rho^{2}} \to
-\sqrt{\omega^{2} - \rho^{2}}) \right]. \nonumber \\
\label{spectral.density1}
\end{eqnarray}
Using the quadratic transformation for the hypergeometric function
\cite{Bateman}
\begin{equation}
{}_2F_1(a, b; a-b+1;z) = (1 - z)^{-a} {}_2F_1
\left(\frac{a}{2}, -b+\frac{a+1}{2}; 1+a-b; -\frac{4z}{(1-z)^2} \right)
\end{equation}
the expression (\ref{A,alpha}) for $A$,
Eqs. (\ref{roots.real}) and (\ref{D.real})
one finally obtains
\begin{eqnarray}
A && (\omega, {\bf k}) =
\frac{\Gamma(\alpha)}{\Gamma(1-\alpha)}
\left( \frac{2}{m r_{\mathrm{av}}^{2}} \right)^{\alpha-1}
\mbox{sgn} \omega\,\theta(\omega^2 - \rho^2)\times\nonumber\\
&&\left[\frac{(\mathcal{A}_1)_{11}}{\mathcal{D}^{\alpha/2}}{}_2F_1
\left(\frac{\alpha}{2},\frac{1-\alpha}{2}; 1; -4
\frac{\frac{k^2}{2m}(\mu + \sqrt{\omega^2- \rho^2})}{\mathcal{D}}\right)
\theta(\mu + \sqrt{\omega^2 - \rho^2})\right.\nonumber\\
&&\left.- (\sqrt{\omega^{2} - \rho^{2}} \to
-\sqrt{\omega^2 - \rho^2}) \right],
\label{spectral.density.final}
\end{eqnarray}
where the chemical potential $\mu$ can be, in principle, determined
from the equation (\ref{number.rho}) which fixes the carrier density.
Here, however as stated above, one assumes that the carrier density is
sufficiently high that $\mu = \epsilon_{F}$.

In BCS theory $A(\omega, {\bf k})$ given by
Eq.~(\ref{BCS.spectral.density}) consists of two pieces which are the
spectral weights of adding and removing a quasi-particle from the system
respectively. Note that our splitting of $A(\omega, {\bf k})$
is different since each term in (\ref{spectral.density.final})
corresponds to both the addition and the removal of a fermion.

\subsubsection{The sum rule for the spectral density}

It is well-known that for the exact Green's function $G(\omega,
{\bf k})$ the spectral function (\ref{spectral.definition})
must satisfy the sum rule
\begin{equation}
\int_{- \infty}^{\infty} d \omega A(\omega, {\bf k}) = 1.
\label{sum.rule}
\end{equation}
The Green's function (\ref{Green.final.real}) calculated here is, of
course, approximate. One reason for this is the use of
the long-distance asymptotic result (\ref{asymptotic}) for the phase
correlator (\ref{correl.2}). This means that its Fourier
transformation (\ref{Fourier.D.final}) is, strictly speaking, valid
for small ${\bf k}$ only, while the expressions have been
integrated out to infinity.  Another approximation that has been  made
is the restriction to static phase fluctuations.  Thus it
is important to check whether the sum rule (\ref{sum.rule}) is
satisfied with sufficient accuracy.

It is remarkable that for (\ref{spectral.density.final}) the sum
rule (\ref{sum.rule}) can be tested analytically (see the details
in \cite{Gusynin.new}). One can derive the following result,
\begin{equation}
\int_{- \infty}^{\infty} d \omega A(\omega, {\bf k}) = \frac{\Gamma
(\alpha)}{\Gamma(2-\alpha)}.
\label{devsumrule}
\end{equation}

The numerical value of the integral at the temperatures
of interest may be estimated in the following way. At
$T=T_{\mathrm{BKT}}$ the phase stiffness is given by
$J = 2/\pi T_{\mathrm{BKT}}$, so that for $T$ near to
$T_{\mathrm{BKT}}$ the value $\alpha$ from (\ref{A,alpha}) is
\begin{equation}
\alpha \simeq 1 - \frac{1}{32} \frac{T}{T_{\mathrm{BKT}}},
\qquad T \sim T_{\mathrm{BKT}}.
\label{alpha.estimation}
\end{equation}
In particular, $\alpha(T=T_{\mathrm{BKT}}) =31/32$ yields the following
estimate for the right hand side of (\ref{devsumrule}),
$\Gamma(\alpha)/\Gamma(2-\alpha) \simeq 1.037$. This shows that
for $T \sim T_{\mathrm{BKT}}$ the spectral density
(\ref{spectral.density.final}) is reasonably good.

The parameter $\alpha$ can however differ strongly from unity
at $T>T_{\mathrm{BKT}}$ and in the underdoped regime.

\subsubsection{Main results for the spectral density.
The problem of gap filling. Limitations of numerical
analytical continuation}
\label{sec:continuation}

\begin{figure}[h]
\centering \resizebox{0.5\textwidth}{!}{%
  \includegraphics[bb=20 355 275 530]{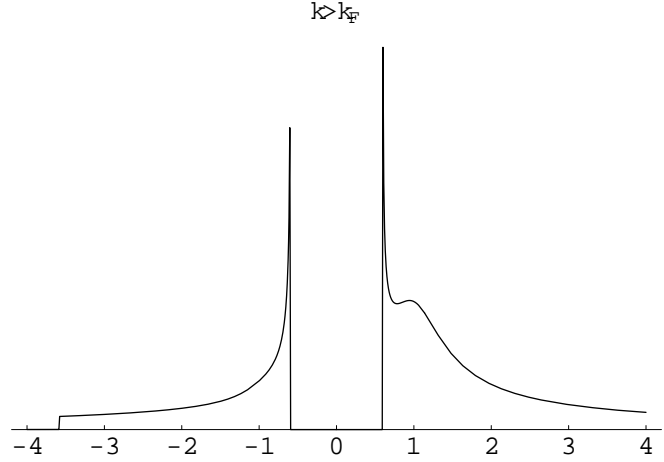}
}
\caption{
The plot of the spectral function $A(\omega , {\bf k}) $ as a
function of $\omega$ in units of the zero temperature gap $\Delta$
for $k > k_{F}$ at $T = 1.12 T_{\mathrm{BKT}}$ as taken from
\protect{\cite{Gusynin.new}}.}
\label{fig:24}
\end{figure}

An example  of the  plots of the spectral density
$A(\omega, {\bf k})$ given by (\ref{spectral.density.final}) at
$T > T_{\mathrm{BKT}}$ is presented in Fig.~\ref{fig:24}.
To draw these plots the value of $\alpha$ from
Eq.~(\ref{alpha.estimation}) and the mean-field value of $\rho$
obtained from Eq.~(\ref{rho.bound}) were used.

The temperature dependence of the spectral function
(\ref{spectral.density.final}) is discussed in detail in
\cite{Gusynin.new}. It is shown, in particular, that
below $T_{\mathrm{BKT}}$ the quasi-particle peaks
at $\omega = \pm E(\mbox{\bf k})$ have a finite
temperature dependent width which is, of
course, related to the spin-wave (longitudinal) phase fluctuations.
When $T \to 0$ the width goes to zero, but this limit cannot be
correctly derived from (\ref{spectral.density.final}) because this is
an ordinary function, while the BCS spectral density
(\ref{BCS.spectral.density}) is a distribution.
This sharpening of the peaks
with decreasing $T$ in the superconducting state was experimentally
observed \cite{Campuzano} and represents a striking difference
from the BCS ``pile-up'' (\ref{BCS.spectral.density}) which is
present for all $T < T_{c}$.

It was pointed out in \cite{Franz} (see Sec.~\ref{sec:Franz}) that the
broadening of the spectral function caused by these fluctuations
can be greater than the experimental data permits.

For $T > T_{\mathrm{BKT}}$ the quasi-particle peaks become less pronounced as
the temperature increases. Indeed, the value of $A(\omega, \mbox{\bf k})$ at
$\omega = \pm E(\mbox{\bf k})$ is, in contrast to the case $T < T_{\rm BKT}$,
already finite. This is caused by the fact that $\mathcal{D} \neq 0$ since
$\xi_{+}(T)$ is now finite due to the influence of vortex fluctuations. As the
temperature increases further, $\xi_{+}(T)$ decreases so that the
quasi-particle peaks disappear (see Fig.~\ref{fig:24}).  This behaviour
qualitatively reproduces the ARPES studies of the cuprates for the anti-node
direction \cite{Campuzano} (see also \cite{Norman.1999}) which show that the
quasi-particle spectral function broadens dramatically when passing from the
superconducting to normal state.

It is important to stress that due to the very smooth
dependence of $\xi_{+}^{-1}(T)$ on $T$ (see Eq.~(\ref{BKT.length})) as
the temperature varies from $T < T_{\mathrm{BKT}}$ to
$T > T_{\mathrm{BKT}}$ there is no sharp transition at the point
$T = T_{\mathrm{BKT}}$. In particular, there is a smooth evolution of
the superconducting (excitation) gap $\Delta_{\mathrm{SC}} = \rho$ into
the gap $\Delta_{\mathrm{PG}}$ which is also equal to $\rho$ and in fact
can be called a pseudogap because at $T>T_{\mathrm{BKT}}$ the system
is not superconducting. This qualitatively fits experiment
\cite{Timusk,Campuzano,Norman.1999,Renner} and appears to be completely
different from BCS theory \cite{Schrieffer}, where the gap
vanishes at $T=T_{c}$.

We, however, would like to stress the main reason why further studies are
essential. While the temperature behaviour of the quasi-particle peaks is
quite similar to experiment \cite{Timusk,Campuzano}, the gap in the
spectrum remains unfilled, i.e. the spectral density is identically zero
inside the gap ($A(\omega , {\bf k}) = 0$ for $|\omega| < \rho$).
Furthermore, there is also an excess of the spectral weight on the gap
edges which is seen as the extra peaks \cite{Gusynin.new}.
This is obviously related to the facts that the classical phase
fluctuations were treated in the static approximation and that the coupling
between the spin and charge degrees of freedom was neglected. One
therefore needs to consider how these effects could lead to the
filling of the gap.

The $T$-matrix approximation seems to result in the opposite
result. For example, the classical fluctuations do lead to a
pseudogap, as shown by the Hubbard model at half-filling
\cite{Tremblay.1997}. We note, however, that our statement
is applicable to the uncoupled static phase fluctuations only and
neglects entirely modulus fluctuations.
Further studies are necessary to make a comparison of this analytical
result for the phase fluctuations with other complementary approaches,
which we believe should reproduce our result when the same
assumptions are made.
Here we make only one further comment related to the numerical
analytical continuation.

\begin{figure}[h]
\centering
\resizebox{0.6\textwidth}{!}{%
  \includegraphics[bb=110 280 460 495]{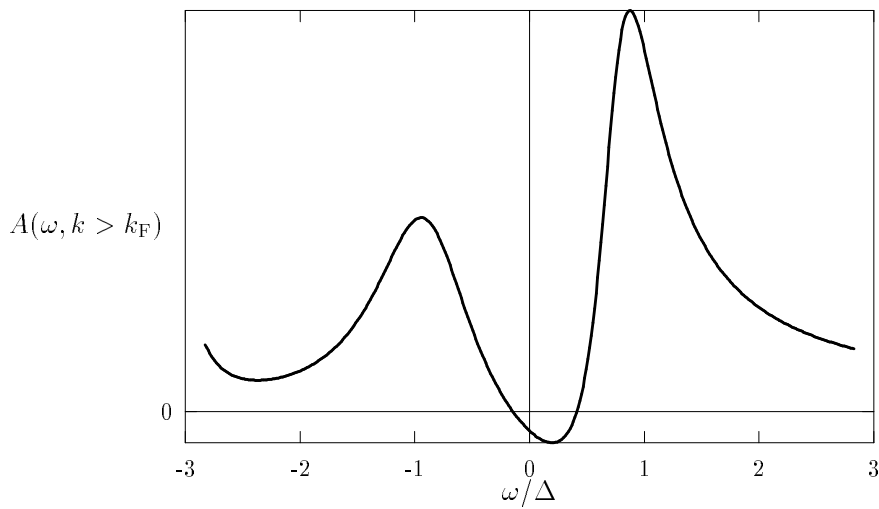}
}
\caption{
Plot of the spectral function $A(\omega , {\bf k}) $ as a
function of $\omega$ in units of the zero temperature gap $\Delta$
for $k > k_{F}$ at $T = 1.12 T_{\mathrm{BKT}}$ obtained by means
of numerical analytical continuation of
Eq.~(\ref{Green.final.Matsubara}).}
\label{fig:25}
\end{figure}

It is very instructive to estimate the accuracy one would expect
if a similar problem were to be investigated numerically.
Using the analytical expression
(\ref{Green.final.Matsubara}) obtained for the Green's function on
the imaginary axis one can perform its numerical analytical
continuation by means of Pad\'e approximants \cite{Vidberg}
(see also \cite{Carbotte} and a recent comprehensive discussion of the
method in \cite{Beach}). Using Eq.~(\ref{Green.final.Matsubara})
calculated to ten digits of precision we tried to recover the known
spectral function (\ref{spectral.density.final}) shown in
Fig.~\ref{fig:24}, but even for 110 Matsubara frequencies the
numerical result presented in Fig.~\ref{fig:25} cannot indicate
whether the gap is filled or not.
As already mentioned in Sec.~\ref{sec:self}, numerical
analytical continuation is often used
for the subsequent extraction of the spectral function in the methods where
the $T$-matrix equations are solved numerically on the imaginary axis. It is
clear, however, that the precision of the numerical solution is even less
than that used above for Eq.~(\ref{Green.final.Matsubara}) and thus
numerical results alone are insufficient for definitive conclusions about
pseudogap filling. There is however a real-time
technique \cite{Gooding,Kornilovitch,Dagotto} which allows one to
avoid numerical analytical continuation, but demands some other
assumptions.

The results discussed in this section show that, while the physical
picture based on non-interacting static phase fluctuations
is sufficient to derive at least the qualitative phase diagram,
additional consideration is necessary to understand the
mechanisms responsible for gap filling.
Thus the next two sections will be devoted to two possible ways
that the pseudogap can be filled.

\subsection{Gap filling by static phase fluctuations due to
quasi-particle vortex interactions. The phenomenology of ARPES}
\label{sec:Franz}

As shown by Franz and Millis \cite{Franz} (see also \cite{Dorsey})
one way of filling the gap can be obtained for
static phase fluctuations via the coupling of the spin and charge
degrees of freedom which leads to a Doppler shift for the
quasi-particles.

As already discussed in Sec.~\ref{sec:phase.green}, for this
mechanism the disordered state above $T_{\mathrm{BKT}}$ can be
thought of as a ``soup'' of fluctuating vortices with positive
and negative topological charges and with total vorticity constrained
to zero. Each of these vortices is surrounded by a
circulating supercurrent which decays as $1/r$ with the distance
from the core. Such supercurrents, within a semiclassical approximation,
lead to a Doppler-shifted local quasi-particle excitation spectrum
of the form \cite{deGennes.book,Tinkham}
\begin{equation}
E_{\bf k}=E_{\bf k}^{(0)}+\hbar{\bf k}\cdot{\bf v}_s({\bf r}),
\label{en}
\end{equation}
where ${\bf v}_s({\bf r})$ is the local superfluid velocity and
$E_{\bf k}^{(0)} =\sqrt{\epsilon_{\bf k}^2+|\Delta_{\bf k}|^2}$
is the usual BCS spectrum.
The change in the local excitation spectrum will affect the
spectral properties of the superconductor in that the physically
relevant spectral function must be averaged over the positions of
fluctuating vortices.

This effect will be particularly pronounced
in a $d$-wave superconductor since  Eq.~(\ref{en})
implies the formation of a region around a nodal point on the Fermi
surface with $E_{\bf k}<0$ for arbitrarily small ${\bf v}_s({\bf r})$.
Physically this corresponds to a region of gapless excitations on
the Fermi surface which leads to a finite  density of states (DOS)
at the Fermi level.
As first discussed by Volovik \cite{Volovik}, a similar situation arises
in the {\em mixed\/} state of a $d$-wave superconductor
where the superflow around the field-induced vortices
leads to a residual DOS proportional to $\sqrt{H}$.

\subsubsection{Phenomenological and phase fluctuation coupled
Green's functions}

In the mean field approximation (neglecting, among other things, phase
fluctuations)
the diagonal Green's function was written in \cite{Franz}
(compare with Eq.~(\ref{K2}))
\begin{equation}
\mathcal{G}_0^{-1}(\omega, {\bf k}) =
\omega-\epsilon_{\bf k}+i\Gamma_1-
\frac{\Delta_{\bf k}^2} {\omega+\epsilon_{\bf k}}\,,
            \label{g0}
\end{equation}
where the angular dependence of the gap function for
$d$-wave superconductor is
$\Delta_{\bf k} = \Delta_{d} \cos 2 \theta$ and
following \cite{Norman.1998} the single particle scattering rate
$\Gamma_1$ was added to the usual mean-field solution to describe
the ARPES data in the overdoped samples where there are no strong pseudogap
effects.
It is instructive  to use Dyson's equation (\ref{Dyson})
to extract from (\ref{g0}) the expression
for self-energy as done in \cite{Norman.1998,Norman.1999}
\begin{equation}
\Sigma(\omega, {\bf k}) =
- i\Gamma_1 +
\frac{\Delta_{\bf k}^2} {\omega+ i0 + \epsilon_{\bf k}} \,.
                  \label{self1.Norman}
\end{equation}
Note that the form of the scattering rate in $\mathcal{G}_0$
($\Sigma$) constitutes a non-trivial assumption. It is not
pair-breaking, in the sense that it is ineffective at small
$\omega$ and $\epsilon_{\bf k}$; i.e. in the  region
$\omega< E^{(0)}_{\bf k}$.
By contrast in a $d$-wave superconductor, a conventional scattering
rate enters via the replacement $\omega\to\omega+i\Gamma$, leading to a
broadening which is effective even at low $\omega$ and $\epsilon_{\bf k}$.
As shown by Norman {\em et al.} \cite{Norman.1998} the form given in
Eqs.~(\ref{g0}) and (\ref{self1.Norman}) agrees with the ARPES data at
$T<T_c$.  It is demonstrated in \cite{Franz} that it also agrees with
STS.

At $T>T_c$ Norman {\em et al.} \cite{Norman.1998} showed that additional
{\em pair-breaking\/} scattering is needed to account for the ARPES data
in the underdoped case where the gap is filled well above $T_c$,
They modelled this phenomenologically by introducing another
scattering rate $\Gamma_0\ne\Gamma_1$, and making the
replacement $\omega\to\omega+i\Gamma_0$ in the last term of
Eqs.~(\ref{g0}) and (\ref{self1.Norman}):
\begin{equation}
\Sigma(\omega, {\bf k}) =
- i\Gamma_1 +
\frac{\Delta_{\bf k}^2} {\omega+ \epsilon_{\bf k} + i \Gamma_0} \,.
                  \label{self2.Norman}
\end{equation}
They suggested that $\Gamma_0$ could arise from exchange of pair
fluctuations and precisely this kind of effect has been intensively
studied within $T$-matrix approximation (see Sec.~\ref{T.matrix}),
but as argued in \cite{Franz,Tremblay.1997}
this proposed mechanism does not account for
the observed magnitude of $\Gamma_0$.

As mentioned above  another likely source of the pair-breaking
scattering leading to gap filling has been suggested in
\cite{Franz}, namely supercurrents induced by phase fluctuations.
In order to determine how $\mathcal{G}_0$ is changed in the presence of
superflow it is useful to recall the origin of the energy shift in
Eq.~(\ref{en}). This can be  derived \cite{deGennes.book,Tinkham}
by assuming a state of {\em uniform} superflow with
${\bf v}_s=\hbar{\bf q}/m$ induced by an order parameter of the form
$e^{2i{\bf q}\cdot{\bf r}} \Delta_{\bf k}$.  By solving the appropriate set
of Bogolyubov-de Gennes equations and retaining only terms to linear order in
${\bf q}$, one finds that the energy is modified as indicated in (\ref{en})
while the coherence factors are to the same order unchanged. This result is
then semiclassically extended to non-uniform situations by assuming slow
spatial variations of ${\bf v}_s({\bf r})$.

One can follow this exact procedure and solve the appropriate
Gor'kov equations for $\mathcal{ G}_{\bf q}$ in the presence of superflow.
One finds the following intuitively plausible result which is exact for
uniform flow up to terms linear in ${\bf q}$:
\begin{equation}
\mathcal{G}_{\bf q}(\omega, {\bf k})
= \mathcal{G}_0( \omega - \eta, {\bf k}-{\bf q}) \,,
        \label{gq}
\end{equation}
where $\eta\equiv\hbar{\bf v}_F({\bf k})\cdot{\bf q}\simeq\hbar{\bf k}\cdot
{\bf v}_s$. Here
${\bf v}_F({\bf k})=(\partial\epsilon_{\bf k}/\partial{\bf k})_{k=k_F}\simeq
\hbar{\bf k}_F/m$
is the Fermi velocity and the last equality holds when the Fermi surface is
approximately isotropic. In the following
it is assumed that Eq.~(\ref{gq}) can be applied locally when
${\bf v}_s({\bf r})$ varies slowly in space.
Applying the above prescription to (\ref{g0}) one finds, again to
the leading order in ${\bf q}$,
\begin{equation}
\mathcal{G}_{\bf q}^{-1}(\omega, {\bf k}) =
\omega-\epsilon_{\bf k}+ i \Gamma_1-
\frac{(\Delta_{\bf k}-\zeta)^2}{\omega+\epsilon_{\bf k}-2\eta}\,,
                      \label{gqq}
\end{equation}
where $\zeta\equiv{\bf v}_\Delta({\bf k})\cdot{\bf q}$ with
${\bf v}_\Delta({\bf k})=(\partial\Delta_{\bf k}/\partial{\bf k})_{k=k_F}$.
One can easily estimate $v_\Delta/v_F\sim
(\xi_{\mathrm{coh}}k_{\mathrm{F}})^{-1}\sim\Delta_d/\epsilon_F$ which is
typically a small number in superconductor.
One therefore expects that $\zeta\ll\eta$.  A more detailed numerical
analysis indeed shows that, as long as $\Delta_d/\epsilon_F$ is small
compared to unity, the effect of $\zeta$ on the spectral lineshape is
negligible compared to that of $\eta$, and will be dropped in the
following.

A typical experimentally measured quantity, such as the ARPES or STS
lineshape, will provide information on the spectral function
associated with $\mathcal{G}_{\bf q}$ {\em averaged} over the phase
fluctuations (see also \cite{Franz.1999}). Thus, one needs to evaluate
\begin{equation}
\bar{\mathcal{G}}_{\bf q}(\omega, {\bf k})
=\int d\eta P(\eta)\mathcal{G}_{\bf q}(\omega, {\bf k}) \,,
\label{avg}
\end{equation}
where $P$ is the probability distribution of $\eta$ given by
\begin{equation}
P(\eta)=
\langle \delta[\eta-\hbar{\bf k}\cdot{\bf v}_s({\bf r})]\rangle \,.
\label{p2}
\end{equation}
The angular brackets indicate thermodynamic averaging over the phase
fluctuations in the ensemble specified by the 2D $XY$ Hamiltonian
(\ref{XY.Hamiltonian}).

Further details of the precise derivation and the comparison with the ARPES
and STS  experimental data can be found in \cite{Franz}, but will not be
discussed here since our main goal is simply to discuss possible mechanisms
for gap filling. We note only that since the calculations reported in
\cite{Franz}, new and controversial ARPES data has been  obtained
\cite{Chuang,Feng}.

As already discussed the last term in Eq.~(\ref{g0}) can be thought
of as a superconducting self-energy given by Eq.~(\ref{self1.Norman}).
Eqs.~(\ref{gqq}) and (\ref{avg}) then imply that the primary effect of
the phase fluctuations is to smear the functional dependence of
$\Sigma (\omega,{\bf k})$ on the energy variable, broadening the
spectral lineshape. The detailed analysis performed in \cite{Franz}
shows that $\eta$ acts primarily to {\em fill in}
the gap, in a way similar to the inverse pair lifetime
$\Gamma_0$ introduced by phenomenological considerations
in Ref.\cite{Norman.1998}. $\Gamma_1$, on the other hand,
does not affect the lineshape at low energies: notice that
since the self-energy given by Eq.~(\ref{self1.Norman}) diverges
at the point ${\bf k} ={\bf k}_F$, $\omega=0$  if
$\Delta_{\bf k} \ne 0$, the corresponding Green's function
$\mathcal{G}_0({\bf k}_F,\omega=0)=0$ for any $\Gamma_1$.

\subsubsection{Discussion}

The qualitative behaviour of ARPES and STS lineshapes (see
\cite{Timusk,Campuzano,Renner,Norman.1998}) in underdoped BiSCCO clearly
establishes the existence of a scattering mechanism which becomes operative at
$T>T_c$ and which acts primarily to fill in the gap at low energies.

It is shown by Franz and Millis \cite{Franz} that vortex phase
fluctuations associated with proliferation of unbound vortex-antivortex
pairs in the system interacting with the quasi-particles
provide a reasonable explanation for this scattering.

Analysis of \cite{Franz} also indicates that longitudinal (spin
wave) phase fluctuations are almost completely suppressed, above and
below $T_c$.
As we mentioned in Sec.~\ref{sec:mapping}
(see also Sec.~\ref{sec:full.superfluid}) it has been proposed in
\cite{Roddick} that in high-$T_c$ materials, longitudinal phase
fluctuations governed by the $XY$ Hamiltonian (\ref{XY.Hamiltonian})
are important in that they significantly contribute to the observed
temperature dependence of the magnetic penetration depth.
The broadening of the spectral function
which would be caused by these fluctuations was calculated in
\cite{Franz}, and it was found that it is much greater than the
experimental  data would permit. They therefore conclude that longitudinal
fluctuations are suppressed, perhaps by the Coulomb interaction.

Quantitatively there exists considerable discrepancy between
the parameters describing the ARPES and STS lineshapes, in particular
the single particle scattering rate $\Gamma_1$. It is possible that
this discrepancy can be resolved using new ARPES data \cite{Chuang,Feng}.

Sizable transverse phase fluctuations implied by this work
will also affect other properties of the underdoped systems, such as the
electronic specific heat, fluctuation diamagnetism and transport
\cite{Franz.1999}. Vortices existing above $T_c$ should also generate local
magnetic fields which are zero on average but have a non vanishing variance.
If such fields could be detected, e.g. by muon spin rotation experiment,
this would constitute direct evidence for the phase fluctuation model of
the pseudogap phase.

Finally we mention the recent paper \cite{Westfahl} where the effect
of phase fluctuations on inelastic neutron scattering and NMR
experiments was studied using an approach similar to that of
\cite{Franz}.

\subsection{Gap filling due to dynamical phase fluctuations without
quasi-particle vortex interactions}
\label{sec:Capezzali}

Another mechanism of the pseudogap filling
due to  vortex-vortex dynamics has been suggested
by Capezzali and Beck \cite{Capezzali.1999} where the same $s$-wave
pairing model as in Sec.~\ref{chap:our.Green} was considered.
This model is treated by the Stratonovich-Hubbard
transformation (see Sec.~\ref{sec:formalism}), decoupling the interaction
term by the complex pairing field $\Phi$.
The one-electron Green's function is then approximately given
by \cite{Pedersen,Capezzali.thesis}:
\begin{equation}
G(i \omega_{n},{\bf k})^{-1}= i\omega_n
-\varepsilon({\bf k}) + \mu- \sigma( i\omega_n, {\bf k})-
\frac{\langle \Phi \rangle^{2}}{
i\omega_{n} + \varepsilon({\bf k})- \mu + \sigma(- i\omega_n, {\bf k})} \,.
                               \label{Green.Capezzali}
\end{equation}
The expression for the self-energy in the next to leading order approximation
\begin{equation}
\sigma (i \omega_{n}, {\bf k})
=-\sum_{m =-\infty}^{\infty} \int \frac{d^2 q}{(2 \pi)^2}
\langle |\Phi(i\Omega_m, {\bf q})|^{2}
\rangle G(i \omega_n - i \omega_m, {\bf k}-{\bf q})
                    \label{self.Capezzali}
\end{equation}
involves the dynamic correlation function of the pairing field.
Note that Eq.~(\ref{self.Capezzali}) does not contain the leading order of the
self-energy which is included into the Green's function (\ref{Green.Capezzali})
via the term with pairing field $\langle \Phi \rangle$
(compare with Eqs.~(\ref{self1.Norman}) and (\ref{self2.Norman})).
In diagrammatic language Eq.~(\ref{self.Capezzali}) corresponds
to a two-vertex bubble, one of the lines of which represents the propagator
of the complex bosonic Hubbard-Stratonovich field, while the other line
corresponds to the dressed propagator of the fermions.

In \cite{Capezzali.1999} rather than aiming at a self-consistent solution
as described in Sec.~\ref{sec:T.matrix}, a simple form for the pairing
correlation function was assumed and its influence on the
one-electron properties were investigated.
Introducing the amplitude and phase in the pairing field
(see Eq.~(\ref{parametrization.Hubbard})) {\em but} without the
corresponding replacement for the Fermi field
(see  Eq.~(\ref{parametrization.fermi}),
Capezzali and Beck made the following assumptions:

\noindent
$(i)$ The mean-field transition temperature $T_{c}^{\mathrm{MF}}$
is, following \cite{Oda}, identified with the temperature $T^{\ast}$.
Below this temperature the amplitude (modulus) fluctuations
are assumed to be space- and time-independent, i.e. they are
treated in the mean-field approximation in a way similar to that discussed in
Secs.~\ref{sec:our.diagram} and \ref{sec:form}. They are thus approximated by
a typical mean-field  (BCS-like) analytical form
(see Eq.~(\ref{rho.dependence})),
\begin{equation}
\langle |\Phi(t, {\bf r})|^2 \rangle \equiv \Delta^2(T) \simeq \Delta_0^2
\left(1 - \frac{T}{T_{c}^{\mathrm{MF}}} \right)
\label{ampl}
\end{equation}
where $\Delta_0^2$ is treated as a free parameter. For BCS theory it is given
by $\Delta_0^2 = 3.02 \Delta^2(T=0)$. Note that once more the average
$\langle \Phi(t, {\bf r})\rangle$ is zero. Using the definition of the Fourier
transform (\ref{Fourier.momentum}),
and space and time independence of the amplitude, one can obtain that
\begin{equation}
\begin{split}
\langle |\Phi(i \Omega_m, {\bf q})^2| \rangle & \equiv
\int_{0}^{\beta} d \tau \int d^2 r \exp(i \Omega_m \tau - i {\bf q} {\bf r})
\langle \Phi^{\ast} (\tau, {\bf r}) \Phi(0, {\bf 0}) \rangle
                 \\
=  \Delta^2(T) \int_{0}^{\beta} & d \tau \int d^2 r
\exp(i \Omega_m \tau - i {\bf q} {\bf r})
\langle \exp (i \theta(\tau, {\bf r}) - \theta(0, {\bf 0}) ) \rangle \,.
\end{split}
\label{Phi.Capezzali}
\end{equation}
The Fourier transform of the phase correlator in (\ref{Phi.Capezzali}) was
taken in the form given by Eq.~(\ref{correlator.dynamic}) (see the details in
Sec.~\ref{sec:phase.green}, in particular the last paragraph of the section).
Thus the main difference between the first assumption of \cite{Capezzali.1999}
and the assumptions used in Chap.~\ref{chap:our.Green} is the consideration
of the more general case of the dynamical phase fluctuations. As
will soon become apparent it is their presence that leads to gap filling.

\noindent
$(ii)$ Below the critical temperature $T_{c}$ one keeps the
same form (\ref{correlator.dynamic}) for the phase correlations, but with
$\xi_{+}^{-1}(T)=0$, corresponding to the algebraic decay of correlations.
However, in contrast to the calculations in Chap.~\ref{chap:our.Green},
a non-zero coupling between the planes (in the third dimension) is taken into
account in \cite{Capezzali.1999}. This is achieved by introducing a non-zero
value for the average of $\Phi(x)$,
$\langle \Phi(x) \rangle^2 = \lambda \langle |\Phi(x)|^2 \rangle$
with a variable parameter $\lambda\leq 1$. Due to the presence of the third
direction the nonzero value of $\langle \Phi \rangle$ is not forbidden
by the CMWH \cite{Coleman} theorem.

\subsubsection{Discussion}

For numerical computation the value of $\sigma(i\omega_n, {\bf k})$
from (\ref{self.Capezzali}) was evaluated in \cite{Capezzali.1999,Capezzali.thesis}
to lowest order by using the noninteracting $G(i \omega_n, {\bf k})$ with
an isotropic spectrum $\varepsilon({\bf k})= {\bf k}^2/2m$.

\begin{figure}[h]
\centering
\resizebox{0.7\textwidth}{!}{%
  \includegraphics{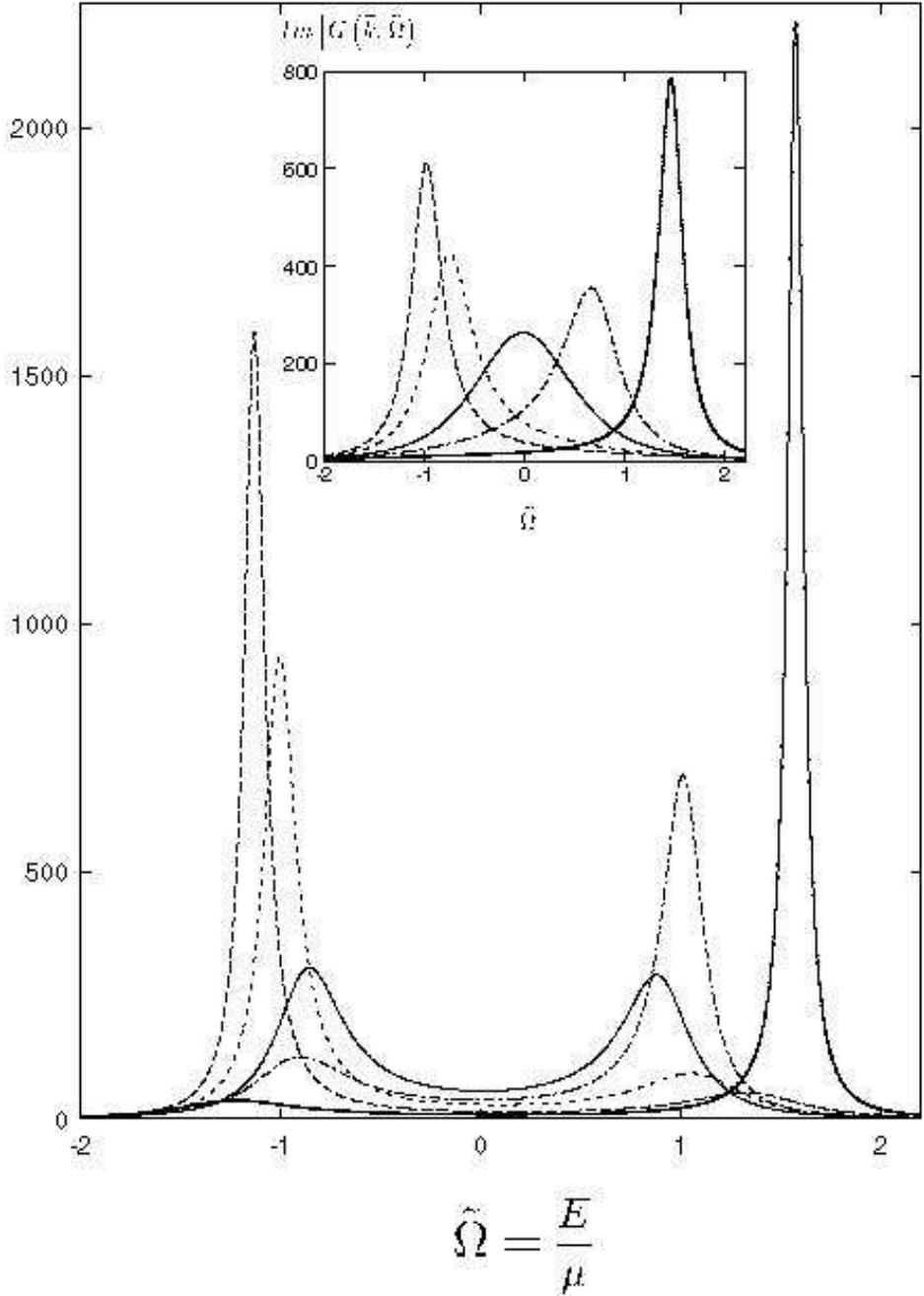}
}
\caption{
The spectral function $A(\omega, {\bf k})$ taken from
\protect{\cite{Capezzali.1999}}
for $k=0.5k_{F}$ (dashed line),
$k=0.7k_{F}$ (dotted line),
$k=k_{F}$ (full line),
$k=1.2k_{F}$ (dash-dotted line),
$k=1.5k_{F}$ (heavy full line), for $T=1.01T_{c}$ and (inset)
$T=1.5T_{c}$.}
\label{fig:26}
\end{figure}
In Fig.~\ref{fig:26} we show the spectral function from \cite{Capezzali.1999}
obtained for the decay constant (see Eq.~\ref{correlator.dynamic})
$\Gamma=0.5 \epsilon_{F}$ and the modulus strength (see Eq.~\ref{ampl})
$\Delta_0^2 = 0.76 \epsilon_F^2$.
The following observations can be made on the basis of this figure and the
other results presented in \cite{Capezzali.1999} :

\noindent
$(i)$ Above $T_{c}$, a pseudogap opens around $\mu \simeq \epsilon_{F}$. Its
effective width is almost $T$-independent, in spite of the $T$-dependence of
$\langle|\Phi|^{2} \rangle$ in the self-energy.

Below $T^{\ast}$, but well above $T_c$, one sees a one-peak structure in the
spectral function for all wave-vectors $k$, centered below (above) the
chemical potential for $k < (>) k_F$. As one approaches $T_{c}$, a two peak
structure emerges for wavevectors $\sim k_F$ where the peak below (above)
the chemical potential has a greater spectral weight for $k < (>) k_F$. This
structure becomes increasingly visible near $T_c$ where the density of states
at the chemical potential, $N\left(\mu\right)$, becomes practically zero for
all $k \ne k_F$. Thus in this limit the pseudogap is delimited by two rather
pronounced peaks.

It is important that, in contrast to the previous section, the filling is
achieved without {\em any\/} vortex-quasi-particle interactions. This, of
course, does not exclude the fact that there are phase fluctuations. If one
could  divide the total contribution of the fluctuations in
(\ref{Phi.Capezzali}) into the classical ($m=0$) and other (in some sense
non-classical with $m \neq 0$) fluctuations, it can be shown that the latter
contribution also results in some gap filling \cite{GLQSh.new}.

\noindent
$(ii)$ Below $T_{c}$, the presence of $\langle \Phi \rangle \neq 0$ produces
two new peaks: a true "superconducting gap" emerges out of the pseudogap. Its
width is given by the geometrical superposition of average and fluctuating
part of $\Phi(x)$. The fluctuations of the latter remain visible in the form of
secondary shoulders inside the superconducting gap which approach
each other about in the same proportion as the main peaks move away
from each other.

\noindent
$(iii)$ The value for the decay constant $\Gamma \simeq 0.5 \mu$ is relatively
large. For a smaller $\Gamma$, and in particular for the case of critical
slowing down of phase fluctuations ($\Gamma\left(T_{c}^{+}\right)=0$)
\cite{Huber} the two shoulders would become secondary peaks inside the
superconducting gap.

\noindent
$(iv)$ Near $T_{c}$, as seen above, the spectral functions are doubly peaked
in some wave-vector domain around $k_{F}$. In this case the width of the
pseudogap is given by the separation between these two peaks, which is
essentially determined by $\langle|\Phi|^{2}\rangle$. At higher temperatures,
approaching $T^*$, the spectral functions have only one peak. However their
width is enhanced over essentially the same $k$-domain. Due to this fact, the
pseudogap fills up gradually, as $T^{\ast}$ is approached (from below), but
with little change in its width.

Thus the mechanism for gap filling due to vortex-vortex interactions
which are described by dynamical phase fluctuations can be also regarded as
a strong candidate. Both mechanisms are based on the same physical premise,
namely the existence of a ``soup'' of vortices above $T_{\rm BKT}$. Indeed
both mechanisms ought to be present simultaneously, but it is not yet
clear which is the dominant mechanism for gap filling, and more
studies are certainly needed.

On the other hand, it must be pointed out that the existence of
fluctuating vortices is not a necessary condition for the appearance of
a pseudogap.  Indeed, systems with $SO(n>2)$ symmetry
(where $n$ is the number of the order parameter components)
do not have topologically stable vortices, but still may
possess a pseudogap-like regime. As discussed in \cite{Kleinert.2000}
there is a situation when the second order phase transition in
such a system is driven by long-wavelength transverse
(or {\em  directional} in terms of \cite{Kleinert.2000}) fluctuations,
and not by  longitudinal (or {\em size\/}) fluctuations as happens in more
conventional scenarios. During this sort of transition the ordering
field acquires firstly as the temperature decreases its size (modulus)
while the direction is still strongly fluctuating and only a further
temperature decrease leads to the formation of the ordered state.
Thus one can concede this scenario of a second order phase
transition leaves room for pseudogap-like behaviour even in
more general theories where the stable vortex excitations are absent.
Nevertheless, their presence in the particular case of $SO(2)$
theory considered here suggests that vortices can play an essential
role and their specific contribution may even be dominant in this
case.

\section{Concluding remarks}

We have concluded the description of the physical properties of 2D
metals in which conductivity is the result of doping.
We paid particular attention to the normal and superconducting
properties directly related to phase fluctuations
of the complex ordering field which increase significantly
as the carrier density decreases. The vital role of
phase fluctuations for phase transitions in low dimensional systems, as
mentioned above, has been clearly outlined in the classical papers of
Berezinskii, Kosterlitz and Thouless and the many other physicists who
followed.
However for many years physicists have ignored the important fact that
even above $T_c$ ($T_{\mathrm{BKT}}$), in the so called normal phase,
there is a remaining non-zero parameter, which drastically changes
the spectrum of the phase and leads to pseudogap behaviour.
While in theoretical papers this feature was simply overlooked,
experimental techniques probably did not have sufficient
precision to observe the pseudogap. During the last 3-5
years the situation has drastically changed and the resolution of ARPES
and thermodynamical experiments permit one to measure quite subtle
spectral features, and thus leading to the discovery of the pseudogap.
According to Abrikosov ``the pseudogap in the
quasi-particle spectrum above $T_c$ and its amazing stability
in strongly underdoped samples was the most spectacular among
phenomena'' observed in high-$T_c$ compounds \cite{Abrikosov.1997}.
The theory discussed above, however, does not relate the pseudogap
appearance directly to the superconducting properties of oxides and
stresses the possible main role of the low dimensionality of their
magnetic and electron spectra. And indeed, the experimental
data is not in contradiction with this conclusion. Pseudogap
features have already been found in quasi-1D superconductors
\cite{Terrasi,Zurick} as well as in the HTSC systems mentioned above.
It is interesting that pseudogap behaviour close to
the Fermi level is also seen in 3D manganite systems \cite{Park}. However,
in the last case its connection with some kind of fluctuation is
under question \cite{Joynt}. Nevertheless even if the pseudogap in
manganites is caused by different effects, it is remarkable,
that in all these cases the systems are bad metals, i.e. for
the systems under consideration,
the undoped state is an insulator and conductivity
only results from doping.

Throughout this review we discussed mainly the simplest 2D
fermi-system with attraction. There is, however, a general
theoretical question as to what are the generic effects that
lead to a pseudogap. It is clear that to gain greater insight
into this problem a wider class of the models should
be considered.
Nevertheless, some general arguments can be given already.
First of all, the lowered dimensionality of the space is an
essential component. This results in the CMWH theorem that allows one to
state that there is a new region between the mean-field transition
temperature $T_{c}^{\mathrm{MF}}$ and a true transition
temperature which is in $d=2$ for the systems with two-component
($SO(2)$) order parameter may be well approximated by the
of topological transition temperature, $T_{\mathrm{BKT}}$. This
case of $SO(2)$ symmetry, which is relevant for superfluids and
superconductors, led us to the rich BKT physics of vortices that
undoubtedly contributes to the pseudogap properties of these systems.
However, as we saw even in these 2D systems the pseudogap region appears
to be small if the carrier density is too high, or in other words the
coherence length is too big and size fluctuations dominate over
directional \cite{Kleinert.2000}. In $d=3$ this coherence length should
be even smaller \cite{Kleinert.2000} and therefore the conditions for the
pseudogap in 3D is difficult if not impossible to satisfy.

What would happen if one considers a multi-component
($SO(n>2)$) order parameter? Since there is no stable vortices in
this case, the topological transition is also absent. Thus
as mentioned, for example, in \cite{Tremblay.1999} the higher
symmetry of the order parameter space leads to an even larger region
with directional fluctuations than in the case of $SO(2)$ symmetry.
Therefore the larger dimensionality of the space of the order parameter
also favours pseudogap formation. Further study of these types of models
would be helpful for a deeper understanding of the pseudogap origin.

A series of experiments, in particular, NMR, neutron scattering
and spin relaxation demand the introduction of the concept of a spin gap.
This is not surprising because the undoped magnetic subsystems, which are
antiferromagnetic insulators  have spins localized on the copper ions
and are characterized by the magnetic anisotropy of ``easy plane''
type. This shows that the average magnetic moment of every site
is nonzero, and therefore the average antiferromagnetic vector
in the basic CuO$_2$ plane is also nonzero.\footnote{It is
remarkable that in the paper \cite{Kresin} the role of these planes
in the physics of HTSC is compared with the role of the hydrogen atom
in atomic physics.}
Doping which suppresses the long-range magnetic order, does not,
however, suppress the average value of the modulus of the spin on
the lattice point. This spin can also be parametrized via its
modulus and phase. Then the physical picture in the spin spectrum
will be very similar to that for superconductivity, when the gap
in the spectrum is related to the average spin (modulus of the
magnetisation) and the absence of the long-range order is the result
of ``disorder'' by angular (directional) fluctuations which can also
be described by a correlation function. Thus the system may have an
additional crossover temperature related to modulus formation for
another -- spin -- physical value.
The consideration of this system, which will be close to that
described above, is given either by the 2D sigma model or by
the model of nearly antiferromagnetic liquid. We did not however
describe them in detail because our goal was to present simple
analytical (but at the same time, we hope, non-trivial) models
which can lead to pseudogap character in the one-particle spectrum.
The spin aspect inevitably demands in the absence of long-range order
a separate treatment and the explanation is therefore less clear,
and was avoided for simplicity.

Another important and new feature we wanted to stress is the role
of impurities. All bad metals are such that each carrier has
its ``own'' dopant. Thus it is necessary to show that impurities
do not lead to pseudogap suppression. It is clear, however, that
this direction needs substantial development to include more fully
the physically relevant case of strong disorder and the effects of
localization.
As seen above, these effects are probably in fact responsible for the
development of an inhomogeneous state.

Finally, it is very important to develop the theory for the case
of anisotropic pairing. Although, as we saw, the BCS-Bose crossover
theory has already been generalized to the case of $d$-wave
pairing, the formation of an anisotropic pseudogap and an
anisotropic BKT transition remain serious problems to be
addressed.\footnote{Here we refer a reader to a very recent paper
\cite{Randeria.action} which exploits a discrete
version of the variable transformation (\ref{parametrization.Hubbard}),
(\ref{parametrization.fermi}) to study the low temperature properties
even for $d$-wave pairing.}
Unfortunately, the analytical treatment of the problem is very limited and,
as we saw in Sec.~\ref{sec:continuation}, numerical methods
have their own limitations.

Therefore, there is no doubt that physicists studying
pseudogap properties have many problems to solve and
the results described above are only one step in this process.

\renewcommand{\theequation}{\Alph{section}.\arabic{equation}}
\appendix

\section{Calculation of the effective potential}
\setcounter{equation}{0}
\label{sec:A}

Here we sketch the derivation of the effective potential. To obtain it one
must write Eq.~(\ref{Omega.Potential.modulus}) in the momentum representation:
\begin{eqnarray}
&& \Omega_{\rm pot}(v, \mu, T, \rho) = v \left\{\frac{\rho^{2}}{U}
- T \sum_{n = -\infty}^{+\infty}\int \frac{d^2  k}{(2
\pi)^{2}}\,\mbox{Tr} [\ln \mathcal{G}^{-1} (i \omega_{n},  {\bf k})
e^{i \delta \omega_{n} \tau_{3}}]\right.\nonumber\\
&& \qquad + \left. T \sum_{n = -\infty}^{+\infty}
\int \frac{d^2 k}{(2 \pi)^{2}}\,
\mbox{Tr} [\ln G_{0}^{-1} (i \omega_{n},  {\bf k})
e^{i \delta \omega_{n} \tau_{3}}] \right\}, \quad
\delta \to +0,
\label{A1}
\end{eqnarray}
where
\begin{equation}
\mathcal{ G}^{-1} (i \omega_{n},  {\bf k}) =
i \omega_{n} \hat I - \tau_{3} \xi( {\bf k}) +
\tau_{1} \rho,\quad
G_{0}^{-1} (i \omega_{n},  {\bf k}) =
\left. \mathcal{ G}^{-1} (i \omega_{n},  {\bf k})
\right|_{\rho = \mu =0}
\label{A2}
\end{equation}
are the inverse Green's functions.
Note that the derivation of the effective potential
(\ref{Omega.Potential.expr})
in the ``old'' $\Phi$ and $\Phi^{\ast}$ variables differs
only in the replacement of the $\tau_{1} \rho$ term by
$\tau_{+} \Phi + \tau_{-} \Phi^{\ast}$.

The exponential factor  $e^{i \delta \omega_{n} \tau_{3}}$ is added
to (\ref{A1}) (see also Eqs.~(\ref{number.Green}), (\ref{Omega}),
(\ref{number.Green.T}), (\ref{number.Green.cons}) and
(\ref{number.impure})) to provide
the correct regularization which is necessary to perform the calculation
with the Green's functions \cite{Abrikosov.book}.
For instance, one obtains
\begin{eqnarray}
&& \lim_{\delta \to +0} \sum_{n = -\infty}^{+\infty}
\mbox{Tr} [\ln \mathcal{ G}^{-1} (i \omega_{n},  {\bf k})
e^{i \delta \omega_{n} \tau_{3}}] =
\lim_{\delta \to +0} \left\{ \sum_{n = -\infty}^{+\infty}
\mbox{Tr} [\ln \mathcal{ G}^{-1} (i \omega_{n},  {\bf k})]
\cos \delta \omega_{n} + \right.
\nonumber\\
&&\left.
i \sum_{\omega_{n} > 0} \sin \delta \omega_{n}
\mbox{Tr} [(\ln \mathcal{ G}^{-1} (i \omega_{n},  {\bf k}) -
\ln \mathcal{ G}^{-1} (- i \omega_{n},  {\bf k})) \tau_{3}] \right\}
\nonumber\\
&& =\sum_{n = -\infty}^{+\infty}
\mbox{Tr} [\ln \mathcal{ G}^{-1} (i \omega_{n},  {\bf k})] -
\frac{\xi( {\bf k})}{T},
\label{A4}
\end{eqnarray}
where
\begin{displaymath}
\ln\frac{ \mathcal{ G}^{-1} (i \omega_{n},  {\bf k})}{i\omega_n}\simeq
\frac{- \tau_3\xi( {\bf k})+\tau_1\rho}{i \omega_{n}}, \quad
\omega_{n} \to \infty
\end{displaymath}
and
\begin{displaymath}
\sum_{\omega_{n} > 0} \frac{\sin \delta \omega_{n}}{\omega_{n}}
\simeq \frac{1}{2 \pi T} \int_{0}^{\infty} dx\,
\frac{\sin \delta x }{x} = \frac{1}{4 T}\, \mbox{sign}\, \delta.
\end{displaymath}

To calculate the sum in (\ref{A4}), one must first use
the identity $\mbox{Tr} \ln \hat A = \ln \det \hat A$, so that
(\ref{A1}) takes the form
\begin{eqnarray}
\Omega_{\rm pot}(v, \mu, T, \rho) & = & v \left\{
\frac{\rho^{2}}{U} -  T \sum_{n = -\infty}^{+\infty}
\int \frac{d^2 k}{(2 \pi)^{2}}\,
\ln \frac{\det \mathcal{ G}^{-1} (i \omega_{n},  {\bf k})}
{\det G_{0}^{-1} (i \omega_{n},  {\bf k})}
\right.
\nonumber        \\
&& - \left.  \int \frac{d^2 k}{(2 \pi)^{2}}\,
[-\xi ( {\bf k}) + \varepsilon ( {\bf k})] \right\}.
\label{A5}
\end{eqnarray}
Calculating the determinants of the Green's functions (\ref{A2})
one obtains
\begin{eqnarray}
\Omega_{\rm pot}(v, \mu, T, \rho) & = & v \left\{
\frac{\rho^{2}}{U} -  T \sum_{n = -\infty}^{+\infty}
\int \frac{d^2 k}{(2 \pi)^{2}}
\ln \frac{\omega_{n}^{2} + \xi^{2}( {\bf k}) + \rho^{2}}
{\omega_{n}^{2} + \varepsilon^{2}( {\bf k})}
\right.
\nonumber                \\
&& -  \left. \int \frac{d^2  k}{(2 \pi)^{2}}
[-\xi ( {\bf k}) + \varepsilon ( {\bf k})] \right\},
\label{A6}
\end{eqnarray}
where the role of $G_{0} (i \omega_{n},  {\bf k})$ in the
regularization of $\Omega_{\rm pot}$ is now evident . The summation
in (\ref{A6}) can be done if one uses the representation
\begin{equation}
\ln \frac{\omega_{n}^{2} + a^{2}}{\omega_{n}^{2} + b^{2}} =
\int_{0}^{\infty} dx
\left(\frac{1}{\omega_{n}^{2} + a^{2} + x} -
      \frac{1}{\omega_{n}^{2} + b^{2} + x} \right),
\label{A7}
\end{equation}
and then
\begin{equation}
\sum_{k = 0}^{\infty}\frac{1}{(2k+1)^{2} + c^{2}} =
\frac{\pi}{4c} \tanh \frac{\pi c}{2}.
\label{A8}
\end{equation}
We find
\begin{equation}
\ln \frac{\omega_{n}^{2} + a^{2}}{\omega_{n}^{2} + b^{2}} =
\int_{0}^{\infty} dx   \left(
\frac{1}{2 \sqrt{b^{2} + x}} \tanh \frac{\sqrt{b^{2} + x}}{2 T} -
\frac{1}{2 \sqrt{a^{2} + x}} \tanh \frac{\sqrt{a^{2} + x}}{2 T}
\right).                     \label{A9}
\end{equation}
Integrating (\ref{A9}) over $x$, one thus obtains
\begin{equation}
T \sum_{n = -\infty}^{+\infty}
\int \frac{d^2 k}{(2 \pi)^{2}}
\ln \frac{\omega_{n}^{2} + \xi^{2}( {\bf k}) + \rho^{2}}
{\omega_{n}^{2} + \varepsilon^{2}( {\bf k})} =
2 T  \int \frac{d^2 k}{(2 \pi)^{2}}
\ln \frac{\cosh [\sqrt{\xi^{2}( {\bf k}) + \rho^{2}}/2T ]}
{\cosh [\varepsilon( {\bf k})/2T ]}.       \label{A10}
\end{equation}
Finally, substituting (\ref{A10}) into (\ref{A6}),
\begin{eqnarray}
&& \Omega _{\rm pot}(v, \mu, T, \rho)  =
\nonumber           \\
{v} && \left\{
\frac{\rho^2}{U} -
\int \frac{d^2 k}{(2 \pi)^{2}}
\left[2T
\ln\frac{\cosh[{\sqrt{\xi^2( {\bf k}) + \rho^2}/{2T}}]}
{\cosh[{\varepsilon( {\bf k})/{2T}}]}
 - [\xi( {\bf k})- \varepsilon( {\bf k})] \right] \right\}.
                                         \label{A11}
\end{eqnarray}
It is easy to show that at $T=0$, the expression (\ref{A11}) reduces to
that obtained in Sec.~\ref{sec:effective}.

Finally, we give formulas for the summation over the Matsubara
frequencies used in Secs.~\ref{sec:derivation}:
\begin{eqnarray}
& & T \sum\limits_{n=-\infty}^\infty\mbox{Tr}
[\mathcal{ G}(i\omega_{n}, {\bf k}) \tau_3
\mathcal{ G}(i \omega_{n},  {\bf k}) \tau_3] =
2T \sum\limits_{n=-\infty}^\infty
\frac{\xi^{2}( {\bf k})-\rho^2 - \omega_ {n}^{2}}
{[\omega_{n}^{2} + \xi^{2}( {\bf k}) + \rho^{2}]^{2}}
\nonumber                \\
  = &&   - \frac{\rho^2}{[\xi^2( {\bf k}) + \rho^2]^{3/2}}
\tanh \frac{\sqrt{\xi^2( {\bf k}) +\rho^2}}{2T}
\nonumber                \\
&& -
\frac{\xi^2( {\bf k})}{2T[\xi^2( {\bf k})+\rho^2]}
\frac{1}{\cosh^2 \frac{\ds \sqrt{\xi^2( {\bf k})+\rho^2}}{\ds 2T}},
\label{A12}
\end{eqnarray}

\begin{eqnarray}
& & T \sum \limits_{n=-\infty}^\infty \mbox{Tr}
[\mathcal{ G} (i\omega_{n}, {\bf k}) \hat I
\mathcal{ G}(i \omega_{n},  {\bf k}) \hat I] = 2T
\sum \limits_{n=-\infty}^\infty \frac{\xi^{2}( {\bf k}) + \rho^2 -
\omega_ {n}^{2}}{[\omega_{n}^{2} +
\xi^{2}( {\bf k}) + \rho^{2}]^{2}}
\nonumber              \\
& & =  - \frac{1}{2T} \frac{1} {\cosh^{2}
\frac{\ds \sqrt{\xi^{2}( {\bf k}) + \rho^{2}}}{\ds 2T}},
                     \label{A13}
\end{eqnarray}
where the Green's function $\mathcal{ G}(i \omega_{n},  {\bf k})$ is
given by (\ref{A2}). Both formulas can easily be calculated using
Eq.~(\ref{A8}) and its derivative with respect to $c$.

\section{Another representation for the retarded Green's function}
\setcounter{equation}{0}
\label{sec:B}

Here we shall obtain another representation for the retarded
fermion Green's function which is more convenient for the derivation
of the spectral density. Recall that when the imaginary part of
$G(\omega,
{\bf k})$ is nonzero $\mu +\sqrt{\omega^{2}
- \rho^{2}} > 0$ and $v_1 > 0$, $v_2 < 0$. This allows one to
transform the analytically continued (by means of Eq.
(\ref{roots.real})) integral
\begin{equation}
L \equiv \int_{0}^{\infty} du
\frac{[u (u + 1)]^{\alpha - 1}} {[(u + v_1)(u + v_2)]^{\alpha}}
\label{L}
\end{equation}
from Eq.~(\ref{Green.final.Matsubara}) in the following manner
($\alpha<1$)
\begin{eqnarray}
 L & = & (-1)^{\alpha} \int_{0}^{|v_2|} du
\frac{[u (u + 1)]^{\alpha - 1}} {[(u + v_1) ( |v_2| - u)]^{\alpha}} +
\int_{|v_2|}^{\infty} du
\frac{[u (u + 1)]^{\alpha - 1}} {[(u + v_1) (u - |v_2|)]^{\alpha}}
\nonumber               \\
& = &
\frac{(-1)^{\alpha}}{u_1^{\alpha}} \Gamma(\alpha)
\Gamma(1-\alpha)
F_1 \left(\alpha, \alpha, 1-\alpha; 1;
\frac{v_2}{v_1}, u_2 \right)
\nonumber                \\
&& +
\frac{1}{|v_2|}\frac{\Gamma(1-\alpha)}{\Gamma(2-\alpha)}
F_1\left(1, \alpha, 1-\alpha; 2-\alpha;
\frac{v_1}{v_2}, \frac{1}{v_2} \right).
\label{L.transformation}
\end{eqnarray}
The first Appell function in (\ref{L.transformation}) can be
reduced to the hypergeometric function using the identity
\cite{Bateman} which is valid for $\gamma = \beta + \beta^{\prime}$
\begin{equation}
F_1 (\alpha, \beta, \beta^{\prime}, \beta + \beta^{\prime}; x,y ) =
(1-y)^{-\alpha} {}_2F_{1} \left( \alpha, \beta;
\beta+\beta^{\prime}; \frac{x - y}{1 - y} \right).
\label{appelltohyper}
\end{equation}
Thus one gets
\begin{eqnarray}
&&L = \frac{(-1)^{\alpha} \Gamma(\alpha)\Gamma(1-\alpha)} {[u_1
(1-u_2)]^{\alpha}}{}_2F_{1} \left(\alpha, \alpha;
1;\frac{u_2(1-u_1)} {u_1(1-u_2)} \right) + \frac{1}{|u_2|}
\frac{\Gamma(1-\alpha)}{\Gamma(2-\alpha)}\times\nonumber\\
&&F_1 \left(1, \alpha, 1-\alpha; 2-\alpha;\frac{u_1}{u_2},
\frac{1}{u_2} \right);
\nonumber             \\
&& \qquad \frac{u_2(1-u_1)}{u_1(1-u_2)} < 1,
\qquad \frac{u_1}{u_2} < 0,
\qquad \frac{1}{u_2} < 0.
                      \label{L.final}
\end{eqnarray}
This finishes the derivation of Eq.~(\ref{Green.real.spectral}).

\end{document}